\documentclass[1p,compress]{elsarticle}

\usepackage{booktabs}
\usepackage{multirow}
\usepackage{sectsty}
\usepackage{graphicx}
\usepackage{amsmath}
\usepackage{amssymb}
\usepackage{subcaption}
\usepackage{caption}
\usepackage{lastpage}
\usepackage{enumitem}
\usepackage{lipsum}
\usepackage{titlesec}
\usepackage{wrapfig}
\usepackage{xcolor}
\usepackage{mdframed}
\usepackage{slashed}
\usepackage{lineno}

\usepackage{ifmtarg}
\usepackage{xifthen}
\usepackage{environ}
\usepackage{multido}

\usepackage{todonotes}

\usepackage{hyperref}

\journal{Physics Reports}

\makeatletter%
\newcommand{\theoremhang}{
  \begingroup%
  \setlength{\unitlength}{.005\linewidth}
    \begin{picture}(0,0)(1.5,0)%
      \linethickness{0.45pt} \color{black!50}%
      \put(-3,2){\line(1,0){206}}
      \multido{\iA=2+-1,\iB=50+-10}{5}{
        \color{black!\iB}%
        \put(-3,\iA){\line(0,-1){1}}
        \put(203,\iA){\line(0,-1){1}}
      }%
    \end{picture}%
  \endgroup%
}%
\newcommand{\theoremhung}{
  \nobreak
  \begingroup%
    \setlength{\unitlength}{.005\linewidth}
    \begin{picture}(0,0)(1.5,0)%
      \linethickness{0.45pt} \color{black!50}%
      \put(-3,0){\line(1,0){206}}
      \multido{\iA=0+1,\iB=50+-10}{5}{
        \color{black!\iB}%
        \put(-3,\iA){\line(0,1){1}}
        \put(203,\iA){\line(0,1){1}}
      }%
    \end{picture}%
  \endgroup%
}%

\newcounter{note}
\newcommand{\thetheorem}{\arabic{note}}
\NewEnviron{note}[1][]{%
    \par\noindent\theoremhang\par\nobreak\noindent
    \refstepcounter{note}\postdisplaypenalty=10000 %
    {\sffamily\bfseries\upshape Note \thetheorem\@ifnotmtarg{#1}{\ (#1)}}\ \ \itshape\ignorespaces%
    \BODY 
    \par\addvspace{-1ex}\nobreak\noindent\theoremhung\par\addvspace{.4ex}%
}
\makeatother

\newcommand{\as}{\alpha_s}

\newcommand{\msbar}{$\overline{\rm MS}$}
\newcommand{\pslash}{\slashed{p}}
\newcommand{\ie}{{\em i.e.}}
\newcommand{\eg}{{\em e.g.}}

\newcommand{\simas}[1]{\raisebox{-.1ex}{
            $\stackrel{\small{#1}}{\sim}$}}

\begin{document}

\begin{frontmatter}

\title{Lattice determinations of the strong coupling}
\author[ed]{Luigi Del Debbio}
\ead{luigi.del.debbio@ed.ac.uk}
\address[ed]{School of Physics and Astronomy, University of Edinburgh, Edinburgh EH9 3JZ, UK}

\author[tcd,ific]{Alberto Ramos}
\ead{alberto.ramos@ific.uv.es}
\fntext[ific]{Present address: Instituto de Física Corpuscular (IFIC),
  CSIC-Universitat de Valencia 46071 - Valencia, SPAIN}
\address[tcd]{School of Mathematics and Hamilton Mathematics Institute, Trinity College Dublin, Dublin 2, Ireland}




\begin{abstract}
Lattice QCD has reached a mature status. State of the art lattice
  computations include $u,d,s$ (and even the $c$) sea quark effects,
  together with an estimate of electromagnetic and isospin breaking
  corrections for hadronic observables. This precise and first
  principles description of the standard model at low energies allows
  the determination of multiple quantities that are essential inputs
  for phenomenology and not accessible to perturbation theory.

  One of the fundamental parameters that are determined 
  from simulations of lattice QCD is the strong coupling constant,
  which plays a central role in the quest for precision at the
  LHC. Lattice calculations currently provide its best determinations,
  and will play a central role in future phenomenological studies.
  For this reason we believe that it is timely to provide
  a pedagogical introduction to the lattice determinations of the
  strong coupling. Rather than analysing individual studies, the
  emphasis will be on the methodologies and the systematic errors that
  arise in these determinations. We hope that these notes will help
  lattice practitioners, and QCD phenomenologists at large, by
  providing a self-contained introduction to the methodology and the
  possible sources of systematic error. 

  The limiting factors in the determination of the strong coupling
  turn out to be different from the ones that limit other lattice
  precision observables. We hope to collect enough information here to
  allow the reader to appreciate the challenges that arise in order to
  improve further our knowledge of a quantity that is crucial for LHC
  phenomenology.
\end{abstract}


\end{frontmatter}

{\emph{Preprint:} IFIC/20-56}

\newpage
\tableofcontents

\section{Introduction}

Nowadays lattice QCD is a mature field. Several low energy tests of the strong
interactions involve careful lattice QCD computations, which in the last 10
years have acquired the status of precision physics. Lattice determinations of
the strong coupling constant are amongst these precise measurements; they
actually represent the most precise results available in the literature for this
fundamental parameter of the Standard Model. This situation will continue to
improve as the combined result of the increase in computer power and the
ingenuity of new methodologies. The next few years will probably see several
new and very precise lattice determinations of the strong
coupling. With the world average of $\alpha_s$ already dominated by lattice
determinations this work seems fully justified.

The reader might not feel comfortable with this situation. But one of the points
that we want to emphasize in this review is that this is not necessarily a bad
situation. The reason is that there are several methods \emph{within} lattice
QCD to extract the strong coupling, and they are affected by systematic effects
in different ways. These methods differ among themselves at least as much as the
different extractions from phenomenological data. Determinations of the strong
coupling from lattice QCD are, as a matter of fact, a vast subject.  The main
objective of this review is to present the different techniques to determine the
strong coupling on the lattice. We do not aim to review the individual papers,
but instead to present the different methods and their general characteristics,
advantages and limitations. Along the way we hope to clarify what each method
needs in order to improve their current results substantially. Hopefully this
review should provide enough information so that the non-experts in the field
can understand and be critical when reading the specialized lattice literature.
In this respect, this work is a complement, and not an alternative, to the
excellent FLAG review~\cite{Aoki:2019cca}, where a detailed description of each
published work can be found, together with quality criteria on the
results. For instance, we will not aim to provide here a {\em best}
value for the strong coupling, for which we refer to FLAG, but we will
try to insist on the systematic errors in lattice determinations.

The systematic errors that affect most lattice QCD calculations are
quite different from those that impact on the determinations of the
strong coupling. We live in an era where state of the art lattice QCD
computations include electromagnetic and charm effects with the aim of
reaching a sub-percent precision in many observables.  But these
effects represent a very small contribution to the uncertainty in the
strong coupling, well below our current precision. Instead, when it
comes to the determinations of the strong coupling, the limiting
factor in lattice analyses is in fact very similar to those that
limit the precision of many phenomenological studies: the use of
perturbation theory at relatively low scales makes it difficult to
estimate the uncertainty associated with the truncation of the
perturbative series.

Let us now summarize the material in the review. In
section~\ref{sec:standard-model-at} we introduce some elementary facts about the
strong interactions. We focus on the $\Lambda$ parameter, whose knowledge is
equivalent to the direct determination of the value of the strong coupling. In
this respect, it is important to realise that, in the absence of quark masses,
the determination of the coupling in QCD is equivalent to setting the scale of
theory by specifying the value of {\em just one} dimensionful quantity. This
point will be discussed in detail below. Section~\ref{sec:phys-defin-strong}
focus on how the strong coupling is determined. We will see that in fact lattice
methods share the same basic strategy as other phenomenological determinations,
and face the same challenges. In section~\ref{sec:lattice-field-theory} we
provide an introduction to lattice field theory. Special emphasis is put on the
topics that enter the determination of the strong coupling: continuum
extrapolations, scaling violations and scale setting are some of the topics that
are explained in detail. Contrary to other lattice computations, that are
intrinsically low-energy computations, the determination of the strong coupling
requires to make contact with the Standard Model at the electroweak scale. In
section~\ref{sec:deco-heavy-quark} we focus on the effects that the heavy charm
and bottom quarks have in this peculiar situation.
Section~\ref{sec:conv-observ-coupl} introduces the different
techniques used on 
the lattice to determine the strong coupling. The focus will be on how they
address the systematic uncertainties inherent in these computations. Finally
section~\ref{sec:pres-future-latt} will discuss the present status and
our anticipation for the future of
lattice determinations of the strong coupling, with an emphasis on the role of
the different methods. 

The authors have their own (possibly sometimes divergent) opinions
about the topics covered in this review. Of course we find our
position well founded and we are happy to defend it, but in a review
work like this it is important to keep in mind that there can be some
controversies.  This can only be positive in a field that is
still an active area of research. We have tried to explain our point
of view as clearly as possible, and when necessary, we have explicitly
stated that we are exposing our own point of view. We hope that this
work becomes a useful reference even for those who disagree with some
of our opinions.


\section{The standard model at low energies}

\label{sec:standard-model-at}

The standard model (SM) of particle physics classifies all known fundamental
particles and describes their interaction via three of the four fundamental
forces. Its predictions agree with experiments with an astonishing precision.
The gravitational force is the only interaction that is unaccounted for by the
SM. Of the three fundamental interactions in the SM, the weak and
electromagnetic interactions are two aspects of a unified electroweak force. At
temperatures below the electroweak scale ($\sim 100\, {\rm GeV}$) two different
interactions (weak and electromagnetic) emerge.  The weak interactions become
relevant only at very short distances, less than the diameter of a proton, due
to the massive nature of the weak $W^\pm, Z$ bosons that mediate the
interaction. On the other hand the electromagnetic force, being mediated by
massless photons, is a long range interaction. It describes the interactions
between electrically charged particles, and is responsible of many phenomena in
different areas, from optics, to radiation or the structure of the atoms. From
the theoretical point of view \emph{quantum electrodynamics} (QED) is a
relativistic quantum field theory amenable to precise computations by using
perturbation theory, due to the smallness of the coupling between charged
particles ($\alpha_{\rm EM}\sim 1/137$). Some of the theoretical predictions of
QED have been confirmed by experiments with a precision better than one part in
a million~\cite{Parker:2018vye,Hanneke:2008tm}.

The remaining SM interaction, the strong nuclear force, is responsible for
binding protons and neutrons together, forming the atomic nucleus, and for
binding the more fundamental quarks and gluons together inside protons, neutrons
and other hadrons. The strong nuclear force is also a short range interaction,
but in contrast with the case of the weak interactions, the reason is not that
the mediators of the interaction are massive, since gluons are massless. Instead
the reason is a dynamical feature of the theory of strong interactions,
\emph{Quantum Chromodynamics} (QCD), called confinement. The
force
between \emph{color charged particles} remains constant and different
from zero at large separations between the charges. Pulling two
quarks apart requires an increasing amount of energy, until 
eventually new pairs of quarks are created. Particles charged under the strong
interactions are therefore \emph{confined} in color ``neutral'' hadrons (like
the proton). On the other hand, the strong interactions become asymptotically
weaker at very short distances, a phenomenon called \emph{asymptotic
freedom}~\cite{Politzer:1973fx,Gross:1973id}. Quarks behave as almost free
particles at distances much smaller than the size of a proton.

This qualitative picture of the strong interactions explains many experimental
phenomena, from scaling in deep inelastic scattering (DIS) experiments to the
fact that not a single free quark has ever been observed. Because of asymptotic
freedom perturbative predictions of QCD can be compared with high-energy
experiments. Quantitative comparisons for processes like vector boson
production, event shape observables at the Large Electron-Positron collider
(LEP) or scaling violations in DIS -- just to name a few -- remain the most
stringent tests of QCD as the theory of the strong interactions, although none
of them reaches the precision of the tests of QED. Low-energy predictions for the
strong interactions are more elusive; as the coupling increases, computations
based on perturbation theory are no longer adequate. Accurate predictions in
this regime require a non-perturbative formulation of the theory, and have
become possible only recently thanks to large scale lattice QCD simulations.

There is currently clear evidence supporting the idea that QED and QCD
is all that is needed to explain most experimental results of particle
physics at scales below the electroweak scale with very high
accuracy. From photoproduction in proton-proton collision, to the mass
of the proton or the energy binding of the atomic nucleus or the
formation of the atom.

But the attentive reader should have noted that at the core of this
picture for the strong interactions (free quarks at ``high'' energies and
confinement at ``low'' energies) lies a fundamental question to be
asked: high and low energies compared with what? how does a scale
arise in QCD? 

\subsection{QCD and the scale of the strong interactions}
\label{eq:QCDScale}

The strong interactions are described by a relativistic quantum field theory. It
describes the interactions between color charged particles: the 6 quark flavors
and the gluons. It is a non-abelian gauge theory with symmetry group
$SU(3)$.
Matter content and symmetries is all that is needed to write down the action of
QCD, that reads~\footnote{We are going to work in 4-dimensional Euclidean space.
The gauge field $A_\mu(x)$ lives in the Lie algebra $\mathfrak{su}(3)$, and
therefore, for matter in the fundamental representation of the gauge group, it is an anti-hermitian
traceless $3\times 3$ matrix.}
\begin{equation}
  \label{eq:QCDAction}
  S[A] =\int {\rm d}^4x\, \left\{
    -\frac{1}{2g^2} {\rm Tr}\left( F_{\mu\nu} F_{\mu\nu}\right) +
    \sum_{i=1}^6\bar \psi_i(\gamma_\mu D_\mu + m_i)\psi_i
  \right\}\,,
\end{equation}
where $D_\mu = \partial_\mu + A_\mu$, $m_i$ is the bare mass of quark
flavor $i$, and $g$ is the bare gauge coupling. The field strength is
defined by
\begin{equation}
  \label{eq:FieldStrengthDef}
  F_{\mu\nu} = \partial_\mu A_\nu - \partial_\nu A_\mu + [A_\mu,A_\nu]
\end{equation}
It is worth noting that quark masses are the only dimensionful
parameters of the QCD action, since the gauge coupling $g$ is
dimensionless in 4 dimensions. At the classical level quark masses are
the only source of breaking of scale invariance.

QCD predictions are made by computing expectations values
of fields in the Euclidean theory as path integral averages with
partition function 
\begin{equation}
  \label{eq:PathIntegralDef}
  \mathcal Z = \int \mathcal{D}A\, e^{-S[A]}\,.
\end{equation}
All physical information is then extracted from these correlators. The
path integral written above is, naively, ill defined. A simple
perturbative calculation for instance shows that the path integral is
plagued by ultraviolet (UV) divergences, \ie\ divergences that arise
when summing over the high-energy modes in the
theory. Expectation values can be made finite by modifying
the theory at short distances. There are several possibilities for such a
\emph{regularization} of the theory, the most natural consists in
defining the theory on a four dimensional Euclidean lattice with
spacing $a$. When performing Fourier transforms in a discretized
spacetime, momenta are limited to the first Brillouin zone, which
implies that the inverse lattice spacing provides a UV
cutoff. There are other possibilities to make expectation
  values finite, like defining the theory in an arbitrary number of
dimensions (dimensional regularization), that are more convenient in the
context of perturbative computations.

Independently of the details of the regularization procedure, any
physical quantity $P(Q)$, measured at a typical scale $Q$, computed
from some expectation value in the regularized theory, will depend not
only on $Q$ and the particular values of the gauge coupling and quark
masses ($g,m_i$), but also on the short distance scale (denoted $a$)
at which QCD is modified. Denoting the mass dimension of $P$ by $d_P$,
we have: 
\begin{equation}
  \label{eq:PhysRegulatedQuantity}
  a^{d_P} P(Q) = {\mathcal P}(aQ,g,am_i)\,.
\end{equation}
Note that the quantity on the left-hand side of
Eq.~\eqref{eq:PhysRegulatedQuantity} is the dimensionless product
$a^{d_P} P(Q)$, and that accordingly the function ${\mathcal P}$ only
depends on dimensionless quantities. The problem is how to make any
solid prediction when the arbitrary value of the short distance $a$
appears in all determinations of physical quantities. The answer comes
under the name of \emph{renormalization}.  Even if determinations in
the regularized theory depend on the particular choice of ultraviolet
cutoff ($a$), the physics at large distances compared with the cutoff
(the regime $aQ\ll 1$) is universal if it is parametrized in terms of
the \emph{renormalized} coupling ($\bar g(\mu)$) and
\emph{renormalized} quark masses ($\bar m_i(\mu)$). The
renormalization scale $\mu$ is an arbitrary scale that is introduced
in the renormalization procedure. A more precise relation would then
read
\begin{equation}
  \label{eq:PhysRenormQuantity}
  \mathcal P(aQ,g,am_i) 
  = \bar{\mathcal P}(Q/\mu,\bar g(\mu),\bar m_i(\mu)/\mu) +
  \mathcal O\left((aQ)^p, (a\mu)^p, (am)^p\right) + \dots\, . 
\end{equation}
Note that the arbitrary scale $a$ does not show up in the first term
on the right-hand side. Moreover in the limit where the short-distance
scale $a$ is much smaller than the physical ($Q$) and renormalization
($\mu$) scales a \emph{precise} prediction for any physical observable
emerges
\begin{equation}
  \label{eq:PhysRenormLimit}
  \frac{P(Q)}{M^{d_P}} = \frac{1}{(aM)^{d_P}} \, 
  \bar{\mathcal P}(Q/\mu,\bar g(\mu),\bar m_i(\mu)/\mu) \,.
\end{equation}
In the equation above we have expressed $P(Q)$ in units of some
physical mass scale $M$, which in turn can be obtained from a lattice
simulation in units of the cutoff $a$ -- this the quantity in the
denominator in the RHS of the expression above.

The renormalized quantities $\bar g(\mu), \bar m_i(\mu)$ are functions
of the quark masses and coupling constant of the finite theory (the
bare parameters $g,m_i$), the cut-off $a$ and the renormalization
scale $\mu$. The physics content of this renormalization process is
that at low energies the theory is sensitive to the particular choice
of cutoff \emph{only} via the relation between bare and renormalized
parameters. This relation is not observable and remains an arbitrary
choice needed in order to make physical predictions.  The set of
prescriptions that are necessary to fully specify the relation between
bare and renormalized quantities is called a {\em renormalization
  scheme}.

Note that in the renormalization procedure, we have introduced a new
scale $\mu$. This is not an accident, and is unavoidable,
independently of the chosen regularization and/or renormalization
schemes. The renormalization scale $\mu$ is arbitrary and physical
quantities must be independent on $\mu$. This requirement can be
expressed as a set of mathematical conditions, which go under the name of
Callan-Symanzik~\cite{Callan:1970yg,Symanzik:1970rt} equations:
\begin{equation}
  \label{eq:CallanSymanzikObs}
  \mu \frac{{\rm d}}{{\rm d}\mu} 
  \bar{\mathcal P}(Q/\mu,\bar g(\mu),\bar m_i(\mu)/\mu) = 0\, .
\end{equation}
These equations can be used to determine how the renormalized coupling $\bar
g(\mu)$ and the renormalized quark masses $\bar m_i(\mu)$ change (``run'') with
the renormalization scale. One of the main characters of this review is the
$\beta$-function, which dictates the dependence of the renormalized coupling on
the renormalization scale~\cite{Politzer:1973fx, Gross:1973id}\footnote{In this
section we will use massless renormalization schemes, where the $\beta$ function
is independent of the values of quark masses. See
section~\ref{sec:deco-heavy-quark} for a discussion of massive renormalization
schemes.}
\begin{equation}
  \label{eq:beta}
  \mu\frac{{\rm d}}{{\rm d}\mu} \bar g(\mu) = \beta(\bar g)\,.
\end{equation}

This renormalization group (RG) equation is a first order equation,
and therefore its solution depends on exactly one integration
constant. Moreover the solution to this equation has to respect the
correct boundary condition given by the asymptotic behavior of the
$\beta$-function determined in perturbation theory~\footnote{In
  general perturbative expansions in quantum field theories are
  asymptotic. Through this work a function $f(x)$ having an asymptotic
  expansion will be denoted by $f(x) \simas{x\to 0}$ \dots}
\begin{equation}
  \label{eq:BetaFunAsymp}
  \beta(\bar g) \simas{\bar g\to 0}  -\bar g^3 \sum_{k=0}^\infty b_k
  \bar g^{2k}\, ,
\end{equation}
where
\begin{subequations}
  \label{eq:beta_univ}
  \begin{eqnarray}
    b_0 &= \frac{1}{(4\pi)^2}\left(11-\frac{2N_f}{3} \right) \, ,\\
    b_1 &= \frac{1}{(4\pi)^4}\left(102-\frac{38N_f}{3} \right) \, ,
  \end{eqnarray}
\end{subequations}
and $N_f$ is the number of fermions in the fundamental representation (\ie\
quarks).  Note that for $\bar g \to 0$ the $\beta$-function is
\emph{negative} (at least for $N_f<17$), which implies asymptotic
freedom, \ie\ the decrease of the coupling with increasing energy.

It is instructive to discuss in some detail the integration of the RG
equation, Eq.~\eqref{eq:beta}. We can readily see that
\begin{equation}
  \label{eq:IntStepOne}
  \frac{d\mu}{\mu} = \frac{d\bar{g}}{\beta(\bar{g})}
  \quad \Longrightarrow \quad
  \log\left(\frac{\mu_1}{\mu_2}\right) =
  \int_{\bar{g}_2}^{\bar{g}_1} \frac{dx}{\beta(x)}\, ,
\end{equation}
where $\bar{g}_1=\bar{g}(\mu_1)$, $\bar{g}_2=\bar{g}(\mu_2)$. The
logarithmic divergence on the left-hand side of
Eq.~\eqref{eq:IntStepOne} when $\mu_1$ (resp. $\mu_2$) tends to
infinity is reflected in the divergence of the integral on the
right-hand side when $\bar{g}_1$ (resp. $\bar{g}_2$) tend to zero. The
asymptotic behaviour of the integrand is
\begin{align}
  \frac{1}{\beta(x)} 
  &= -\frac{1}{b_0 x^3} \frac{1}{1 + \frac{b_1}{b_0} x^2 + O(x^4)} \\
  & \underset{x\to 0}{\simeq} -\frac{1}{b_0 x^3}
    \left[
    1 - \frac{b_1}{b_0} x^2 + O(x^4)
    \right]\, ,
\end{align}
and therefore the integral can be rewritten as
\begin{align}
  \int_{\bar{g}_2}^{\bar{g}_1} \frac{dx}{\beta(x)}
  &= \left[
    \frac{1}{2b_0\bar{g}_1^2} - \frac{1}{2b_0\bar{g}_2^2}
    + \frac{b_1}{b_0^2} \log \bar{g}_1 -
    \frac{b_1}{b_0^2} \log \bar{g}_2 
    \right] + \nonumber \\
  & \quad + \int_{\bar{g}_2}^{\bar{g}_1}{\rm d}x\, \left[\frac{1}{\beta(x)} +
    \frac{1}{b_0x^3} - \frac{b_1}{b_0^2x}\right]\, .
\end{align}
Note that, when rewritten in this form, the integrand that appears on
the right-hand side is $\mathcal O(x)$ when $x\to 0$, and hence the integral is
finite when the integration limit tends to zero. After some algebraic manipulations
Eq.~\eqref{eq:IntStepOne} yields
\begin{align}
  &\mu_1
  \left[b_0\bar{g}_1^2\right]^{-\frac{b_1}{2b_0^2}}\,
  e^{-\frac{1}{2b_0\bar{g}_1^2}}\,
  \exp\left\{-
    \int_{0}^{\bar{g}_1}{\rm d}x\, \left[\frac{1}{\beta(x)} +
    \frac{1}{b_0x^3} - \frac{b_1}{b_0^2x}\right]\right\} = \nonumber
  \\
  &\quad \mu_2
  \left[b_0\bar{g}_2^2\right]^{-\frac{b_1}{2b_0^2}}\,
  e^{-\frac{1}{2b_0\bar{g}_2^2}}\,
  \exp\left\{-
    \int_{0}^{\bar{g}_2}{\rm d}x\, \left[\frac{1}{\beta(x)} +
    \frac{1}{b_0x^3} - \frac{b_1}{b_0^2x}\right]\right\} \, .
    \label{eq:IntConst}
\end{align}
The equality holds for any value of $\mu_1$ and $\mu_2$, showing that
the combination in Eq.~\eqref{eq:IntConst} has units of mass, and is
independent of $\mu$. It is called the $\Lambda$-parameter and can be
understood as the \emph{intrinsic scale} of QCD that we were looking
for. Note that the integration of the renormalization group equation
Eq.~\eqref{eq:beta} is exact, and the $\Lambda$-parameter can be
defined as:
\begin{equation}
  \label{eq:lam}
  \Lambda = \mu
  \left[b_0\bar g^2(\mu)\right]^{-\frac{b_1}{2b_0^2}}\,
  e^{-\frac{1}{2b_0\bar g^2(\mu)}}\,
  \exp\left\{-
    \int_{0}^{\bar g(\mu)}{\rm d}x\, \left[\frac{1}{\beta(x)} +
    \frac{1}{b_0x^3} - \frac{b_1}{b_0^2x}\right]\right\}\,.
\end{equation}
This expression is valid beyond perturbation theory. Hadron masses, meson decay
constants, or any other dimensionful quantity in QCD, can be measured in units
of $\Lambda$, and are given by dimensionless functions of the renormalized
coupling, and the renormalized quark masses (also expressed in units of
$\Lambda$). It is in this respect that we like to think of the $\Lambda$
parameter as an {\em intrinsic scale}\ of QCD.

The renormalized theory is defined by specifying the value of
the renormalized coupling at a given scale, or equivalently by
specifying the value of the $\Lambda$ parameter. Note that
Eq.~\eqref{eq:lam} is an implicit equation for $\bar g (\mu)$,
and therefore the running coupling is a function of $\Lambda/\mu$; at
high energies compared with $\Lambda$ (i.e. $\mu/\Lambda\gg 1$) the
running of the coupling is given by
\begin{equation}
  \label{eq:RGEHighEnergy}
  \bar g^2(\mu) \simas{\mu/\Lambda\gg 1}   
  \frac{1}{2 b_0 \log(\mu/\Lambda) + b_1 \log\log(\mu/\Lambda)}
  + \dots\,.
\end{equation}
At scales \emph{much larger than} $\Lambda$, $\bar g(\mu)$ is small,
QCD is weakly coupled and quarks behave as almost free particles.

\subsection{The determination of the intrinsic scale of QCD}

There is quite some freedom when renormalizing QCD. In the framework of
perturbative computations there are many valid ways to subtract the divergent
parts of Feynman diagrams that differ by finite terms. Non-perturbatively there
are also multiple conditions to use as a definition for renormalized coupling
and quark masses. This freedom is called \emph{choice of scheme}.

The value of the strong coupling constant at high energies is a
necessary input for the study of all QCD cross sections at the Large
Hadron Collider (LHC) and many other high-energy experiments. For this
reason it is convenient to quote its value in a scheme that can be
easily used for phenomenological input. The so-called \emph{modified
  minimal subtraction} ($\overline{\rm MS}$)
scheme~\cite{Bardeen:1978yd} is by far the most widely-used
choice. This scheme is defined in the context of perturbative
computations; however the $\Lambda$-parameter extracted in this
convenient scheme still has a non-perturbative meaning, as discussed
below.

\subsubsection{Scheme dependence}

Most of the time we are going to deal with \emph{mass-independent}
renormalization schemes. Any modification of the theory that is performed in
order to regularize and renormalize QCD can always be made at energies much
larger than the quark masses.\footnote{In some cases massive
  renormalization schemes might be more convenient, like for example
  heavy quarks regulated on the lattice. In practice the lattice spacings that are
currently accessible to simulations provide a UV cutoff that is not much larger
than the mass of the heavy quarks $c$ and $b$. In this context mass-dependent
renormalization schemes might have some advantageous properties. See
\eg~\cite{Boyle:2016wis,Fritzsch:2018kjg}.} From a perturbative point of view
we can say that the UV divergences of Feynman diagrams are independent of any
quark mass. A nice property of mass-independent schemes is that the RG functions
(like the $\beta$-function Eq.~\eqref{eq:beta}) are independent of the quark
masses. Of course there is nothing fundamentally wrong with renormalization
schemes that are not mass-independent, and we will study in detail the
relation between mass-independent and mass-dependent renormalization schemes in
chapter~\ref{sec:decoupling}, but first let us state some basic relations
between massless renormalization schemes.

By convention we normalize couplings in different schemes so that they
agree to leading order. This implies that renormalized couplings in
two schemes $s$ and $s'$ are related perturbatively by
\begin{equation}
  \label{eq:gssp}
  \bar g^2_{s'}(\mu)\,  \simas{\bar g_s\to 0}\,  \bar g^2_s(\mu)  +   c_{ss'}\bar g^4_s(\mu) +\dots
\end{equation}
with $c_{ss'}$ a finite number. The $\beta$-function for the couplings $\bar
g_{s}(\mu)$ and $\bar g_{s'}(\mu)$ are different (i.e. $\beta$-functions are
scheme dependent), but it is easy to check that the two leading terms in its
asymptotic expansion~\eqref{eq:beta_univ} are scheme independent: $b_0$ and
$b_1$ are universal. Higher order coefficients $b_n$ with $n>1$ are scheme
dependent. For the case of the $\overline{\rm MS}$ scheme, the $\beta$-function
is known up to five
loops~\cite{vanRitbergen:1997va,Czakon:2004bu,Baikov:2016tgj,Luthe:2016ima,Herzog:2017ohr}
(see table~\ref{tab:bnms}). 
\begin{table}
  \centering
  \begin{tabular}{lc}
    \toprule
    $N_f$& $\bar \beta(\alpha_{\overline{\rm MS}})$\\
    \midrule
    3 &1.0 \, +\,  $0.565884\, \alpha_{\overline{\rm MS}} \, +\,  0.453014\, \alpha_{\overline{\rm MS}}^2 \, +\, 
                                                     0.676967\, \alpha_{\overline{\rm MS}}^3
                                                   \, +\, 
                                                     0.581082\, \alpha_{\overline{\rm MS}}^4$ \\
    4 &1.0 \, +\,  $0.490197\, \alpha_{\overline{\rm MS}} \, +\,  0.308790\, \alpha_{\overline{\rm MS}}^2 \, +\, 
                                                     0.485901\, \alpha_{\overline{\rm MS}}^3
                                                   \, +\, 
                                                     0.280899\, \alpha_{\overline{\rm MS}}^4$ \\
    5 &1.0 \, +\,  $0.401347\, \alpha_{\overline{\rm MS}} \, +\,  0.149427\, \alpha_{\overline{\rm MS}}^2 \, +\, 
                                                     0.317223\, \alpha_{\overline{\rm MS}}^3
                                                   \, +\, 
                                                     0.081429\, \alpha_{\overline{\rm MS}}^4$ \\
    6 &1.0 \, +\,  $0.295573\, \alpha_{\overline{\rm MS}} -
        0.029401\, \alpha_{\overline{\rm MS}}^2 \, +\, 
        0.177980\, \alpha_{\overline{\rm MS}}^3 \, +\,  0.002360\, \alpha_{\overline{\rm MS}}^4$\\
    \bottomrule
  \end{tabular}
  \caption{$\beta$-function in the $\overline{\rm MS}$ scheme. Here we
  write a series in $\alpha_{\overline{\rm MS}} = \bar
  g^2_{\overline{\rm MS}}/(4\pi)$ and
  divide out the leading order behaviour according to $\bar
  \beta(\alpha_{\overline{\rm MS}}) =
  -\beta(\bar g_{\overline{\rm MS}})/(b_0\bar g^3_{\overline{\rm MS}})$. Surprisingly, the
  perturbative coefficients remain 
  small up to five-loops.}
\label{tab:bnms}
\end{table}

On the other hand the $\Lambda$-parameter, defined in Eq.~\eqref{eq:lam}, is
also scheme dependent. It is easy to see by using the one-loop relation between
couplings, Eq.~\eqref{eq:gssp}, that
\begin{subequations}
  \begin{eqnarray}
    \left[\frac{\bar g_{s'}^2(\mu)}{\bar g_s^2(\mu)}\right]^{-\frac{b_1}{2b_0}}
    &\,  \simas{\bar g_s\to 0}\, &
      1  - \frac{b_1}{2b_0}c_{ss'}\bar g_{s}^2(\mu) + \dots\,, \\
    \frac{1}{\bar g_{s'}^2(\mu)} - \frac{1}{\bar g_s^2(\mu)}
    &\,  \simas{\bar g_s\to 0}\, &
      c_{ss'} + \dots\,.
  \end{eqnarray}
\end{subequations}
Since the integral in Eq.~\eqref{eq:lam} is $\mathcal O(\bar
g^2)$, one can obtain an exact relation between $\Lambda$-parameters
by taking the limit $\bar g_s\to 0$~\cite{Hasenfratz:1980kn}
\begin{equation}
  \frac{\Lambda_{s'}}{\Lambda_s} = 
  \exp\left(\frac{-c_{ss'}}{2b_0}\right)\,.
\end{equation}

In other words, the relation of $\Lambda$-parameters in different
schemes is \emph{exactly} known via a one-loop computation, as
reported in Eq.~\eqref{eq:gssp}. This observation, together with
Eq.~\eqref{eq:lam} allows a precise non-perturbative definition of the
$\Lambda$-parameter even for schemes that are intrinsically defined in
a perturbative context: even if $\overline{\rm MS}$ is a
``perturbative scheme'', $\Lambda_{\overline{\rm MS}}$ is a meaningful
quantity beyond perturbation theory.

\subsubsection{Quark thresholds}

Quarks are not massless particles. Every physical process in QCD
depends not only on the intrinsic scale of the strong interactions,
but on the quark masses. In particular we expect that if the process
takes place at a some energy scale much lower than the mass of some
quark, this quark should ``decouple'' from all physical processes.

How this decoupling takes place in mass-independent renormalization
schemes is in fact not trivial. The RG functions (and therefore the
renormalized parameters of the theory) do not depend on any
quark mass: at any scale, the top and the up quarks give the very same 
contribution to the $\beta$-function and therefore to the running
coupling.  Still, physical observables written in terms of these
renormalized parameters should ``know'' when the energy scale of the
physical process is large or small compared with some quark masses.  

We will return to these problems in detail in section~\ref{sec:decoupling}, here
it is sufficient to mention that decoupling in mass-independent renormalization
schemes can be understood as a matching between different theories. At energy
scales below the top quark mass, it is more convenient to use an effective
5-flavor QCD theory, without the top quark. The effects of the top quark at low
energies can be conveniently reabsorbed in a redefinition of the coupling and
quark masses of the 5-flavor theory (these can be computed perturbatively), with
further corrections being power suppressed $\sim\mathcal O\left((\Lambda/m_{\rm
t})^2\right), \mathcal O\left((Q/m_{\rm
t})^2\right)$.

In a similar way, at energies much below the bottom (respectively charm)
quark mass threshold, 4- (respectively 3-) flavor QCD is an excellent
description of nature. Each theory has its own set of fundamental
parameters, so \eg\ the 4-flavor theory is completely defined by the
values of the 4-flavor coupling constant and of the quark
masses. These effective theories can be used to describe any physical
process at energy scales much below the corresponding thresholds
$m_{\rm b}\sim 4\, {\rm GeV}$ and $m_{\rm c}\sim 1.4\, {\rm GeV}$. 


Following the notation in Ref.~\cite{Chetyrkin:2005ia}, the coupling in the
effective theory with $N_{\rm f} -1$ active flavors is related to the coupling
in the fundamental theory with $N_{\rm f} $ active flavors by the relation 
\begin{equation}
  \label{eq:gdec}
  \bar g^2_{\overline{\rm MS}, N_{\rm f}-1 }(m^\star) = \bar g^2_{\overline{\rm
      MS},N_{\rm f}}(m^\star) \times \xi(\bar g^2_{\overline{\rm MS},N_{\rm f}}(m^\star))\,.
\end{equation}
This expression neglects power corrections in the matching between
theories, and  $\xi(x)$ is a just a polynomial.
In Eq.~(\ref{eq:gdec}) $m^\star = m_{\overline{\rm MS} }(m^\star)$ is
the $\overline{\rm MS} $ mass at its own scale. In this case the
one-loop term vanish and we have
\begin{equation}
  \label{eq:xi}
  \xi(\bar g ) = 1 + c_1 \bar g ^4 + c_2 \bar g ^6 + c_3 \bar g ^8 + \mathcal O(\bar g ^{10})\,.
\end{equation}
with the first 4 terms
known~\cite{Chetyrkin:2005ia,Weinberg:1980wa,Bernreuther:1981sg,Grozin:2011nk,Schroder:2005hy}. 

Eq.~(\ref{eq:gdec}) allows to relate the values of the
$\Lambda$ parameters with a different number of flavors. 
For example, for processes at energy scales above the top mass
threshold we would need the value of $\Lambda^{(6)}_{\overline{\rm
    MS}}$. This can be obtained from $\Lambda^{(5)}_{\overline{\rm
    MS}}$ by first determining the value of the five flavor coupling
at the top mass threshold\footnote{We will not discuss any subtleties in the determination
  of the top quark mass here.} ($m_{\rm t}^\star \approx 163$
GeV)
\begin{equation}
  \bar g _{\rm t,5} = \bar g_{\overline{\rm MS},5 }(m^\star_{\rm t})\,,
\end{equation}
from the implicit equation
\begin{equation}
  \frac{\Lambda^{(5)}_{\overline{\rm MS}}}{m^\star_{\rm t}} = 
  \left[b_0\bar g _{\rm t,5}^2\right]^{-\frac{b_1}{2b_0^2}}\,
  e^{-\frac{1}{2b_0\bar g _{\rm t,5}^2}}\,
  \exp\left\{-
    \int_{0}^{\bar g _{\rm t,5}}{\rm d}x\, \left[\frac{1}{\beta^{(N_{\rm f} =5)}(x)} +
    \frac{1}{b_0x^3} - \frac{b_1}{b_0^2x}\right]\right\}\,.
\end{equation}
Now this value of the coupling is transformed into the 6 flavor
coupling by using the decoupling relations Eq.~(\ref{eq:gdec}), to
obtain
\begin{equation}
  \bar g^2_{\overline{\rm MS}, 6}(m^\star_{\rm t}) =
  \frac{\bar g_{\rm t, 5}^2}{\xi(\bar g_{\rm t, 5})}\,.
\end{equation}
Finally the value of $\bar g _{\rm t, 6} = \bar g _{\overline{\rm MS}, 6}(m^\star_{\rm
  t})$ can be used to determine the value of the 6-flavor $\Lambda$
parameter
\begin{equation}
  \Lambda^{(6)}_{\overline{\rm MS} } = m^\star_{\rm t}
    \left[b_0\bar g _{\rm t, 6}^2\right]^{-\frac{b_1}{2b_0^2}}\,
  e^{-\frac{1}{2b_0\bar g _{\rm t, 6}^2}}\,
  \exp\left\{-
    \int_{0}^{\bar g _{\rm t, 6}^2}{\rm d}x\, \left[\frac{1}{\beta^{(N_{\rm f} =6)}(x)} +
    \frac{1}{b_0x^3} - \frac{b_1}{b_0^2x}\right]\right\}\,.
\end{equation}
All this procedure can be summarized by defining
\begin{equation}
  \varphi_{(N_{\rm f}) }(t) =
    \left[b_0t^2\right]^{-\frac{b_1}{2b_0^2}}\,
  e^{-\frac{1}{2b_0t^2}}\,
  \exp\left\{-
    \int_{0}^{t}{\rm d}x\, \left[\frac{1}{\beta^{(N_{\rm f} )}(x)} +
    \frac{1}{b_0x^3} - \frac{b_1}{b_0^2x}\right]\right\}\,.
\end{equation}
and using
\begin{equation}
  \label{eq:lamrat}
  \frac{\Lambda^{(N_{\rm f} +1)}}{\Lambda^{(N_{\rm f})}} =
  \frac{\varphi_{(N_{\rm f} +1)}\left( \frac{\bar g ^\star}{\xi(\bar g^\star)} \right) }
  {\varphi_{(N_{\rm f})}(\bar g ^\star)}\,.
\end{equation}

Of course such a conversion suffers from several uncertainties. 
The expressions for the $\beta$-functions (\ie\, $\beta^{(N_{\rm f}
  =5,6)}$) and the  decoupling relations Eq.~(\ref{eq:gdec}) are only
known to a certain order in perturbation theory. 
This implies that the conversion of $\Lambda$-parameters
Eq.~(\ref{eq:lamrat}) carries a perturbative uncertainty. 
On top of that there are power corrections that have been neglected in
the matching between the effective $N_{\rm f}$ and fundamental
$N_{\rm f} + 1$ theories (see
section~\ref{sec:deco-heavy-quark}). However, we have now strong
numerical 
evidence~\cite{Athenodorou:2018wpk, Korzec:2016eko} showing that both 
the perturbative and power corrections are very small in the ratio
Eq.~(\ref{eq:lamrat}). 
Even for the case of the decoupling of the charm quark (at a rather
low energy scale $m^\star_{\rm c} \approx 1.4$ GeV), these
effects seem to be too small to affect the current determinations of
$\Lambda$. The interested reader is encouraged to read
section~\ref{sec:deco-heavy-quark} and consult the original
reference~\cite{Athenodorou:2018wpk} where these issues are discussed
in detail. 

\subsubsection{Challenges in the determination of
  $\alpha_{\overline{\rm MS}}(M_Z)$}
\label{sec:chall-determ-alph}

From the previous discussion it seems logical to quote the intrinsic scale of
QCD by giving the value of $\Lambda_{\overline{\rm MS}}^{(5)}$, which is a well
defined quantity, even beyond perturbation theory. Together with
Eq.~\eqref{eq:lam} and the coefficients of the $\beta$-function in the
$\overline{\rm MS}$ scheme reported in table~\eqref{tab:bnms}, it can be used
for high energy phenomenology. Moreover if one is interested in a process at an
energy scale above the top quark threshold (or below the bottom/charm
thresholds), the procedure described in the previous section can be used to
determine the 3,4 or 6 flavor $\Lambda$-parameters.

For historical reasons it is now standard to quote the intrinsic
scale of the strong interactions in an indirect way by referring to
the value of the strong coupling in the $\overline{\rm MS}$ scheme at
a reference scale $\mu=M_Z$ (\ie\ the mass of the $Z$ vector boson
$M_Z\approx 91.19\, {\rm GeV}$). The current world
average for 
\begin{equation}
  \alpha_{\overline{\rm MS} }(\mu) = {\bar
  g^2_{\overline{\rm MS}}(\mu)}/{4\pi}
\end{equation}
quoted in the PDG~\cite{pdgtbp:2020} is: 
\begin{equation}
  \label{eq:alpha_pdg}
  \alpha_{\overline{\rm MS} }(M_Z) = 0.1179(10)\, ,
\end{equation}
with an uncertainty of around $\sim 1\%$ obtained by combining the uncertainty
in the determination from several processes. Note that this coupling refers to
the 5 flavor theory since $m_{\rm b} < M_Z < m_{\rm t}$. By using the five-loop
asymptotic expansion of the $\beta$-function, the world average
Eq.~\eqref{eq:alpha_pdg} is equivalent to\footnote{Note that due to the
logarithmic running of the strong coupling a $\sim 6\%$ uncertainty in
$\Lambda^{(5)}_{\overline{\rm MS}}$ translates into an $\sim 1\%$ uncertainty in
$\alpha_{\overline{\rm MS} }(M_Z)$.}
\begin{equation}
  \Lambda_{\overline{\rm MS}}^{(5)} = 0.207(12)\, {\rm GeV}\,.
\end{equation}

But, what are the challenges in a precise determination of the strong
coupling? To understand this subtle point, we have to look carefully
at the fundamental equation used to determine $\Lambda$ in units of
some reference scale $\mu_{\rm ref}$:
\begin{align}
  \label{eq:lam_again}
  \Lambda = \mu_{\rm ref}
  \left[b_0\bar g^2(\mu_{\rm ref})\right]^{-\frac{b_1}{2b_0^2}}\,
  & e^{-\frac{1}{2b_0\bar g^2(\mu_{\rm ref})}} \times \nonumber \\
  & \times \exp\left\{-
    \int_{0}^{\bar g(\mu_{\rm ref})}{\rm d}x\, \left[\frac{1}{\beta(x)} +
    \frac{1}{b_0x^3} - \frac{b_1}{b_0^2x}\right]\right\}\,.
\end{align}
As already discussed above, this is the solution of a first-order differential
equation, and $\Lambda$ can be understood as an integration constant. In other
words, knowing the value of the coupling at some reference scale is sufficient
to determine $\Lambda$ and hence to fix the coupling value at all energies
according to the RG running. In principle only {\em one}\ number is needed to
fully determine the value of the strong coupling. It could be for instance the
value of $\bar g(\mu_{\rm ref})$. The systematic errors in the determination of
the strong coupling are better understood by discussing the determination of
$\Lambda$. As we can see, \eqref{eq:lam_again} involves the integral of the
$\beta$-function from $0$ to $\bar g(\mu_{\rm ref})$. Since
\begin{equation}
  \lim_{\mu\to\infty}\bar g (\mu) = 0\,,
\end{equation}
the lower limit of the integral corresponds to an infinite energy scale.
Therefore the determination of $\Lambda$ requires the knowledge of the
nonperturbative beta function for all energies between $\mu_\mathrm{ref}$ and
infinity. In practice this limit is never reached, neither in experimental
processes nor in lattice simulations. 
At most we can
  determine the non-perturbative running in a limited range of scales,
  say from $\mu_{\rm ref}$ to $\mu_{\rm PT}$. 
  At energies higher than $\mu_{\rm PT}$ one uses the perturbative
  approximation of the non-perturbative $\beta$-function. 
  If we denote $\beta^{(\ell)}$ the perturbative $\beta$
  function to $\ell$-loops, we have that
  \begin{align}
  \label{eq:PTerrorOne}
  \int_{0}^{\bar g(\mu_{\rm ref})}{\rm d}x\, \left[\frac{1}{\beta(x)} -
  \frac{1}{\beta^{(\ell)}(x)} \right] = 
  \mathcal O\left(\left(\bar g^2(\mu_{\rm ref})\right)^{\ell-1}\right)\,.
  \end{align}
And therefore this uncertainty propagates to the determination of $\Lambda$:
\begin{align}
	\nonumber
	\int_{0}^{\bar g(\mu_{\rm ref})}{\rm d}x\, \left[\frac{1}{\beta(x)} +
	  \frac{1}{b_0x^3} - \frac{b_1}{b_0^2x}\right]
    \simas{\bar g (\mu_{\rm PT}) \to 0}
  & \int_{\bar g(\mu_{\rm PT})}^{\bar g(\mu_{\rm ref})}{\rm d}x\, 
  \left[\frac{1}{\beta(x)} +
	\frac{1}{b_0x^3} - \frac{b_1}{b_0^2x}\right] \\
	\nonumber
	& + \int_{0}^{\bar g(\mu_{\rm PT})}{\rm d}x\,
	\left[\frac{1}{\beta^{(\ell)}(x)} +
	\frac{1}{b_0x^3} - \frac{b_1}{b_0^2x}\right] \\
  & + \mathcal O(\bar g^{2\ell -2}(\mu_{\rm PT}))\,.
    \label{eq:intpt}
\end{align}

As shown in table~\ref{tab:bnms} the $\beta$-function is known up to five loops
in the $\overline{\rm MS}$ scheme. Note, nevertheless, that in practice one
never reaches this level of accuracy. In order to apply Eq.~(\ref{eq:intpt}) in
the $\overline{\rm MS} $ scheme, the value of the coupling $\bar
g_{\overline{\rm MS} } (\mu_{\rm PT})$ is needed. The latter is determined by
matching an experimental quantity with its asymptotic perturbative expansion,
typically known up to 3-4 loops. In this case the accuracy in the extraction of
$\Lambda_{\overline{\rm MS}}$ will be limited by the limited knowledge in the
perturbative expression of the physical observable, and not by the perturbative
knowledge in $\beta_{\overline{\rm MS} }(\bar g)$. 

This phenomenon is present in one form or another in any extraction of the
strong coupling, not only the ones from lattice QCD, but also in
phenomenological extractions. Even if $\Lambda_{\overline{\rm MS} }^{(5)}$ is
defined non-perturbatively, perturbation theory is needed for its determination.
Of course this does not mean that a truly non-perturbative determination of the
$\Lambda$-parameter is impossible. The situation is conceptually very similar to
many other systematic effects present in any lattice determination; for example
lattice calculations are always performed at non-zero lattice spacing (and on a
finite volume) and this does not prevent us to obtain values in the continuum
(and in infinite volume). We need to simulate several lattice spacing (and
several volumes) and \emph{perform an extrapolation}. The situation here is very
similar: the determination of $\Lambda$ has to be understood as an extrapolation
in $\bar g^{2n}(\mu_{\rm PT})$, with perturbation theory as a guide.

The scale $\mu_{\rm PT}$ is usually called the \emph{scale of matching with
perturbation theory}. One crucial point to note is that the size of the missing
terms is $\mathcal O(\bar g^{2(\ell-1)}(\mu_{\rm PT}))$, where $\ell$ is the
number of loops included in the computation of the beta function. In order to
have a significant change in the contribution of the missing terms, the matching
scale $\mu_{\rm PT}$ has to be changed substantially due to the slow logarithmic
running of the strong coupling (cf. Eq.~\eqref{eq:RGEHighEnergy}). These issues
play a central role in the determination of the systematic error presented in
section~\ref{sec:syst-extr-alph}.


\section{Physical definitions of the strong coupling}




\label{sec:phys-defin-strong}

\subsection{Determinations of $\alpha_{\overline{\rm MS}}(M_Z)$}

How is the value of the strong coupling constant extracted from experimental
data? The generic procedure can be sketched as follows. Broadly speaking, the
experimental results for a physical process $P(Q)$ at high energies $Q$ are
compared with the perturbative prediction (typically available up to some order
$n$), 
\begin{equation}
  \label{eq:Pseries}
  P(Q) = \sum_{k=0}^n c_k(s) \alpha_{\overline{\rm MS}}^k(\mu) + \mathcal
  O(\alpha_{\overline{\rm MS}}^{n+1}(\mu)) +
  \mathcal O \left( \frac{\Lambda^p}{Q^p} \right) \,, 
  \qquad (s=\mu/Q)\,.
\end{equation}
Several subtle points are involved in this comparison.  First we
should notice that the coefficients $c_k(s)$ grow logarithmically with
$s$, and therefore the renormalization scale $\mu$ has to be chosen
close to the physical scale of the process $Q$, in order to avoid
large logarithms and a poorly converging perturbative series. We
should also note that once $\alpha_{\overline{\rm MS} }(\mu)$ is known
at some energy scale $\mu\sim Q$, one can use the 5-loop
$\beta$-function in the $\overline{\rm MS}$ scheme to ``run'' this
result either to a common reference scale (i.e. $M_Z$), or up to
infinite energy and quote the value of the
$\Lambda_{\overline{\rm MS}}$ parameter. The considerations raised in
section~\ref{sec:chall-determ-alph} also apply to the determinations
that follow this approach. In this case the renormalization scale
$\mu = sQ$ plays the role of $\mu_{\rm PT}$: the energy scale at which
we match with perturbation theory. One would like to extract
$\alpha_{\overline{\rm MS} }(M_Z)$ (or $\Lambda_{\overline{\rm MS}}$)
by using data at several values of $\mu = sQ$, and take as the final
result a suitable extrapolation $sQ\to\infty$. Since the value of $s$
cannot be taken to be arbitrarily large, such a procedure requires
data at several values of the physical scale $Q$ in order to have a
real constraining power on the value of the coupling.

In Eq.~\eqref{eq:Pseries} we show the two types of corrections present in the
perturbative expansion of a physical quantity. First the {\em missing higher
orders}, due to the fact that we only know a finite (typically $n = 2, 3$)
number of terms in the perturbative expansion of the observable. Second,
non-perturbative corrections (usually called \emph{power corrections}). These
are of the form $e^{-A/\alpha_{\overline{\rm MS}}(Q)} \sim \mathcal
O\left(\frac{\Lambda^p}{Q^p} \right)$ with $p=2A\, b_0$ and decrease faster
than any power of $\alpha_{\overline{\rm MS}}$.  In order to keep the truncation
and non-perturbative corrections small, the chosen process should be ideally
inclusive and defined at high enough energies. High energy scales ensure that
$\alpha_{\overline{\rm MS}}(Q)$ is small. Inclusive measurements do not require
a quantitative description of the strong interactions of hadronic states and
therefore are less affected by systematic errors coming from models of
hadronization and parton showers. 

\begin{figure}
  \centering
  \includegraphics[width=\textwidth]{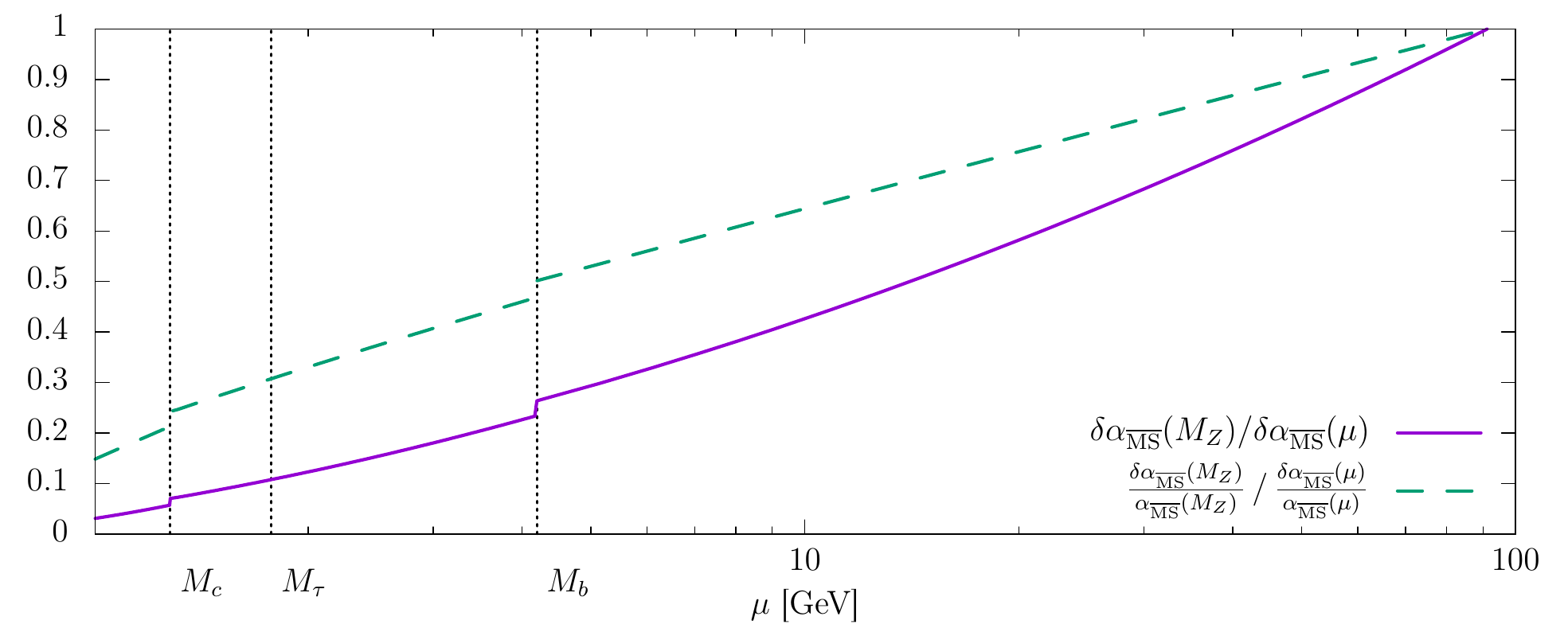} 
  \caption{Error of the coupling at a scale $\mu$ compared with the
    error propagated to the reference scale $M_Z$. 
    Note that when the strong coupling is determined at low energy
    scales the result at the reference scale $M_Z$ becomes more precise. 
    For example, the error in $\alpha_{\overline{\rm MS} }(M_\tau)$ is
    reduced by almost one order of magnitude when the result is
    evolved to the scale $M_Z$ (solid purple line). 
    This effect is \emph{not} only due to the reduction in the coupling
    itself: the relative error in the coupling is reduced
    approximately by a factor three (dashed green line).}
  \label{fig:alpha_err_scale}
\end{figure}

Obtaining a precise value for the strong coupling with high energy
experimental input has its own challenges. At high energies the strong
coupling is small, which is just what is needed to have the truncation
and power corrections under control, but at the same time the effect
that one is trying to measure is small. This usually translates in
larger uncertainties in $\alpha_{\overline{\rm MS}}(M_Z)$ from
determinations based on data at high energies (see
figure~\ref{fig:alpha_err_scale}). Extracting the value of 
the strong coupling at lower energies usually leads to smaller
uncertainties, although the estimation of the truncation uncertainties
and the non-perturbative effects become more challenging: clearly the
extrapolation $sQ\to\infty$ is more difficult without data at large
$Q$.
 
Another point to take into account is that in contrast with the
perturbative computations, quarks are not the observed final states of
any physical process. Hadronization and other non-perturbative effects
have to be taken into account when comparing experimental data with
perturbative predictions, usually by using Monte Carlo
generators. 

\paragraph{Extraction from data}

A well-know example of the extraction of the strong coupling from
experimental data is the extraction of
$\alpha_{\overline{\rm MS}}(M_Z)$ using data for $\tau$ decaying into
hadrons. We briefly summarise the procedure here, in order to
highlight the main steps and the sources of uncertainties, we refer
the reader to an extensive review like e.g. Ref.~\cite{Pich:2013lsa}
for a detailed discussion. The physical processes considered in this
case are the decays of $\tau$ leptons. More specifically, the ratio of
the hadronic and leptonic decay widths can be written as
\begin{equation}
  \label{eq:tau_decays}
  R_{\tau, V+A} = \frac{\Gamma(\tau \to \nu_\tau + \text{hadrons})}
  {\Gamma(\tau \to \nu_\tau e^-\bar\nu_e)} = 
  3|V_{ud}|^2 S_{\rm EW}\left(1 + \delta_{P} + \delta_{\rm NP}\right)
  \, .
\end{equation}
In this case the typical energy scale of the process is set by the
$\tau$ mass $Q=M_\tau = 1.77682(16)$ GeV.  In
Eq.~\eqref{eq:tau_decays} $S_{\rm EW}(Q)$ is the electroweak
contribution to $R_{\tau, V+A}$, and $\delta_{\rm P}, \delta_{\rm NP}$
are the QCD perturbative and non-perturbative corrections to the
process respectively. The non-perturbative (i.e. power) corrections
are estimated to be very small
$\delta_{\rm NP}\sim 10^{-4}$~\cite{Pich:2013lsa}. On the other hand
the perturbative prediction
\begin{equation}
  \label{eq:Rpt}
  \delta_{\rm P} = \sum_{k=1}^4 r_n(s)\,
  \left(\frac{\alpha_{\overline{\rm MS}}(\mu)}{\pi}\right)^k +
  \mathcal O(\alpha_{\overline{\rm MS}}^5) 
  \, , \qquad (s=\mu/M_\tau)\,.
\end{equation}
is known up to four
loops~\cite{Baikov:2008jh,Baikov:2012er,Baikov:2012zn}. The impressive
perturbative knowledge in 
the ratio $R_{\tau, V+A}$ makes this quantity a good candidate to
determine $\alpha_{\overline{\rm MS}}$. In fact such determinations
are one of the most precise phenomenological determinations. On the other
hand the scale at which $\alpha_{\overline{\rm MS}}$ is determined is
relatively low, and it cannot be changed, since $M_\tau$ is what it
is.

The procedure with other observables is basically the same, although
some details, like the number of known terms in the perturbative
expansion or the size of the non-perturbative effects
(i.e. $\delta_{\rm NP}$ in Eq.~\eqref{eq:tau_decays}), change from one
observable to another.

Combining multiple collider observables in a global fit provides a
better lever-arm to constrain $\alpha_s$ together with the parton
distribution functions (PDFs). Global fits that include the wider
ranges of data provide determinations of the strong coupling constant
with good statistical accuracy, see \eg
Refs.~\cite{Ball:2011us,Ball:2018iqk,Harland-Lang:2015nxa}. The
challenges here stem from controlling the systematic errors (both
theoretical and experimental) in fits that involve very large and
diverse datasets and the relatively low energies involved (see for
example the recent review\cite{Salam:2017qdl}). Moreover,
as recently discussed in 
Ref.~\cite{Forte:2020pyp}, determinations of the strong coupling from
hadronic processes should entail a simultaneous determination of the
parton distribution functions. 

\paragraph{Extraction from lattice simulations}

Lattice QCD offers an interesting alternative to phenomenological
determinations. Being a non-perturbative formulation of QCD, one can
combine input from well-measured QCD quantities -- like for example
the proton mass, or a meson decay constant -- with the perturbative
expansion of a short distance observable that does not need to be
directly observable (like the quark anti-quark force). The advantage
of this approach is that the experimental input comes from the hadron
spectrum with a negligible uncertainty. Hadronization corrections are
not needed, since we are working directly in a non-perturbative
framework.

\begin{figure}[t!]
  \centering
  \includegraphics[width=\textwidth]{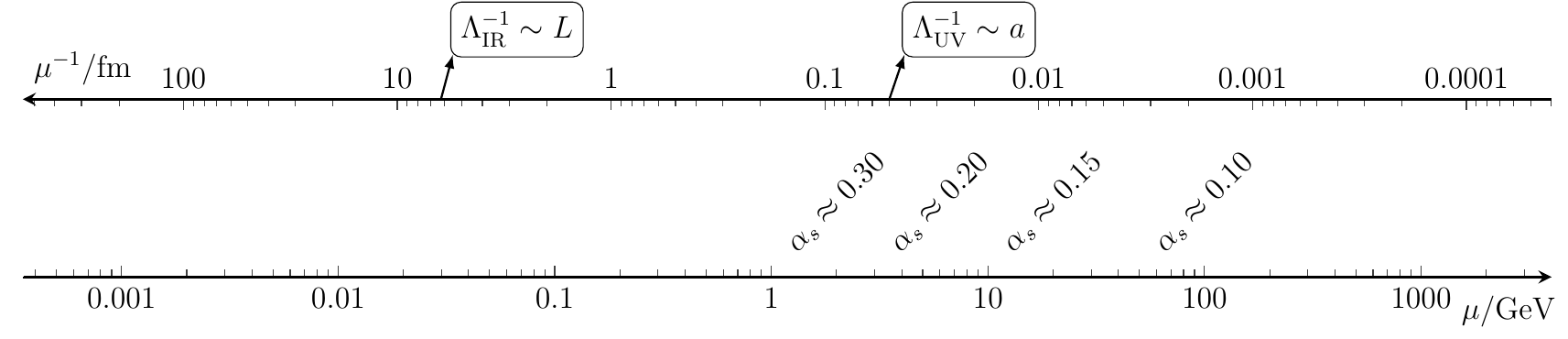}
  \caption{Scales in a typical state of the art lattice
    computation. The volume of the simulation is a few fm, while the
    cutoff $1/a$ is a few GeV. Lattice QCD can resolve observables in
    this window of scales, where the strong coupling is $\alpha_{\overline{\rm MS}} \sim
    0.25$.}
  \label{fig:LQCDscales}
\end{figure}

Despite being very different approaches, both the phenomenological and
the lattice QCD methods have to overcome similar challenges. Lattice
extractions of the strong coupling must be done at sufficiently high
energies so that truncation and power corrections are well under
control when matching to perturbative expansions. Due to the finite
nature of computer resources, every lattice QCD simulation has two
intrinsic scales: the total physical volume simulated $L$ (IR cutoff),
usually of a few fermi in order to keep finite-volume corrections well
under control, and the lattice spacing $a$ (the UV cutoff $\sim 0.04$
fm in the most challenging present day simulations, which corresponds
roughly to a cutoff of 5~Gev in energies). Any lattice QCD simulation
can only resolve a process if it is defined at a scale between these
IR and UV cutoffs (see figure~\ref{fig:LQCDscales}).  The number of
lattice points in each direction is given by the ratio $L/a$, viz. the
separation of the UV and IR cutoffs determines the memory footprint
and computing power,
and hence the computational cost, of the corresponding simulations,
putting in practice a limit on the energy scales that can be studied
in any lattice simulation. While in principle lattice techniques can
be used to compute non-perturbatively the running of the coupling
until the perturbative regime is reached, in practice the range of
scales that can be studied \emph{in a single lattice simulation} is
limited by computer resources. Reaching scales higher than a few GeV
requires a dedicated approach.

\subsection{Systematics in the extraction of
  $\alpha_{\overline{\rm MS}}$}
\label{sec:syst-extr-alph}

\begin{figure}
  \centering
  \includegraphics[width=\textwidth]{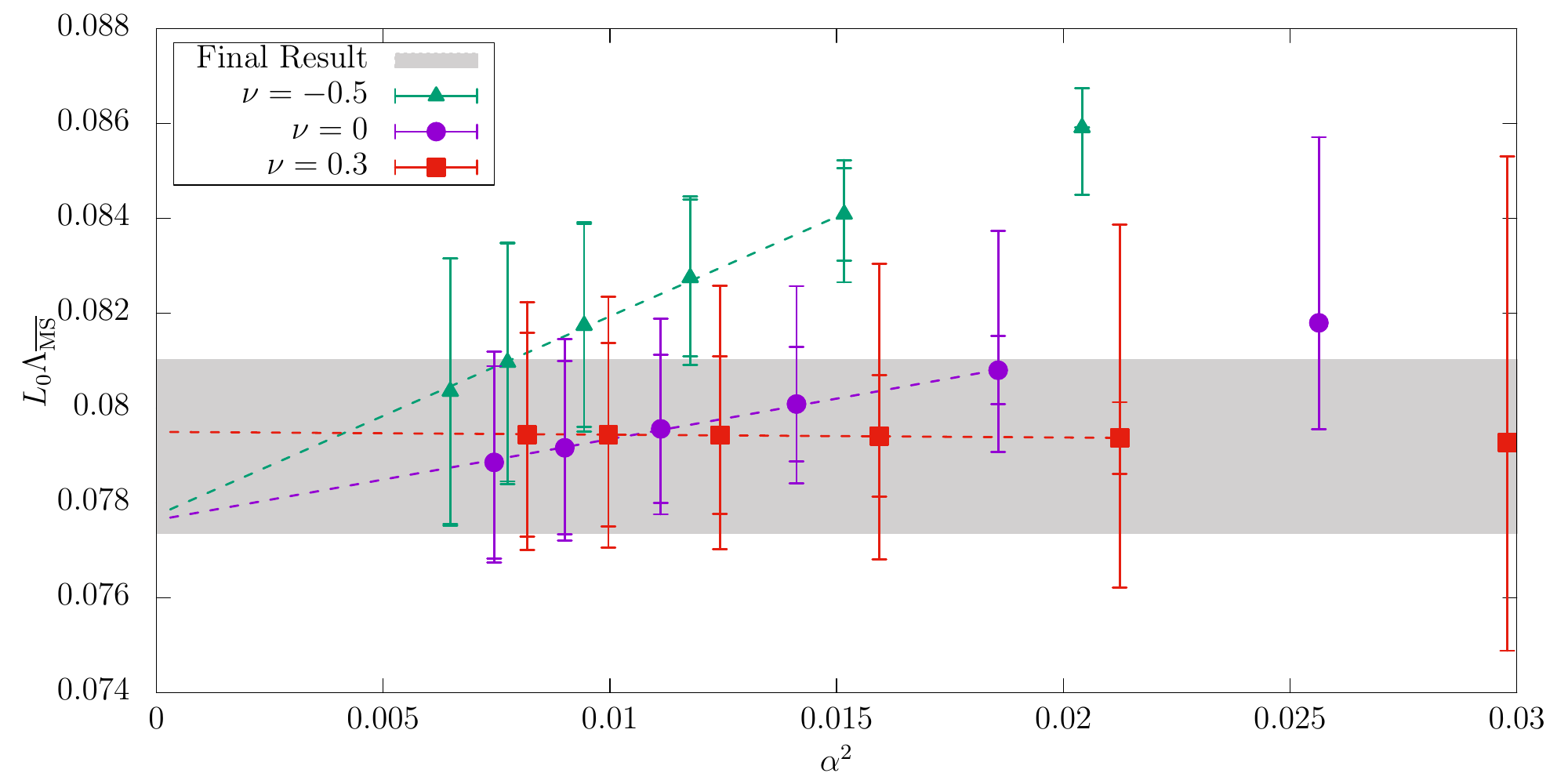} 
  \caption{Determination of the $\Lambda$-parameter in units of a
    scale $L_0\sim 1/(4 {\rm GeV})$. Different values of $\nu$
    represent different choices of observable, and each point
    corresponds to a determination of the dimensionless product
    $\Lambda_{\overline{\rm MS}}  L_0$ using a perturbative expansion
    (like Eq.~(\ref{eq:Pseries})). 
    The horizontal axes label the scale of matching with perturbation
    theory ($\mu$ in Eq.~(\ref{eq:Pseries})) parametrized in terms of
    the leading corrections $\mathcal O(\alpha_{\overline{\rm MS}
    }^2(\mu))$. The first error bar shows the
    statistical uncertainty. The second error bar shows the total
    systematic plus statistic added in quadratures. The systematic
    uncertainty is determined by varying the renormalization scale a
    factor of two above and below some specific value. For $\nu=-0.5$
    the systematic uncertainty determined with this method is unable
    to account for the difference with the final results at values of
    $\alpha\sim 0.13-0.2$, while in other cases ($\nu = 0.3$) the
    systematic uncertainty overestimates the true
    difference. See text for more details. 
    (source~\cite{Brida:2016flw, DallaBrida:2018rfy}). }
  \label{fig:msbar}
\end{figure}

The truncation performed in Eq.~\eqref{eq:Pseries} neglects higher
order terms (i.e. perturbative corrections) and non-perturbative power
corrections. When performing an extrapolation to $Q\to\infty$, these
effects only affect \emph{how fast} we approach the extrapolated
value. An example of this behaviour can be seen in
figure~\ref{fig:msbar}, where different observables (labeled by $\nu$)
are used to estimate $\Lambda_{\overline{\rm MS} }$ by matching with
perturbation theory at different physical scales $Q$, which are
translated in different values of $\alpha_{\overline{\rm MS} }$ in the
plot. Different observables predict compatible results for
$\Lambda_{\overline{\rm MS} }$ when the extrapolation $Q\to\infty$,
corresponding to $\alpha\to 0$ in the plot, is performed. The results
of these extrapolations agree well within errors with the result
quoted in Ref.~\cite{Brida:2016flw} (gray error band in the
plot). Note however that some observables ({it viz.}\, $\nu=-0.5$)
show a slow approach to the extrapolated value, with significant
discrepancies even at energy scales $Q\sim 8-10$ GeV.

In practice performing the extrapolation $Q\to\infty$ is very
difficult. 
Data over a large range of energy scales is required in order to
perform such an analysis. 
For example the data in figure~\ref{fig:msbar} involves precise lattice
determinations of the target observables for energy scales $Q\in
2-140$ GeV. This is only possible with a dedicated approach (see
section~\ref{sec:finite-size-scaling}).  

How do we estimate the systematics effects in the
extraction of $\alpha_{\overline{\rm MS}}$ when the data does not
allow an extrapolation $Q\to \infty$?
Of course this is a complex subject in itself, that is of much
relevance not only for the extraction of $\alpha_{\overline{\rm MS}}$,
but also in the interpretation of many experimental data from hadron
colliders. A possible estimate of the uncertainty due to the
missing terms is given by the last known term in the series
$c_n(s) \alpha_{\overline{\rm MS}}^n(\mu)$. Generally this results in
significant theoretical uncertainties, and, perhaps more
interestingly, correlations between experimental data. 

A \emph{common} approach to estimate these uncertaintites 
exploits the fact that the truncated perturbative expansion to $n^{\rm
  th}$ order
\begin{equation}
  \label{eq:TruncObsP}
  P^{(n)}(Q, s) = \sum_{k}^n c_k(s) \alpha_{\overline{\rm MS}}^k(\mu)\,,
  \qquad (s=\mu/Q)\,.
\end{equation}
still depends on $\mu$, while the true value of the observable $P(Q)$
does not. The truncation of the perturbative expansion introduces a
spurious dependence on the renormalization scale, which is in general
an unphysical, arbitrary quantity. Higher-order effects are estimated
by looking at the variation of $P^{(n)}(Q, s)$ when the
renormalization scale $\mu$ is changed by a factor two around some
preferred value (for example $\mu=Q$). In principle the relation
between a variation in the renormalization scale $\mu$ and the size of
the missing higher-order terms given by
$\delta_n=|P(Q) - P^{(n)}(Q,s)|$ is unclear, beyond the fact that the
scale dependence in Eq.~\eqref{eq:TruncObsP} is due to the truncation
of the perturbative expansion. Under some assumptions on the size of
the coefficients of the perturbative expansion
($c_{n+2}(s)\alpha(\mu)\ll c_{n+1}(s)$), it is possible to show 
that the scale variation yields a sensible estimate of $\delta_n$ (see
for example Ref.~\cite{Cacciari:2011ze}).  Formally,
\begin{equation}
  \label{eq:truncation_1}
  \mu \frac{{\rm d} P^{(n)}(Q,s)}{{\rm d} \mu} \propto \alpha_{\overline{\rm MS}}^{n+1}(\mu)\,,
\end{equation}
which implies that, at least parametrically, changes in $\mu$ capture the
correct size of the missing terms. As an example, a recent comprehensive study
of the theoretical uncertainties for numerous observables, based on scale
variations, can be found in
Refs.~\cite{AbdulKhalek:2019bux,AbdulKhalek:2019ihb}.

What about power corrections? they are not captured by this kind of analysis.
Estimating them requires to have access to different physical scales $Q$.
Ideally one would like to work at sufficiently high energies so that they are
negligible compared with the accuracy of the data. In practice this is not
always the case. Note that the perturbative running is logarithmic, and
distinguishing this perturbative running from a power-like behaviour requires
data that span large energy ranges.

The assumptions that underlie the scale variation procedure constrain both the
non-perturbative effects and the character of the perturbative series. In
particular, the assumption that the first unknown term of the
perturbative series is smaller than the last known one is implicit in any
estimate that uses Eq.~\eqref{eq:truncation_1}. Also the value of
$\alpha_{\overline{\rm MS}}(\mu)$ is assumed to be small enough so that these
uncertainties are meaningful. These assumptions might seem reasonable and mild,
and often yield sensible estimates, but there are examples in the literature
where they have been shown not to be accurate. Let us mention here
three relevant cases. 
\begin{itemize}
\item The convergence of the perturbative series in practice is not as
  good as we would like. Due to the asymptotic nature of the PT
  series, one expects that at some point the coefficients in
  the perturbative series will grow factorially, see
  Ref.~\cite{Beneke:1998ui} for a review.
  
\item Extractions of the strong coupling from $\tau$ decays can be done by
  applying two frameworks in perturbation theory, called fixed order
  perturbation theory (FOPT) and contour improved perturbation theory (CIPT).
  Using $\alpha_{\overline{\rm MS}}(m_\tau)=0.34$ as the typical value of the
  strong coupling at the scale set by the mass of the $\tau$, the contributions
  to both perturbative series look as follows\footnote{The
    perturbative series has the form
    \begin{equation}
      \delta = \sum_n(K_n + g_n)\left(\frac{\alpha}{\pi}\right)^n
    \end{equation}
    where the $K_n$ coefficients are the same in FOPT and CIPT, while
    the $g_n$ coefficients are different in both formulations. 
    The coefficient $K_5$ is unknown. Here we use an estimate $K_5=275$. 
  Note that this does not affect the difference in $\delta$
  between both formulations.}
  \begin{eqnarray}
    \delta_{\rm FOPT} &=&
                          0.1082 + 0.0609 + 0.0334 + 0.0174 + 0.0087 = 0.2286\,,\\
    \delta_{\rm CIPT} &=&
                          0.1479 + 0.0297 + 0.0122 + 0.0086 + 0.0037 = 0.2021\,.
  \end{eqnarray}
  The terms in both series decrease, and each of the expansions by themselves
  seem reasonable. Taking the last term as a measure of the uncertainty in the
  truncation, these values should be accurate with a precision $\sim 0.01$. But
  both approaches result in values that differ by more than twice this amount.
  What is more worrisome, for the highest orders the difference
  between both estimates \emph{grows} as more terms are included in
  the expansion.  

\item The scale variation approach to estimate truncation uncertainties
  (changing the value of the renormalization scale $\mu$ by a factor two around
  a preferred value) has been compared with non-perturbative data in a careful
  study for values of $\alpha_{\overline{\rm MS}}\sim
  0.1-0.2$~\cite{DallaBrida:2018rfy} (see figure~\ref{fig:msbar}). The error
  bars in figure~\ref{fig:msbar} include an estimate of the truncation
  uncertainty using this approach. It is clear from the plot that for $\nu=-0.5$
  the scale variation underestimates the systematic error. Note that the known
  terms in the perturbative series for all these observables suggest an apparent
  good perturbative behavior. 
\end{itemize}

These examples show that estimating the truncation uncertainties within
perturbation theory is difficult. In the absence of a theorem, our attempts to
quantify these uncertainties remain exploratory, and caution should be exercised
in interpreting the results. One should never forget the asymptotic nature of
perturbative series in QCD~\cite{tHooft:1977xjm}. Eventually a \emph{factorial
growth} of the size of the terms in the series is expected, which is deeply
related to the non-perturbative effects of the theory. Note however that the
structure of the corrections have been investigated at length (see
\eg~Ref.~\cite{Beneke:1998ui}). It is clear
that there are at least two ingredients in the quality of any
extraction of the strong coupling.  
\begin{enumerate}
\item The value of $\alpha_{\overline{\rm MS}}$ at which perturbation theory is
  used. Non-perturbative (power) corrections decrease very quickly with
  $\alpha_{\overline{\rm MS}}$. In order to make them negligible one needs to
  have access to high energy scales, hence, small values of
  $\alpha_{\overline{\rm MS}}$.

\item The extraction has to be performed over a range of values of
  $\alpha_{\overline{\rm MS}}$. This allows the unknown terms in the series to
  vary substantially, so that one can check that indeed they are negligible. A
  reasonable requirement would be that the first unknown term
  $\alpha_{\overline{\rm MS}}^{n+1}$ varies significantly, say by a factor four.
\end{enumerate}

These are two criteria that are usually relevant in any extrapolation. As
shown in the examples above, they will impact the quality of any extraction of
the strong coupling. Of course some determinations cannot really change the
value of the momentum scale at which perturbation theory is used. A
good example is the above 
mentioned extraction from $\tau$ decays, since the mass of the $\tau$ is what
it is and sets the overall energy scale to the process. For the case of
lattice simulations, changing the values of $\alpha_{\overline{\rm MS}}$ at
which perturbation theory is used is challenging, yet feasible. We shall keep
these two criteria in mind when describing any lattice computation/method, and
not only the quoted theoretical uncertainties.

We would like to end this section recommending the reader the recent
contribution to the Lattice Field Theory Symposium by M. 
Dalla Brida~\cite{DallaBrida:2020pag}, where these issues are also
discussed in detail.


\section{Lattice field theory}
\label{sec:lattice-field-theory}
This section summarizes, briefly, the ideas that underlie lattice
QCD. While it does not provide an extensive discussion of lattice QCD, it
is intended to present the framework of lattice simulations for the
non-expert reader in a self-consistent form, setting the notation for the
following sections, and providing references for further
reading. Hoperfully it will yield the foundations to better understand the
sources of systematic errors that are discussed in what follows. It can
safely be skipped unless the reader is actually interested in the details
of the lattice simulations.

The formulation on a discretized lattice provides a non-perturbative
definition of a Quantum Field Theory (QFT). The starting point is the
path integral of the theory in Euclidean space
\begin{equation}
  \label{eq:path_integral}
  \mathcal Z = \int \mathcal D\bar \psi \mathcal D\psi \mathcal D
  A_\mu\,
  e^{-S_{\rm QCD}[\bar \psi,\psi,A_\mu]}\,.
\end{equation}
where
\begin{equation}
  \label{eq:EuclAction}
  S_{\rm QCD}[\bar\psi,\psi,A_\mu] = \int {\rm d} ^4x\, \left\{- \frac{1}{2g^2}{\rm
    Tºr}\{F_{\mu\nu}F_{\mu\nu}\} + \sum_{f=1}^{N_{\rm f}}\bar\psi_f
  (D_\mu\gamma_u + m_f)\psi \right\}
\end{equation}

Lattice field theory gives a precise definition to the path integral in
Eq.~\eqref{eq:path_integral} by discretizing the spacetime in a hyper-cubic
lattice with spacing $a$. In this approach matter fields are defined at the
lattice sites $x_\mu=an_\mu$ for $n_\mu=1,\dots,L_\mu/a$, where $L_\mu$ is the
physical size in direction $\mu$. After this discretization the path integral
Eq.~\eqref{eq:path_integral} is simply an integral over very many degrees of
freedom (the value of each of the fields at each spacetime point). For example
the measure for the fermion fields becomes
\begin{eqnarray}
  \mathcal D \bar \psi(x)
  &\longrightarrow& \prod_{n_\mu=1}^{L_\mu/a} d\bar \psi(an_\mu)\,,\\
  \mathcal D \psi(x)
  &\longrightarrow& \prod_{n_\mu=1}^{L_\mu/a} d\psi(an_\mu)\,.
\end{eqnarray}
As already discussed the lattice spacing $a$, provides the UV cutoff of the
theory. 

The most appealing characteristic of the lattice formulation of QCD is that it
allows quantitative computations to be performed using numerical simulations.
Field correlators are computed as integrals in spaces of very large (but finite)
dimensions using Monte Carlo techniques.

A crucial step in lattice field theory consists in defining a ``discretized''
version of the continuum action $S_{\rm QCD}$.  When constructing lattice
actions, one has to pay special attention to the symmetries of the theory.
Ideally the lattice action should preserve \emph{exactly} as many of the
symmetries of the continuum action as possible. The case of gauge symmetry plays
a special role, since it is crucial to guarantee the renormalizability of the
theory. Another common requirement for the lattice action is to reduce to the
continuum action in the naive classical limit $a\to 0$\footnote{ Recently
several works have pointed out that universality might still give the correct
continuum results, even in cases where the lattice action does not reproduce the
continuum action in the naive limit $a\to 0$. The interested reader should
consult the seminal work~\cite{Bietenholz:2010xg}.}.   

\subsection{Gauge fields on the lattice}

A naive discretization of the pure gauge action, obtained by
substituting the derivatives in the continuum action by finite
differences, results in discretization effects that break gauge
invariance. The way to construct lattice actions for gauge theories is
rooted in the geometric interpretation of gauge invariance, and was
first proposed by Wilson~\cite{PhysRevD.10.2445}. Since the gauge
field acts as an affine connection in the continuum theory, its
lattice counterpart is the parallel transporter along the links of
discretized spacetime. Hence the key idea is to work with link
variables
\begin{equation}
  \label{eq:LinkVarsDef}
  U(x,\mu) = e^{aA_\mu(x)} \,,
\end{equation}
where the pair $(x,\mu)$ uniquely identifies the link that originates
from point $x$ in the positive $\mu$ direction.  These link variables
can be seen as a discretization of a continuum Wilson line, the
parallel transporter mentioned above, linking the points $x$ and
$x+a\hat \mu$~\footnote{The four-vector $\hat \mu$ has all components
  equal to zero, except the coordinate $\mu$, that has a value 1.}
\begin{equation}
  \label{eq:ParallelTransp}
  U(x,\mu) = \mathcal P\exp \left\{ a\int_0^1{\rm d}t\,
    A_\mu(\gamma(t)) \right\} + \mathcal O(a^2)\,.
\end{equation}
Here $\gamma(t)$ is a path that links the point $x$ with $x+a\hat\mu$,
\eg
\begin{equation}
  \label{eq:SimplePath}
  \gamma(t) = x + at\hat \mu\, , \quad t\in [0,1]\, .
\end{equation}
A product of link variables along a closed loop is called a {\it
  Wilson loop}.
\begin{figure}[h!]
  \centering
  \begin{subfigure}{0.3\textwidth}
    \centering
    \includegraphics[height=0.2\textheight]{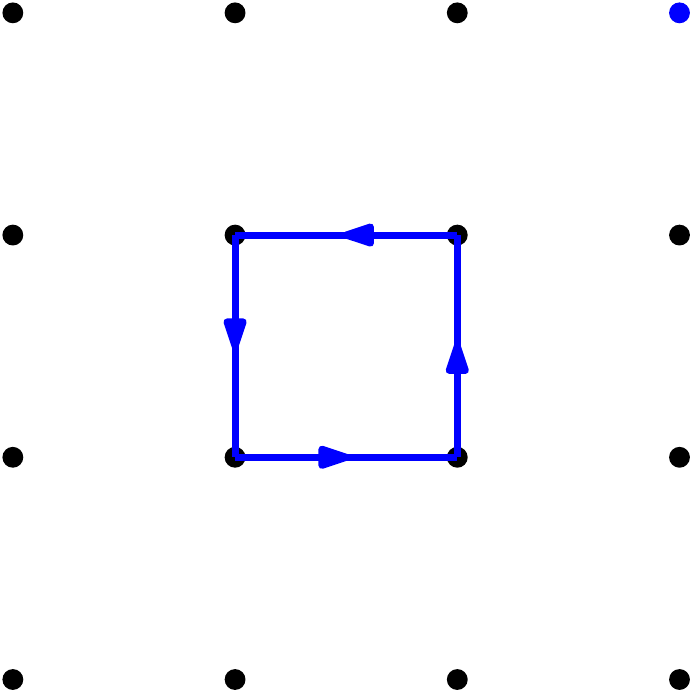}
    \caption{$\mathcal S_0$}
  \end{subfigure}\hspace{0.2\textwidth}
  \begin{subfigure}{0.3\textwidth}
    \centering
    \includegraphics[height=0.2\textheight]{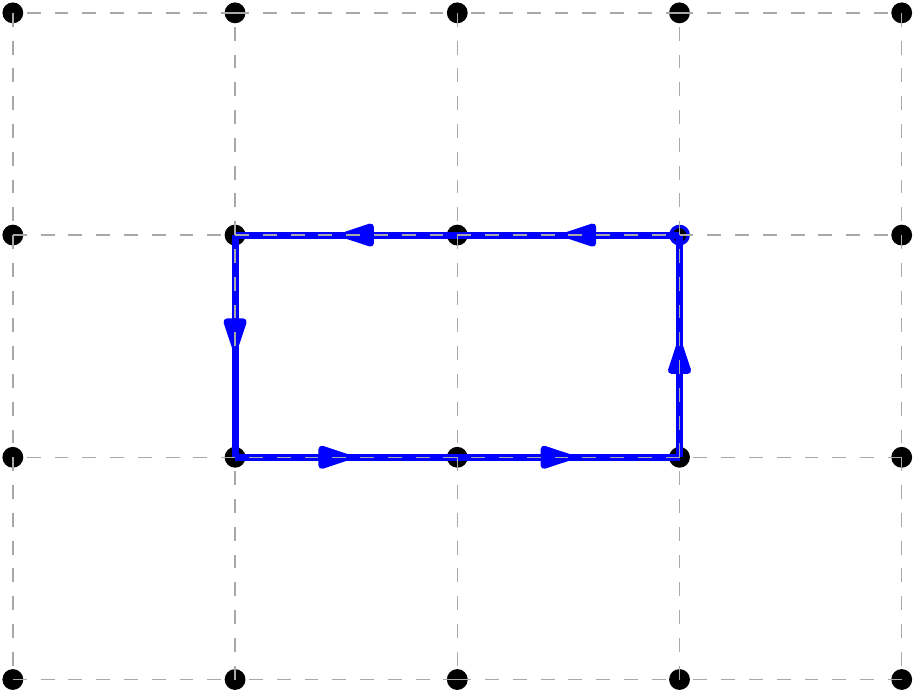}
    \caption{$\mathcal S_1$}
  \end{subfigure}
  
  \caption{Traces of the plaquette ($\mathcal S_0$) and the $2\times 1$ Wilson
    loops ($\mathcal S_1$) can be used to construct different lattice
    actions.}
  \label{fig:wloops}
\end{figure}

Link variables transform under a gauge transformation $\omega(x)\in SU(3)$ as
\begin{equation}
  U(x,\mu) \to \omega(x) U(x,\mu) \omega(x+a\hat \mu)^\dagger\,,
\end{equation}
which implies that the trace of a Wilson loop is gauge invariant. Lattice
actions are constructed by combining these traces, with different choices
yielding the same low-energy physics (compared to the cutoff scale), with
different lattice artefacts. The simplest choice, first proposed by Wilson,
consists in using the smallest possible loop (the plaquette)
\begin{equation}
  \label{eq:Sw}
  S_W = \frac{1}{g_0^2}\sum_{\mathcal W \in \mathcal S_0}{\rm
    tr}\{1-U(\mathcal W)\}\,,
\end{equation}
where the sum runs over all oriented Wilson loops of type $\mathcal S_0$ (see
figure~\ref{fig:wloops}). 
It is easy to check that for a given plaquette $\mathcal W_x$ in the $\mu,\nu$
plane with left lower corner at point $x$, we have
\begin{equation}
  {\rm Re}\, {\rm tr}\{1-U(\mathcal W_x)\} = \frac{1}{2} a^4{\rm
    tr}\{F_{\mu\nu}(x)F_{\mu\nu}(x)\} + \mathcal O(a^5)
\end{equation}
And therefore the Wilson action reduces to the continuum YM action in the naive
limit $a\to 0$.

\paragraph{Improved actions}

This is not the only option. By including in the definition of the lattice
action larger loops the rate of convergence to the continuum can be
improved, \ie\ the size of lattice artefacts can be reduced. This is a
general programme that goes under the name of {\em improvement}, and
extends to the fermionic part of the action. Improved actions play an
important role in reducing lattice artefacts and therefore providing more
precise extrapolations to the continuum limit. Although the literature
covers a wide range of lattice actions, most lattice simulations are
performed using some particular choice of the one-parameter family that can
be constructed from the plaquette and the $2\times 1$ loops (see
figure~\ref{fig:wloops})
\begin{equation}
  \label{eq:ImprovedGaugeAction}
  S_{\rm latt} = \frac{1}{g_0^2}\sum_{i=0,1} c_i \sum_{\mathcal W \in \mathcal S_i}{\rm
    tr}\{1-U(\mathcal W)\}\,.
\end{equation}
The constants $c_0,c_1$ have to obey the constraint
\begin{equation}
  c_0+8c_1 = 1\, ,
\end{equation}
in order to recover the classical continuum limit, but otherwise can be chosen
at will.  Clearly the simple Wilson action, $S_W$ in Eq.~\eqref{eq:Sw}, is
recovered by choosing $c_0=1, c_1=0$. Other popular choices include the Symanzik
tree-level improved action ($c_0=5/3, c_1=-1/12$), or the Iwasaki action
($c_0=3.648, c_1=-0.331$). A detailed discussion of the improvement of
lattice actions is beyond the scope of this review. However the reader
should keep in mind that there are discretization effects when
reading the lattice literature. These discretization effects induce a
systematic error in the lattice observables, which in turn impacts on
the determination of the strong coupling. We shall return to this
issue later in this paper, when discussing the different lattice
extractions of $\alpha$.  

For the path integral to be fully specified, it is also necessary to
define the integration measure of the gauge link variables
$U_\mu(x)$. In order to ensure the gauge invariance of the path
integral, the measure ${\rm d} U$ of each link variable needs to be
invariant under both left and right multiplication by elements of the
group:
\begin{equation}
  \label{eq:MeasureInv}
  {\mathrm d} U = \mathrm{d}(g U) = \mathrm{d}(U g)\, ,
\end{equation}
where $g$ is a generic element of the gauge group. Imposing the
normalization condition
\begin{equation}
  \label{eq:MeasureNorm}
  \int dU = 1\, ,
\end{equation}
the integration measure is uniquely defined to be the Haar measure on
the group. The reader interested in more details can consult any
standard reference on compact topological groups. When integrating
over all the lattice link variables, we will use the shorthand
notation
\begin{equation}
  {\rm d} U = \prod_{x,\mu} {\rm d} U(x,\mu) \,.
\end{equation}

\subsection{Fermions on the lattice}

\subsubsection{Fermionic Path Integral}
\label{sec:grassmann-variables}

Fermions in the functional integral language are represented by
Grassmann (anti-commuting) variables
\begin{equation}
  \{\psi_i,\psi_j\} =   \{\psi_i,\bar \psi_j\} =   \{\bar \psi_i,\bar
  \psi_j\} = 0\,.
\end{equation}
Obviously $\psi_i^2 = \bar \psi_i^2 = 0$, so that any function of
Grassmann variables is defined by its Taylor expansion up to second
order. Integrals over Grassmann variables are defined by
\begin{equation}
  \begin{split}
    \int\, {\rm d} \psi = \int\, {\rm d} \bar \psi = 0\,, \\
    \int\, {\rm d} \psi \psi  = \int\, {\rm d} \bar \psi \bar\psi = 1\,, \\
  \end{split}
\end{equation}
Computing integrals of a function of Grassmann variables
$f(\psi,\bar\psi)$ is therefore a problem in combinatorics. 
When computing the integral over several Grassmann variables we will use the
shorthand notation
\begin{equation}
  {\rm d} \psi = \prod_i {\rm d} \psi_i\, , \qquad
  {\rm d} \bar \psi = \prod_i {\rm d} \bar \psi_i\,.
\end{equation}
A key role is played by the integrals
\begin{eqnarray}
  \label{eq:grassmann_int}
  \int {\rm d} \psi{\rm d} \bar \psi\,
  e^{-\bar \psi_i M_{ij}\psi_j} &=& (-1)^{\frac{n(n-1)}{2}}\det M\,,\\ 
  \nonumber
  \int {\rm d} \psi{\rm d} \bar \psi\,
  \psi_{k_1}\psi_{k_2}\cdots\bar\psi_{l_1}\bar\psi_{l_2}\cdots
  e^{-\bar \psi_i M_{ij}\psi_j} &=&
                                    \sum_{P}(-1)^{\sigma_P}
                                    (M^{-1})_{k_{a_1} l_{b_1}}\cdots(M^{-1})_{k_{a_2} l_{b_2}}
\end{eqnarray}
with the sum is over all permutation of the $k,l$ indices. 

As discussed above, matter fields on the lattice are associated to sites,
denoted by the suffices in the equations above. The fermionic action is
quadratic in the fermion fields, and different discretizations can be cast into
different choices for the matrix $M$. 
A discretized version of the derivative 
\begin{eqnarray}
  \partial_\mu f(x) = \frac{f(x+a\hat\mu) - f(x)}{a}\,,\\
  \partial_\mu^* f(x) = \frac{f(x) - f(x-a\hat\mu)}{a}\,,
\end{eqnarray}
enters in the non-diagonal elements of this matrix $M$. Moreover, for fermions
minimally coupled to the gauge field, the discrtized derivative needs to be
replaced by its covariant version. Hence in QCD the matrix $M$ depends on the gauge field
configuration $U_{\mu}(x)$ and the mass of the fermions $m_\mathrm{f} $. We will
use the notation
\begin{equation}
  \label{eq:LatDiracOp}
  D_\mathrm{f} = D\left[U, m_\mathrm{f} \right]\, ,
\end{equation}
to denote the lattice Dirac operator for one fermion species, the latter being
identified by the index $\mathrm{f}$.  

Although the simulation of Grassmann variables on a computer is
possible, it is computationally very inefficient. Instead, the
previous relation is used to directly define the path integral of
lattice QCD as (cf. 
Eq.~(\ref{eq:grassmann_int}))
\begin{equation}
  \mathcal Z = \int {\rm d} U\,
  \, \left[\prod_{\mathrm{f}=1}^{N_{\rm f}}\det(D_{\rm f})\right]
  e^{-S_G[U]}\,.
\end{equation}
Note that by integrating out the fermions fields exactly, one is
effectively simulating a non-local theory.

\subsubsection{Chiral symmetry and lattice fermions}

The Euclidean action for a single free fermion in the continuum reads
\begin{equation}
  S_F[\psi,\bar\psi] =
  \int {\rm d}^4x\, 
  \bar\psi (\gamma_\mu\partial_\mu  + m)\psi\,.
\end{equation}
A naive attempt to discretize this action leads to the so-called \emph{doubling
problem:} instead of describing a single fermion, the lattice action describes
$2^4$ fermion flavors. In fact this phenomenon is intimately related with chiral
symmetry. In the absence of a mass term the fermion action is invariant under
chiral transformations
\begin{equation}
  \psi \mapsto e^{\imath \theta \gamma_5}\psi\, , \quad
  \bar{\psi} \mapsto \bar{\psi}  e^{\imath \theta \gamma_5}\, .
\end{equation}
This is just a consequence of the kernel of the fermion bilinear being
proportional to $\gamma_\mu$. The Nielsen-Ninomiya
theorem~\cite{Nielsen:1981hk,Karsten:1981gd,Pelissetto:1987ad} shows that any
local lattice hermitian action that preserves translational invariance and
chiral symmetry describes an equal number of positive- and negative-chirality
fermions. In the case of the naive fermion action, the Nielsen-Ninomiya theorem
is satisfied because of the 16 fermions, 8 have positive chirality and the
remaining 8 have negative chirality. It is possible to reduce the number of
\emph{doublers} from 15 to just 1, but in order to describe a single fermion one
has to break one of the hypotheses of the Nielsen-Ninomiya theorem. Giving up
locality leads to serious difficulties for the renormalization of the theory.
Therefore most efforts have focused on four particular approaches.

\begin{description}
\item[Wilson fermions] One of the most popular choices follows
  Wilson's original proposal to break chiral symmetry at finite
  lattice spacing by an irrelevant operator~\cite{Wilson:1975id}. This
  is implemented by adding a dimension 5 term to the Lagrangian that
  is just a suitable discretization of
  \begin{equation}
    \label{eq:wterm}
    \mathcal L_5 = \frac{r}{2} \bar \psi \partial^2 \psi\, .
  \end{equation}
  The addition of this irrelevant operator has nevertheless an
  important impact in the spectrum of the theory by removing all the
  doublers.  
    
  There are two {\em unpleasant}\ effects of this extra term in the
  action. Firstly the fermion mass is no longer protected by chiral
  symmetry in the regularised theory, and therefore acquires an
  additive renormalization. As a consequence the massless theory can
  only be obtained by fine tuning the bare mass in the
  action. Secondly the scaling violations are linear in the lattice
  spacing $\mathcal O(a)$. The massless theory can be
  \emph{non-perturbatively improved} by adding the so called
  Sheikholeslami-Wohlert term~\cite{Sheikholeslami:1985ij}
\begin{equation}
  \label{eq:csw}
  \mathcal L'_5 = c_{\rm SW} \bar \psi F_{\mu\nu}[\gamma_\mu,\gamma_\nu]\psi\,.
  \end{equation}
  If the coefficient $c_{\rm SW}$ is chosen appropriately
  (see~\cite{Luscher:1996sc}), all remaining 
  linear cutoff effects are proportional to the quark masses
  $\mathcal O(am)$.~\footnote{These terms can be further
    eliminated. See~\cite{Sommer:1997xw} for a comprehensive review of the
    improvement programme.}  This last term is usually called \emph{clover}
  term in the lattice jargon, and the discretization is referred as
  Wilson-clover fermion action. 
  
\item[Twisted mass fermions] A close relative of Wilson clover
  fermions are twisted mass fermions. In this case one uses the same
  5-dimensional operator to break chiral symmetry, except that in this
  case the mass term is of the form
  \begin{equation}
    m + \imath \mu \gamma_5\tau_3\,,
  \end{equation}
  where $\tau_3$ is the third Pauli matrix acting in flavor
  space. Twisted mass lattice QCD always describes multiples of two fermion
  flavors with a mass given by a combination of $m$ and
    $\tau$. Note that in the continuum one can always set 
    $\mu=0$ (with the help of a non-anomalous chiral transformation),
    recovering the usual mass term.

  On the other hand, at non-zero values of the lattice spacing, the
  twisted mass term cannot be reduced to the standard Wilson form,
  because of the explicit breaking of chiral symmetry of the Wilson
  term Eq.~(\ref{eq:wterm}). 
  The main advantage of this formulation is that, for a specific
  choice of the mass parameter $m$ (called \emph{maximally twisted})
  all \emph{physical observables} are automatically $\mathcal
  O(a)$-improved~\cite{Frezzotti:2003ni}. In 
  twisted mass lattice QCD there is no need of tuning the $c_{\rm SW}$
  term of Eq.~\eqref{eq:csw}. On the other hand, parity and flavor
  symmetries are broken at
  finite lattice spacing, and, as already mentioned, one can only
  simulate an even number of quarks. 
  The reader might be interested in the review~\cite{Shindler:2007vp}.

\item[Staggered fermions] One can live with some of the doublers and
  in fact use them in one's favor~\cite{Kogut:1974ag,
    Susskind:1976jm}. This is the approach taken by the 
  staggered formulations of QCD: some of the doublers are used to
  represent the 4 spin components of the fermion. The staggered
  fermion formulation therefore reduces the amount of doublers to
  4. The main advantage of this approach is that it preserves an
  exact $U(1)$ symmetry that is enough to guarantee, among other
  things, that scaling violations are proportional to $a^2$. Moreover
  they are computationally cheap.

  The main drawback of this particular fermion formulation is that the 4
  remaining doublers are degenerate in mass, while the 4 lightest quarks in
  nature have very different masses. In order to describe single flavors, the
  staggered fermion formulation uses a \emph{rooting prescription:} the fermion
  determinant that describes the 4 doublers is replaced with its fourth root
  with the hope that this will describe a single
  flavor~\cite{Marinari:1981qf}. This rooting 
  prescription has the unpleasant effect of breaking locality. It has been
  argued that locality is recovered in the continuum.
  If this is actually the case or not has been the subject of many heated
  discussions in the past, although the issue has never been completely
  resolved~\cite{Sharpe:2006re,Bernard:2006vv,Bernard:2007eh,
    Creutz:2007yg}.  
  
\item[Domain Wall fermions] One way to circumvent the Nielsen-Ninomiya
  theorem is to require that the fermion action, at finite lattice
  spacing, is invariant under a set of modified chiral
  transformations,
  \begin{equation}
    \label{eq:ModifiedChiralTransf}
    \psi \mapsto e^{i \alpha \gamma_5 [1 - 2\zeta aD]} \psi\, , 
    \quad \mathrm{and}\ \quad
    \bar\psi \mapsto \bar\psi e^{i \alpha \gamma_5 [1 - 2(1-\zeta) aD]}\, , 
  \end{equation}
  where $\zeta$ is a free parameter~\cite{Luscher:1998pqa}. A lattice
  Dirac operator that is invariant under these transformation
  satisfies
  \begin{equation}
    \label{eq:GWRelation}
    \left\{\gamma_5, D\right\} = 2 a D \gamma_5 D\, .
  \end{equation}
  This relation is known as Ginsparg-Wilson relation, and was derived
  from renormalization group arguments in Ref.~\cite{Ginsparg:1981bj}
  long before the symmetry above was suggested.

  One solution of Eq.~\eqref{eq:GWRelation} is the overlap operator
  obtained in Ref.~\cite{Neuberger:1997fp}:
  \begin{equation}
    \label{eq:NeubOp}
    a D_N = \frac12 \left[
      1 + \gamma_5 \mathrm{sgn}\left(H\right)
      \right]\, ,
  \end{equation}
  where $H = \gamma_5 (D_W - M)$, where $D_W$ is a lattice Dirac
  operator that has the correct naive continuum limit, and $M\sim 1/a$ is a
  mass parameter.

  The operator $D_N$ is a representation in terms of four-dimensional fields of
  five-dimensional domain-wall fermions (DWF)~\cite{Kaplan:1992bt}. In DWF
  formulations, a fermionic zero-mode with definite chirality, localised on the
  boundary of the (semi-infinite) extra dimension, plays the role of a
  four-dimensional chiral fermion, while the other states decouple from the
  low-energy dynamics. As it turns out, the five-dimensional formulations of
  DWFs presented in Refs.~\cite{Shamir:1993zy,Brower:2012vk} has become the
  method of choice for simulating chiral fermions on the lattice. Without
  entering into a detailed explanation of the DWF construction on the lattice,
  we write down explicitly the action for DWF, and discuss some of its features,
  while referring the interested reader to the literature for detailed
  discussions~\cite{Kaplan:2009yg,Kennedy:2006ax}. Denoting the coordinate in
  the fitfth direction by $s \in [0,\infty)$, the DWF action in
  Ref.~\cite{Shamir:1993zy} reads
  \begin{align}
    \label{eq:ShamirDWF}
    S =& \sum_{x,s,\mu} \bar\psi(x,s) \gamma_\mu \partial_\mu \psi(x,s)
    + M \bar\psi(x,s) \psi(x,s) + \frac{r}{2} \sum_{x,s,\mu}
         \bar\psi(x,s) \partial_\mu^2 \psi(x,s)  \, + \nonumber \\
    &+ \sum_{x,s>0} \bar\psi(x,s) \gamma_5\partial_5 \psi(x,s) +
    \frac{r}{2} \sum_{x,s>0} \bar\psi(x,s) \partial_5^2
      \psi(x,s) \, + \nonumber \\
       &+ \frac12 \sum_x \bar\psi(x,0) \gamma_5 \psi(x,1) +
         \frac{r}{2} \sum_x \bar\psi(x,0) \left[
         \psi(x,1) - 2 \psi(x,0)
         \right]\, .
  \end{align}
  The index $\mu$ runs over the first four Euclidean directions, and
  the derivatives $\partial_\mu$ become covariant derivatives when the
  fermions are coupled to a four-dimensional gauge field. In actual
  simulations the fifth dimension has a finite size, while the lattice chiral
  symmetry is only recovered when the size of the fifth dimension goes
  to infinity. In practice a balance must be found between the breaking
  of chirality due to the finite fifth dimension and the growing cost
  of the simulations as the size of this dimension is increased.

\end{description}


\subsection{Masses, correlators and all that}

Quantum Field Theories provide the machinery to evaluate field
correlators, \ie\ the expectation value of functions of the elementary
fields that appear in the definition of the path integral. Specific
physical properties can then be extracted from these correlators by
means of dedicated analyses. It is interesting to discuss a few
explicit examples in order to introduce some of the quantities that
are used later in this review. 

Let us begin with a two-point correlator
\begin{equation}
  \label{eq:TwoPtCorr}
  C_{\Gamma\Gamma'}(t) = \sum_{\mathbf{x},\mathbf{y}}
  \langle O_\Gamma(t,\mathbf{x})\,
   O_{\Gamma'}^\dagger(t,\mathbf{y}) \rangle\, ,
 \end{equation}
 where $O_\Gamma$ is a quark bilinear
 \begin{equation}
   \label{eq:QuarkBilinear}
   O_\Gamma(t,\mathbf{x}) = \bar{q}(t,\mathbf{x}) \Gamma
   q'(t,\mathbf{x})\, . 
 \end{equation}
 The matrix $\Gamma$ determines the spin structure of the bilinear, which
 we assume to be a color singlet. In Eq.~\eqref{eq:QuarkBilinear} we have
 suppressed the flavor indices, for simplicity, we assume the bilinear to
 be a non-singlet with respect to flavor transformations.

 Using Wick's theorem, the correlator can be rewritten in terms of
 quark propagators:
 \begin{equation}
   \label{eq:TwoPtProps}
   C_{\Gamma\Gamma'}(t) = \sum_{\mathbf{x},\mathbf{y}}
   \mathrm{tr}\
   \left[
     \Gamma' S(0,\mathbf{y}; t,\mathbf{x})
     \Gamma S'(t,\mathbf{x}; 0,\mathbf{y})
   \right]\, ,
 \end{equation}
 where $S(x;y)$ and $S'(x,y)$ are the propagators of the quarks $q$
 and $q'$ respectively, computed as the inverse of the Dirac operator
 in the gauge background. The expectation value is computed by
 averaging the trace above over an ensemble of gauge configurations
 generated by Monte Carlo methods.

 In order to extract physical quantities from a two-point function, we
 insert a complete set of hadronic states,
 \begin{equation}
   \label{eq:UnitarityId}
   1 = \sum_n \int \frac{d^3\mathbf{p}}{(2\pi)^3}
   \frac{1}{2E_n} |n,\mathbf{p}\rangle \langle n,\mathbf{p}|\, .
 \end{equation}
A few lines of algebra show that the sum over the spatial coordinates
implements a projection over zero-momentum states, and therefore
\begin{equation}
  \label{eq:TwoPtSpectral}
  C_{\Gamma\Gamma'}(t) =
  \sum_n \frac{1}{2E_n}\,
  \langle 0 | O_\Gamma(0) | n, {\mathbf p}=0\rangle 
  \langle n, {\mathbf p}=0 | O_{\Gamma'}(0)^\dagger | 0\rangle e^{-E_n t}\, .
\end{equation}
The two-point function can be trivially extended to project on eigenstates of the spatial momentum. 

Eq.~\eqref{eq:TwoPtSpectral} shows the main features that allow the
extraction of physical observables from field correlators. In Euclidean
space the time dependence of correlators is a sum of exponentials, whose
decay rates are determined by the energies of the states that have a
non-vanishing matrix element
$\langle 0 | O_\Gamma(0) | n, {\mathbf p}=0\rangle$. For this reason these
operators are often called {\em interpolating operators}. For large time
separations, the correlators are dominated by the lowest energy state, and
the time dependence becomes a simple exponential. The prefactors
multiplying the exponential yield various combinations of matrix elements
of interest.

The so-called effective mass is defined as the time-derivative of the 
two-point function:
\begin{equation}
  \label{eq:EffMassDef}
  M_\mathrm{eff}(t) = -\frac{d}{dt} \log C_{\Gamma\Gamma'}(t) =
  E_0 + c\, e^{-(E_1-E_0)t} + \ldots\, .
\end{equation}
For large times $t$, $M_\mathrm{eff}(t)$ tends to a constant, which is the mass
of lowest-energy state. However for the range of separations that can be
achieved in practice, it is always important to check that the contamination
from excited states is sufficiently small, or otherwise under control, since
this is one of sources of systematic error that affect the computation of
physically interesting observables.

Another interesting example is the computation of the PCAC mass. Specialising
the above interpolating operators to the case of two degenerate light quarks
($u$ and $d$), and following the notation in Ref.~\cite{DelDebbio:2007pz}, we
define
\begin{equation}
  \label{eq:PandA}
  P^{ud}(x) = \bar{u}(x) \gamma_5 d(x)\, , \quad \mathrm{and} \quad
  A^{ud}_\mu(x) = \bar{u}(x) \gamma_\mu \gamma_5 d(x)\, ,
\end{equation}
and the two-point correlators
\begin{align}
  C_{PP}(t) &= \sum_\mathbf{x} \langle P^{ud}(t,\mathbf{x})
              P^{du}(0)\rangle\, ,\\
  C_{AP}(t) &= \sum_\mathbf{x} \langle A^{ud}_0(t,\mathbf{x})
              P^{du}(0)\rangle\, .
\end{align}
Following the arguments above, $C_{PP}(t)$ at large distances is dominated by a
a single pseudoscalar state. Denoting by $M_\mathrm{PS}$ the mass of the
pseudoscalar state, and by $G_\mathrm{PS}$ its vacuum-to-meson matrix element,
we obtain
\begin{align}
   C_{PP}(t) &= - \frac{G_\mathrm{PS}^2}{M_\mathrm{PS}}\,
               e^{-M_\mathrm{PS} t} + \ldots\, .
\end{align}
Interestingly the ratio
\begin{align}
  \label{eq:mPCAC}
  m_\mathrm{\tiny PCAC} = \left(
  \frac12 (\partial_0 + \partial_0^*) C_{AP}(t) + c_A a \partial_0^*
  \partial_0 C_{PP}(t)
  \right) / C_{PP}(t)\, ,
\end{align}
tends to the PCAC mass defined in the continuum theory through the
axial Ward identity. The interest in the PCAC mass is two-fold. For
fermionic formulations that break explicitly chiral symmetry at finite
lattice spacing, like \eg\ the Wilson fermions described above, the
bare parameters in the action need to be tuned to approach the chiral
limit. In particular an additive renormalization of the bare fermion
mass is required. The chiral theory is defined by requiring the PCAC
mass to vanish. After renormalization, the rate of convergence of
$m_\mathrm{\tiny PCAC}$ will be proportional to the lattice spacing
$a$ if the theory is not improved, while it becomes proportional to
$a^2$ when $c_{\rm SW}$ in Eq.~\eqref{eq:csw} and $c_A$ in
Eq.~\eqref{eq:mPCAC} are properly tuned.

Note that the decay constant of the PS state, defined as
\begin{equation}
  \label{eq:DecayConst}
  \langle 0 | A_\mu(0) | \mathrm{PS}\rangle = i Z_AF_\mathrm{PS} p_\mu\, , 
\end{equation}
can be computed directly from fitting\footnote{The renormalization
  constants $Z_A$ and $Z_m$ need also to be computed. 
We point the reader to the reviews~\cite{Sommer:1997xw,Vladikas:2011bp}
for more details on the topic.} $C_{AP}(t)$ and $C_{PP}(t)$,
which yield the matrix elements
$\langle 0 | A_0(0) | \mathrm{PS}\rangle$ and
$\langle 0 | P(0) | \mathrm{PS}\rangle$. Alternatively, the decay
constant can be evaluated from the quantities defined above using
\begin{equation}
  \label{eq:DecayConstTwo}
  F_\mathrm{PS} = Z_m\frac{m_\mathrm{\tiny PCAC}}{M_\mathrm{PS}^2}\,
  G_\mathrm{PS}\, .
\end{equation}
It is interesting to remark that the two definitions differ by lattice
artefacts. While in general this is not necessarily a cause for
concern, it shows that definitions that are equivalent in the
continuum limit do differ at finite lattice spacing. This is
something to keep in mind in choosing observables when aiming for high
precision measurements.

\subsection{Systematic effects in lattice QCD}

Any lattice QCD computation has several sources of systematic uncertainties
that have to be kept under control in order to be able to quote accurate
results.  Since lattice QCD is a first principle definition of QCD, these
sources of systematic uncertainties reflect the current limitations in
computer power, or in our knowledge of efficient algorithms. Since both
computer power and our knowledge of efficient algorithms are constantly
improving, lattice QCD is able to solve now problems that were basically
impossible just a few years ago.

The most basic limitation that computer power sets in any lattice
computation is related to the number of points $(L/a)^3\times (T/a)$
simulated (the lattice volume). The computer cost increases \emph{at
  least} linearly with this lattice volume, and sets a basic
compromise between simulating a large physical volume, and a small
lattice spacing. Computing resources also limit the range of quark
masses that can be simulated, even though it has become common to have
simulations with physical quark masses. In summary we have the
following main sources of systematic error.
\begin{description}
\item[Finite volume corrections: ] QCD quantities determined on a
  large but finite box suffer from finite volume effects. These are
  exponentially suppressed with the smallest mass present in the
  spectrum of the theory (i.e. the pion $M_\pi$)~\cite{Luscher:1985dn}
  \begin{equation}
    \propto e^{-M_\pi L}\,.
  \end{equation}
  Usually $M_\pi L > 4$ is sufficient for finite volume effects to be a
  small sub-percent correction to the quantities determined on a finite
  volume box. $M_\pi L>3$ is the bare minimum to keep these effects under
  control.

  For some hadronic quantities (\eg\ meson decay constants) chiral
  perturbation theory yields an estimate of the size of the finite
  volume corrections, allowing to subtract them from the data in some
  cases. 

  Since one typically uses a short distance observable to determine
  the strong coupling, these determinations are normally affected very
  little by finite volume effects. Nevertheless every determination of
  the strong coupling needs a determination of the scale (see
  section~\ref{sec:scale-sett-latt}), where finite volume corrections
  can be substantial.
  
\item[Continuum extrapolation: ] All determinations on the lattice require
  a continuum extrapolation to reproduce QCD results. Lattice artefacts are
  small only if we can achieve a significant separation between the scales
  at which observables are defined and the lattice cutoff. We will examine
  in detail the process of taking the continuum limit in
  section~\ref{sec:cont-limit-scale}. Here we just mention that this point
  is particularly delicate for the determinations of the strong
  coupling. As we try to use observables computed at high energies in order
  to define the strong coupling, we necessarily need to face the issue of
  larger cutoff effects. Typical current large volume simulations use
  $a\approx 0.04-0.1~{\rm fm}$.
  
\item[Chiral extrapolation: ] Many lattice QCD computations used to be
  performed at nonphysically heavy values of the quark masses. There
  are two reasons for this. First, lattice QCD simulations become more
  expensive at lighter quark masses. The gap in the spectrum of the
  Dirac operator depends on the mass of the lightest quark in the
  simulation. This has the effect of making simulations close to
  physical values of the quark masses computationally very
  expensive. Second, close to physical values for the quark masses
  finite volume effects are larger, and therefore there is an extra
  cost due to the need to simulate in larger physical volumes. 

  For example, simulating physical values for the quark masses
  ($M_\pi \approx 135\, {\rm MeV}$) with sub-percent finite volume
  effects $M_\pi L = 4$ and a very fine lattice spacing
  $a\approx 0.05\, {\rm fm}$ requires a lattice with $L/a\approx 120$
  points in each direction. At the time of writing this report this is
  right at the edge of current capabilities for most choices of lattice
  action. Algorithmic developments have made it possible to simulate 
  directly at the physical point, but these physical regimes of the
  parameters are usually simulated on coarser lattices, making the
  chiral extrapolation a crucial ingredient of any lattice QCD
  computation.
\end{description}

\subsection{The continuum limit and scale setting}
\label{sec:cont-limit-scale}

Any lattice action has $N_{\rm f}+1$ free parameters: the bare quark
masses in lattice units $am_0$, and the bare coupling $g_0$ (usually
the lattice community uses $\beta = 6/g_0^2$ as input parameter for the
simulations). While the role of the bare quark masses is clear (they
directly affect the values of the quark masses), the role of the bare
coupling is less obvious. In fact the bare coupling is tuned in order
to approach the continuum limit. Naively the continuum limit amounts
to take $a\to 0$. But in the lattice action there is nowhere any
reference to the lattice spacing $a$ (or any other parameters with
dimensions), raising the question about how to actually take the
continuum limit.

Since all lattice input parameters are dimensionless, lattice QCD by itself only
gives predictions of dimensionless quantities. For example, the study of the
proton correlator yields the proton mass in lattice units $aM_p$. The key idea
in order to make contact with physical, dimensionful, quantities is to choose
\emph{one} quantity as a reference scale. Every other dimensionful quantity is
computed in units of this reference scale. For example, one can take as
reference scale the proton mass $M_p$. Any other quantity, say for instance the
$\Omega$ baryon mass is measured in units of this reference mass -- in practice
in a lattice simulation we determine the dimensionless ratio $(aM_\Omega)/(aM_p)
= M_\Omega/M_p$.

If we focus on the case of $N_{\rm f}=2+1$ simulations (two degenerate
light quark masses plus the strange quark), a prediction for the value
of $M_{\Omega}$ would conceptually proceed as follows:
\begin{enumerate}
\item Choose a value of the bare coupling $g_0$. Measure the values of the
  masses of the $\pi$ and $K$ mesons and the reference scale in lattice units
  (i.e. $aM_\pi, aM_K, aM_p$). Tune the bare quark masses such that the ratios
  of the pseudo goldstone bosons masses to the reference scale $(aM_\pi)/(aM_p)$
  and $(aM_K)/(aM_p)$ are equal to the physical values $M_{\pi}^{\rm
  exp}/M_p^{\rm exp}$ and $M_k^{\rm exp}/M_p^{\rm exp}$, see \eg\ the values
  reported by the PDG~\cite{Eidelman:2004wy}. This procedure fixes the values of
  the bare quark masses $am_0$ for each choice of $g_0$. The lattice spacing is then $a=(a M_p)/M_p^{\rm exp}$, where the numerator is the output of the numerical simulation, and the denominator is the reference scale.  
\item Repeat the process for several values of $g_0$. The final prediction has
  to be taken as the limit where the reference scale in lattice units is much
  smaller than one, \ie\ $aM_p \ll 1$: 
  \begin{equation}
    \label{eq:cont_limit}
    \frac{M_\Omega}{M_p} =  \lim_{aM_p\to 0} \frac{aM_\Omega}{aM_p}\,.
  \end{equation}
  By using the experimental value of the proton mass $M_p \approx 940$ MeV, this
  last prediction of a dimensionless ratio can be translated in a prediction for
  $M_\Omega$. 
\end{enumerate}

The first step sets up a \emph{line of constant physics} (LCP). It requires
one experimental input per quark mass. Pseudo-goldstone bosons are the
natural candidates, since their mass depends strongly on the values of
quark masses. The value of the reference scale ($M_p$ in the example) is an
extra input, used to convert lattice dimensionless predictions in
dimensionful quantities. Note that although usually one takes the values of
the quantities to fix from experiment, one can follow the same procedure to
set up a line of constant physics at arbitrary non-physical values of the
quark masses (for example to investigate a world with mass degenerate
$u,d,s$ quarks): lattice QCD also allows to make \emph{unambiguous}
predictions of dimensionless quantities for \emph{unphysical} values of the
quark masses. The second step \emph{takes the continuum limit}.  Different
LCP (for example by choosing a different reference scale) would result in
different approaches to \emph{the same universal continuum values}.

These two steps together define a non-perturbative renormalization scheme for
the theory. The bare parameters are tuned in order to reproduce some physical
world (values of the quark masses and reference scale) and quantities are
computed in the limit where the UV cutoff ($1/a$) is much larger than the energy
scales of interest ($M_p, M_\Omega, \dots$). Asymptotic freedom can be used
as a guidance in order to take this continuum limit. At short distances the
theory is weakly coupled, which suggests that a series of decreasing values of
the bare coupling $g_0$ will successively approach the continuum limit.

Even though the procedure sketched above is perfectly correct from a conceptual
point of view, it misses some fine details that are crucial in order to obtain
the precision achieved nowadays. Let us briefly mention them here.
\begin{itemize}
\item As mentioned above lattice QCD simulations become very expensive at the
  bare parameters that correspond to the physical values of the quark masses.
  There are two reasons for this. First is the increasing numerical cost of
  simulating quarks at small values of the bare quark mass. Second, lighter
  values of the quark masses result in lighter values of the $\pi$ meson masses.
  Since $M_\pi L$ is the quantity that dictates the size of finite volume
  effects, simulations at lighter quark masses requires to simulate larger
  physical volumes. In many practical situations the physical point is only
  reached by extrapolating from simulations at heavier quark masses, although
  this situation is changing fast with the increasing computer power and
  improved simulation algorithms.

\item The experimental values of the physical hadronic quantities are affected
  by the electromagnetic interactions. This has implications for the
  determination of the LCP, since the use of the experimental values as input
  for a lattice QCD computation has to be done with care, usually correcting
  them for isospin breaking effects. One has to make sure that the size of the
  electromagnetic effects has a negligible effect in the determination of 
  physical quantities. When this is not the case, a first-principles prediction
  requires to simulate both the strong and the electromagnetic interactions in
  order to make contact with the physical world. 
  We anticipate here that for the case of the the
    determination of $\alpha_s$, isospin breaking effects are 
  not particularly relevant at the current level of precision
  (cf. section~\ref{sec:pres-future-latt}).
\end{itemize}

\subsubsection{Systematics in the continuum extrapolation}

Since short-distance observables are more susceptible to show large
cutoff effects, the continuum extrapolation plays a key role in the
extraction of the strong coupling. Here we want to show in some detail
how to asses the quality of these continuum extrapolations. 

Most lattice QCD simulations choose an improved discretization where the
leading scaling violations that we discussed above are $\mathcal
O(a^2)$. This is achieved either by choosing a fermion formulation where
any $\mathcal O(a)$ violations are forbidden by symmetry arguments (like
domain-wall fermions, twisted mass at maximum twist, or staggered
fermions), or by tuning the parameters of the action (the so-called
Wilson-clover fermions). In this situation any
dimensionless quantity $D$ has an asymptotic expansion of the form
\begin{equation}
  D \simas{a\to 0} D_{\rm cont} + \dots\,,
\end{equation}
where the dots represent the scaling violations, with an asymptotic
expansion with leading term  $\mathcal
O(a^2)$. But the functional form of these scaling violations is in
general very complicated. The corrections include
logarithmic terms of the form (i.e. 
are of the form $a^n\log^k a$). How does
these generic statements affect in practice the extrapolation of
lattice data? It is clear that corrections need to be
  small in order for the precise functional form to have a negligible
  effect on the extrapolation.

\begin{figure}[ht!]
  \centering
  \includegraphics[width=\textwidth]{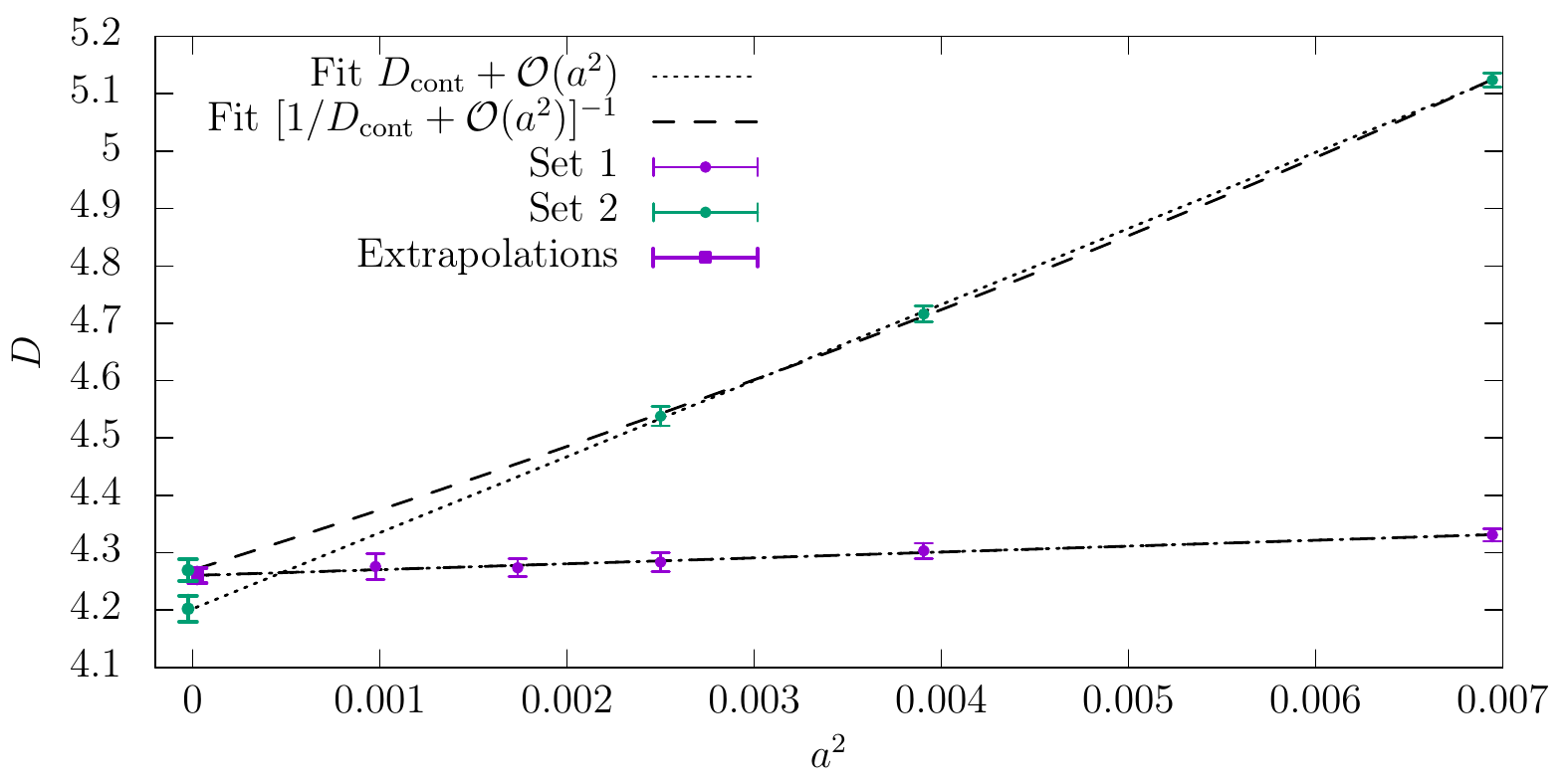} 
  \caption{Continuum extrapolation of a dimensionless quantity. Without
    entering in the details of the definition of the specific quantity in
    the plot, it is worth noting that these data correspond to actual lattice
    simulations, see~\cite{ramos:19lat} for more details.}
  \label{fig:cont}
\end{figure}
In order to be more precise it is best to look at one example.
Figure~\ref{fig:cont} shows the continuum extrapolation of a dimensionless
quantity
using two different discretizations. These are labeled 1 and 2 respectively
in the legend. The extrapolations of both datasets have very different
scaling violations, and this has an impact in the accuracy of the continuum
value. These differences are:
\begin{description}
\item[Range of lattice spacings: ] Dataset 1 includes simulations that
  span a factor 
  \[
    a_{\rm max}/a_{\rm min}\approx 2.5
  \]
  in lattice spacings, while dataset 2 has data that span only a factor 
  \[
    a_{\rm
    max}/a_{\rm min}\approx 1.7\, .  
  \]

\item[Size of the extrapolation: ] Dataset 1 has very small scaling violations:
  the finest lattice spacing is compatible within errors with the continuum
  value. On the other hand dataset 2 shows large scaling violations: the data at
  the finest lattice spacing is about 5 standard deviations away from the
  extrapolated value.
\end{description}
These two points are crucial to understand the assumptions that are behind the
extrapolation of these two datasets. What we need is a way to measure the impact
of the terms \emph{beyond} the leading $\mathcal O(a^2)$ scaling violations. In
order to do so, it is useful to consider not only the extrapolation of $D$, but
also the extrapolation of different functions $f(D)$. In principle this function
seems completely redundant, since 
\begin{equation}
  f(D) \simas{a\to 0} f_{\rm cont} + \dots\,,\qquad (f_{\rm cont} =
  f(D_{\rm cont}))\,,
\end{equation}
one can, in principle, recover \emph{the same} $D_{\rm cont}$ from any choice of
function by using
\begin{equation}
  D_{\rm cont} = f^{-1}(f_{\rm cont})\,.
\end{equation}
However, when the range of lattice spacings available is not very large and/or
the extrapolation is quite sizable, the choice of function can change
substantially the result of the extrapolation. This is illustrated in
figure~\ref{fig:cont}, with two simple choices $f(x) = x$ (i.e. extrapolating
the data itself) and $f(x)=1/x$ (i.e. extrapolating the inverse of the data). In
set 1, these different choices result in practically identical values for the
extrapolated value. On the other hand in set 2 the different choices of $f$
result in extrapolations that differ by several standard deviations. Obviously
the difference between these two choices for the extrapolation are higher order
terms, since
\begin{equation}
  \frac{1}{A + Ba^2} = \frac{1}{A} - \frac{B}{A^2}a^2 +
  \frac{B^2}{A^3} a^4 + \dots \,.
\end{equation}
These results suggest that \emph{the dependence of the extrapolation
  on the choice of function $f$ is equivalent to a dependence on the
  non-leading scaling violations}. It is important to note that
discriminating among different functions cannot always be done by
looking at the quality of the fit. In the quoted example of
figure~\ref{fig:cont} the fits for the two choice of $f$ are equally good.

Another important point worth mentioning is that this kind of analysis
usually only explores the non-leading power corrections (\ie\ $a^4$
in the example).  Logarithmic corrections are much more difficult to
estimate, although one should be very careful with neglecting them (the
reader interested in this topic will enjoy the puzzle described
in Ref.~\cite{Balog:2009np}, and the recent discussion in the context
of lattice QCD~\cite{Husung:2019ytz}). 
These topics are also very nicely discussed in a recent contribution to
the Lattice Field Theory Symposium~\cite{DallaBrida:2020pag}, a
reference that we recommend to the reader will enjoy.

Summarising these results, it is crucial to remember that any extrapolation
is based on some assumptions.  In the case of the continuum extrapolations,
these assumptions depend crucially on three characteristics of the dataset:
the number of lattice spacings simulated, the ratio of the coarser and the
finest lattice spacings, and the size of the extrapolation.

\subsubsection{Scale setting in lattice QCD}
\label{sec:scale-sett-latt}

This process of taking the continuum limit is usually seen by the lattice
community from an equivalent perspective, which comes under the name of
\emph{scale setting}. As discussed above some reference scale~\footnote{For the
sake of the presentation we will assume that the reference quantity has units of
mass.} $M_{\rm ref}$ is used to determine the value of the lattice spacing $a$.
This is simply done by setting up a line of constant physics and declaring that 
\begin{equation}
  \label{eq:a_determination}
  a \equiv \frac{\left(aM_{\rm ref}\right)}{M_{\rm ref}^{\rm exp}}\,,
\end{equation}
where the parentheses emphasise that $\left(aM_{\rm ref}\right)$ is
the quantity that is actually computed in a lattice simulation, and
$M_{\rm ref}^{\rm exp}$ is the experimental value of the reference
scale. Now any other quantity $Q$ with units of mass can be determined
by the expression
\begin{equation}
  Q = \lim_{a\to 0} \frac{(aQ)}{a}\,.
\end{equation}
Obviously this is just a rephrasing of Eq.~\eqref{eq:cont_limit}. Note that
determinations of $a$ are intrinsically entangled with the LCP and the
experimental quantity that is used to set the scale (the particular choice of
$M_{\rm ref}$). Different LCP and/or physical quantities will result in
different values for the lattice spacing. This is however not a problem since
\emph{any} such determinations will give the same predictions for any physical
quantity \emph{after the continuum limit is taken}. This procedure explains the
usual jargon of the field: one experimental input is used to determine the
lattice spacing.

Scale setting is a key ingredient in any lattice determination of the
strong coupling. The $\Lambda$ parameter is a quantity with units of
mass, and therefore the error in the scale translates into an equal
relative error in the determination of $\Lambda$. What is more
important \emph{any unaccounted systematic in the determination of the
scale propagates into an effect of the same relative size to}
$\Lambda$. What quantities are used as reference scales? Ideally one would like
some quantity that has a clean and precise experimental determination,
and that can be determined with high precision and accuracy on the
lattice. The reader might be surprised to read that no such clean
quantity exist. For example the above mentioned proton mass, that
would look as a natural quantity, has a large signal to noise problem
(see appendix~\ref{ap:challenges}), especially close to the physical
point. In practice different quantities are used, with each choice
having pros and cons. Let us briefly review the characteristics of
some of them. 

\bigskip
\noindent
{\em Meson decay constants ($F_\pi, F_K$)}
\medskip

These quantities are clean from a lattice point of view, which explains why they
are a popular choice in the lattice community. They are determined from meson
2-point functions, as discussed above. For example $F_\pi$ is obtained thanks to
the relation
\begin{equation}
  \label{eq:fp_corr}
  \left\langle 0|\bar u\gamma_5\gamma_0 d(0)[\bar u\gamma_5\gamma_0 d(x)]^\dagger |0 \right\rangle \sim
  aM_\pi(aF_\pi)^2e^{-aM_\pi\, x_0} + \dots\, ,
\end{equation}
which is free from the infamous signal-to-noise problem, leading to these
decay constants in lattice units $aF_\pi, aF_K$ being determined with a
precision of a few permille. One problem is that the chiral
corrections on the decay constants (especially $F_\pi$)
  is not small. Recent
determinations are performed at values of the quark masses very close to
its physical values, so in principle this has become a lesser issue in
state of the art calculations.

On the other hand decay constants are not that clean from the
experimental point of view. First, there is a theoretical issue in the
definition of these quantities. They are unambiguous in QCD, but the
electromagnetic interactions render these quantities ill defined --
the quantity in Eq.~\eqref{eq:fp_corr} is not even invariant under a
$U(1)$ gauge transformation. The actual experimental observable is the
photon-inclusive decay rate $\Gamma_{Pl2}$, defined as
\begin{equation}
  \Gamma_{Pl2} = \Gamma(P\rightarrow l\nu) +
  \Gamma(P\rightarrow l\nu\gamma)\, ,
\end{equation}
with $P=K^\pm,\pi^\pm$. Most lattice QCD calculations~\footnote{... albeit not
all!} compute the isospin symmetric quantity $F_P$ which parametrizes the decay
of the pseudoscalar in a world where electromagnetic interactions are switched
off, and isospin is unbroken: 
\begin{equation}
\Gamma(P\rightarrow l\nu)\Big|_{\alpha_{\rm EM} = 0, m_u = m_d} = \frac{G_F
    |V_P|^2 F_{P}^2}{4\pi} \, 
  M_P m_l^2\left[ 1-\frac{m_l^2}{M_P^2}\right]^2\,.
\end{equation}
In order to relate these two quantities we need to estimate the EM
effects. The master formula, as discussed in Ref.~\cite{Marciano1993},
is
\begin{eqnarray}
  \Gamma_{Pl2} = \Gamma(P\rightarrow l\nu)
  &\times&
           \left[1 + \delta_{\rm EW}
           \right] \times \left( 1 + \delta_{\rm EM}^P \right)\, ,
\end{eqnarray}
where the first bracket term is the universal electroweak correction
\begin{equation}
  \delta_{\rm EW} = 
  0.0232 \approx
  \frac{2\alpha_{\rm EM}}{\pi}\log\left(\frac{m_Z}{m_\rho}\right)\,,
\end{equation}
a short distance contributions that affects all semileptonic charged current
amplitudes when expressed in terms of the Fermi constant. Finally the
$\delta_{\rm EM}^P$ piece can be written as
\begin{equation}
  \begin{split}
  \delta_{\rm EM}^P = -\frac{\alpha_{\rm EM}}{\pi}&\left\{ 
           - F\left(\frac{m_l^2}{m_P^2}\right) + 
           \frac{3}{2}\log\left(\frac{m_\rho}{m_P}\right) + c_1^P
         \right.\\
         &+\left. 
           \frac{m_l^2}{m_\rho^2}\left[c_2^P
             \log\left(\frac{m_\rho}{m_l}\right) + 
           c_3^P + c_4^P\left(\frac{m_l}{m_P} \right)\right]
         - \tilde c_2^P \frac{m_P^2}{m_\rho^2}\log\left(\frac{m_\rho^2}{m_l^2}\right)
         \right\}\,.
  \end{split}
\end{equation}
In this expression, the first term is the universal long distance
electromagnetic correction, computed to one loop assuming that the $\pi,K$
are point-like particles, while the terms proportional to
$c_{1,2,3,4}^P, \tilde c_2^P$ parametrize the structure dependent part,
$c_1^P$ being the leading one. A conservative estimate of the
electromagnetic uncertainties consists in taking the full difference
between the point particle approximation -- where all $c_i^P, \tilde c_i^P$
are zero -- and the best phenomenological determination available in the
literature, based on $\chi$PT. This gives an uncertainty $\sim 0.27\%$ for
the case of $F_\pi$ and $\sim 0.21\%$ for $F_K$. These figures set a
conservative limit on the precision that can be quoted for these decay
constants, and therefore a limit on the precision of the scale
determination for any lattice computation that relies on them to set the
scale. Going beyond this precision requires the inclusion of
electromagnetic effects on the lattice and a direct computation of the
decay rate, something that nowadays is challenging from the theoretical
point of view, but an impressive progress has been achieved
recently~\cite{Lubicz:2016xvp, DiCarlo:2019thl}.

For the particular case of $F_K$ one has also to take into account the
strong isospin breaking effects, which vanish at leading order for the case
of $F_\pi$. This is only a practical problem, since the leading corrections
$\propto (m_u-m_d)$ can in principle be determined. Finally, the relation
between the experimentally measured decay rates and the decay constants
involves some CKM matrix elements. In the case of $F_\pi$ the relevant term
is $V_{ud}$, that is very well determined experimentally from super-allowed
$\beta$-decays, but the case of $F_K$ needs a determination of $V_{us}$,
which usually involves lattice input and assumes CKM unitarity relations.

All in all, the pion decay rate ($F_\pi$ in pure QCD) remains an attractive
quantity for scale setting, especially nowadays that we can simulate quark
masses close to their physical values. It can be determined very accurately
on the lattice. The model dependent electromagnetic corrections are below
the 0.3\%, the leading strong isospin breaking effects vanish, and the
necessary CKM matrix element is cleanly determined experimentally.

\bigskip
\noindent
{\em The $\Omega$ mass}
\medskip

The mass of the $\Omega^{-}$ baryon is also a common choice for scale
setting. This particle is stable under the strong interactions and its mass
is known very precisely. Being made of three strange valence quarks, the
dependence on the value of the light quark masses is only induced via loop
effects, which translates in a mild chiral extrapolation if simulations are
performed at constant value of the strange quark mass.

Contrary to the case of the decay constant, what is measured by the
experiment  is directly related with what is  measured on the
lattice. Strong isospin breaking and electromagnetic corrections on
the $\Omega$ mass are also small, with some recent lattice studies
pointing to corrections below the 0.3\%. 

Unfortunately the determination of the
$\Omega^-$ mass on the lattice is challenging. Like
all baryons, the $\Omega^-$ is affected by the signal-to-noise
problem (see appendix~\ref{ap:challenges}). 

The current precision in the scale determined from $M_\Omega$ ranges from
$2\%$ to $0.3\%$. The main difference in the claimed precision stems from
the time range used to extract the mass from the two-point function along
the lines of what we presented above. The most precise determinations use
the values of the correlator at small Euclidean times, where the signal to
noise problem is less severe. On the other hand at this early Euclidean
times there is substantial contamination from excited states, that has to
be disentangled from the real signal of the $\Omega$ mass.

We can summarise the state of affairs by saying that the $\Omega$ mass is a
theoretically clean quantity with a mild dependence on the values of the
light quark masses, and $\le 0.5\%$ isospin breaking corrections, which makes it
an attractive quantity for scale setting. Nevertheless, very precise
determinations of the scale require to control the excited states
contamination in a correlator affected from a strong signal to noise
problem. How to achieve this in practice without assumptions on these
excited states is currently another hot research topic.

\subsubsection{Theory scales}
\label{sec:theory-scales}

An ideal quantity to be used as a reference scale must have some particular
characteristics. First it must have a weak dependence on the quark
masses. Having a simple chiral dependence is crucial for those lattice QCD
simulations that are performed at unphysical values of the quark masses,
and reach the physical point only by extrapolation. Second, the quantity
must be clean from the computational point of view. A quantity that is
complicated to compute on the lattice, as a result of several
extrapolations or some involved fits of the lattice raw data are better
left as predictions, and not as reference scales. Third, the quantity must
have a clear experimental determination, ideally with a weak dependence on
strong isospin breaking effects.

In the previous sections we have seen the typical quantities used for
scale setting (decay constants and $M_\Omega$), and the pros and cons
of each choice. There exists interesting alternatives, with a very
weak chiral dependence and that are straightforward to compute from
the lattice point of view. The drawback is that they are not
quantities that can be accessed by experiments (hence the name
``theory scales''). Nevertheless they are
very useful as intermediate reference scales (i.e. to ``determine the
lattice spacing'' as in Eq.~\eqref{eq:a_determination}).

\bigskip
\noindent
{\em Scales derived from the static potential}
\medskip

One example is the theory scale $r_0$, which is derived from the force
between static quarks, as discussed in
section~\ref{sec:static-potential}. This force $F(r)$ has dimension of
mass squared, and therefore the quantity
\begin{equation}
  r^2 F(r)\,,
\end{equation}
is a dimensionless function of the distance between the static
quarks. This suggests to define a reference scale $r_0$ by the
condition~\cite{Sommer:1993ce}
\begin{equation}
  \label{eq:rZeroDef}
  r^2F(r)\Big|_{r=r_0} = 1.65\,.
\end{equation}
The particular value $1.65$ is chosen so that $r_0\sim 0.5$ fm,
although the shorter distance version $r_1 \sim 0.3$ fm, defined by
$r_1^2F(r_1^2) = 1$ is also commonly used in order to improve the
precision, at the cost of larger cutoff
effects~\cite{Bernard:2000gd}.

The extraction of the scales $r_0,r_1$ from lattice data is not completely free
of challenges. The quantity from which $r_0,r_1$ is extracted has a divergent
signal-to-noise ratio when one approaches the continuum, these challenges have
been overcome by a combination of several techniques
(see~\cite{Sommer:2014mea} for an overview). At the current values of the
simulation parameters a precision $<1\%$ can be achieved in these quantities.

The advantages of scales derived from the static potential are clear. Being
gluonic quantities, their dependence on the value of the quark masses is very
small. The chiral dependence of these quantities is very mild. Even if its
extraction is not completely trivial\footnote{In fact the literature has seen
discrepancies in the values of these scales, although it is not clear that the
fault was the evaluation of the static force, and not the conversion of
$r_0,r_1$ to physical units. See~\cite{Sommer:2014mea}.}, the lattice community
has vast experience determining the static potential.

\bigskip
\noindent
{\em Scales derived from the gradient flow}
\medskip

Recently even better theory scales have been proposed. They are
derived from the gradient flow~\cite{Luscher:2010iy,
  Narayanan:2006rf}, a diffusion like process for the gauge field in a
fictitious time coordinate $t$ called flow time. The gauge field
evolves in flow time according to the equation
\begin{equation}
  \partial_t B_\mu(t,x) = D_\nu G_{\nu\mu}(t,x),\qquad
 B_\mu(0,x) = A_\mu(x), 
 \label{eq:YMflow}
\end{equation}
where $D_\mu = \partial_\mu + [B_\mu,\cdot]$ is the 
covariant derivative with respect to the field $B_\mu$, and 
\begin{equation}
  G_{\mu\nu}= \partial_\mu B_\nu - \partial_\nu B_\mu + [B_\mu,B_\nu],
\end{equation}
is the corresponding field strength tensor. Note that here $x$ denotes the
four-dimensional spacetime coordinates, while the flow time $t$ has units
of length squared. The field $B_\mu(t,x)$ can be seen as a smoothed version
of the original gauge field $A_\mu(x)$ over a length scale
$\sim\sqrt{8t}$. Gauge invariant quantities constructed from the flow field
$B_\mu(t,x)$ do not need renormalization at $t>0$, beyond the usual
renormalization of the bare parameters of the
Lagrangian~\cite{Luscher:2011bx}. For example, the action density
\begin{equation}
  \langle E(t,x) \rangle =
  - \frac{1}{4} \langle {\rm tr}\{G_{\mu\nu}G_{\mu\nu}(t,x)\} \rangle
\end{equation}
is a renormalized observable - \ie\ it has a finite continuum
limit. By dimensional analysis, the quantity
$t^2 \langle E(t,x) \rangle$ is dimensionless, but its value depends
on the scale $\sqrt{8t}$. Similar to what is done for $r_0$, a
convenient scale can be defined by the condition
\begin{equation}
  t^2 \langle E(t,x) \rangle \Big|_{t=t_0} = 0.3\,,
\end{equation}
which results in a hadronic scale ($\sqrt{8t_0} \sim 0.5$ fm). A close
relative to $t_0$ is the $w_0$ scale~\cite{Borsanyi:2012zs}, initially
introduced with the aim of reducing the size of the cutoff effects in
$t_0$ and defined by the condition
\begin{equation}
  t\frac{{\rm d} }{{\rm d}t} t^2 \langle E(t,x) \rangle \Big|_{t=w_0^2} = 0.3\,.
\end{equation}

Flow scales have many advantages. First, as for the
case of $r_0,r_1$ they are
simple gluonic observables, so that their chiral dependence is very
mild. But in contrast with the scales $r_0,r_1$ 
they are given directly by an expectation
value. Their computation in lattice simulations only involves
integrating the flow equation~(\ref{eq:YMflow}), something that can be
done in practice with arbitrary precision. There is no need to look
to the large Euclidean time behavior of a correlator, to perform any
fit or to deal with any signal to noise issue. Moreover flow
observables have a very small variance, making the 
statistical errors in the computation of such quantities very
small. Recent scale determination of $t_0$ in the pure gauge case have
reached a precision $\sim 0.2\%$ in very fine lattice
spacings~\cite{Giusti:2018cmp}. 

Of course the drawback of any of these theory reference scales is that
ultimately they need to be computed in terms of a real experimental
observable if one aims at making a full prediction. But they are
invaluable as intermediate reference scales, especially if one takes
into account that they can be quoted at quite unphysical values of the
quark masses. 

\subsection{Data analysis in lattice QCD}
\label{sec:analysis-lattice-qcd}

As was discussed at the beginning of this section, numerical lattice QCD is
based on the fact that the path integral, after discretization, is an integral
in a large, but finite, dimensional space
\begin{equation}
  \mathcal Z_{\rm latt} = \int\, e^ {-S[U]}\, {\rm d} U \,.
\end{equation}
In typical state of the art current simulations, expectation values
are integrals in $d \approx 10^9$ dimensions. They are computed with a
sub-percent precision using a few ($N \sim \mathcal O(1000)$) gauge fields
($U^{(1)}, \dots, U^{(N)}$ ) that are drawn with probability distribution
\begin{equation}
  {\rm d}\mathcal P(U^{(k)}) \sim \frac{e^{-S[U]}}{\mathcal Z_{\rm latt}}{\rm d}U\,,
\end{equation}
where ${\rm d}U$ is the Haar measure on $SU(3)$.

Drawing representative ensembles in lattice QCD uses the techniques of
Markov chain Monte Carlo. 
For the case of the pure gauge theory very efficient local link update
algorithms exists, but almost every lattice QCD simulation is
performed with some variant of the Hybrid Monte Carlo (HMC)
algorithm~\cite{Duane:1987de}. 
Once a representative ensemble $\{U^{(k)} \}_{k=1}^N$ is available,
estimates of any observable are determined by averaging over the
ensemble
\begin{equation}
  \langle O \rangle = \frac{1}{N}\sum_{k=1}^NO(U^{(k)}) + \mathcal O \left( 1/\sqrt{N} \right)\,.
\end{equation}

A crucial step in any lattice QCD work is the estimate of the
statistical uncertainty. 
\ie\, how much does the estimate
\begin{equation}
  \bar O = \frac{1}{N}\sum_{k=1}^NO(U^{(k)})
\end{equation}
deviates from the exact value of the expectation value $\langle O
\rangle$. An estimate of this uncertainty $\delta \bar O$ is given by
the variance of the mean
\begin{equation}
  \label{eq:varobs}
  (\delta \bar O)^2 = {\rm Var}\left[ \frac{1}{N}\sum_{k=1}^NO(U^{(k)}) \right]\,.
\end{equation}
There are two key points in estimating this uncertainty
\begin{enumerate}
\item The properties of Markov chains ensure that the variance of each
  of the terms in the sum Eq.~(\ref{eq:varobs}) is the same for all
  samples $U^{(k)}$ and in fact given by an expectation value with the
  same probability distribution 
  \begin{equation}
    {\rm Var}\left[ O(U^{(k)}) \right] = \langle (O - \langle O \rangle)^2 \rangle = \sigma^2\,.
  \end{equation}

\item Any Markov chain Monte Carlo algorithm works by producing the
  \emph{next} sample ($U^{(k+1)}$) from the \emph{current} one ($U^{(k)}$). 
  Subsequent measurements of an observable $O(U^{(k)})$
  are \emph{correlated}.  
  Note that this correlations have the unpleasant effect of increasing
  the uncertainties: 
  the error estimate of an observable (Eq.~(\ref{eq:varobs})) is given by 
  \begin{equation}
    \label{eq:varobs2}
    (\delta \bar O)^2 = \frac{\sigma^2}{N}\left[ 1 + \frac{2}{N}\sum_{i>j} \frac{\Gamma(i-j)}{\sigma^2} \right]\,.
  \end{equation}
  where
  \begin{equation}
    \label{eq:acf}
    \Gamma(i-j) = {\rm Cov}\left( O(U^{(i)}), O(U^{(j)}) \right)\,.
  \end{equation}
  The first term in the bracket of Eq.~(\ref{eq:varobs2}) accounts for
  the error estimate if the data were uncorrelated. 
  The second term is due to the correlations. 
  Usually the previous formula for the error of an observable is
  written using the \emph{integrated autocorrelation time}
  \begin{equation}
    \label{eq:tau_rho}
    \tau_{\rm int} = \frac{1}{2} + \frac{1}{N}\sum_{i>j} \rho(i-j)\,,\qquad \left( \rho(t) = \frac{\Gamma(t)}{\sigma} \right)\,,
  \end{equation}
  as
  \begin{equation}
    \label{eq:varobs_tauint}
    (\delta \bar O)^2 = \frac{\sigma}{N}(2\tau_{\rm int})\,.
  \end{equation}
\end{enumerate}

\begin{figure}
  \centering
  \includegraphics[width=\textwidth]{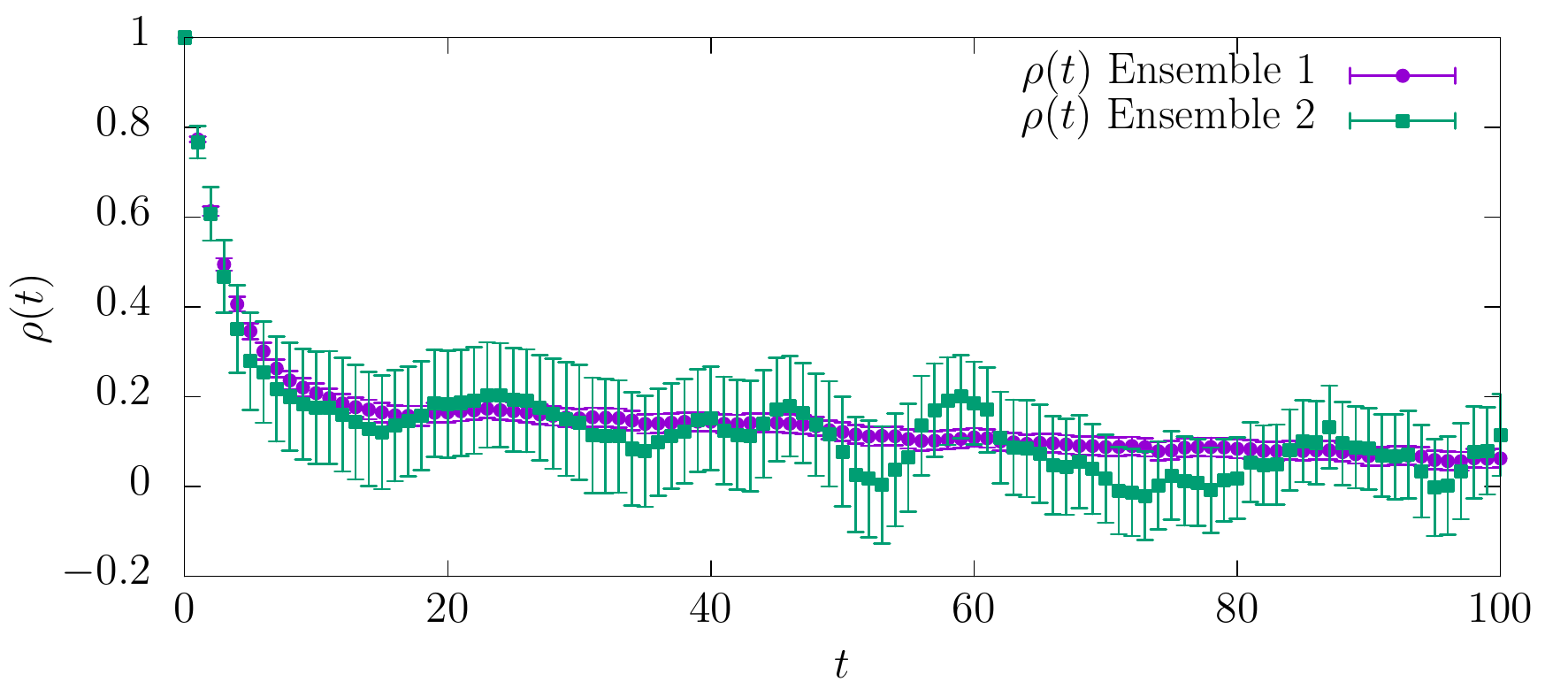} 
  \caption{Two \emph{replica} of a Monte Carlo simulation with
    $\tau_{\rm exp} = 100$. Ensemble 1 has length 20000, while
    ensemble 2 has length 500. A long Monte Carlo run allows to
    determine the normalized autocorrelation function $\rho(t)\equiv
    \Gamma(t)/\sigma^2$ more precisely (see Eq.~(\ref{eq:acf})). This
    results in a more solid estimate of the statistical uncertainties.}
  \label{fig:auto}
\end{figure}

It is clear that   $\tau_{\rm int} = 1/2$ characterizes uncorrelated
data (see Eq.~(\ref{eq:varobs_tauint})). 
The larger the value of $\tau_{\rm int}$ for a particular observable
$O$, the larger the uncertainty \emph{for some ensemble of fixed
  length} $N$. The problem is that $\tau_{\rm int}$ has to be estimated from the data itself. 
The sum in Eq.~(\ref{eq:tau_rho}) has no upper limit $(i-j) \to \infty$,
while in practice it has to be \emph{truncated} at some finite value. 
This means that the estimated value of $\tau_{\rm int}$ will naively
be systematically lower than the correct value (see Fig.~\ref{fig:auto}).

This is not a mere academic observation. 
The properties of Markov chain Monte Carlo ensures that the
autocorrelation function Eq.~(\ref{eq:acf}) is a sum of exponentials
\begin{equation}
  \Gamma (t) = \sum_{k} A_k e^{-t/\tau_k }\,  \xrightarrow[t\to \infty]{}\,
  A' e^{-t/\tau_{\rm exp}}\,.
\end{equation}
At large MC times $t\to\infty$, the dominant contribution to the
autocorrelation function is given by the slowest mode of the Markov operator. 
This is usually called \emph{exponential autocorrelation time}
($\tau_{\rm exp}$). 
Clearly the number of measurements must be large compared with
$\tau_{\rm exp}$ in order to have sensible error estimates and ensure
the ergodicity of the simulation. 
What do we know about $\tau_{\rm exp}$ that is relevant for the
lattice determinations of the strong coupling?
\begin{enumerate}
\item At fixed physical volume, one expects $\tau_{\rm exp}$ to
  increase proportional to $1/a^2$. 
  At fine lattice spacing, one need large statistics in order to
  estimates the uncertainties correctly\footnote{In practice the
    situation might be even more delicate due to a phenomena called
    \emph{topology freezing}. 
  see appendix~\ref{ap:challenges} and the original
  works~\cite{DelDebbio:2004xh, Schaefer:2010hu}.}.

\item The values of $\tau_{\rm exp}$ are not very sensitive to the
  fermion masses. 
  In fact even pure gauge simulations show similar values of
  $\tau_{\rm exp}$ as simulations with dynamical fermions when similar
  algorithms are used.

\item A reasonable estimate of the order of magnitude for $\tau_{\rm
    exp}$ can be made by taking $\tau_{\rm exp} \sim 70$
  MDU\footnote{MDU stands for Molecular Dynamics Unit, and measure
    simulation ``Monte Carlo'' time in HMC simulations. 
  The typical spacing between measurements in realistic simulations is
between 1 and 2 MDU's. A simulation of 1000 MDUs would allow to have
between 500-1000 measurements.
} at $a\approx$ 0.065 fm for \emph{simulations without topology
  freezing} (see~\cite{Bruno:2016plf, Bruno:2014jqa}). 
Values at finer lattice spacing can be estimated using an approximate
$a^2$ scaling. 
This simple estimate of the order of magnitude is already telling us
that estimating statistical uncertainties for $a\sim 0.03$ fm, where
$\tau_{\rm exp}\sim 350$ requires substantial statistics (\ie\,
one would not feel comfortable with less than 4000MDU's).

\end{enumerate}

Let us end this section commenting that the most common analysis
techniques in lattice QCD involves \emph{binning} the data: the Monte
Carlo measurements are averaged in groups of size $N_{\rm bin}$. The
data bins are treated as independent measurements and a naive 
error estimate is performed (usually by resampling). This
alternative analysis method does not improve the determinations of the
statistical uncertainties over the methods that directly determine the
autocorrelation function. 
It is clear that bins of data are less correlated than the data
itself, but it has been  shown that the 
decrease in the correlations is slow $\sim 1/N_{\rm
  bin}$~\cite{Wolff:2003sm}.  
Moreover, in the fairly common case that $N_{\rm bin}$ cannot be taken
much larger than the exponential autocorrelation time $\tau_{\rm
  exp}$, there are no known methods to explicitly include the slow
modes of the Markov operator in the binning analysis. 
On the other hand a direct analysis of the autocorrelation function
allows to include these effects in the error
estimates~\cite{Virotta2012Critical, Schaefer:2010hu} (see also the
summary in~\cite{Ramos:2018vgu}).  

In summary, it is important to point out that in contrast with other
numerical fields where the number of MC samples is very large, lattice
QCD simulations are performed in the uncomfortable situation that the
number of samples is not much larger than the relaxation time of the
Markov operator. In this situation the estimates of statistical
uncertainties can be challenging. 
This observation is especially relevant for the determinations of the
strong coupling, since $\tau_{\rm exp}$ scales like $1/a^2$, and fine
lattice spacing are needed in order to study observables at short
distances.


\section{Decoupling of heavy quark and matching accross thresholds}
\label{sec:deco-heavy-quark}
\label{sec:decoupling}

\subsection{Decoupling Theorem}
\label{sec:hist-remarks-will}

Following a Wilsonian approach to field theory, it is natural to imagine that
the low-energy dynamics is not sensitive to the details of the theory at high
energies -- the high-energy degrees of freedom are {\em integrated out}. More
precisely, when considering observables defined at some low-energy scale $\mu$,
the effects of heavy states of mass $m$ are expected to be encoded in a
redefinition of the couplings, or suppressed by powers of $\mu/m$ in the limit
where $\mu/m\ll 1$. 

\newcommand{\osmass}{\bar{m}_\mathrm{os}}
\newcommand{\oscoup}{\bar{g}_\mathrm{os}}

The original idea of decoupling of massive states dates back to the seminal
paper by Appelquist \& Carazzone~\cite{Appelquist:1974tg}, where decoupling was
proven by analysing the behaviour of Feynman diagrams containing heavy quark
loops in perturbation theory. In order to be able to discuss decoupling, the
relevant quantities need to be precisely defined. It is therefore useful to
summarise, briefly, the results in Ref.~\cite{Appelquist:1974tg}, paying
specific attention to the subtleties due to the regularization and
renormalization of the theory. Since we are ultimately interested in QCD, we can
follow the arguments in~\cite{Appelquist:1974tg} starting from a non-Abelian
gauge theory coupled to a massive fermion as defined in Eq.~\ref{eq:QCDAction}.
Working in the Landau gauge, we choose an {\em on-shell} renormalization scheme,
and denote the renormalized mass and coupling constant $\osmass$ and
$\oscoup$ respectively. The gluon propagator, the three-gluon vertex
and the fermion propagator are given respectively by
\begin{align}
  \label{eq:GluonProp}
  & D_{\mu\nu}^{ab}(k) = \delta^{ab} \,
    \frac{1}{k^2} \left(g_{\mu\nu}-\frac{k_\mu k_\nu}{k^2}\right)\,
    d\left(\frac{k^2}{\mu^2},\frac{\osmass^2}{\mu^2},\oscoup(\mu)\right)\, ,\\
  \label{eq:ThreeGluonVert}
  & i \Gamma^{abc}_{\mu\nu\sigma}(p,q,r) =
    f^{abc}\, \left[
    \left(p-q\right)_\mu g_{\nu\sigma} +
    \left(q-r\right)_\nu g_{\sigma\mu} +
    \left(r-p\right)_\sigma g_{\mu\nu}
    \right] \times \nonumber \\
    &\quad \times 
    G\left(\frac{k^2}{\mu^2},\frac{\osmass^2}{\mu^2},\oscoup(\mu)\right)\, , \\
  \label{eq:FermionProp}
  & S(p) = \frac{1}{p^2-m^2} \left[
    a\left(\frac{p^2}{\mu^2},\frac{\osmass^2}{\mu^2},\oscoup(\mu)\right) 
    \pslash +
    b\left(\frac{p^2}{\mu^2},\frac{\osmass^2}{\mu^2},\oscoup(\mu)\right) m
    \right]\, ,
\end{align}
where in Eq.~\ref{eq:ThreeGluonVert}, the 1PI vertex is evaluated at a
symmetric point $p^2=q^2=r^2=k^2$. The mass counterterm is adjusted by
imposing that the fermion propagator has a pole for $p^2=\osmass^2$, and the
remaining counterterms are defined by providing renormalization
conditions at the scale $\mu$, {\em e.g.}
\begin{align}
  \label{eq:AFieldRenorm}
  &d\left(-1,\frac{\osmass^2}{\mu^2},\oscoup(\mu)\right) = 1\, , \\
  \label{eq:gmuRenorm}
  &G\left(-1,\frac{\osmass^2}{\mu^2},\oscoup(\mu)\right) = 1\, , \\
  \label{eq:PsiFieldRenorm}
  &a\left(-1,\frac{\osmass^2}{\mu^2},\oscoup(\mu)\right) = 1\, .
\end{align}
The running coupling constant at a generic scale $k^2$ is defined as
\begin{align}
  \label{eq:AppRunningCoupling}
  &\oscoup\left(\frac{k^2}{\mu^2},\frac{m^2}{\mu^2},\oscoup(\mu)\right) =
    \oscoup(\mu) \, 
    G\left(\frac{k^2}{\mu^2},\frac{m^2}{\mu^2},\oscoup(\mu)\right)
    \,
    d\left(\frac{k^2}{\mu^2},\frac{m^2}{\mu^2},\oscoup(\mu)\right)^{3/2}\, , 
\end{align}
and its dependence on the scale is described by the beta function:
\begin{align}
  \label{eq:FullBetaFunc}
  k^2 \frac{d}{dk^2} \oscoup = 
  \beta\left(\frac{\osmass^2}{-k^2},\oscoup\right)\, .
\end{align}
The specific value of the coupling at the scale $\mu$ that was used in the
renormalization process serves as the initial condition for integrating the beta
function. Note that, as pointed out in the original derivation in
Ref.~\cite{Appelquist:1974tg}, the beta function depends on the renormalized
mass $\osmass$ through the ratio $\osmass^2/k^2$ -- this is a direct consequence
of working in an on-shell scheme, where the mass of the particle enters
explicitly in the renormalization conditions and therefore in the running of the
coupling constant.

In this specific scheme, Appelquist \& Carazzone show that diagrams containing
heavy propagators are suppressed by powers of $k/\osmass$ or $\mu/\osmass$,
where $k$ is the typical scale of the external momenta of the diagram. The
second, important, result is that, in the decoupling limit, the beta function of
the theory in Eq.~\ref{eq:FullBetaFunc} reduces to the beta function of the
theory where the heavy particle has been integrated out, {\em i.e.} in this
particular case the beta function of pure Yang-Mills theory; the effects of the
heavy states simply show up as power corrections that interpolate between the
two theories.

\subsection{Matching Theories}
\label{sec:matching-theories}

It was noted in Refs.~\cite{Binetruy:1979hc,Binetruy:1980xn} that the decoupling
theorem does not apply in minimal subtraction (MS), since all loops contribute
to the beta function independently of the mass of the state. The same problem
exists for generic mass-independent schemes~\cite{Weinberg:1951ss}. The solution
to this problem is found by matching at low energies the theory with heavy
particles to an effective theory containing only the light degrees of freedom,
{\em i.e.} by tuning the couplings of the effective theory so that it reproduces
the field correlators of the full theory at low energies up to corrections that
are suppressed by the ratio of the scales that characterise the light and heavy
degrees of freedom respectively. Matching gauge theories across thresholds is
first discussed in Ref.~\cite{Weinberg:1980wa}, then analysed in detail in
Refs.~\cite{Wetzel:1981qg,Bernreuther:1981sg}. It has become the method of
choice to define the coupling constant at energies that span a wide range and
hence cross several mass thresholds.

Once again we find it useful to give a pedagogical review of the main
steps in the procedure. We consider a theory with $n_f$ light fermions
and only 1 heavy fermion~\footnote{The argument can be readily
  extended to the case where more than one particle is integrated
  out.}, whose dynamics is specified by a lagrangian $\mathcal{L}$ and
a set of bare couplings and fields:
\begin{align}
  \label{eq:EqandDfields}
  & g, m, m_h, \xi, \psi, A_\mu, c\, . 
\end{align}
Using a familiar notation for QCD, $g$ is the gauge coupling, $m$ the
mass of the light fermions, $m_h$ the mass of the heavy fermion, and
$\xi$ is the gauge parameter. The fields $\psi, A_\mu, c$ describe the
fermions, the gauge field and the ghosts respectively.

The effective low-energy theory will be a theory defined by an effective
lagrangian $\mathcal{L}'$ which only involves the light degrees of freedom, and
a set of rescaled couplings and fields. Following Ref.~\cite{Chetyrkin:1997un},
the bare couplings and fields of the effective theory are denoted by primed
letters, and are connected to the bare couplings and fields of the full theory
through the so-called {\em decoupling constants} $\zeta_i$:
\begin{align}
  \label{eq:DecConst}
  &g' = \zeta_g g\, , \quad m' = \zeta_m m\, , \quad \xi'-1 =
  \zeta_3\left(\xi-1\right)\, , \\
  &\psi' = \sqrt{\zeta_2} \psi\, , \quad A'_\mu = \sqrt{\zeta_3}
    A_\mu\, \quad c' = \sqrt{\tilde\zeta_3} c\, .
\end{align}
It is easy to argue on symmetry grounds that $\mathcal{L}'$ must have
the same form as $\mathcal{L}$, but contain only the light degrees of
freedom:
\begin{align}
  \label{eq:MatchLPrime}
  \mathcal{L}'\left(g,m,\xi,\psi,A_\mu,c; \zeta_i\right)
  = \mathcal{L}\left(g',m',\xi',A'_\mu, c'\right)\, .
\end{align}
Higher-dimensional operators can appear in $\mathcal{L}'$, but are
suppressed by inverse powers of the heavy mass.  The decoupling
constants $\zeta_i$ are determined by computing field correlators in
both theories, and matching them up to power contributions.

The matching procedure yields the relation between the renormalised coupling $\bar\alpha$ in
the two theories:
\begin{align}
  \label{eq:AlphaMatching}
  &\bar\alpha'(\mu) = \left(\frac{Z_g}{Z'_g} \zeta_g\right)^2
    \bar\alpha(\mu) = \zeta_{Rg}^2 \bar\alpha(\mu)\, ,
\end{align}
where $Z_g$ and $Z'_g$ are respectively the renormalization constants for the
coupling in the full and in the effective theory. The decoupling constant
$\bar\zeta_{g}$ has a perturbative expansion
\begin{align}
  \label{eq:ZetaGExp}
  \bar\zeta_{Rg} = 1 + \sum_{\ell=1}^\infty \bar\alpha(\mu)^\ell C_\ell(x)\, ,
\end{align}
where the coefficients $C_\ell$ are functions of the logarithm of the
ratio of scales $x=\log\left(\mu^2/\bar{m}_h(\mu)^2\right)$, and
$\bar{m}_h(\mu)$ is the mass of the heavy fermion in \msbar.

It is interesting to recall here that the functional dependence of the
coefficients $C_\ell$ on $x$ is dictated by the renormalization group equations,
as discussed {\em eg.} in Ref.~\cite{Bernreuther:1981sg}. Keeping in mind that
\begin{align}
  \mu^2 \frac{d}{d\mu^2} x = 1 + \gamma = 1 + \sum_{k=1}^\infty
  \gamma_k \bar\alpha(\mu)^k\, ,
\end{align}
where $\gamma$ is the mass anomalous dimension, we can take the derivative of
Eq.~\ref{eq:AlphaMatching} with respect to the logarithm of $\mu^2$, and solve
the resulting equation order by order in $\bar\alpha(\mu)$. This procedure
yields a set of differential equations that the functions $C_\ell$ must satisfy,
{\em viz.}
\begin{align}
  \label{eq:DiffEqCone}
  &\frac{d}{dx}C_1(x) = \beta_0'-\beta_0\, ,\\
  \label{eq:DiffEqCTwo}
  &\frac{d}{dx}C_2(x) = 2\left(\beta_0'-\beta_0\right) C_1 +
    \beta_1'-\beta_1 -\gamma_0 \frac{d}{dx}C_1\, , \\
  &\dots 
\end{align}
The structure of these equations implies that $C_\ell$ is a polynomial
of degree $\ell$. The coefficients of these polynomials are functions
of the coefficients $\beta_k$, $\beta_k'$ and $\gamma_k$. On top of
that, for each differential equation, an integration constant
$C_{\ell,0}$ appears, which is determined by matching a vertex
function at $\ell$ loops. The simplest example is the integration of
Eq.~\ref{eq:DiffEqCone}, which yields
\begin{align}
  \label{eq:DiffEqOneIntegrate}
  C_1(x) = \left(\beta_0'-\beta_0\right) x + C_{1,0}\, .
\end{align}
As discussed above, $C_1$ is a linear function of $x$, the slope of the function
is given by the difference of the first coefficients of the beta functions in
the two theories, $\beta_0$ and $\beta_0'$, while the integration constant
$C_{1,0}$ needs to be computed from a one-loop matching. Recent matching
calculations up to four loops are
available~\cite{Schroder:2005hy,Chetyrkin:2005ia}.

The final result for the matching of couplings across mass thresholds
is reported in the PDG~\cite{Patrignani:2016xqp}; using the PDG
notation and for the particular case $\mu=m_{\rm h}^\star $, we have
\begin{equation}
  \label{eq:MassThreshMatching}
  \as^{(n_f+1)}(\mu^2) = \as^{(n_f)}(\mu^2) 
  \left(
    1 + \sum_{\ell=2}^\infty c_\ell \left[\as^{(n_f)}(\mu^2)\right]^\ell
  \right)\, ,
\end{equation}
here $\as^{(n_f)}$ is the coupling in the \msbar\ scheme with $n_f$ massless
fermions, and $m_{\rm h}^\star$ is the mass of the heavy fermion, also defined in the
\msbar\ scheme, at the energy scale given by the mass itself. Note that for this
particular choice of scales the $\mathcal O(\alpha^2)$ term in the relation
between couplings vanish. The coefficients $c_{n}$ for $n\leq 4$ and the
physically relevant cases $3\to 4$ and $4\to 5$ are available in
table~\ref{tab:cdec}. We also quote the effect that the last term of the series
$c_4\left[\as^{(n_f)}(\mu^2)\right]^4$ has in the determination of the strong
coupling. As the reader can see, the truncation of the perturbative series in
the decoupling relations has a completely negligible effect on the extraction of
$\alpha_s$ ($\lesssim 0.2\%$ for the case of the charm quark). Note that this
analysis does not exclude potentially large non-perturbative corrections in the
matching between theories (see next section).

\begin{table}
  \centering
  \begin{tabular}{lllllc}
    \toprule
    & $m_{\rm h}$ [GeV]& $c_2\times 10^2$ &$c_3\times 10^2$ &$c_4\times 10^2$ & $\delta\alpha_{\overline{\rm MS} }(M_Z)$ [\%] \\
    \midrule
    $3\to 4$ & $m_{\rm c}\approx 1.3$& -1.547963 &  -2.315990 &  -1.922535 & 0.18\% \\
    $4\to 5$ &$m_{\rm b}\approx 4.2$ & -1.547963 &  -2.042976 & -0.7278004  & 0.02\% \\
    \bottomrule
  \end{tabular}
  \caption{Coefficients for the decoupling of the charm quark ($3\to
    4$) and the bottom quark ($4\to 5$) for the case $\mu=m_{\rm h}$
    (see
    Eq.~(\ref{eq:MassThreshMatching}))~\cite{Chetyrkin:2005ia,Schroder:2005hy,Herren:2017osy}.  
  The last column quotes the effect of the last known term in the
  series in the value of the strong coupling at the scale $M_Z$. }
  \label{tab:cdec}
\end{table}

\subsection{Nonperturbative decoupling}
\label{sec:nonp-deco}

As noted before in Sect.~\ref{eq:QCDScale}, the running of the
coupling constant with the energy scale is determined by the knowledge
of the beta function and the $\Lambda$-parameter of a given theory. Hence
the matching of the coupling between theories described in the
sections above can be reformulated in terms of the matching of the
$\Lambda$-parameters, which leads naturally to a framework where
decoupling can be discussed beyond perturbation theory. We present
here a brief summary of the ideas that were originally developed in
Ref.~\cite{Bruno:2014ufa}.

Following the notation introduced above, we denote quantities in the
theory with heavy particles with unprimed variables, while primed
variables always refer to the effective theory that includes the light
degrees of freedom only. The relation between the $\Lambda$-parameters
is fixed by requiring that low-energy quantities are matched up to
power corrections; without loss of generality we can write
\begin{align}
  \Lambda' = f\left(\Lambda, M\right)\, ,
\end{align}
where $M$ is the RGI mass of the heavy particle in the full
theory. Furthermore, we can argue on dimensional grounds that
\begin{align}
  \label{eq:LamRatioMassThres}
  \Lambda'/\Lambda = P\left(M/\Lambda\right)\, .
\end{align}
The dependence of $P$ on the mass $M$ is encoded in
\begin{align}
  \label{eq:DefEtaFun}
  \eta(M) = \left. \frac{1}{P} M \frac{\partial}{\partial M} P
  \right|_\Lambda\, ,
\end{align}
which can be reliably evaluated in perturbation theory only for large
values of $M$
\begin{align}
  \label{eq:EtaPertTh}
  \eta(M) \simas{M\to\infty} \eta_0 + \eta_1 \bar{g}^2(M) + \ldots \, .
\end{align}
Introducing the variable $\tau=\log\left(M/\Lambda\right)$, and
integrating Eq.~\ref{eq:EtaPertTh}, yields
\begin{align}
  P\left(M/\Lambda\right) =
  \frac{1}{k}\, \exp\left(\eta_0\tau\right)\,
  \tau^{\eta_1/(2b_0)} \times
  \left(1 + O\left(\frac{\log\tau}{\tau}\right)\right)\, ,
\end{align}
where $k$ is a constant that can be computed given the conventions for
$\Lambda$ and $M$.

A nonperturbative matching condition requires that some hadronic scale
remains the same in the two theories, {\em ie.}
\begin{align}
  \label{eq:NPMatch}
  m_\mathrm{had}'\left(\Lambda'\right) =
  m_\mathrm{had}\left(\Lambda, M\right) + O\left(\frac{\Lambda^2}{M^2}\right)\, .
\end{align}
In the equation above $\Lambda$ and $M$ are dimensionful quantities that {\em
define} the coupling and the masses in the full theory, and are given. The
matching condition then determines $\Lambda'$. As a consequence we find:
\begin{align}
  \frac{m_\mathrm{had}\left(\Lambda, M\right)}{\Lambda} =
  \frac{m_\mathrm{had}'\left(\Lambda'\right)}{\Lambda'}
  \frac{\Lambda'}{\Lambda}
  + O\left(\frac{\Lambda^2}{M^2}\right)\, ,
\end{align}
and hence
\begin{align}
  \label{eq:MassRatio}
  \frac{m_\mathrm{had}\left(\Lambda,
  M\right)}{m_\mathrm{had}\left(\Lambda, 0\right)} =
  Q_\mathrm{had} \,  P\left(M/\Lambda\right)
  + O\left(\frac{\Lambda^2}{M^2}\right)\, ,
\end{align}
where
\begin{align}
  Q_\mathrm{had} = \frac{m_\mathrm{had}'\left(\Lambda'\right)}{\Lambda'}
  \frac{\Lambda}{m_\mathrm{had}\left(\Lambda, 0\right)} \, .
\end{align}
The left-hand side of Eq.~\ref{eq:MassRatio} can be measured by
performing MC simulations of the full theory with different values of
the mass $M$, while $Q_\mathrm{had}$ requires an extra simulation in
the effective theory. 

References~\cite{Bruno:2014ufa, Athenodorou:2018wpk} study the effect
of heavy quarks along the lines described above. 
Making a long story short, they estimate the non-perturbative effects
in the matching between theories with the conclusions that for the
most important case of the charm quark, they affect the extraction of
$\alpha_s$ below the $0.4\%$ level. The interested reader is invited to
consult the original works.


\section{Convenient observables for a coupling definition}

\label{sec:conv-observ-coupl}

In principle there are very many possibilities to define the strong
coupling on the lattice. The only thing that is needed is a
\emph{dimensionless finite observable that depends on a single scale}. Our
discussion on scale setting (in particular the discussion about theory
scales in section~\ref{sec:theory-scales}) has already introduced some
observables with these characteristics. In this section we will
introduce the observables that are more commonly used to compute the
strong coupling from the lattice, with special emphasis on the
systematic effects associated with the truncation of the perturbative
series.

We recall that the general method to extract the value of the strong
coupling consists in comparing a perturbative theoretical prediction
with lattice measurements (i.e. eq.~(\ref{eq:Pseries})). It is
convenient to introduce non-perturbative definitions of the coupling
constant, where \emph{the value of the observable} is used for the
definition of the strong coupling.  We take a dimensionless observable
with perturbative expansion
\begin{equation}
  O(\mu) \simas{\alpha_{\overline{\rm MS} }\to 0} \sum_{k=1}
  c_k(\mu/\mu')\alpha_{\overline{\rm MS} }(\mu')\,,
\end{equation}
and define
\begin{equation}
  \label{eq:O2gsq}
  \alpha_O(\mu) = \frac{O(\mu)}{c_1(1)}\,.
\end{equation}
This equation defines a particular renormalization
scheme, which we refer to as $O$-scheme. When computed on the lattice,
$O(\mu)$ can be determined at all energy scales, thus providing a
non-perturbative renormalization scheme.

\begin{figure}[t!]
  \centering
  \includegraphics[width=\textwidth]{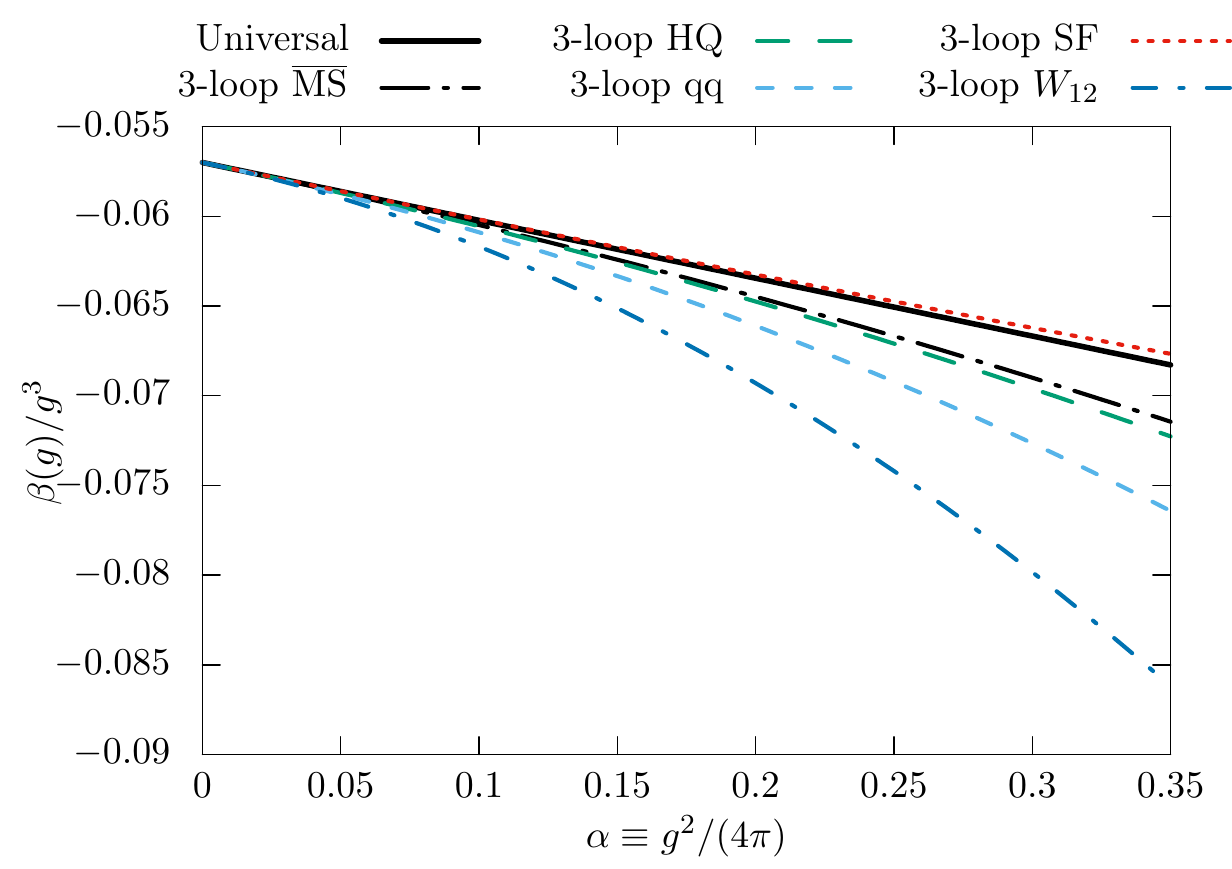} 
  \caption{$\beta$-function in three flavor QCD in different schemes (see
    Eq.~\ref{eq:beta_O}). The 2-loop universal behavior is compared with the
    3-loop $\beta$-function in different schemes:
    $\alpha_{\rm HQ, 4}$ (Eq.~(\ref{eq:alphahq})),
    $\alpha_{\rm qq}$ (Eq.~(\ref{eq:qqpt})),
    $\alpha_{\rm SF}$ (Eq.~(\ref{eq:alphasf})) and
    $\alpha_{W_{12}}$ (Eq.~(\ref{eq:alphaW})). With the exception of the 
    scheme based on Wilson loops, where the 3-loop coefficient is significantly
    larger than in the $\overline{\rm MS} $ scheme, all schemes seem well
    behaved from a perturbative point of view. 
  Note that for the case of the static potential, one more order is
  known perturbatively. } 
  \label{fig:beta}
\end{figure}

Note that the RG equation for the coupling in the $O$-scheme defines a
$\beta$-function, that can also be computed beyond perturbation theory 
\begin{equation}
  \label{eq:beta_O}
  \mu \frac{{\rm d} \bar g_{O}(\mu)}{{\rm d} \mu} =
  \beta_O(\bar g_O)\simas{g_O\to 0} -g_O^3(b_0 + b_1g_O^2) + \dots\,. 
\end{equation}
This $\beta$-function has all the properties that one would expect,
starting from a perturbative expansion with the usual first two universal
terms. Examining the convergence of the perturbative series of this
beta function provides a piece of
information on the size of the truncation effects in the determination
of the strong coupling. If these are small, the beta function
$\beta_O(x)$ has to be well approximated by its perturbative
expansion. Note also that our fundamental relation Eq.~\ref{eq:lam}
allows the determination of the $\Lambda$-parameter in this
scheme ($\Lambda_O$). Figure~\ref{fig:beta} shows the $\beta$-function
for the typical choices of observables that we are going to explore in
detail later. 

\begin{table}[t]
	\centering
	\begin{tabular}{llllc}
		\toprule
		Observable& $l$ & 
		$\alpha_{\overline{\rm MS}}(\mu_{\rm PT}) $ &
		$\mu_{\rm PT}$ [GeV] & Power corrections\\
		\midrule
		QCD vertices        & 3 & $0.20-0.30$         & $2-6$                & ${\sim 1/\mu^2,1/\mu^6}^{+}$  \\
		Static Potential    & 3 & $0.19-0.36^{\star}$ & $1.5-8^{\star}$      & -$^{\star}$                   \\
		HQ correlators      & 2 & $ 0.20-0.36^{\dagger}$   & $\bar m_c^{\dagger} - 4\bar m_c^{\dagger}$ & -                             \\
		Wilson loops        & 2 & $ 0.22-0.40$        &  $1/a = 1.1-4.4$ & ${\sim 1/\mu^2}^{\ddagger}$                             \\
		Vacuum polarization & 4 & $ 0.22-0.31$        & $1.6-4$              & $\sim 1/p^{2k}\, (k=1,...,4)$ \\
		Finite size scaling & 2 & $ 0.11-0.23$        & $4-140$              & -                             \\
		\bottomrule
	\end{tabular}
	\caption{We summarize some key numbers in different techniqes
          used to determine 
          the strong coupling. Column labeled $l$ gives the number of loops
          known in the perturbative expansion of the $\beta$-function. The
          second and third columns shows the range of couplings that
          have been explored in the literature, 
          and the corresponding energy scale (note however that \emph{not all
            individual works explore the same energy scales}). Finally we show if
          power corrections are needed to describe the data, and how many
          terms are used. \\
          $^{+}$ Most of the determinations find it neccesary to add these
          power corrections in order to describe the lattice
          data. Nevertheless, see~\cite{Sternbeck:2012qs, Sternbeck:2010xu}.\\
          $^{\star}$ Most of the works expore this range of scales without
          any need for power corrections.
          In~\cite{Takaura:2018lpw} power corrections are added
          to the analysis, allowing to use data down to 0.6 GeV.\\
          $^{\dagger}$ Most determinations are performed at the charm
          mass, but some works~\cite{Petreczky:2019ozv, McNeile:2010ji}
          explore larger masses. See
          section~\ref{sec:heavy-quark-corr} for more
          details. \\
          $^{\ddagger}$ In this approach cutoff effects manifest
          as power corrections. 
          See~\ref{sec:observ-defin-at} for more details.
	}
\end{table}

We have explained in detail (see
sections~\ref{sec:chall-determ-alph},~\ref{sec:syst-extr-alph}) the particular
systematic effects that affect the lattice determinations of the strong
coupling. Let us now focus on explaining how the different methods to extract
the strong coupling face these challenges. In particular we will pay special
attention to the following issues. 
\begin{description}
\item[Perturbative truncation effects:] We already insisted on the fact that the
  determination of the strong coupling has to be understood as an extrapolation
  (cfr. section~\ref{sec:chall-determ-alph} and~\ref{sec:phys-defin-strong}). An
  important criterion is the range of scales $\mu_{\rm PT}$ at which one matches
  with perturbation theory. More precisely, we are interested in the values of
  the coupling $\alpha_{\overline{\rm MS} }^{(l-1)}(\mu_{\rm PT})$ at such
  scales, where $l$ is the number of analytically known loops in the asymptotic
  expansion of the $\beta$ function, since they parametrize the leading
  perturbative truncation effects. 

\item[Non perturbative effects:] Power corrections are understood to
  arise from the Operator Product Expansion (OPE). The short distance
  observable used to extract the strong coupling $O(\mu)$, typically
  has an OPE that schematically can be written as
  \begin{equation}
    \label{eq:ope}
    O(\mu) = \sum_{k=0} d_k(\mu)\frac{O^{(k)}}{\mu^k}\,, 
  \end{equation}
  where $O^{(k)}$ are operators of dimension $k$. Note that the
  perturbative expansion of the observable Eq.~(\ref{eq:O2gsq})
  neglects all power corrections (terms with $k>0$).
  One problem with these non-perturbative corrections in lattice
  determinations of the strong coupling is that they
  are not computed\footnote{The higher dimensional operators
    $O^{(k)}$ in Eq.~(\ref{eq:ope}) (i.e. ``condensates'') are
    composite operators that typically mix other lower dimensional
    operators. Computing them (i.e. their non-perturbative
    renormalization) is probably more complicated than the
    determination of the strong coupling.}, but estimated from the
  same data for $O(\mu)$ by fitting them. Distinguishing the
  perturbative running from the non-perturbative corrections given the
  limited range of scales that are available in many extractions is
  always challenging. 

  Ideally one would like to work with observables and energy scales
  where the power corrections are negligible. In practice this is not
  always the case.
\end{description}

\subsection{The ghost-ghost-gluon vertex}
\label{sec:quark-gluon-vertex}

The definition of the strong coupling using QCD vertices is the one that is more
similar to the type of computation that is usually done in the context of
perturbation theory. There are several issues in the extraction of these QCD
vertices from lattice QCD simulations. In practice the coupling is extracted
from the gluon/ghost two-point functions, but these are not gauge invariant, and
therefore this scheme can only be implemented by fixing the gauge of the lattice
configurations. The problems of gauge fixing \emph{beyond} perturbation theory
({\em i.e.}\ Gribov ambiguities~\cite{Gribov:1977wm}), have been discussed at length in the
literature (see for example~\cite{Vandersickel:2012tz}), and we can
add very little to the discussion, except 
pointing out that the issue of Gribov ambiguities is also present in other
lattice QCD calculations. One of the most widely used methods of
non-perturbative renormalization (i.e. ``RI/MOM'' schemes,
see~\cite{Martinelli:1994ty}) also 
requires to fix the configurations (typically the 
Landau gauge is used). It is believed that at the relatively high
energy scales where $\alpha_s$ is extracted, this is not a serious issue. 

In principle there are several options to extract the strong coupling, since it
can be defined from different three- and four-point functions.  All the methods
define the coupling by requiring that some vertex is equal to its tree-level
value. 
 The momenta entering in the vertex are part of the definition of the
scheme. The most popular choices set one of the momenta to zero, and
are usually labeled $\widetilde{\rm MOM}$ schemes. 

Nowadays the most common coupling definition uses the un-renormalized
ghost-ghost-gluon vertex, which can be constructed from the gluon and
ghost two-point functions. The non-perturbative coupling definition
reads 
\begin{equation}
  \label{eq:alpha_Taylor}
  \alpha_T(\mu) = \lim_{a\to 0} F_{\rm lat}(p,a)
  D_{\rm lat}(p,a)\frac{g_0^2}{4\pi} \Big|_{\mu = p}\,.
\end{equation}
This is usually referred to as the Taylor scheme, as indicated by the suffix.
Here $D_{\rm latt}(p,a)$ and $F_{\rm latt}(p,a)$ are the ``dressing functions''
of the lattice gluon and ghost two-point functions~\footnote{In the continuum
the relation of the propagators $F^{ab}(p), D^{ab}_{\mu\nu}(p)$ with the
dressing functions $F, D$ is 
\begin{eqnarray}
	F^{ab}(p) &=& -\delta^{ab} \frac{F(p)}{p^2}
	\,,\\
	D^{ab}_{\mu\nu}(p) &=& -\delta^{ab} \left( \delta_{\mu\nu} -
	\frac{p_\mu p_\nu}{p^2} \right)
	D(p)\,.
\end{eqnarray}
On the lattice the relation is similar, but the momentum $p$ is substituted by
a function of $p$ and the lattice spacing $a$ that
only reduces to $p$ in the continuum limit and 
depends on the particular choice of discretization.} and $g_0^2$ is the bare
coupling used for the simulation. The main advantage of this scheme is that one
does not need to determine any three- or four-point function, since the coupling
is directly defined from the computation of the propagators. We are going to
focus the discussion on this particular choice, although most of what we are
going to state is also valid for other, similarly defined schemes. The
perturbative expansion for $\alpha_T(\mu)$,
\begin{equation}
\label{eq:alphaTpt}
  \alpha_T(\mu) \simas{\mu\to\infty} \alpha_{\overline{\rm MS} }(\mu) +
  t_1\alpha_{\overline{\rm MS} }^2(\mu) + t_2\alpha_{\overline{\rm MS} }^3(\mu) 
  + t_3\alpha_{\overline{\rm MS} }^4(\mu)
  + \mathcal O(\alpha_{\overline{\rm MS} }^5)\, , 
\end{equation}
is known up to three-loops~\cite{Chetyrkin:2000dq}.

\begin{figure}
  \centering
    \begin{subfigure}[t]{0.45\textwidth}
    \centering
    \includegraphics[width=\textwidth]{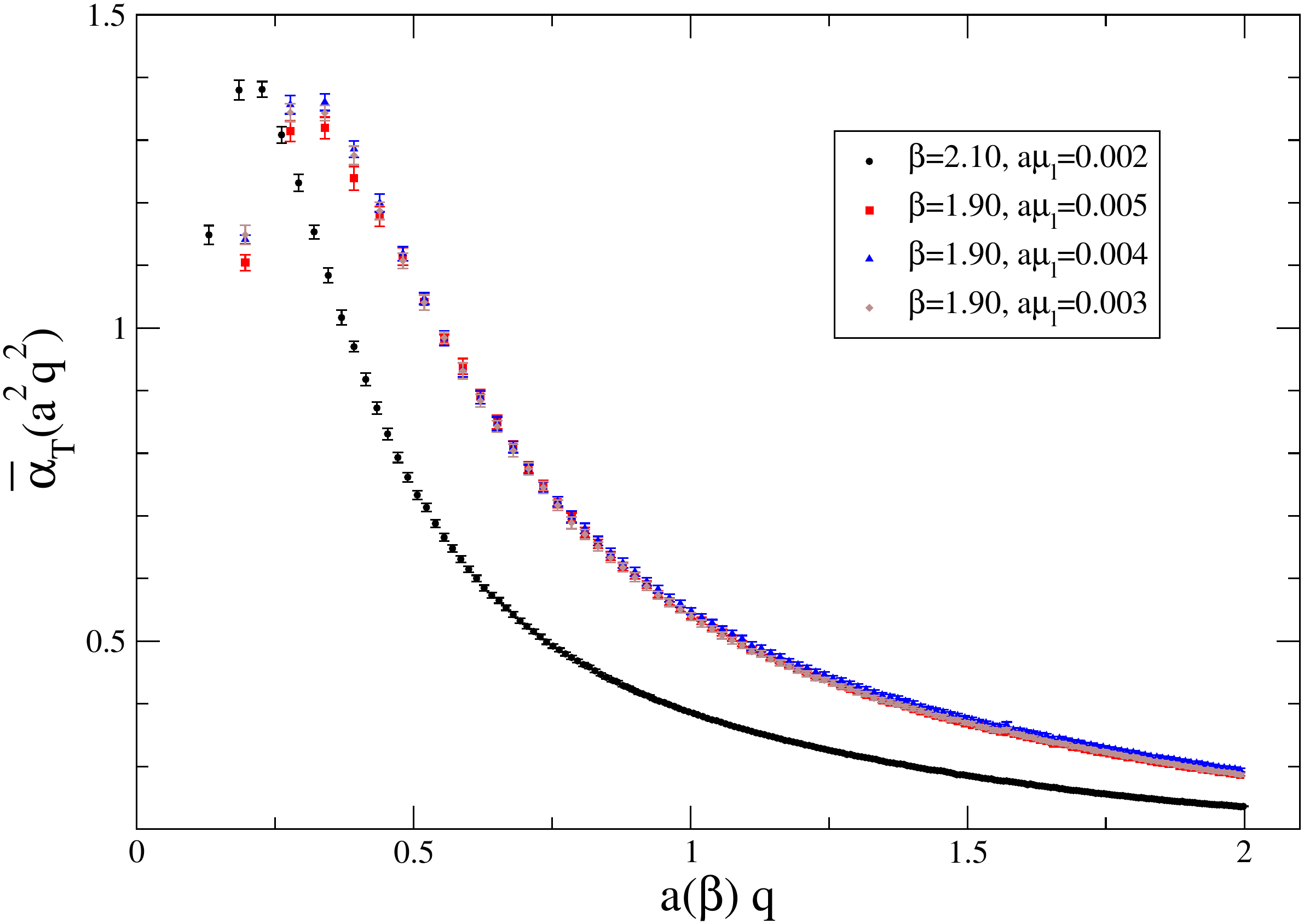}
    \caption{}
  \end{subfigure}
  \begin{subfigure}[t]{0.45\textwidth}
    \centering
    \includegraphics[width=\textwidth]{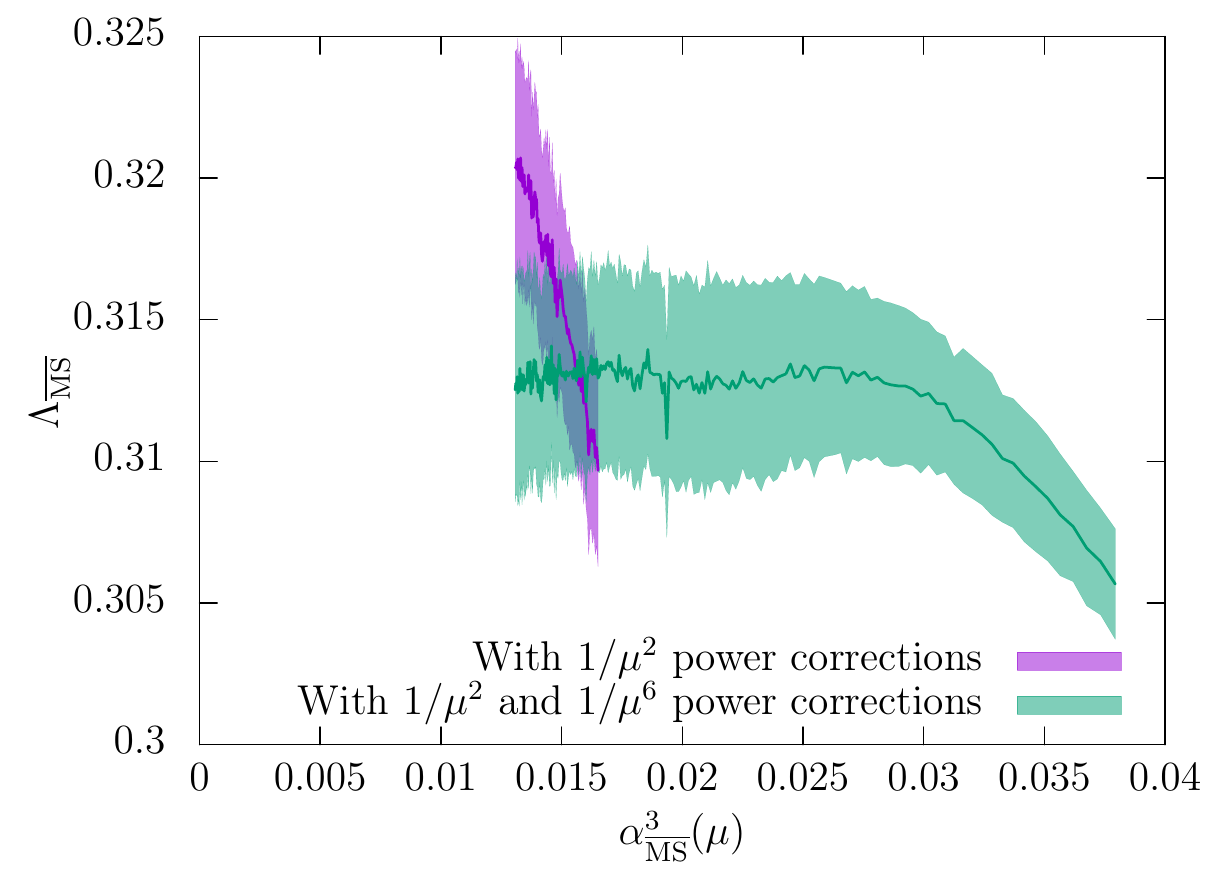}
    \caption{}
  \end{subfigure}

  \caption{Extraction of the strong coupling from QCD vertices.
    (a) Determination of the strong coupling in the Taylor
      scheme. The figure shows the raw values of the coupling for
      two different lattice spacing ($a\approx
      0.089$ fm for $\beta=1.90$ and $a\approx 0.06$ fm for
      $\beta=2.10$) and different values 
      of the quark masses. As the reader can see, the chiral
      dependence is mild, but the cutoff effects are substantial. 
      (source~\cite{Blossier:2013ioa}). 
      (b)
      Extraction of the $\Lambda$-parameter in the Taylor scheme. 
      Including only $1/\mu^2$ power corrections is not enough to
      to reach the perturbative running. 
      In this particular work~\cite{Zafeiropoulos:2019flq} , the
      extraction of $\Lambda$ seems 
      consistent in the region $0.012<\alpha^3<0.025$ by fitting the
      data with an additional $1/\mu^6$ power correction term.}
    \label{fig:alpha_taylor}
\end{figure}

There are two important issues that affect generally these extractions and
that are worth mentioning.
\begin{description}
\item[Non perturbative corrections] Several studies have concluded that
  including non perturbative corrections in the analysis is mandatory to find
  consistent results among observables and range of scales used to determine the
  $\Lambda$-parameter (see reference~\cite{Boucaud:2005xn} for a detailed study
  in pure gauge and energy scales in the range $3-5$ GeV). 

  The leading correction comes from 
  \begin{equation}
    \sim \frac{g_0^2 \langle A^2 \rangle }{p^2}\,.
  \end{equation}
  These corrections are included in the analysis either by using an estimate for
  $\langle A^2 \rangle$ (like for example~\cite{Zafeiropoulos:2019flq}) or by
  fitting their data including such a term. Higher order non-perturbative
  corrections (i.e. $\sim p^{-x}$ for different values of $x$) are also
  typically needed to match the lattice data with the perturbative
  running~\cite{Blossier:2011tf,Blossier:2012ef,
  Blossier:2013ioa,Zafeiropoulos:2019flq} (see Fig.~\ref{fig:alpha_taylor}).

\item[Cutoff effects] The range of scales where $\alpha_T(\mu)$ can be described
  by its perturbative expansion (and the non-perturbative contributions) is
  typically about $\mu \sim 3$ GeV. Since the typical lattice spacings in
  state-of-the-art lattice simulations are in the range $2-5$ GeV, it is really
  challenging to make a continuum extrapolation with several lattice spacings in
  order to extract the value of the coupling at some scale $\mu$. Therefore the
  common approach is to include several terms to either subtract or fit the
  cutoff effects (see~\cite{Blossier:2011tf,Blossier:2012ef,
  Blossier:2013ioa,Boucaud:2018xup}).
\end{description}

Figure~\ref{fig:alpha_taylor} shows these two points exemplified in the results
of one particular work~\cite{Blossier:2013ioa}. Panel (a) shows that the raw
measurements for the Taylor coupling at two different values of the lattice
spacing differ by approximately $50\%$ after subtracting the $H(4)$ breaking
cutoff effects (see~\cite{Blossier:2013ioa} for details). The remaining scaling
violations have an asymptotic expansion that includes terms $\sim\mathcal O(a^2
p^2)$ and are noticeable compared with the statistical precision of the data.
Panel (b) shows the extraction of the $\Lambda$-parameter after matching with
its perturbative expansion at the scale $\mu$~\cite{Zafeiropoulos:2019flq}. The
plot shows the values of $\Lambda_{\overline{\rm MS} }$ as a function of
$\alpha_{\overline{\rm MS} }^3$ (the leading correction to the extraction). The
data with only $1/\mu^2$ corrections subtracted shows a correction compatible
with a large $\alpha^3$ perturbative term. But if another non-perturbative term
$1/\mu^6$ is included as a fit parameter, the data seem to be well described by
the perturbative expression in Eq.~(\ref{eq:alphaTpt}). 

In summary, the extraction of $\alpha_s$ from QCD vertices is hampered
by a slow rate of convergence to the perturbative behavior. Several
non-perturbative corrections need to be fitted at the same time, since
they are noticeable. Even after fitting for the non-perturbative
corrections the range of energies that can be used to determine the
strong coupling is limited. Large energies scales are needed, where
most data come typically from a single lattice spacing.

These extractions also show that distinguishing the perturbative and
non-perturbative corrections is, in practice, very difficult.
Figure~\ref{fig:alpha_taylor} (b) shows the difficulty in
distinguishing a correction of order $\alpha^3(\mu)$ from a $1/\mu^6$
non perturbative correction when we have only access to a limited
range of scales. Indeed we see that a variation of the order of 10 MeV
in $\Lambda$ can be reabsorbed by higher-order power corrections.

We believe that a dedicated study in pure gauge theory with the aim of reaching
energy scales where the non-perturbative data is described by the perturbative
prediction \emph{without} fitting any non-perturbative terms would be very
interesting. A pure gauge simulation would also allow a detailed investigation
of the continuum extrapolations using several fine lattice spacing. 

\subsection{The static potential}
\label{sec:static-potential}

The force between static color charges has been traditionally one of
the first observables to be studied in lattice
QCD~\cite{Wilson:1974sk}. The potential at distance $r$ can be
extracted from Wilson loops, that behave asymptotically as
\begin{equation}
  \mathcal W_{r\times T} \sim \lambda_0^2e^{-V(r)T} + \sum_k
  \lambda_k^2 e^{-V_n(r)T}\,.
\end{equation}
The potential $V(r)$, given by the ground state 
(i.e. the leading decaying exponential), is formally computed as 
\begin{equation}
  V(r) = \lim_{T\to \infty} \frac{1}{T} \log \langle \mathcal
  W_{r\times T} \rangle\,.
\end{equation}
In practice it is extracted at large, but finite, values of $T$, and therefore
several techniques are needed in order to enhance the overlap with the ground
state and to distinguish the leading exponential from the excited state
contamination. When computed on the lattice, the static potential is power
divergent $\sim 1/a$. This is related to the ambiguity in the overall magnitude
of $V(r)$: only energy differences are physical. The cleanest way to deal with
this linear divergence is to define the coupling via the static force
\begin{equation}
  F(r) = \frac{{\rm d} V(r)}{{\rm d} r}\,.
\end{equation}
The derivative with respect to $r$ removes the linear divergence but requires to
perform a numerical derivative of the potential. This is implemented by some
finite difference expression. It is convenient to define the force by
\begin{equation}
  F(r_I) = \frac{V(r) - V(r-a)}{a}\,,
\end{equation}
with $r_I$ chosen so that the force has no cutoff effects to leading 
order in perturbation theory. This has been shown to reduce the cutoff
effects in the force~\cite{Necco:2001xg}.  

A renormalized coupling constant can be defined non-perturbatively
using the static force. The non-perturbative definition of the
coupling and its perturbative expansion in powers of the coupling in
the $\overline{\rm MS}$-scheme reads
\begin{eqnarray}
  \label{eq:qqpt}
  \nonumber
  \alpha_{qq}(\mu) = \frac{4}{3} r^2F(r)\Big|_{\mu = 1/r}
                       &\simas{r\to 0}&\, 
  \alpha_{\overline{\rm MS} }(\nu) +
                                        c_{qq}^{(1)}(s)\alpha_{\overline{\rm MS} }^2(\nu)
  + c_{qq}^{(2)}(s)\alpha_{\overline{\rm MS} }^3(\nu)\\
  &+&
      c_{qq}^{(3)}(s)\alpha_{\overline{\rm MS}
      }^4(\nu) + \dots \qquad (s=\mu/\nu)\,.
\end{eqnarray}
The relation of this coupling to the $\overline{\rm MS} $ scheme is known up to
three loops, but the observable suffers from IR divergences, that manifest
themselves in the naive perturbative expansion of eq.~(\ref{eq:qqpt}) being
divergent. These so-called \emph{soft} and \emph{ultra-soft} divergences can be
re-summed and produce logarithmic corrections to the perturbative series. The
leading one is $\propto \alpha_{\overline{\rm MS} }^4(\nu)\log \alpha_{\overline{\rm
MS}}(\nu_{\rm us})$. This re-summation process introduces
an arbitrary energy scale (so called \emph{ultra-soft scale}), its
natural value being $\nu_{\rm us} = \alpha(\nu)/r$.  
 In principle this means that not only the scale $\nu\approx 1/r$ has
 to be large, but also $\nu_{\rm us} = \alpha(\nu)/r$ has
to be large. The additional factor $\alpha$ is not negligible
taking into account that these extractions take place at a few GeV.

All in all, the perturbative expansion, including terms
$\alpha^4,\alpha^4\log\alpha,\alpha^5\log\alpha,\alpha^5\log^2\alpha$, is
known~\cite{Fischler:1977yf,Billoire:1979ih,Peter:1997me,Schroder:1998vy,Brambilla:1999qa,Smirnov:2009fh,Anzai:2009tm,Brambilla:2009bi}
(see~\cite{Tormo:2013tha} for a summary on the perturbative expressions).

Regarding the perturbative behavior, figure~\ref{fig:beta} shows that
the $\beta$-function in this scheme is well behaved, with a
2-loop coefficient of a similar size as in the $\overline{\rm MS} $
scheme. 

Several works extract the value of the strong coupling not
from the force, but directly from the potential. In particular they
examine the dimensionless quantity $rV(r)$. Basically the same
considerations apply for these works: the perturbative expansion in
the $\overline{\rm MS} $ scheme is known up to order $\mathcal
O(\alpha_{\overline{\rm MS} }^4)$, and the \emph{soft} and
\emph{ultra-soft} gluons give rise to logarithmic corrections in the
perturbative expansion, starting at order $\alpha_{\overline{\rm MS}
}^4$. The main difference is in the role of the linear divergence in
the potential $V(r)\sim 1/a$.

\begin{figure}
  \centering
  \begin{subfigure}[t]{0.45\textwidth}
    \centering
    \includegraphics[width=\textwidth]{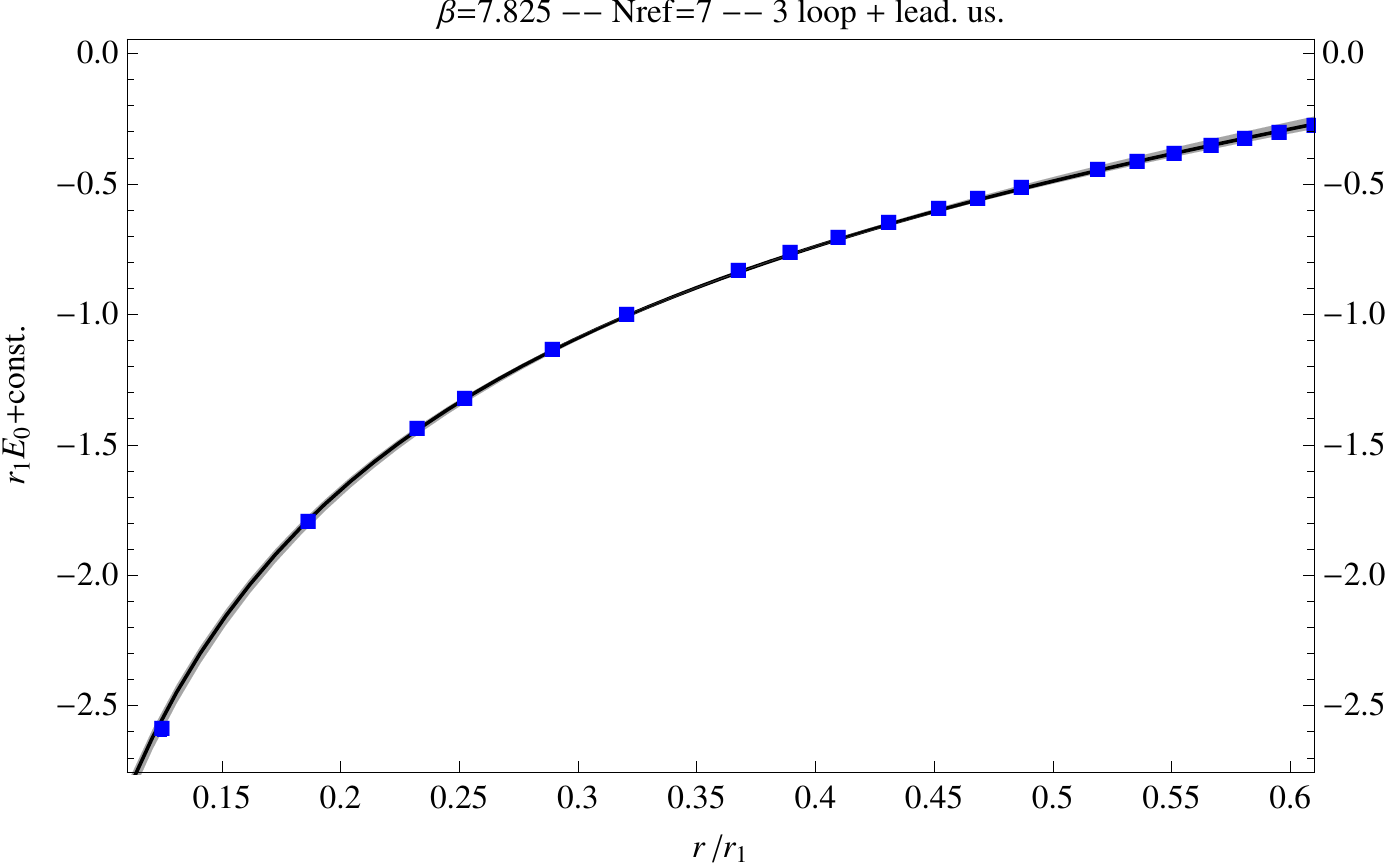}
    \caption{}
  \end{subfigure}
  \begin{subfigure}[t]{0.45\textwidth}
    \centering
    \includegraphics[width=\textwidth]{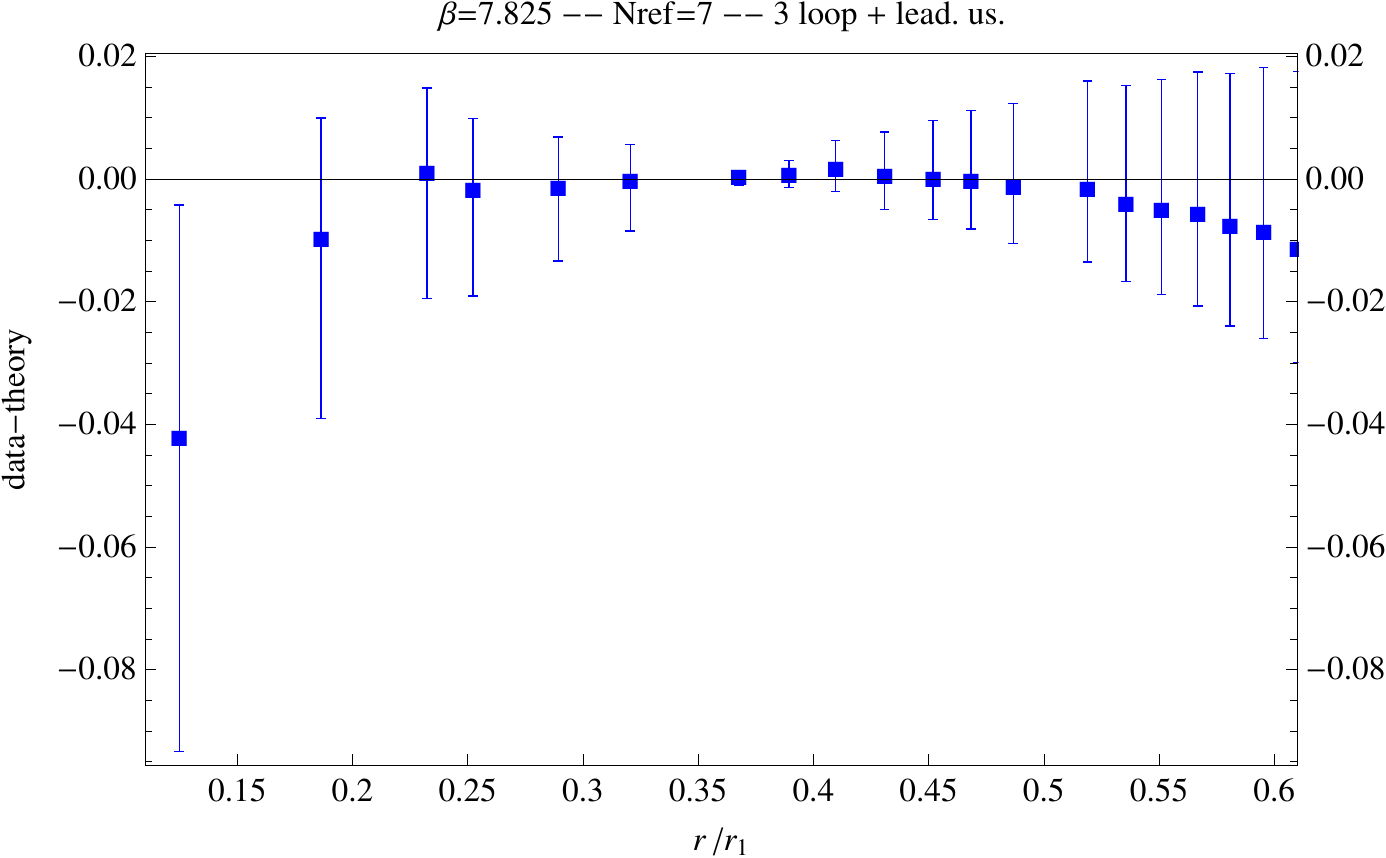}
    \caption{}
  \end{subfigure}

  \caption{(a) Comparison of the perturbative prediction and the
    lattice data at the finest lattice spacings ($a\approx 0.04$
    fm) for the static energy $E_0(r)$. (b) Residuals of the fit.}
  \label{fig:alpha_stat}
\end{figure}

Of course there are many different ways to use the perturbative expansion
Eq.~(\ref{eq:qqpt}) (or a similar expression for the potential
$rV(r)$) to extract $\alpha_s$. Different works, although similar in
spirit, use different approaches to fit the lattice data and deal
with the additive renormalization (in case they use $V(r)$ for the
extraction). Here we will focus on one particular 
work~\cite{Bazavov:2014soa} to show the details of such
analysis. Ref.~\cite{Bazavov:2014soa} uses the expression of
the static force and determines the integral up to a reference
distance $r_{\rm ref}$,
\begin{equation}
  \label{eq:E0}
  E_0(r) = \int_{r_{\rm ref}}^{r}{\rm d} x\,  F(x)\,.
\end{equation}
The perturbative expansion for the force gives a similar perturbative expansion
for $E_0(r)$. Note that this quantity is similar to the static potential $V(r)$,
with the difference that it is free of the linear divergence $\sim 1/a$ (the
relation $E_0(r_{\rm ref}) = 0$ is exact for all lattice spacings), and that
depends now on two scales ($r$ and $r_{\rm ref}$). The perturbative expression
for $E_0(r)$ is fitted to the lattice data. Fig.~\ref{fig:alpha_stat} shows the
result of such an analysis. Non-perturbative corrections do not seem to be
needed to describe the data at least for distances in the range $r\lesssim 0.2$
fm. This distance is, nevertheless, large enough to allow the analysis with
several lattice spacings, so the lattice data can be extrapolated to the
continuum. Other recent works include non-perturbative
corrections in the analysis of the lattice data either to claim a
better convergence~\cite{Ayala:2020odx} or to extend the range described by
their fit to $r\lesssim 0.3$ fm~\cite{Takaura:2018lpw}.

In summary, extracting the strong coupling from the static potential
is, in the opinion of the authors, one of the most promising
approaches. The N$^3$LO perturbative knowledge, together with the fact
that data in this scheme seem to follow the perturbative predictions
at scales as low as $1.5$ GeV means that a precise determination can
be achieved. In particular, contrary to other approaches, there is no
need to fit or parametrize any non-perturbative corrections. At least
in theory, the dependence on the lattice spacing, can be accounted
for. Conceptually, the fact that the observable
  used is not IR safe is not ideal. 
  An additional (lower) energy scale enters into the game, and a few
  works discuss the best methods to deal with these 
  corrections. The main criticism to 
such extractions, is that their results 
depends sensibly on the physics at a scale of just a few GeV, in
particular the range of scales where their assumptions can be checked
(role of the IR divergences, size of cutoff effects,
  matching with perturbation theory) is limited. It
would be interesting to check the agreement with the 
perturbative running down to higher energies, at least in the pure
gauge theory, where very fine lattice spacings can be
simulated (see reference~\cite{sommer:19lat} for a recent study). We
postpone to section~\ref{sec:pres-future-latt} a more detailed study. 

\subsection{Heavy quark correlators}
\label{sec:heavy-quark-corr}
The idea of using correlators of heavy quarks to extract the value of
the strong coupling has its origins in a phenomenological
determination of $\alpha_s$: moments of quarkonium correlators in the
vector channel can be compared with experimental data for $e^+e^- \to
\text{hadrons}$.

As was first noted by the HPQCD collaboration~\cite{Allison:2008xk}, the strong
coupling and the charm quark mass can be extracted on the lattice from
correlators of the pseudoscalar density involving two heavy quarks, 
\begin{equation}
  G(x_0) = a^6(am_0)^2 \sum_{\mathbf x} 
  \langle \overline \psi\gamma_5\psi(\mathbf{x},x_0) 
  \, \overline \psi\gamma_5\psi (\mathbf{0},0)
  \rangle\,, 
\end{equation}
where $am_0$ is the bare quark mass. This correlator has a short
distance divergence $\sim 1/x_0^3$, but if one uses a fermion formulation
that preserves some chiral symmetry (like staggered or domain wall
fermions), the PCAC relation ensures that moments of
$G(x_0)$, defined as
\begin{equation}
  G_n = \sum_{x_0} \left( \frac{x_0}{a} \right)^n G(x_0)\,,
\end{equation}
are dimensionless quantities with a well defined continuum limit~\footnote{With
Wilson fermions an extra finite renormalization for the axial current would be
needed.} for $n\ge 4$. The main contribution to these moments comes from
Euclidean times $x_0 \sim 1/m$.

The extraction of the strong coupling is performed using the reduced
even moments
\begin{eqnarray}
  r_4 &=& \frac{G_4}{G_4^{(0)}}\,,\\
  r_n &=& \bar m(\mu) \frac{am_{\eta}}{2am_0}\left[
          \frac{G_n}{G_n^{(0)}} \right]\,,\qquad (n\ge 6)
\end{eqnarray}
where $G_n^{(0)}$ denotes the leading order prediction for $G_n$ in bare lattice
perturbation theory. This normalization is introduced in order to reduce cutoff
effects (i.e. to leading order, $r_n$ is free of lattice artifacts). Here
$m_\eta$ denotes the mass of the $\eta_c$ meson, and $r_n$ admits a perturbative
expansion that allows an extraction of the strong coupling constant:
\begin{equation}
  r_n \simas{\alpha_{\overline{\rm MS}}(\mu)\to 0} 1 +
  r_{n,1} \alpha_{\overline{\rm MS}}(\mu) + 
  r_{n,2}(s) \alpha_{\overline{\rm MS}}^2(\mu) + 
  r_{n,3}(s) \alpha_{\overline{\rm MS}}^3(\mu) + \dots
\end{equation}
where
\begin{equation}
  s = \frac{\mu}{\bar m(\mu)}\,. 
\end{equation}
A non-perturbative definition of the strong coupling at the scale $\bar m(\bar
m)$ is given by
\begin{equation}
\label{eq:alphahq}
  \alpha_{\rm HQ, n}(\mu)\Big|_{\mu = \bar m(\bar m)} =
  \frac{r_n-1}{r_{n,1}} \simas{\alpha_{\overline{\rm 
        MS}}(\mu)\to 0}
  \alpha_{\overline{\rm MS}}(\nu) + 
  \frac{r_{n,2}(s)}{r_{n,1}}  \alpha_{\overline{\rm MS}}^2(\nu) + 
  \frac{r_{n,3}(s)}{r_{n,1}} \alpha_{\overline{\rm MS}}^3(\nu) + \dots\,,
\end{equation}
with $s=\nu/\bar m(\bar m)=\nu/\mu$. The first three coefficients
($r_{n,1}, r_{n,2}(s), r_{n,3}(s)$) are analytically known
(see~\cite{Maier:2009fz, Maier:2007yn, Boughezal:2006px, Chetyrkin:2006xg}. 
Reference~\cite{McNeile:2010ji} has values tabulated for $N_{\rm f} = 3$). 

There are two crucial points in these determinations: the estimate of the
truncation uncertainties, and controlling the continuum extrapolations. Note
that these two points have competing interests: the continuum extrapolation is
more easily kept under control for a quantity measured at larger distances, and
therefore the ``high'' moments $r_{6,8,10}$ have milder continuum
extrapolations. On the other hand the truncation uncertainties are smaller for a
short distance quantity: the first moment $r_4$ has a better behavior. This is
just another manifestation of the ``window problem'' (see
figure~\ref{fig:LQCDscales}).  

\begin{figure}
  \centering
  \begin{subfigure}[t]{0.45\textwidth}
    \centering
    \includegraphics[width=\textwidth]{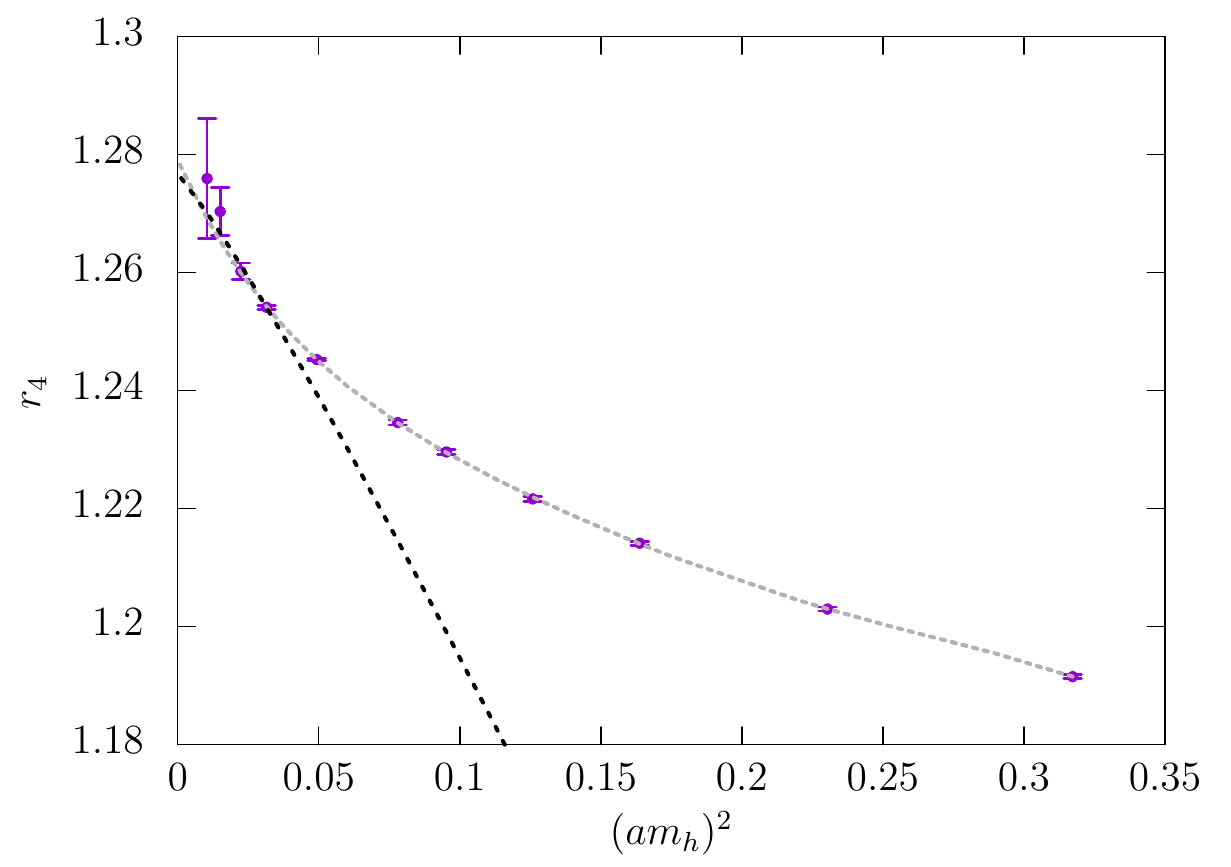}
    \caption{}
  \end{subfigure}
  \begin{subfigure}[t]{0.45\textwidth}
    \centering
    \includegraphics[width=\textwidth]{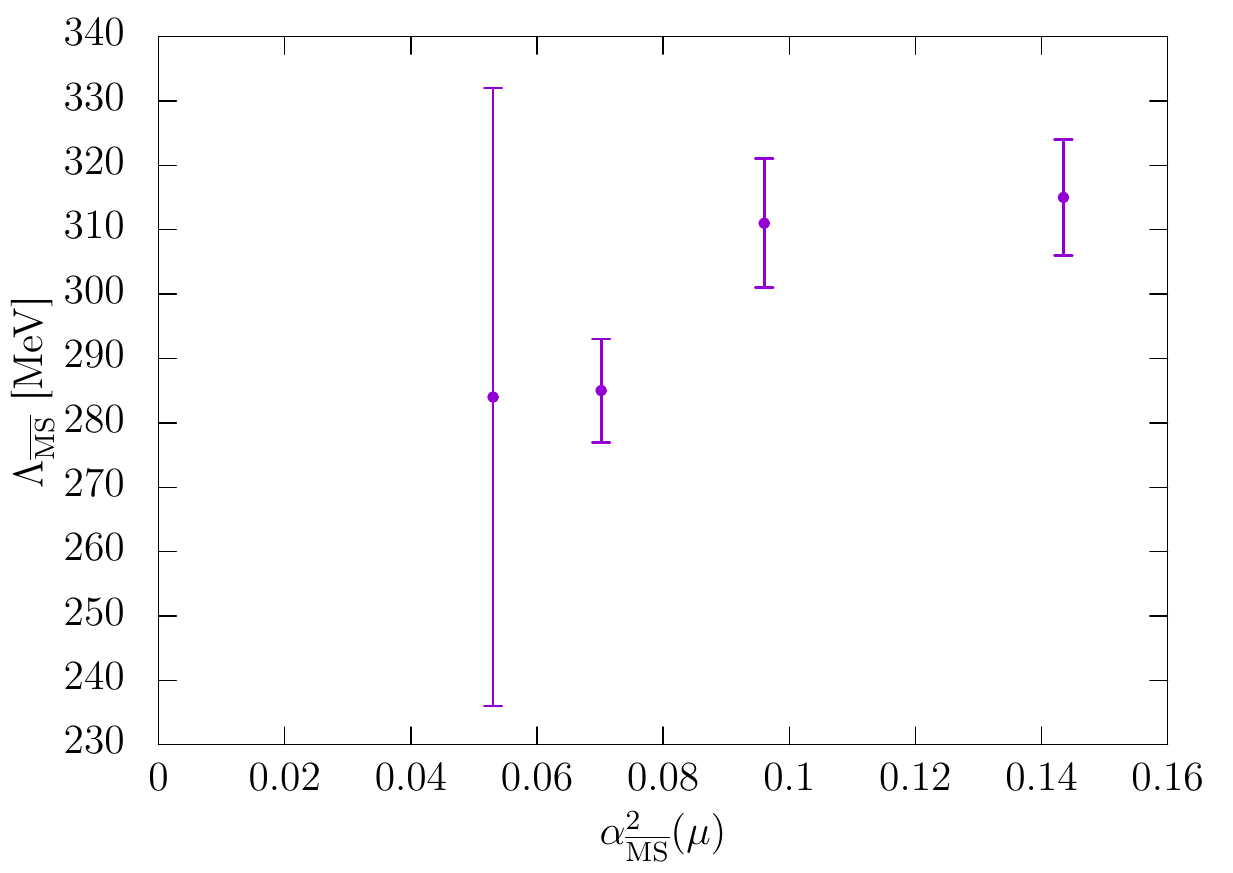}
    \caption{}
  \end{subfigure}
  
  \caption{(a) Continuum extrapolation of $r_4$ at the lower energy
    scale $\mu\sim \bar m_c$. Scaling violations are significant, even
    at the smaller quark masses used in the study. The extrapolations
    is performed using both a 5$^{\underline{th}}$ degree polynomial,
    or a second degree one with a restricted fitting window
    (source~\cite{Petreczky:2019ozv}).  (b)
    Dependence of the 
    $\Lambda_{\overline{\rm MS} }$ 
    parameter extacted using heavy quark correlators on the value of
    the coupling at the matching scale with perturbation theory
    ($\alpha_{\overline{\rm MS} }(\mu)$) (source~\cite{Petreczky:2019ozv}).}
  \label{fig:hqcorr}
\end{figure}

The estimates for these uncertainties vary significantly across different
studies, and we will comment in detail on the issue of the truncation
uncertainties in section~\ref{sec:pres-future-latt}. Here we will focus on the
more technical issue of the continuum extrapolation. Figure~\ref{fig:hqcorr}
shows the continuum extrapolation of $r_4$ at scale $m_{\rm c}$ of the
works~\cite{Maezawa:2016vgv,Petreczky:2019ozv}. As the reader can see, the scaling
violations are significant and have a complicated functional form. Different
works in the literature deal with these complicated cutoff effects in very
different ways
\begin{description}
\item [JLQCD collaboration~\cite{Nakayama:2016atf}:] In this case they
  prefer to only perform extrapolations linearly in $a^2$. 
  Their data for $r_4$ does not allow such an extrapolation, and
  therefore it is excluded from their analysis. 
  
\item [HPQCD collaboration~\cite{McNeile:2010ji,Chakraborty:2014aca}:]
  In these works all moments are used, and masses above the charm quark mass are
  used, including data with $am\sim 0.9$. Cutoff effects are large and the data
  is contaminated by effects $\sim (am_{\rm c})^{2p}$. Therefore their fit
  Ansatz includes terms $a^{2p}$ with $p$ up to 10. These fits typically have
  more terms than data, and require to include an estimate of the size of these
  coefficients as Bayesian priors.

\item [Ref.~\cite{Petreczky:2019ozv}:] Here energy scales
larger than the physical charm quark mass are explored, but the continuum
extrapolations are difficult and the data usually has associated large
uncertainties (see figure~\ref{fig:hqcorr}).
\end{description}

The main drawback of this approach is the large cutoff effects that
affect the quantity used to extract the strong coupling. This makes it
very challenging to explore energy scales larger than the physical
charm quark mass $m_c\sim 1.4$~GeV, which is not particularly
large. The recent work in Ref.~\cite{Petreczky:2019ozv} explores
different energy scales in the range $\bar m_c - 3\bar m_c$, but the
continuum extrapolation is very challenging already at
$\mu\gtrsim 2m_c$~\footnote{Note that the leading cutoff effects in
  this approach are $\mathcal O(a^2\alpha_s)$, while in other
  extractions (like for instance the one based on the static
  potential) they are suppressed by an extra factor of
  $\alpha_s$.}. Even at the scale of the charm quark mass, scaling
violations are significant and have a complicated functional form (see
Figure~\ref{fig:hqcorr}). Together with the fact that the perturbative
relation is known only to NNLO the situation is far from ideal:
truncation uncertainties at the energy scales reached by current
simulations are not small (see detailed discussion in
section~\ref{sec:heavy-quark-corr-1}). This might change in the
future, as smaller lattice spacings can be simulated, allowing a 
reduction of the discretization effects. A detailed study in pure gauge and
reaching energy scales significantly larger than the charm quark mass,
would probably give very important information on the systematics of
this method.

\subsection{Observables defined at the scale of the cutoff}
\label{sec:observ-defin-at}

Lattice QCD offers the interesting possibility of extracting the
strong coupling from expectation values computed at non-zero lattice
spacing. This approach is fundamentally different from the strategies
outlined above, where a quantity is computed in QCD (i.e. extrapolated
to the continuum), and then compared with a perturbative
prediction. The observables that we are going to discuss in this
subsection are defined at a scale given by the lattice spacing
$1/a$. Lattice bare perturbation theory is able to relate these purely
lattice observables with a power series in the renormalized
coupling. The usual problem that a naive approach has to face is that
bare lattice perturbation theory is just terrible. Absurdly small
values of the lattice coupling $\alpha_{\rm latt} = g_0^2/4\pi$ have
to be used in order to reach the domain of apparent convergence. It
has been argued that this apparent failure of lattice perturbation
theory is just a due to the choice of the bare coupling $g_0^2$ as
expansion parameter~\cite{Lepage:1992xa}. If lattice quantities are
expressed as a perturbative series in a \emph{renormalized} coupling,
like $\alpha_{\overline{\rm MS} }(\mu)$, their perturbative behavior
improves substantially (see~\cite{Lepage:1992xa}).

The HPQCD collaboration has pursued the systematic study of several Wilson loops
of size $n\times m$ (denoted $\langle W_{nm} \rangle$) and used them to extract
the value of the strong coupling. In these analyses lattice quantities are
expressed as a perturbative series in terms of the renormalized coupling
$\alpha_V(\mu)$; for an SU(3) gauge theory coupled to $N_f$ fermions in the
fundamental representation, the latter coupling is defined
by~\cite{Schroder:1998vy}
\begin{equation}
  \label{eq:alphaV}
  \alpha_{\rm V}(\mu) = \alpha_{\overline{\rm MS}}(\bar \mu) + \frac{2.6 - 0.3 N_{\rm f} }{\pi}\alpha_{\overline{\rm MS}}^2(\bar \mu)
  + \frac{53.4 - 7.2 N_{\rm f} + 0.2 N_{\rm f} ^2}{\pi^2}
  \alpha_{\overline{\rm MS}}^3(\bar \mu)\,,
\end{equation}
with $\bar \mu = \exp(\gamma-5/6)\mu \approx 0.774\times \mu$. Wilson
loops have a perturbative expansion 
\begin{equation}
  \label{eq:wmn}
  - \log \langle W_{nm} \rangle  \simas{a\to 0}
  w_1 \alpha_{\rm V}(\mu) +
  w_2 \alpha_{\rm V}^2(\mu) + 
  w_3 \alpha_{\rm V}^3(\mu) + \dots \,.
\end{equation}
Alternatively one
can use Creutz ratios or tadpole improved Wilson loops. This latter choice 
\begin{equation}
  \label{eq:wmnu}
  - \log \left( \frac{\langle W_{nm} \rangle}{\sqrt{\langle W_{11}
        \rangle^{n+m}}}  \right)   \simas{a\to 0}
  w_1^{\rm b} \alpha_{\rm V}(\mu) +
  w_2^{\rm b} \alpha_{\rm V}^2(\mu) + 
  w_3^{\rm b} \alpha_{\rm V}^3(\mu) + \dots \,.
\end{equation}
is supposed to lead to smaller truncation uncertainties. In all cases
the perturbative coefficients 
are known for several choices of $n, m$~\cite{Mason:2005zx}. The scale is given
by $\mu = d/a$, where $d\approx \pi$ with the exact value depending on the
choice of Wilson loop. Note that non-perturbative couplings can be defined by
expressions
\begin{equation}
  \label{eq:alphaW}
  \alpha_{W_{nm}}(1/a) = - \frac{\log \langle W_{nm} \rangle }{w_1}
  \simas{a\to 0} \alpha_{\rm V}(\mu) +
  \frac{w_2 }{w_1} \alpha_{\rm V}^2(\mu) + 
  \frac{w_3 }{w_1} \alpha_{\rm V}^3(\mu) + \dots \,.
\end{equation}

Several quantities are fitted to the previous perturbative
expressions, with the value of $\alpha_{\rm V}(\mu_{\rm ref})$ as fit
parameter. The values of $\alpha_{\rm V}(\mu)$ at other scales are
obtained from $\alpha_{\rm V}(\mu_{\rm ref})$ and the RG equation
\begin{equation}
  \mu^2 \frac{{\rm d} \alpha_{\rm V}(\mu)}{{\rm d} \mu^2} =
  -\alpha_{\rm V}^2\sum_{k = 0}^3 \beta_k\alpha_{\rm V}^k\,.
\end{equation}
The final value of $\alpha_{\rm V}(\mu_{\rm ref})$ can be converted to
the more convenient $\overline{\rm MS} $ scheme using Eq.~(\ref{eq:alphaV}).

An example of such extractions are the HPQCD works
(see~\cite{Allison:2008xk}). A
total of 22 quantities (all like Eq.~(\ref{eq:wmn})) are fitted to
their perturbative expression. Unfortunately the truncated
perturbative expression Eq.~(\ref{eq:wmn}) does not describe the data
well, and several extra terms (up to $\alpha_{{\rm
V}}^{10}$) are necessary in order to obtain a sensible fit. The
coefficients of these terms are constrained with Gaussian priors,
which eventually lead to stable fits, with a consistent determination
of the strong coupling using any of the 22 quantities (although some
of them require to also fit several power corrections).

The main advantage of methods based on observables defined at the
cutoff scale, is that high energies can be reached without having to
worry about the continuum extrapolation. The statistical accuracy is
excellent, since the observables entering the determination
Eq.~(\ref{eq:wmn}) have a very small variance. On the other hand the
uncertainty in these determinations is dominated by the truncation of
the perturbative series. The fact that several higher order terms have
to be fitted (and constrained with Gaussian priors) in order to
describe the data is not ideal. It is clear that expressing lattice
quantities as a power series in \emph{renormalized couplings}, as
suggested in~\cite{Lepage:1992xa} greatly improves the predictive
power of perturbation theory (bare perturbation theory is just useless). 
Still these lattice observables are far from ideal from a perturbative
point of view: even if energy scales $1/a\approx 4$ GeV are reached,
perturbation theory does not predict the lattice data and truncation
uncertainties are not small (see detailed discussion in
section~\ref{sec:observ-at-cutoff}).  

Another delicate point in this approach is that cutoff
effects have the same functional form as the non-perturbative effects
(power corrections) in the expansion of $\alpha_P(1/a)$. This can be
easily understood by noting that
\begin{equation}
  a^2 \sim \exp\left\{- \frac{4\pi}{2b_0\alpha_W(1/a)} \right\}\,,
\end{equation}
Of course in any extraction of the strong coupling based on these methods,
these effects are not parametrically the leading ones, since the
truncation of the perturbative series Eq.~(\ref{eq:alphaW}) misses
terms of order $\mathcal O(\alpha_W^n) \sim \log^n a$. However in
practice it 
is not clear which effects dominate (the $\mathcal O(a^2)$ cutoff effects
or the $\log^n a$ from the truncation of the perturbative series), and
this might even depend on the particular observable used to set the scale. 

\subsection{The hadron vacuum polarization}

The hadronic vacuum polarization function (HVP) is defined from two-point
functions of the vector and axial-vector currents
\begin{eqnarray}
  V_\mu^a(x) &=& \bar \psi_a\gamma_\mu \psi_a(x)\,, \\
  A_\mu^a(x) &=& \bar \psi_a\gamma_5\gamma_\mu \psi_a(x)\,,
\end{eqnarray}
after a decomposition in Fourier space (with $J_\mu= V_\mu, A_\mu$)
\begin{equation}
  \int \mathrm{d} ^4x\, e^{\imath p x}\, \langle J_\mu^a(x) J_\nu^a(0)
\rangle =
(\delta_{\mu\nu}p^2 - p_\mu p_\nu)\Pi_J^{(1)}(p^2) - p_\mu
p_\nu\Pi_J^{(0)}(p^2) \,.
\end{equation}
The quantity
\begin{equation}
  \Pi(p^2) = \Pi_{V}^{(0)}(p^2) + \Pi_{V}^{(1)}(p^2) + \Pi_{A}^{(0)}(p^2) + \Pi_{A}^{(1)}(p^2)\,.
\end{equation}
is dimensionless and has a perturbative expansion
\begin{equation}
  \label{eq:hvppt}
  \Pi(p^2) \simas{p\to\infty} c_0
  + \sum_{k=1}^4c_k(s)\alpha^k_{\overline{\rm MS}}(\mu) + \mathcal O(\alpha^5_{\overline{\rm MS} })\,.
  \qquad (s=p/\mu)\,.
\end{equation}
known up to 5-loops. The constant term $c_0(s)$ is divergent, so that the strong
coupling is usually extracted from the difference $\Pi(p^2) - \Pi(p^2_{\rm
ref})$, or the Adler function
\begin{equation}
  D(p^2) = p^2 \frac{{\rm d} \Pi(p^2)}{{\rm d} p^2}\,.
\end{equation}

The recent work~\cite{Hudspith:2018bpz} determines the finite
difference
\begin{equation}
  \Delta(p^2,p^2_{\rm ref}) = \frac{\Pi(p^2) -\Pi(p^2_{\rm
      ref})}{\log(p/p_{\rm ref})} \,.
\end{equation}
at high energies in order to make contact with the perturbative running. They
use several values of $p\sim 2-4$ GeV and different fit procedures, ranges of
$p$ and values of $p_{\rm ref}$ to extract $\alpha_{\overline{\rm MS} }(M_Z)$.

The main issue with extractions based on the HVP is that power
corrections are significant even for large
momenta~\cite{Shintani:2010ph}. In fact, as discussed in detail in
Ref.~\cite{Hudspith:2018bpz}, these corrections show a very poor
convergence. One would expect that higher power corrections become
negligible as the momentum increases, but this is not the case due to
accidental cancellations between condensates of different dimensions
(up to $1/\mu^8$).  Ref.~\cite{Hudspith:2018bpz} pushes the
determination to high energies, so that the data can be described
without any power corrections, but then cutoff effects become larger
and the window of scales to obtain the strong coupling decreases.
Despite the impressive perturbative knowledge in this scheme (5
loops), the authors think that more work is needed in order to
convincingly show that contact with the perturbative running has been
made, and that the continuum extrapolations are under control. Being a
relatively new technique, there is not a single work for the pure
gauge theory. Once again, we would like to stress that a detailed
study in this simpler case, where very fine lattice spacings can be
simulated, would shed some light on many of these issues.

\subsection{Eigenvalues of the Dirac operator}
\label{sec:eigenv-dirac-oper}


Recently a novel approach to extract the strong coupling has been
proposed. 
It uses the spectral density of the continuum Dirac operator
\begin{equation}
  \rho(\lambda) = \frac{1}{V} \left\langle
    \sum_k \left[ \delta(\lambda - \imath \lambda_k) + \delta(\lambda + \imath \lambda_k) \right] 
  \right\rangle\,,
\end{equation}
and its perturbative expansion
\begin{equation}
  \label{eq:rho_pt}
  \rho(\lambda) = \frac{3\lambda^3}{4\pi^2} \left(
    1 - \rho_1(s)  \alpha_{\overline{\rm MS} }(\mu) 
    - \rho_2(s)  \alpha_{\overline{\rm MS} }^2(\mu) 
    - \rho_3(s)  \alpha_{\overline{\rm MS} }^3(\mu) 
    + \mathcal O(\alpha_{\overline{\rm MS} }^4)
  \right)\,.\qquad
  (s = \mu/\lambda)\,.
\end{equation}
A non-perturbative coupling definition can be defined by using
\begin{equation}
  \alpha_D(\lambda) = \rho_1(s)^{-1}\left[\frac{4\pi^2}{3\lambda^3}\rho(\lambda) -
    1\right]\,.
\end{equation}
with a $\beta$-function known up to 3-loops. 

Alternatively one can use the derivative
\begin{equation}
  F(\lambda) = \frac{\partial\rho(\lambda)}{\partial\log\lambda}
  = 3-F_1(s)  \alpha_{\overline{\rm MS} }(\mu) 
    - F_2(s)  \alpha_{\overline{\rm MS} }^2(\mu) 
    - F_3(s)  \alpha_{\overline{\rm MS} }^3(\mu) 
    - F_4(s)  \alpha_{\overline{\rm MS} }^4(\mu) 
    + \mathcal O(\alpha_{\overline{\rm MS} }^4)
\end{equation}
to also define the strong coupling whith a perturbative expansion
known up to 4-loops.   

Only one work in the literature uses this method to extract the strong
coupling~\cite{Nakayama:2018ubk}. Naively the truncation error at the
energy scales used to extract $\alpha_s$ (\ie\
$\lambda \approx 0.8-1.2$ GeV) turns out to be very large ($\sim 20\%$
in $F(\lambda)$).  The renormalization scale is pushed to higher
values by using $\mu/\lambda = 5$. For these values of $\mu$
truncation effects are expected to be reduced to a percent level,
while the perturbative expansion for the strong coupling still shows
good convergence properties.

It is important to note that the dominant uncertainty from the
truncation of the perturbative series is determined by an estimate of
the leading missing coefficient in the perturbative series
Eq.~(\ref{eq:rho_pt}).  Due to the low energy scale of the
determination, the usual procedure of varying the renormalization
scale by a factor 2 below and above $\lambda$ would result in a
substantially larger truncation error.  On the other hand the
work~\cite{Nakayama:2018ubk} does not seem to need any power
corrections to describe the data at energy scales $0.8 - 1.2$ GeV.

The main issue with this novel approach seems to be the very low energy
scales at which the extraction is performed. 
These low energy scales are needed in order to be able to extrapolate
the data to the continuum. 
The work~\cite{Nakayama:2018ubk} shows that a cut $a\lambda < 0.5$
is needed in order to avoid a substantial
deviation from the leading $\mathcal O(a^2)$ scaling violations. 
This restricts the energy scales that can be reached with their data-set
(with lattice spacing $a^{-1} = 2.5, 3.6$ and $4.5$ GeV) to $\lambda <
1.2$ GeV. 
This novel method needs further study, in particular at finer lattice
spacing, in order to convincingly show that contact with the
perturbative running is done and to be able to perform a robust
estimate of the truncation effects.



\subsection{Finite size scaling}
\label{sec:finite-size-scaling}

Here we will describe a theoretical idea to overcome the fundamental
limitation of any of the previous lattice determination of the strong
coupling~\footnote{The ideas of finite size scaling are able to deal
  with many other multiscale problems, like the description of heavy
  quarks in lattice simulations, see Ref.~\cite{Sommer:2010ic} for
  example, or the renormalization of composite operators.}, namely the
compromise between reaching large energy scales to control
perturbative truncation effects, and and having a clear separation
between these energy scales and the lattice cutoff so that cutoff
effects in observables measured at small distances are kept under
control. This fundamental compromise present in any single lattice
computation -- the scales $\mu$ that can be probed have to
obey $1/a\gg \mu \gg m_\pi$ -- originates from the fact that we want to
accommodate in a single lattice computation the scales used to match
with perturbation theory $\mu_{\rm PT}$ and the scales used to
describe hadronic physics. Finite size scaling solves this problem by
adopting a different strategy. Only a finite range of scales is
resolved in any single lattice computation, and a recursive procedure
allows to relate the energy scales explored with different
simulations.

This idea is implemented by using a \emph{finite volume renormalization
scheme}~\cite{Luscher:1991wu}. The observable used to define the coupling
$O(\mu)$ is defined at a scale linked with the finite volume of the simulation
$\mu \propto 1/L$ and the coupling will run with the size of the system. Very
small physical volumes -- the so-called \emph{femtouniverse} -- are simulated in
order to reach high energy scales. The only constraint for these simulations is
that the energy scale explored has to be far away from the cutoff ($1/a\gg \mu
\sim 1/L$). This is easily achieved by using lattices of moderate size $L/a\sim
10-50$. 

\begin{figure}
  \centering
  \includegraphics[width=\textwidth]{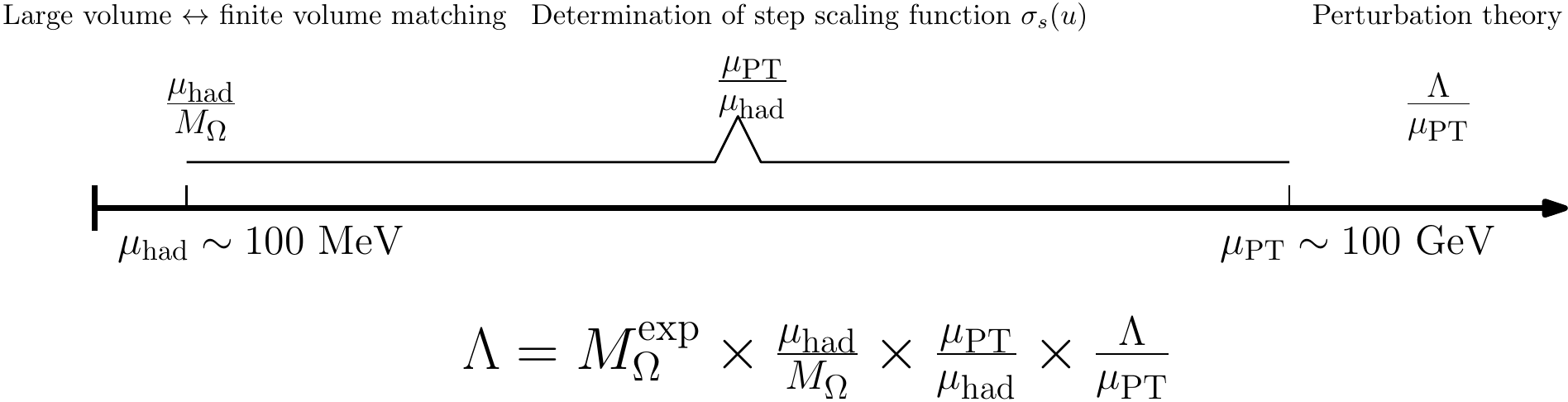} 
  \caption{  A product of dimensionless ratios together with a single
  dimensionful experimental input is used to determine $\Lambda$.
  At low energies an experimental quantity ($M_\Omega$ in
  the example) is computed in terms of $\mu_{\rm had}$, a hadronic
  scale defined in a 
  massless finite volume scheme (see section~\ref{sec:match-an-exper}). 
  The finite volume scheme is used to connect non perturbatively the
  hadronic ($\mu_{\rm had}$) and perturbative ($\mu_{\rm PT}$) regimes
  of QCD by determining the step 
  scaling function (see section~\ref{sec:solving-rg-equations},
  Eq.~(\ref{eq:ston}) and Eq.~(\ref{eq:ratio_beta})). 
  Finally using perturbation theory one can determine the $\Lambda$
  parameter in units of the high energy perturbative scale ($\mu_{\rm
    PT}$). 
  }
  \label{fig:fss_scheme}
\end{figure}

There are three key ingredients in any finite size scaling study:
\begin{itemize}
\item  We need a way to relate our finite volume simulations
with experimental data. 
\item We need a way to match simulations in
different volumes.
\item We need to match our results in
very small volumes (high energies) with a perturbative
computation.
\end{itemize}

Figure~\ref{fig:fss_scheme} shows a schematic representation of the procedure.
Note that we are effectively separating the problem of solving
non-perturbatively the RG equations and determining an overall hadronic scale.
But before explaining in detail these three ingredients, let us discuss some
technical details of QCD in small physical volumes.

\subsubsection{QCD in a finite volume simulations}

Typical lattice simulations aim to explore QCD in the infinite volume
limit. One is usually interested in the determination of hadronic
quantities like the spectrum or some decay form factors, which can be
reliably computed only using large volume simulations, with say
$L m_\pi > 4$. In these simulations the choice of boundary conditions
is not very relevant, since we are supposed to describe the infinite
volume physics within a percent accuracy.

On the other hand, when studying QCD on a finite volume we are exploring a
completely unphysical regime of the theory with then aim of solving
the RG equations. The size of the system is just 
part of the renormalization prescription, and we are free to choose it at will.
But in this particular situation the boundary conditions play a fundamental
role. Different choices of boundary conditions have to be understood
as different renormalization schemes, and not as small corrections. The
choice of boundary conditions is part of the definition of the observable
itself.  

One would naturally think that periodic boundary conditions are a natural
choice, but it was soon found~\cite{GonzalezArroyo:1981vw} that QCD observables
computed in a finite periodic box have complicated perturbative expansions.
Perturbative expansions derive from a saddle point expansion of the path
integral
\begin{equation}
  \label{eq:path_int}
  \mathcal Z = \int\mathcal D A_\mu\, e^{-S[A_\mu]}
\end{equation}
around the minimum. In infinite volume this minimum ($A_\mu=0$) is
unique, up to gauge transformations. On the other hand fields with
periodic boundary conditions on a finite volume do not have a unique
minimum of the action. All zero momentum fields
$A_\mu = \text{constant}$ have zero action. Note that these
configurations, in a finite volume, are not all related to each other
by gauge transformations. Gauge transformations
\begin{equation}
  \Omega(x) = \exp\left(\omega^a(x) T^a\right)
\end{equation}
have to be single valued functions, imposing the condition
$\omega^a(x+2\pi) = \omega^a(x)$. It is easy to see that this implies
that gauge transformation can only shift the zero mode of the gauge
fields by a multiple of $2\pi$, 
\begin{equation}
  A_\mu^a \to A_\mu^a + 2\pi n\,.
\end{equation}

Naively these zero momentum field configurations produce flat
directions in the path integral Eq.~(\ref{eq:path_int}), making the
integral divergent. Expectation values are finite, but in
general the perturbative expansion can no longer be written as a power
series in $\alpha$. Fractional powers (like $\alpha^{3/2}$) or even
logarithmic contributions ($\alpha\log\alpha$) appear in the
perturbative expansion\footnote{The concrete form of these terms
  depends on the number of periodic directions and the rank of the
  group.}. These non-analytic terms in the perturbative expansion of
observables defined in a periodic box result in coupling definitions
(defined via Eq.~(\ref{eq:O2gsq})) that generically do not even share
the universal coefficients of the $\beta$-function. The
$\Lambda$-parameter can not be defined in schemes defined from these
observables (readers interested in this topic can consult the
review~\cite{vanBaal:1988qm}). 

Of course one could still match these observables with the perturbative
expansion, including the non-analytic terms\footnote{The matching with the
asymptotic perturbative behavior for these kind of observables might be
delicate, and require access to substantially larger energy scales.}, but
fortunately there are better solutions. Since these non-analytic perturbative
expansions are just a consequence of our choice of boundary conditions, we can
choose the latter wisely in order to avoid these complications, and there exists
several options to accomplish this goal.

\begin{description}
\item[Twisted boundary conditions] Demanding physical quantities to be periodic
  does not necessarily require a periodic gauge field $A_\mu(x)$. It is enough
  for $A_\mu(x)$ to be
  \emph{periodic modulo a gauge transformation}~\cite{tHooft:1981sz}
  \begin{equation}
    A_\mu(x+L_\mu\hat \mu) = \Omega_\mu(x)A_\mu\Omega^\dagger_\mu(x) +
    \imath \Omega_\mu(x)\partial_\mu\Omega^\dagger_\mu(x)\,.
  \end{equation}
  The matrices $\Omega_\mu(x)$, known as transition matrices, can be chosen in
  order to guarantee that the action has a unique minimum up to gauge
  transformations~\cite{GonzalezArroyo:1981vw}.

  In some sense, twisted boundary conditions are the most natural choice, since
  translational invariance is fully preserved. Fermions can be added without
  major conceptual problems. However, for fermions in the fundamental
  representation of the gauge group, boundary conditions impose that the number
  of flavors to be included in the model have to be a multiple of the rank of
  the group (\ie\ $N_{\rm f} \propto 3$ for the case of $SU(3)$). For QCD
  applications this might be a problem, since only the three and six flavor
  theories can be formulated with this choice of boundary conditions.   

  The first applications of twisted boundary conditions in finite size scaling
  used a ratio of Polyakov loops as definition of the renormalized
  coupling~\cite{deDivitiis:1994yz}. This coupling definition suffers from large
  statistical errors, due to the large variance of this particular observable.
  Most recent works use the gradient flow, which we introduced in
  section~\ref{sec:theory-scales}, to define the renormalized
  coupling~\cite{Ramos:2014kla}.
  
\item[Schr\"odinger Functional boundary conditions]  The most common choice of
  boundary conditions to study QCD on a finite volume are called Schr\"odinger
  Functional boundary conditions~\cite{Luscher:1992an, Sint:1993un}. Dirichlet
  boundary conditions are imposed on the spatial components of the gauge field
  at Euclidean times $x_0=0,T$
  \begin{equation}
    \label{eq:sfbcs}
    A_i(x)\Big|_{x_0=0} = C_i(\mathbf x);\qquad
    A_i(x)\Big|_{x_0=0} = C'_i(\mathbf x)\,.
  \end{equation}
  The time component of the gauge field inherits its boundary
  conditions from the gauge fixing condition. If the boundary fields 
  $C_i(\mathbf x), C'_i(\mathbf x)$ are chosen appropriately, the
  minimum of the action is unique up to gauge transformations. There
  are two common choices in the literature. First one can choose
  $C_i(\mathbf x), C'_i(\mathbf x)$ to be constant diagonal
  matrices. For the case of $SU(3)$, which has two diagonal generators,
  the generic form of the the background field is
  \begin{eqnarray}
    \label{eq:CCp}
    C_k &=& 
            \frac{i}{L}{\rm diag}
            \bigg\{\eta-{\pi\over 3},\,
            \eta\Big(\nu-{1\over2}\Big),\,
            -\eta\Big(\nu+{1\over 2}\Big)+{\pi\over 3}\bigg\}\,,\\
    C_k' &=& \frac{i}{L}{\rm diag}\bigg\{
             -\eta-\pi,\,
             \eta\Big(\nu+{1\over 2}\Big)+{\pi\over 3},\,
             -\eta\Big(\nu-{1\over 2}\Big)+{2\pi\over 3}\bigg\}\,,
  \end{eqnarray}
  where the parameters $\eta,\nu$ can be chosen at will. An important
  advantage of this setup is that derivatives of the effective action
  with respect to these boundary parameters can be used to define a
  renormalized coupling at a scale given by the volume of the system: 
  \begin{equation}
  \label{eq:alphasf}
    \frac{12\pi}{\bar g^2_{{\rm SF},\nu}(\mu)} =  
    \left\langle \frac{\partial S}{\partial
        \eta}\right\rangle\bigg|_{\eta=0}, 
    \qquad 
    (\mu= 1/L)\,.
  \end{equation}
  Different choices of $\nu$ represent different renormalization schemes
  (different coupling definitions). Conveniently, the values of $\bar g_{{\rm
  SF},\nu}$ for all values of $\nu$ can be calculated from expectation values
  evaluated at $\nu=\eta = 0$. These couplings have been the 
  preferred choice for finite size scaling studies in QCD
  (see~\cite{Capitani:1998mq,DellaMorte:2004bc,Aoki:2009tf,Tekin:2010mm}), but
  they are gradually being replaced by the new coupling definitions based on the
  gradient flow~\cite{Fritzsch:2013je}. These are more conveniently defined with
  zero background field (i.e. $C_i(\mathbf x) = C'_i(\mathbf x) = 0$ in
  eq.~(\ref{eq:CCp})). 

  Schr\"odinger functional boundary conditions can be easily simulated on the
  lattice, but breaking translational invariance has the unpleseant effect of
  producing linear $\mathcal O(a)$ cutoff effects (even in the pure gauge
  theory~\cite{Luscher:1992an}). These can in principle be removed by tuning
  boundary counterterms. They are only known in perturbation theory and the
  effect of the higher order corrections have to be studied in detail in any
  step scaling study. 
  
\item[Open-SF boundary conditions] For a long time, topology freezing was
  thought not be an issue in small volume simulations since non-trivial
  topological sectors are highly suppressed on small volumes. But nowadays it is
  clear~\cite{Fritzsch:2013yxa} that these simulations are also suffering from
  this effect when the physical volume is $\sim 1\,{\rm fm}$ and either
  twisted or SF bondary conditions are used~\footnote{There exists an index
  theorem with twisted boundary conditions that guarantees that \emph{when
  simulating massless quarks} only topologically trivial sectors contribute to
  the path integral. This opens the door for three flavor QCD to actually
  determine the running coupling without having to worry about any topology
  freezing problems.\label{foot:twisted}}. A way to overcome this issue is just
  to use definitions in the sector of zero topological charge, as suggested in
  Ref.~\cite{Fritzsch:2013yxa}.

  Using open boundary conditions in Euclidean time provides a solution to the
  problem: the topological charge is no longer quantized and transitions between
  different topological sectors are allowed~\cite{Luscher:2014kea}. As discussed previously, the breaking of translational invariance leads also in this case to linear $\mathcal O(a)$ cutoff effects near the Euclidean time boundaries.

\end{description}

Whatever the choice of boundary conditions, simulations on small physical
volumes have one more crucial ingredient that makes them attractive: the
possibility to directly simulate massless quarks on the lattice. This is
possible because in this regime the size of the system acts as IR cutoff ($\sim
1/L$). The Dirac operator has a gap, even at $m=0$. This possibility removes a
source of systematic effect in the lattice determinations of the strong
coupling, since the matching with the perturbative regime is usually only known
to high accuracy in massless renormalization schemes. There is no need to
perform an extrapolation to zero mass.

\subsubsection{Solving the RG equations}
\label{sec:solving-rg-equations}

Finite volume renormalization schemes work by providing coupling
definitions that depend on a single scale given by the volume of the
system ($\mu\sim 1/L$). It is clear that large energy scales can be
easily reached simulating a small physical volume, requiring only a
modest lattice size ($L/a\sim 10-30$). What is not so clear is how to
relate different finite volume simulations in order to determine the
actual running of the strong coupling.

\begin{figure}
  \centering
  \includegraphics[width=0.7\textwidth]{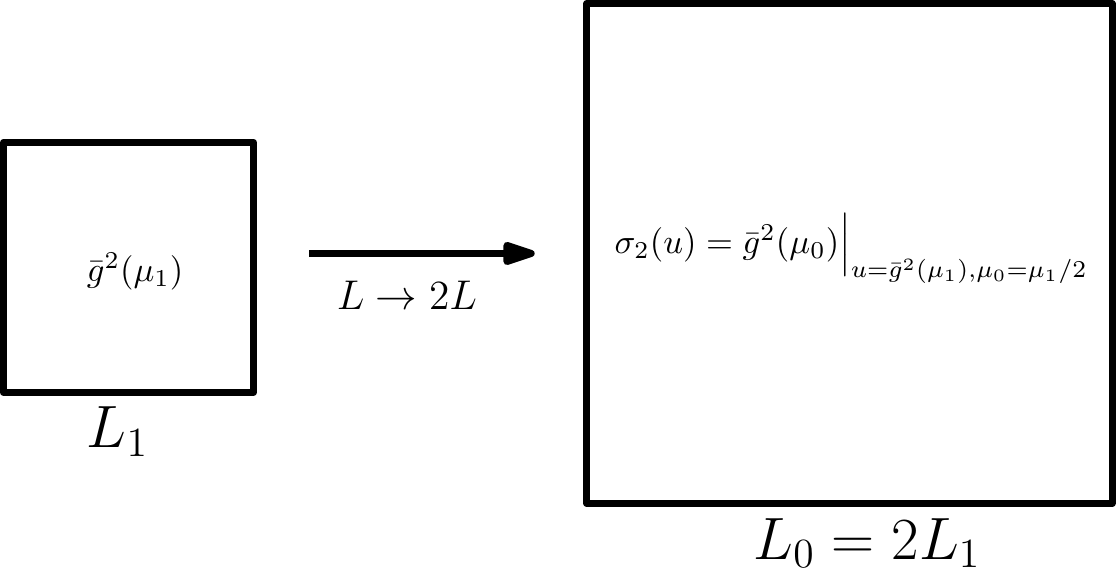} 
  \caption{In finite volume renormalization schemes the coupling $\bar
    g ^2(\mu)$ is defined at a scale given by the physical size  of
    the system ($\mu = 1/L$). If at some scale $\mu_1 = 1/L_1$ the coupling has
    value $u = g ^2(\mu_1)\Big|_{\mu_1 = 1/L_1}$, and we measure the
    coupling in a volume two times larger ($2L_1$), we obtain the step
    scaling function $\sigma_2(u)$ (see Eq.~(\ref{eq:ssf})). This is a
    discrete version of the $\beta$-function that measures how much
    the coupling changes when the scale is changed by a factor 2.}
  \label{fig:step}
\end{figure}

The \emph{step scaling function} $\sigma_s(u)$ is the key quantity that makes it
possible solving the RG equations (see figure~\ref{fig:step}). It measures the
change in the coupling when the renormalization scale changes by a factor $s$,
and therefore can be understood as a discrete version of the $\beta$-function,
\begin{equation}
  \label{eq:ssf}
  \sigma_s(u) = \bar g ^2(\mu)\Big|_{\bar g ^2(\mu/s) = u}\,,\qquad
  (\mu = 1/L)\,.
\end{equation}
The values $s=2,3/2$ are typical in the literature. The advantage of the step
scaling function is that it can be easily computed on the lattice. We have to
recall that there is a one-to-one relationship between the bare parameters used
in a lattice simulation (i.e. $g_0^2,am_0$) and the lattice spacing $a$. In
massless schemes we simulate with $\bar m=0$, and the lattice spacing $a$ is set
solely by the value of the bare coupling $g_0^2$. This means that if we keep the
same bare simulation parameters and just multiply the number of lattice points
by a factor $s$, we will obtain a lattice estimate of the step scaling
function
\begin{equation}
  \label{eq:ssf_comp}
  L/a \to sL/a \Longrightarrow u=\bar g^2 \to \Sigma_s(u,a/L)\, .
\end{equation}
Of course the previous process still depends on the value of the cutoff $a$, but
this result can be repeated with several pairs of lattices in order to obtain a
continuum result
\begin{equation}
  \label{eq:sigma_cont}
  \lim_{a\to 0} \Sigma_s(u,a/L) = \sigma_s(u)\,.
\end{equation}

\begin{figure}
  \centering
  \includegraphics[width=\textwidth]{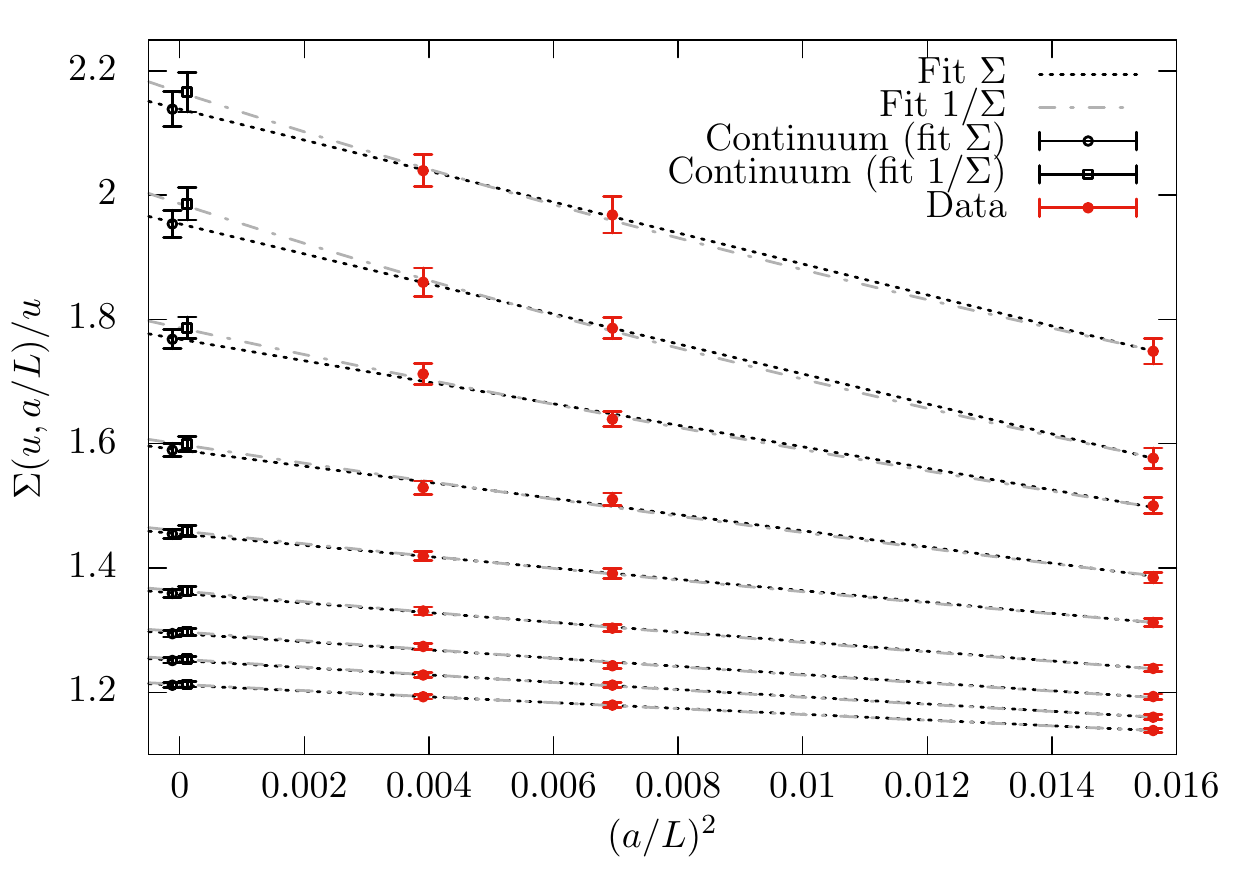} 
  \caption{The bare coupling $g_0^2$ is fixed on several lattice sizes
    $L/a=8,10,12,16$ so that the renormalized coupling is equal to $\bar g ^2 =
    u_i$ for nine values $u_i \approx 2.12, 2.39, 2.73, 3.20, 3.86, 4.49, 5.30,
    5.86, 6.54$. This means that all simulations have the same physical volume
    $L$, up to scaling violations. By computing the coupling on lattices twice
    as large, one determines a lattice approximation $\Sigma_s(u,a/L)$ of the
    step scaling function $\sigma_2(u)$. This plot shows the continuum
    extrapolation of the lattice step scaling function $\Sigma_s(u,a/L)$ (see
    Eq.~(\ref{eq:sigma_cont})). The continuum values can be used to parameterize
    $\sigma_2(u)$ for $u\in[2.12,6.54]$, or to determine the $\beta$-function.
    (Source~\cite{DallaBrida:2016kgh}).}
\label{fig:ssf}
\end{figure}

By repeating a series of continuum extrapolations at different values
of the coupling $u$, the function $\sigma_s(u)$ can be determined. 
Once the step scaling function is known, the RG problem is solved
non-perturbatively. Note that the relation
\begin{equation}
  \label{eq:basic_ssf}
  \log s = -\int_{\sqrt{u}}^{\sqrt{\sigma_s(u)}}\, \frac{{\rm d} x}{\beta(x)}\,,
\end{equation}
is exact. The determination of the $\Lambda$-parameter
Eq.~(\ref{eq:lam}) uses the previous relation to break the fundamental
integral of the $\beta$-function 
\begin{equation}
  \label{eq:basic_int}
    \int_{0}^{\bar g(\mu_{\rm had})}\, \frac{{\rm d}x}{\beta(x)}\,,
\end{equation}
into several pieces. The scale $\mu_{\rm had}$ is a low-energy scale, of the
same order as the reference scale used to ``set the scale'', as discussed in
section~\ref{sec:scale-sett-latt}. With the help of the step-scaling function
one can produce a series of couplings $\bar g_k$ such that 
\begin{equation}
  \bar g_0^2 \equiv \bar g^2 (\mu_{\rm PT});\qquad
  \bar g_{k-1}^2  = \sigma_s(\bar g^2 _{k})\,.
\end{equation}
Note that since the step scaling function changes the scale by a factor $s$, we
have $\bar g^2 _k = \bar g^2 (s^{k}\mu_{\rm had})$. Recalling the basic relation
Eq.~(\ref{eq:basic_ssf}) the integral of Eq.~(\ref{eq:basic_int}) can now be
written as
\begin{equation}
  \label{eq:split_int}
  \int_{0}^{\bar g(\mu_{\rm had})}\, \frac{{\rm d}x}{\beta(x)} =
  \int_{0}^{\bar g(s^n\mu_{\rm had})}\, \frac{{\rm d}x}{\beta(x)} +
  n\log s\,.
\end{equation}
With moderate values of $n$ (\ie $\, n\sim 10$) one can reach high energy scales, since
$s^n\mu_{\rm had}\sim 100$ GeV. The remaining integral in
Eq.~(\ref{eq:split_int}) can be very well approximated by using perturbation
theory (cf. 
section~\ref{sec:syst-extr-alph}).
\begin{equation}
  \int_{0}^{\bar g(s^n\mu_{\rm had})}\, \frac{{\rm d}x}{\beta(x)}
  \simas{n \to \infty}
  \int_{0}^{\bar g(s^n\mu_{\rm had})}\, \frac{{\rm d}x}{\beta_{\rm
      PT}(x)} + \dots\,.
\end{equation}
The scale $s^n\mu_{\rm had}$ is the scale at which the result is
matched with perturbation theory
\begin{equation}
  \label{eq:ston}
  \mu_{\rm PT} = s^n\mu_{\rm had}\,.
\end{equation}
One of the big advantages is that
finite size scaling allows to vary this scale substantially: it can be
pushed to very large values with a moderate computational effort. 
The $\Lambda$-parameter can be determined by matching with
perturbation theory at different energy scales. We have already seen
an example of such analysis in figure~\ref{fig:msbar}.

\bigskip
\noindent
{\em Direct determination of the $\beta$-function}
\medskip

Recent works use the step scaling function to directly determine the
$\beta$-function. This approach has some technical advantages like making the
determination of arbitrary ratios of scales $\mu_1/\mu_2$ possible, or allowing
different scales factors $s$ to be used to fix the same set of coefficients.
Several parametrizations are possible. At high energies the most natural one is
just to write
\begin{equation}
  \beta(x) = -x^3\sum_{n=0}^N b_nx^{2n}\,,
\end{equation}
with the first few coefficients fixed by the perturbative
prediction. In this case, once the $\beta$-function is known, one can
use 
\begin{equation}
  \label{eq:ratio_beta}
  \frac{\mu_{\rm PT}}{\mu_{\rm had}} = \exp \left\{ -\int_{g(\mu_{\rm PT})}^{g(\mu_{\rm had})}\, \frac{{\rm d} x}{\beta(x)} \right\}
\end{equation}
to connect the perturbative and hadronic scales similarly to
Eq.~(\ref{eq:ston}).  

Another advantage is that it allows a direct
comparison of the data with perturbation theory by comparing the
non-perturbatively determined $\beta$-function and its perturbative
expansion (see~\cite{Brida:2016flw, DallaBrida:2016kgh} for some
examples).  

\subsubsection{Matching to an experimental quantity}
\label{sec:match-an-exper}

In the previous section we have seen how finite size scaling techniques are able
to solve non-perturbatively the RG equations: ratios of scales defined
in a massless scheme, like
$\mu_{\rm PT}/\mu_{\rm had}$ can be determined precisely. But what we
really need are the values of these scales in physical units.
Following the discussion of section~\ref{sec:scale-sett-latt}, we need
to determine $\mu_{\rm had}/M_{\rm ref}$, where $M_{\rm ref}$ is a
reference scale (for example the $\Omega$ mass) determined in large
volumes and with physical values of the quark masses (\ie\ in a setup
that can be matched with an experimental input).

In detail, the procedure is as follows: first one fixes the value of
the coupling in a given massless finite volume renormalization scheme
to some particular value 
\begin{equation}
  \bar g ^2(\mu_{\rm had}) = \text{fixed}\,,
\end{equation}
for several values of $L/a$. Since the coupling depends only
on one scale $\mu_{\rm had}\propto 1/L_{\rm had}$, the physical volume
of all these simulations is the same, up to scaling violations
(i.e. the different values of $L/a\propto 1/(a\mu)$ are really
different values of $a$ at the same $L=L_{\rm had}$).

Second, one determines in large volume simulations the value of some low-energy
scale \emph{at the same values of the bare coupling $g_0$} and for physical
values of the quark masses. This low energy scale is typically whatever
reference quantity is used to set the scale. For example let us assume that it
is the mass of the $\Omega$ baryon $M_{\Omega}$. The large volume lattice
simulations yield values of the mass in lattice units, we denote this
dimensionless quantity $\hat{M}_{\Omega}(a)$ where we have written explicitly its
dependence on the lattice spacing $a$. Since the bare coupling has been kept the
same, the values of $a$ in our determination of $a\mu_{\rm had} = a/L_{\rm had}$
are the same, up to scaling violations, as the values of $a$ in our
determinations of $\hat{M}_{\Omega}(a)$. This allows to determine the ratio
\begin{equation}
  \frac{\mu_{\rm had}}{M_{\Omega}} = \lim_{a\to 0} \frac{a\mu_{\rm
      had}}{\hat{M}_{\Omega}(a)}\,. 
\end{equation}
Note that the fact that different values of the quark masses or physical volumes
are used for the determination of $a\mu_{\rm had}$ and $aM_{\Omega}$ is not an
issue once the continuum extrapolation is performed~\cite{Luscher:1996sc}\footnote{If one uses a
fermionic action that violates chiral symmetry, like Wilson fermions, the
scaling violations might naively be $\mathcal O(a)$, unless the bare coupling
$g_0$ is shifted by an term $\propto am_q$. This is a technical issue only of
practical importance, that does not change the validity of the above
statements.}. Once this ratio is known in the continuum, the experimental value
of $M_{\Omega}$ can be used to determine $\mu_{\rm had}$ in physical units
\begin{equation}
  \mu_{\rm had} = M_{\Omega}^{\rm exp} 
  \times \frac{\mu_{\rm had}}{M_{\Omega}}\,.
\end{equation}

We suggest the reader to look again to Fig.~\ref{fig:fss_scheme} for a schematic
summary of the full procedure: the $\Lambda$-parameter, and therefore the strong
coupling constant, is determined from just {\em one} experimental dimensionful
quantity (like $M^{\rm exp}_\Omega$). Perturbation theory is only needed at
scales larger than $\mu_{\rm PT} = s^n\mu_{\rm had}$. This scale can be made
(almost) arbitrarily large with a modest (but dedicated) computational effort.


\section{Present and future of lattice determinations of $\alpha_s$}
\label{sec:pres-future-latt}

\begin{figure} \centering
\includegraphics[width=\textwidth]{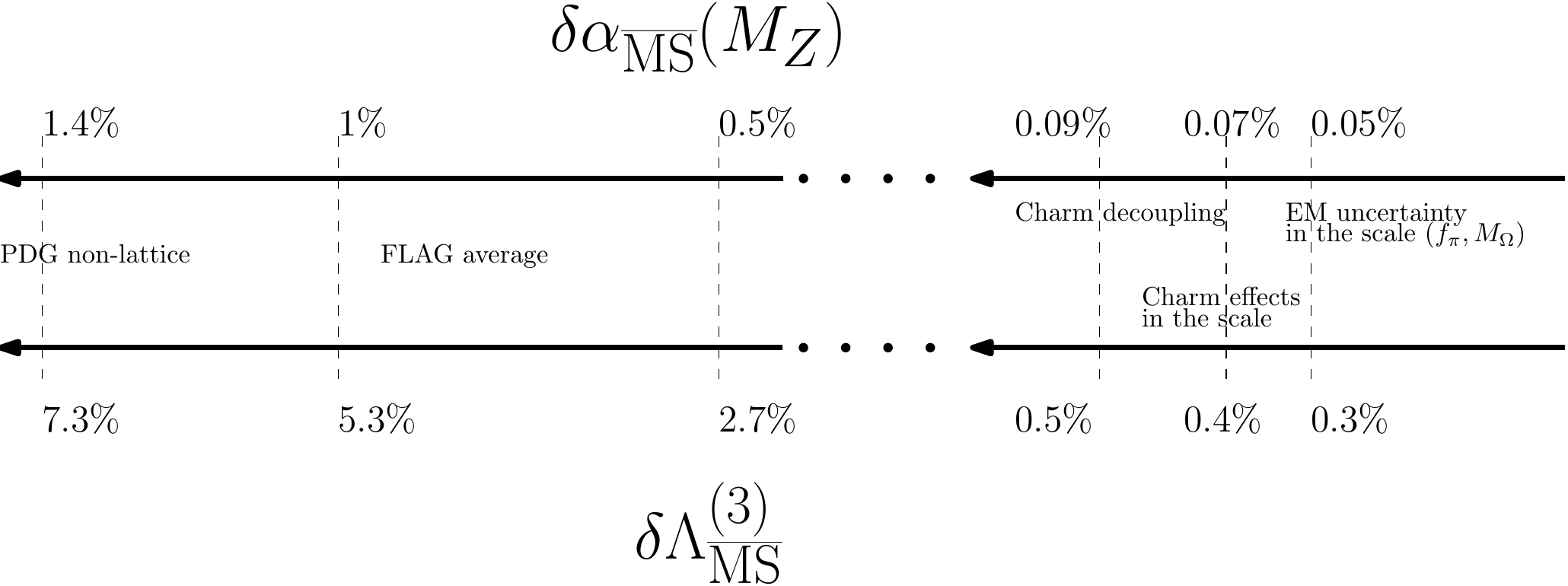}
  \caption{Current uncertainty in the determination of the strong
coupling. A 5\% uncertainty in the three flavor $\Lambda$ parameter
translates in a sub-percent precision in the strong coupling at the
electroweak scale. Electromagnetic uncertainties in the scale
(section~\ref{sec:scale-sett-latt}) and charm quark effects
(section~\ref{sec:decoupling}) are far from being the limiting factor
in the precision (note the gap in the axes). See text for more
details. }
  \label{fig:lamerr}
\end{figure}

In this review we have focused on the determination of the $\Lambda$-parameter,
which in our opinion should be considered the fundamental parameter from which
the strong coupling constant can be deduced. From our point of view it has
several conceptual advantages, first and foremost the fact that it is defined
non-perturbatively, and therefore allows a clean separation of the role of
perturbation theory. Moreover since $\Lambda$ has units of mass, its connection
with scale setting is transparent. Many determinations of the strong coupling
claim a precision below 1\%. This is impressive, but if translated to an
uncertainty in the $\Lambda$-parameter the achievement looks quite modest
compared with other state of the art lattice QCD computations: a 5\% uncertainty
in $\Lambda ^{(3)}_{\overline{\rm MS} }$ is enough to achieve a 1\% precision in
$\alpha_{\overline{\rm MS} }(M_Z)$. In an era where many lattice computations of
decay constants, the hadron spectrum, or the anomalous magnetic moment of the
muon reach a sub-percent precision, the current $\approx 5\%$ uncertainty in the
$\Lambda$-parameter is a clear testimony to the difficulties involved in solving
this multi-scale problem.

As a consequence many of the fine details needed in current state of the art
lattice QCD computations are not crucial for improving the current determination
of $\Lambda_{\overline{\rm MS}}$. Let us illustrate this point in detail by
discussing the effect of electromagnetic corrections and massive sea quarks.
\begin{description}
\item[Electromagnetic effects] They enter in any determination of the
strong coupling in two different ways. 

  First there are the electromagnetic effects in whatever quantity is
  used to set the scale. If the pion decay constant is used to set the
  scale, one can claim that these effects are of the order of $0.3\%$
  in $f_\pi$ (see discussion in
  section~\ref{sec:scale-sett-latt}). The size of this effect has been
  recently confirmed by numerical QCD+QED
  computations~\cite{DiCarlo:2019thl} (although with a smaller
  precision).The other popular choice for 
  scale setting is the $\Omega$ baryon mass.  Electromagnetic
  corrections in the $\Omega$ baryon have been computed recently via
  lattice QCD simulations and found to be significantly below
  0.3\%~\cite{Blum:2018mom}. Since the relative error in the scale
  propagates into an equal relative error in $\Lambda$, the effects of
  the EM uncertainties in the scale will only affect the strong
  coupling at the $0.05\%$ level: well below the precision of any
foreseeable result in the near future.

  The second point where electromagnetic corrections affect the
  lattice determinations of the strong coupling is in the actual
  running. The running of the strong coupling is affected by the fact
  that quarks are electrically charged and couple to photons. These
  effects are estimated in perturbation theory (see for
  example~\cite{Mihaila:2012bt} for a calculaiotn up to three loops)
  with the result that the leading effect is an
  $\mathcal O(\alpha_s\alpha_{EM})$ term that can be interpreted as a
  correction in the leading $b_0$ term of the $\beta$ function. This
  correction is below the percent level and translates in a per-mille
  effect in $\Lambda$, \ie\ a completely negligible effect in
$\alpha_s(M_Z)$

\item[Charm sea quark effects] Some lattice collaborations nowadays include a
dynamical charm quark in their simulations (these simulations are usually
labeled as $N_{\rm f}=2+1+1$), while other collaborations still ignore the
effects of a dynamical charm quark ($N_{\rm f} =2+1$)~\footnote{Note that it is
not clear what setup is better \emph{at the current values of the simulation
parameters}. A dynamical charm quark has the unpleasant effect of enhancing
cutoff effects due to the fact that $am_c$ is typically not very small at the
simulated lattice spacing.}. A naive estimate of the contribution of charm quark
loops can be obtained from the large-$N_c$ counting rules, which suggest that
quark loop effects are $1/N_{\rm c}$ suppressed. Moreover decoupling arguments
show that these are further suppressed by a factor $(E/M_c)^2$ where $E$ is the
typical energy scale of an hadronic quantity, and $M_c$ is the RGI-invariant
charm quark mass. For the sake of the argument we will take here $E\approx 0.4$
MeV (this corresponds to the energy scale of a typical hadronic scale like $r_0,
\sqrt{8t_0}\approx 0.5$ fm). Putting these pieces together we
obtain
\begin{equation}
  \frac{1}{N_{\rm c}} \times \left( \frac{E}{M_c} \right)^2
  \longrightarrow 3\%\, \text{effect}\,.
\end{equation}
One can argue that there is an additional suppression by a
factor $\alpha_{\overline{\rm MS} }(M_c)$. In this case the overall effect will
be $\mathcal O(1\%)$. 

Either way, we have strong evidence that charm quark loop effects are
dynamically suppressed, resulting in overall effects much smaller than
the uncertainty quoted above.
\begin{itemize}
\item The FLAG report does not find any significant difference
  between lattice $N_{\rm f} = 2+1$ and $N_{\rm f} = 2+1+1$
  computations in several quantities (decay constants, quark masses,
  \dots)~\cite{Aoki:2019cca}. Some of these computations agree with a
  sub-percent precision.

\item Charm quark effects in dimensionless ratios of gluonic
  quantities (ie\ $r_0/\sqrt{t_0}, w_0/r_0, \dots$) have been
  estimated recently~\cite{Bruno:2014ufa} by comparing $N_{\rm f} = 2$
  lattice simulations with two heavy quarks and the pure gauge theory. 
  The measured effect is below the 0.4\% level.

\end{itemize}

The previous considerations apply directly to the case of the determination of
the strong coupling. The  relative error in the scale (\ie\  $r_0, t_0, f_\pi,
\dots$) propagates linearly into $\Lambda$, and these effects are well below the
$0.1\%$ uncertainty in $\alpha_s(M_Z)$.  

Charm quark effects also affect the decoupling relations used to
translate
$\Lambda_{\overline{\rm MS}}^{(3)} \to \Lambda_{\overline{\rm
    MS}}^{(4)} \to \Lambda_{\overline{\rm MS}}^{(5)}$, which are
needed to quote the value of the strong coupling at the electroweak
scale. The perturbative relations have non-perturbative corrections
$\mathcal O(\Lambda^2/M_c^2)$, and these effects have been recently
estimated to be about $0.2\%$~\cite{Athenodorou:2018wpk}.  This
estimate is a result of comparing $N_{\rm f} =2$ lattice simulations
with two heavy quarks and the pure gauge theory ($N_{\rm f} =0$). The
same reference adds a factor two to this estimate, resulting in a
conservative uncertainty in $\Lambda$ of about $0.4\%$. Again, this is
well below our current precision.
\end{description}

In summary, it is useful to keep in mind that the precision in lattice
determinations of the strong coupling is limited for reasons that are
very different from the ones that limit other current lattice QCD
computations.

\begin{figure}[t]
  \centering
\includegraphics[width=\textwidth]{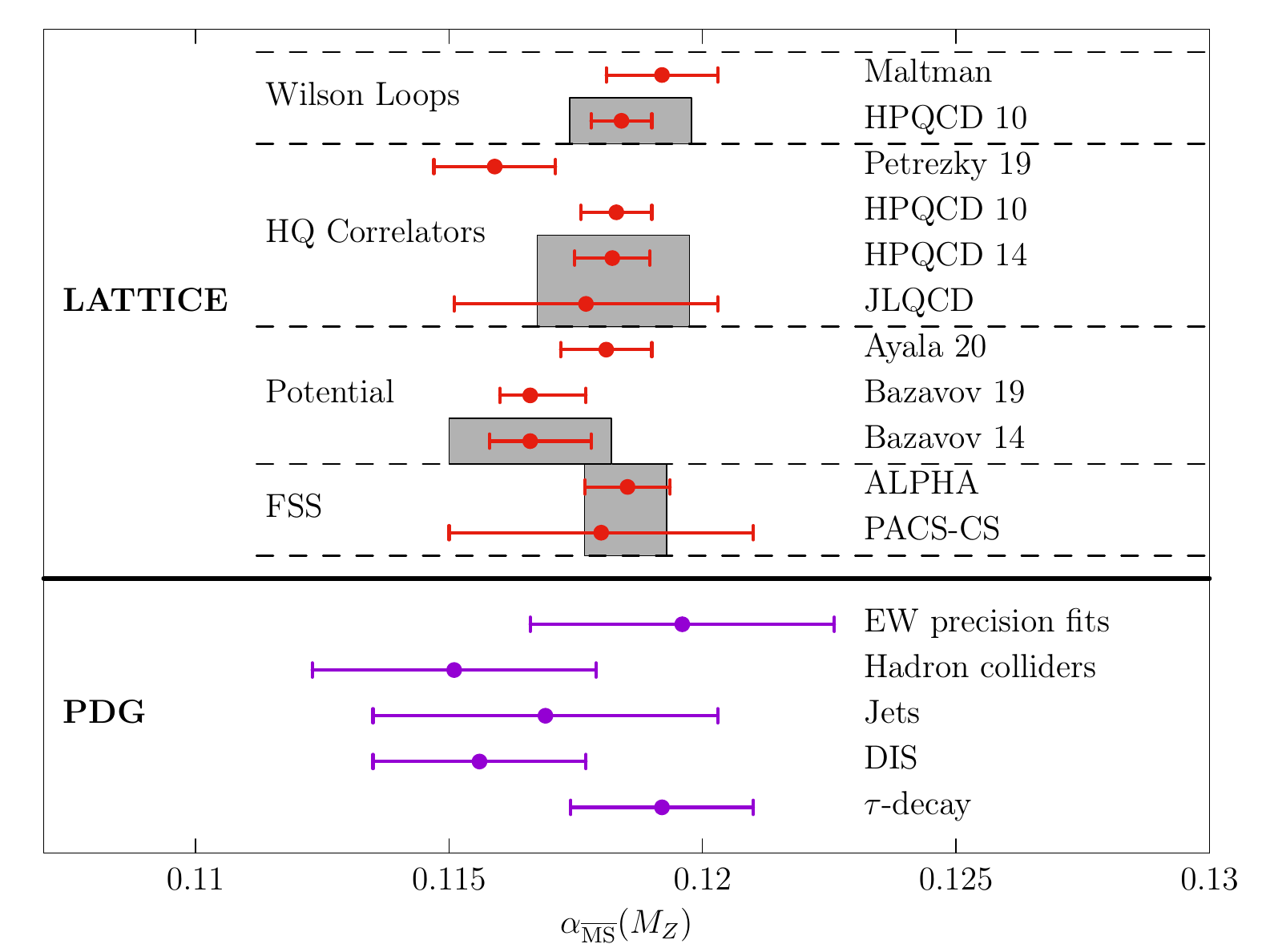}
  \caption{Summary of lattice and non-lattice result for the strong coupling
  using different methods. For non-lattice results we use the PDG
  averages~\cite{Patrignani:2016xqp}. For the lattice results we quote the works
  that enter in the FLAG average~\cite{Aoki:2019cca} and the recent updates that
  have been published after the FLAG review. The gray boxes are the FLAG average
  for each method. The recent updates (Petrezcky 19~\cite{Petreczky:2019ozv} and
  Bazavov 19~\cite{Bazavov:2019qoo}) are not included in the current FLAG
  averages because they were published \emph{after} the FLAG update. See text
  for more details. 
(References: Maltman~\cite{Maltman:2008bx}, HPQCD
10~\cite{McNeile:2010ji}, Petrezky 19~\cite{Petreczky:2019ozv}, HPQCD
14~\cite{Chakraborty:2014aca}, JLQCD~\cite{Nakayama:2016atf}, Ayala
20~\cite{Ayala:2020odx}, Bazavov
19~\cite{Bazavov:2019qoo}, Bazavov 14~\cite{Bazavov:2014soa},
ALPHA~\cite{Bruno:2017gxd}, PACS-CS~\cite{Aoki:2009tf}). 
} 
  \label{fig:summary}
\end{figure}

In figure~\ref{fig:summary} we summarize the current status of the
lattice and non-lattice determinations of the strong coupling. For the
case of the non-lattice determinations we use the
PDG~\cite{Patrignani:2016xqp} averages, which include phenomenological
determinations from $\tau$-decays, DIS, QCD Jets, hadron collider data
and electroweak precision fits. For the lattice results we use the
FLAG averages~\footnote{It is important to point out that each data
point by the PDG in figure~\ref{fig:summary} is the average of several
works. Some of these individual works claim smaller errors than the
average. This is very similar to the situation in the lattice methods
despite the fact that the procedure for averaging is 
very different. In this review we will just take the averages at face
value, leaving the reader with the task of making their own opinion
on each averaging procedure.}.

Without duplicating the detailed work done in the FLAG review, in what follows
we shall try to assess the limiting factors in the different extractions of the
strong coupling. The first naive observation is that figure~\ref{fig:summary}
does not include lattice results for some of the methods reviewed in
section~\ref{sec:conv-observ-coupl}. Let us then begin by discussing these
methods.  

\subsection{Methods not entering the FLAG average}
\label{sec:methods-not-entering}

FLAG applies several quality criteria to determine which works enter in the
average. These quality criteria aim at covering all the possible sources of
systematic effects in the calculation. Data sets that are unlikely to allow for
a reasonable control of systematic effects are excluded from final averages. The
methods that do not qualify to enter in the final average are discussed in this
subsection. 

\begin{description}
\item[QCD vertices] There are three lattice works in the latest FLAG review that
extract the value of the strong coupling from QCD
vertices~\cite{Blossier:2011tf,Blossier:2012ef,Blossier:2013ioa}. In all cases
the data set does not allow for a controlled continuum extrapolation, which
prevents these works from contributing to the average. Another
work~\cite{Zafeiropoulos:2019flq} has appeared since the FLAG review was
published, but the extraction of the strong coupling is still done from
simulations at a single lattice spacing.

The limiting factor in these determinations is the ability to
reach energy scales that are sufficiently high to make contact with perturbation
theory while having the continuum extrapolation under control.

\item[HVP] Two works extract the strong coupling from the vacuum
polarization~\cite{Shintani:2010ph,Hudspith:2018bpz}. Nevertheless these works
do not qualify for the FLAG average because the continuum extrapolation is not
under control.

{Here we face the same issues: the difficulty to reach the
perturbative region while having the continuum extrapolation under
control.}

\item[Eigenvalues of the Dirac operator] There is only a single work that
determines the strong coupling from the eigenvalues of the Dirac
operator~\cite{Nakayama:2018ubk}. The extraction is performed at a very low
energy scale, where $\alpha\approx 0.6 - 0.4$. According to the FLAG criteria,
this does not allow to control the matching with perturbation theory.
\end{description}

All in all, these methods fail to show convincing evidence for a safe contact
with the asymptotic perturbative behavior and/or fail to show that the continuum
limit can be reached at the energy scales needed to extract the value of
$\alpha_s$. At the time of writing, we think that it would be better to use the
value of the strong coupling as an input to investigate the physics related with
these phenomena, instead of using these physics effects to perform a precision
determination of the fundamental parameters of the SM. Of course this situation
might change in the future. More powerful computers might allow to reach higher
energies. Eventually a safe contact with perturbation theory in the continuum
can lead to precise values of $\alpha_s$, but this is not the case at the
moment. 

\subsection{Methods that enter the FLAG average}
\label{sec:methods-that-enter}

Next we will comment critically on the methods that actually enter in the final
FLAG average (see Fig.~\ref{fig:summary}). These methods show compelling
evidence of producing robust results in the continuum limit (by using several
lattice spacing to extrapolate their results) and reach energy scales where the
perturbative matching is convincing. 

A crucial point in the following discussion is the estimate of the
truncation error, \ie\ the uncertainty that is introduced in the
determination of the coupling by working at a given order in
perturbation theory. As we shall see below, this is the main source of
uncertainty in most of the determinations of the strong
coupling~\footnote{The FLAG estimate of the truncation uncertainties
  for the static potential, heavy quark correlators and small Wilson
  loops determinations are larger that the uncertainties quoted by
  some of the works that enter in the average, in some cases by more
  than a factor two. This has in fact produced some controversies (see
  for example~\cite{dEnterria:2015kmd}).}. Here we will look at these
uncertainties in detail using the scale variation method. In
particular we will examine how a given measurement of an observable in
three flavor QCD at a physical scale $Q$ produces different
determinations of $\alpha_ {\overline{\rm MS} }(M_Z)$ when the ratio
between the renormalization and physical scales is varied. Without
loss of generality, we will assume that the observable is determined
in $N_{\rm f} = 3$ QCD and normalized
as a coupling (as discussed at the beginning of
section~\ref{sec:conv-observ-coupl}) 
\begin{equation}
  \label{eq:Pinpt}
  \alpha(Q) = \alpha_{\overline{\rm MS} }(sQ) + \sum_{k=2}^n c_k(s)\alpha_{\overline{\rm MS} }^k(sQ)\,.
\end{equation}
The procedure that we employ to estimate the uncertainty in our
determination of $\alpha_{\overline{\rm MS}}(M_Z)$
is detailed in appendix~\ref{sec:scale-vari-estim}. 
The reader interested in numerical values should consult the freely
available package~\url{https://igit.ific.uv.es/alramos/scaleerrors.jl}. 
In summary we proceed as follows:
\begin{enumerate}
\item Fix the value $\alpha(Q)$ using a canonical value
  $\Lambda^{(3)}_{\overline{\rm MS} } = 341\, {\rm MeV}$ (equivalent
  to $\alpha_{\overline{\rm MS} }(M_Z) \approx 0.1185$\footnote{Note that since we assume that
    the observable is determined in three-flavor QCD, we need to cross
    the charm and bottom thresholds.}) for some value of the
  scale $\mu=s_{\rm ref} Q$. This is achieved by {\em
    reverse engineering}, \ie\ by computing
\begin{equation}
    \alpha(Q) = \alpha_{\overline{\rm MS} }(\mu) + 
    \sum_{k=2}^n c_k(s_{\rm ref})\alpha_{\overline{\rm MS} }^k(\mu)\,,
  \qquad (\mu = s_{\rm ref}Q)\,.
\end{equation}
  where $\alpha_{\overline{\rm MS} }(\mu)$ is the value of the three
  flavor coupling at scale $s_{\rm ref}Q$ obtained from
  $\Lambda^{(3)}_{\overline{\rm MS} } = 341$ MeV.
  
\item Use Eq.~(\ref{eq:Pinpt}) again, in order to determine the values of
  $\alpha_{\overline{\rm MS} }(sQ)$, by solving
  \begin{equation}
    \alpha(Q) = \alpha_{\overline{\rm MS} }(sQ) +
    \sum_{k=2}^n c_k(s)\alpha_{\overline{\rm MS} }^k(sQ)\,.
  \end{equation}
  at the values $s = s_{\rm ref}/2, 2s_{\rm ref}$.
  
\item Use the 4- and 5-loop $\beta$-function to run the values of
  $\alpha_{\overline{\rm MS} }(sQ)$ obtained in step 2. to the
  reference scale $M_Z$ (crossing the charm and bottom thresholds).  A
  comparison between the values of 
  $\alpha_{\overline{\rm MS}}(M_Z)$ is a measure of the truncation
  uncertainty due to the scale variation.
\end{enumerate}

\begin{table}[t!]
\centering
\begin{tabular}{lllcrrr}
  \toprule
Observable & loops &\(Q\) [GeV] & FLAG error [\%] & $\delta_{(4)}^\star\, [\%]$ & $\delta_{(2)}\, [\%]$& \(\delta_{(2)}^\star\, [\%]\)\\
\midrule
          &   & 1.5 & 1.4 &      &  2.6  &  2.7\\
Potential & 4 & 2.5 &        & 0.9  &  1.5  &  1.5\\
          &   & 5.0 &        & 0.4  &  0.8  &  0.8\\
\midrule
HQ \(r_4\) & & \(m_{\rm c}\) & 1.3 &    &   2.7  &  2.8\\
HQ \(r_4\) & 3  & \(2m_{\rm c}\) &       &1.2 &   1.5  &  1.6\\
HQ \(r_6\) & & \(2m_{\rm c}\) &       &    &   2.3  &  1.2\\
HQ \(r_8\) & & \(2m_{\rm c}\) &       &    &   2.8  &  4.8\\
\midrule
$- \log W_{11}$      & 3 & 4.4 & 1.0 & 2.8 &   3.3  &  2.5 \\
$-\log W_{12}/u_0^6$ &   &4.4 &        & 3.5 &   3.2  &  3.1 \\
\midrule
  FSS & 3 & 80 &  & 0.1  &  0.2  &  0.2\\
  \bottomrule
\end{tabular}
\caption{\label{tab:scale_truncation}
Summary of truncation uncertainties on $\alpha_{\overline{\rm MS} }(M_Z)$ estimated by varying the scale. We compare the error quoted by flag with a change of scale by factors two and four around \(s=1\) or \(s=s^\star\) (the value of fastest apparent convergence). 
For each method we quote the number of known loops in the relation
between the observable and the $\overline{\rm MS} $ scheme according
to h counting in section~\ref{sec:conv-observ-coupl}.
Details on the different types of extractions can be found in
section~\ref{sec:static-potential-1} (potential),
section~\ref{sec:heavy-quark-corr-1} (HQ),
section~\ref{sec:observ-at-cutoff} ($\log W_{11}, \log W_{12}$) and
section~\ref{sec:finite-size-scaling-1} (FSS).}
\end{table}

The usual procedure in phenomenology is to vary the renormalization
scale by a factor 2 above and below some reference scale. Estimates of
the truncation uncertainties that use renormalization scales below 1
GeV tend not to be reliable. Since many extractions of the strong
coupling are performed at relatively low scales, the above mentioned
procedure might lead to unreasonable uncertainties. For this reason,
the uncertainty resulting from comparing the change
$s_{\rm ref}\to 2s_{\rm ref}$ provides complementary information on
the size of the truncation uncertainties, specially if one explores
the dependence with $s_{\rm ref}$ in a range $1-2$. In order to get a
more quantitative understanding of these effects, we will use the
following quantities.
\begin{description}
\item [$\delta_{(4)}(s_{\rm ref})$:] Change the renormalization scale by a factor
  two above and below some reference scale $s_{\rm ref}Q$.  We quote a symmetric
  error by averaging the difference between the scales $s_{\rm ref}Q$ and
  $2s_{\rm ref}Q$, and the difference between the scales $s_{\rm ref}/2Q$ and
  $s_{\rm ref}Q$. Note however that in some cases the error is markedly
  asymmetric. 
     
\item[$\delta_{(2)}(s_{\rm ref})$:] Change the renormalization scale by a factor
  two above the reference scale $s_{\rm ref}Q$ only. 
  
\end{description}

We will show explicitly in the computations below how the two measures
$\delta_{(4)}(s_{\rm ref})$ and $\delta_{(2)}(s_{\rm ref})$ depend on
the choice of $s_{\rm ref}$. In principle any number $\mathcal O(1)$ is a
reasonable choice for $s_{\rm ref}$, but there can be significant
differences in the results depending on its actual value. For this
reason we will explore two common choices.
\begin{itemize}
\item Take $s_{\rm ref} = 1$. 
  \ie\, the renormalization and physical scales are the same. 
  In this case the uncertainties will be labeled $\delta_{(2)}, \delta_{(4)}$.
\item Take $s_{\rm ref} = s^\star$ as \emph{the value of fastest apparent
    convergence}. This value is determined with the condition
  \begin{equation}
    c_2(s^\star) = 0
  \end{equation}
  \ie\ the NLO coefficient in the relation between $\alpha(Q)$ and
  $\alpha_s$ vanishes (see Eq.~(\ref{eq:Pinpt})).
  In this case the uncertainties will be labeled $\delta^\star_{(2)},
  \delta^\star_{(4)}$.
\end{itemize}

A summary of the results is presented in table~\ref{tab:scale_truncation}. One
can readily see that the truncation uncertainties obtained with this method are
in the same ballpark as the FLAG uncertainties, except for the case of
the Wilson loops, where our estimates are substantially larger. Once
again we would like to end 
with a warning: estimates of the truncation uncertainties based on the scale
variation can (and have been shown to) fail in some cases (see discussion in
section~\ref{sec:syst-extr-alph} and specifically figure~\ref{fig:msbar}).

\subsubsection{Finite size scaling}
\label{sec:finite-size-scaling-1}

\begin{figure}[t]
  \centering
      \begin{subfigure}[t]{0.48\textwidth}
    \centering
    \includegraphics[width=\textwidth]{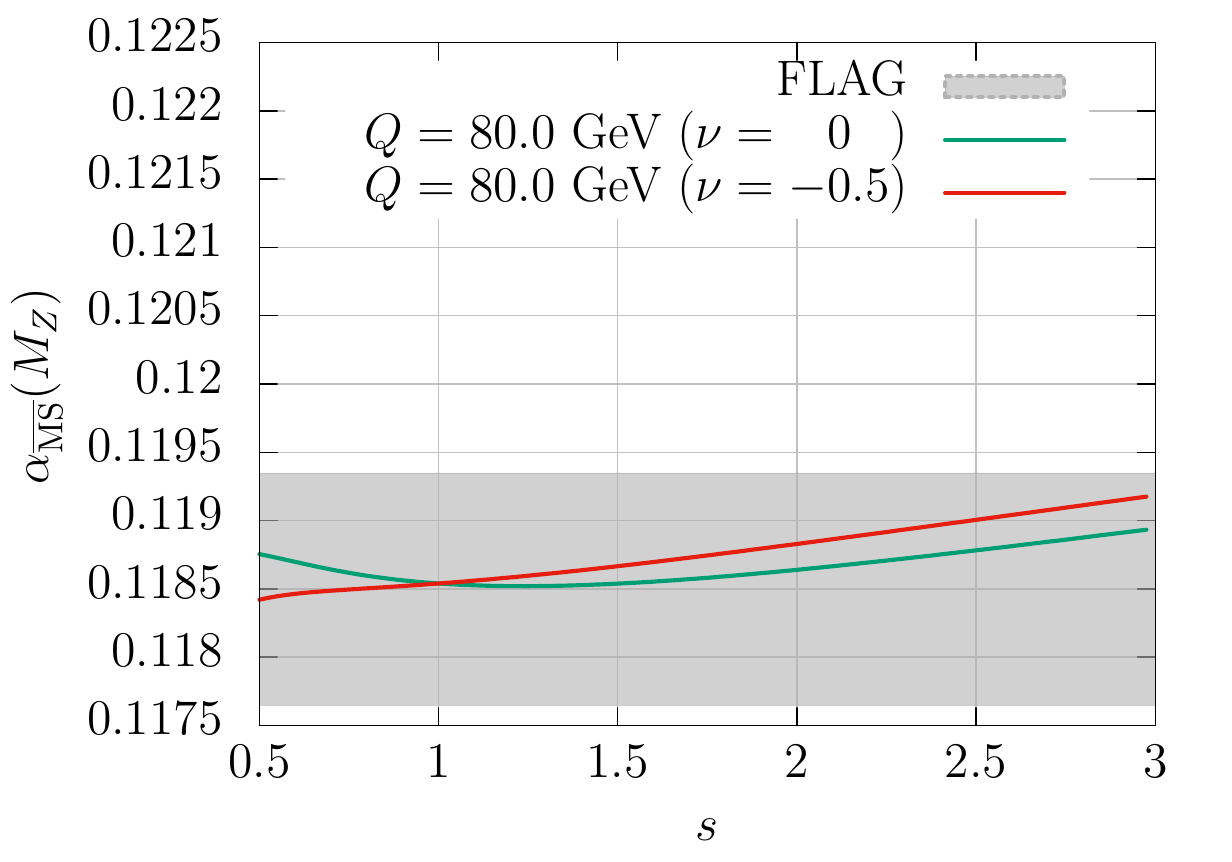}
    \caption{}
  \end{subfigure}
  \begin{subfigure}[t]{0.48\textwidth}
    \centering
    \includegraphics[width=\textwidth]{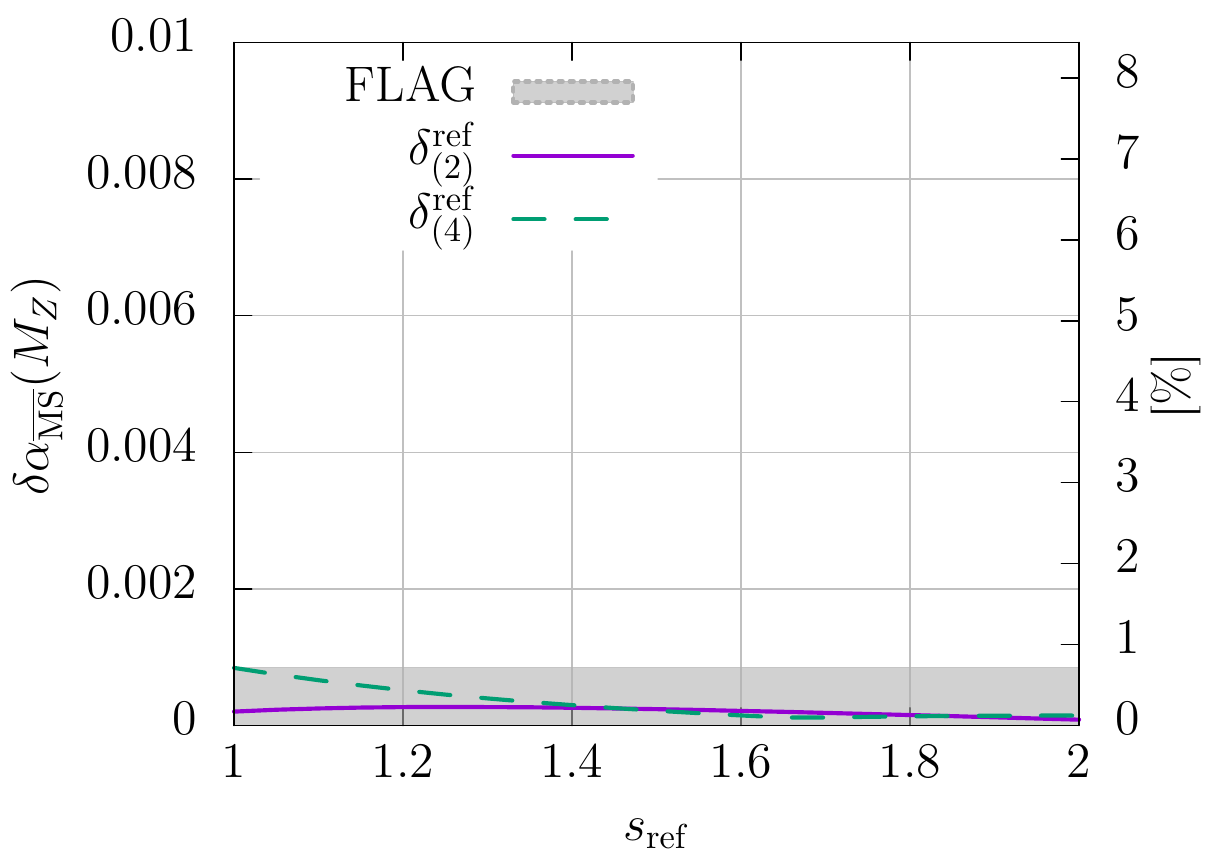}
    \caption{}
  \end{subfigure}
  \caption{Truncation effects in $\alpha_s$ extractions from finite
    size scaling. 
    Note that in this methods the extraction is performed at very high
    energies $Q\approx 80$ GeV. 
    The scale dependence of the coupling at $Q=8$ GeV in plot (a) is
    just for reference. 
    Plot (b) shows that truncation uncertainties are typically much
    smaller than the quoted uncertainties.} 

  \label{fig:fsstruncation}
\end{figure}

The FLAG average is the result of Refs.~\cite{Aoki:2009tf,Bruno:2017gxd}, wich
are in good agreement with each other. Perturbation theory is used at very high
energies ($Q \approx 80$ GeV), where perturbative estimates of the truncation
uncertainties are reliable. They affect the extraction of $\alpha_s$
only at the $0.1-0.2\%$ level (see figures~\ref{fig:fsstruncation}),  well below
the quoted uncertainties.  

The most recent work~\cite{Bruno:2017gxd} explores the dependence of
the extractions of $\Lambda$ on the physical energy scale over a large
range of energy scales $Q\sim 4 - 140$ GeV (see Figure~\ref{fig:msbar}
and the related discussion). This study compares their extraction of
the coupling $\alpha_s$ with the extraction performed with several
observables after extrapolating the renormalization scale at which
perturbation theory is used $\mu \to \infty$ (see discussion in
section~\ref{sec:syst-extr-alph}). All in all, results based on finite
size scaling do not depend on estimates of the perturbative
uncertainties done in perturbation theory, although
figure~\ref{fig:fsstruncation} show that they tend to be rather small,
as expected given that the electroweak scale is reached by
non-perturbative simulations. 
We should also point to the detailed study in~\cite{DallaBrida:2018rfy}.

The continuum extrapolation is the main source of systematic
uncertainty. {According to our
discussions in sections~\ref{sec:cont-limit-scale}
and~\ref{sec:finite-size-scaling}} the method allows to
to perform a controlled continuum extrapolation by using several
values of the lattice spacing at each energy scale. In particular, the
most delicate continuum extrapolations in~\cite{DallaBrida:2016kgh}
uses three  values of the lattice 
spacing that vary by a factor two at each value of the scale.

The statistical precision of the non-perturbative running is the main limiting
factor of these determinations. Note that coupling definitions using the
gradient flow -- see section~\ref{sec:theory-scales}
and~\ref{sec:finite-size-scaling} -- have a very small variance. These couplings
have been used in Ref.~\cite{Bruno:2017gxd}, but not in Ref.~\cite{Aoki:2009tf}
(these techniques were not known at the time), explaining the large difference
in the error between the two works: the result in~\cite{Aoki:2009tf} quotes a
2.5\% error in $\alpha_{\overline{\rm MS} }(M_Z)$, while the
result~\cite{Bruno:2017gxd} quotes a 0.7\% uncertainty. 
  
\subsubsection{Static potential}
\label{sec:static-potential-1}

\begin{figure}[t]
  \centering
      \begin{subfigure}[t]{0.48\textwidth}
    \centering
    \includegraphics[width=\textwidth]{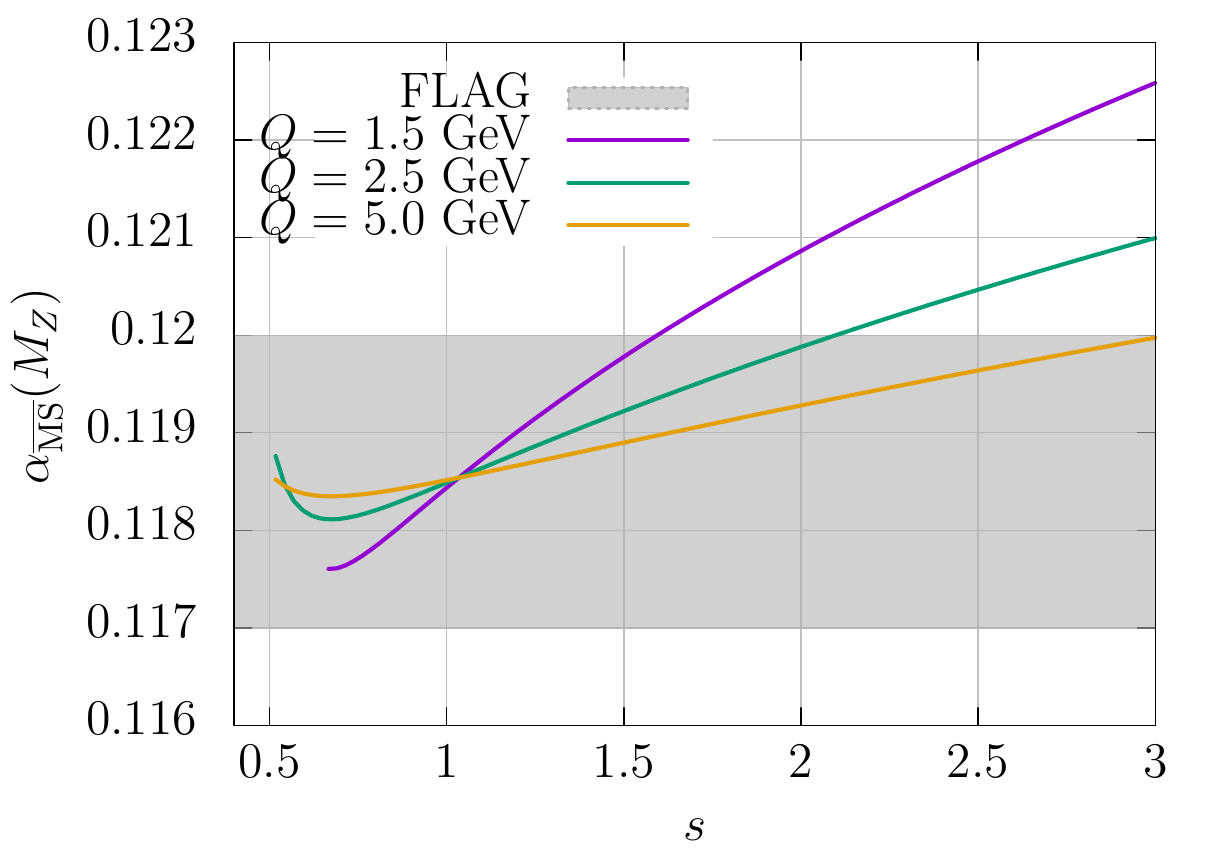}
    \caption{}
  \end{subfigure}
  \begin{subfigure}[t]{0.48\textwidth}
    \centering
    \includegraphics[width=\textwidth]{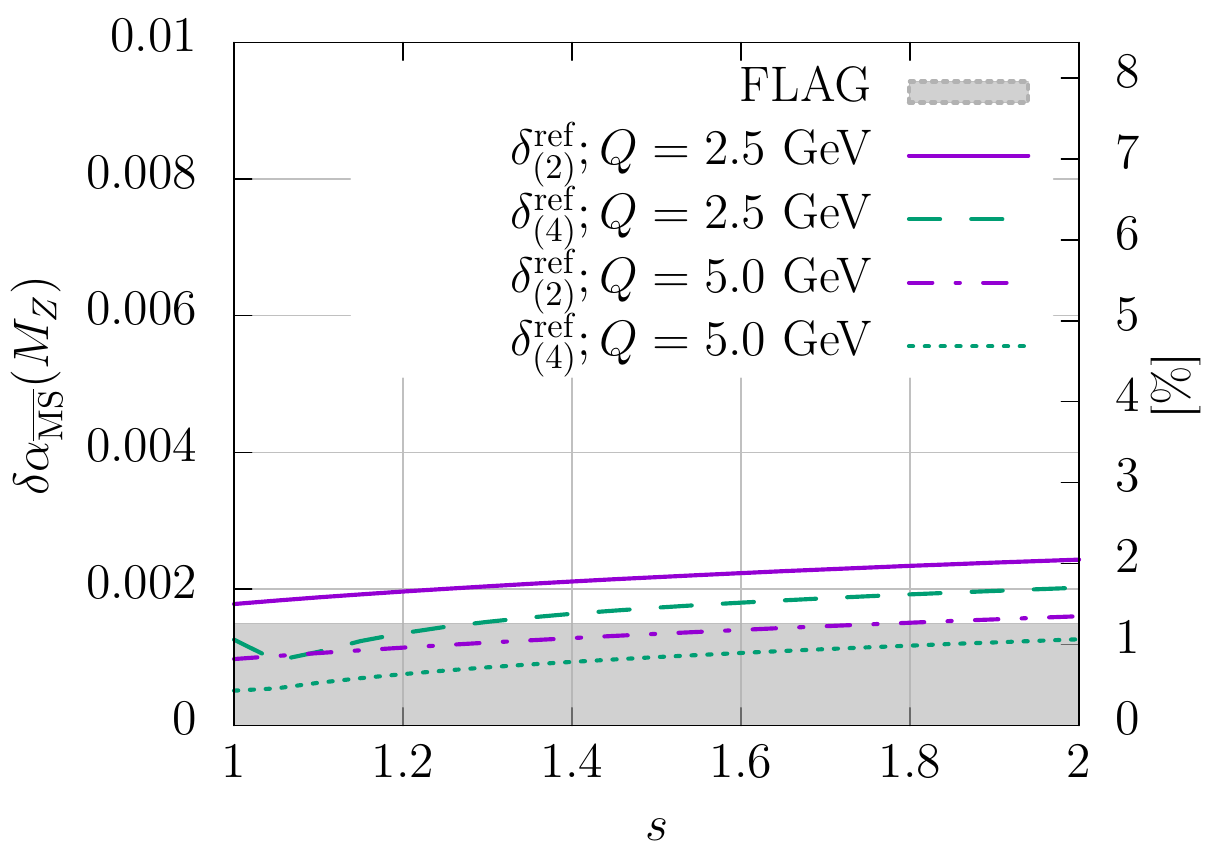}
    \caption{}
  \end{subfigure}
  \caption{Truncation effects in $\alpha_s$ extractions from the
    static potential. 
  Note that the high energy scale $Q=8$ GeV is only reached in the
  last work~\cite{Bazavov:2019qoo}. 
The FLAG error is determined for }
  \label{fig:pottruncation}
\end{figure}

The FLAG average is basically the result of Ref.~\cite{Bazavov:2014soa} with an
added uncertainty because of the perturbative truncation errors. Since the
publication of the FLAG report, two new determination of the strong coupling by
the same group has been published~\cite{Bazavov:2019qoo,
  Ayala:2020odx}. In these last works the
determination of the strong coupling is extracted at energy scales $Q\approx 2.5-8$
GeV\footnote{The same work also extract the
  strong coupling at shortes distances in a finite temperature setup,
  finding good agreement.}, improving significantly the previous
determination, which used $Q\in 
1.4-4.9$ GeV. 
Energy scales $\approx 5$ GeV can be reached with several values of
the lattice spacing. 
On the other hand the largest energy scales are reached at a single
lattice spacing, preventing a proper continuum extrapolation. 
The strong coupling is typically extracted from fits using 
energy scales about $2.5$ GeV and above. 

We should recall that the simulations at very fine lattice
spacing have a relatively small volume and are performed at fixed topology (see
discussion in section~\ref{sec:static-potential}), and in relatively
small physical volumes.

In what follows we will focus on the truncation uncertainties. 
In our analysis we ignore the logarithmic corrections due the IR
divergences of the static potential. This is partially justified: if
one takes the logarithms as constants, truncation uncertainties as
estimated below show only a mild variation (see appendix~\ref{sec:static-potential-app}). 
Nevertheless there is some concern here, since the natural scale at
which the terms $\log\alpha$ are evaluated (the \emph{ultra-soft}
scale) is significantly smaller (see discussion in
section~\ref{sec:static-potential}). 

Figure~\ref{fig:pottruncation} shows the scale dependence of the strong coupling
and the truncation errors $\delta_{(2)}^{\rm ref}, \delta_{(4)}^{\rm ref}$ as a
function of $s_{\rm ref}$. First it is important to point out that the error
estimate by FLAG, even if it comes from a completely different argument is in
the same ballpark as our estimates. For the relevant energy scales used in
present works~\cite{Bazavov:2014soa,Bazavov:2019qoo} ($Q\in 1.4-5$ GeV)
the perturbative estimates of the truncation uncertainties are in 
the range $\delta\alpha_{\overline{\rm MS} }(M_Z) \approx 2.6\%-0.8\%$, while
the FLAG estimate of the truncation uncertainties is 1.4\% (the most recent
work~\cite{Bazavov:2019qoo} had not appeared at the time the FLAG review was
published).

It is interesting to consider in detail the analysis of
Refs.~\cite{Bazavov:2014soa, Bazavov:2019qoo, Ayala:2020odx}.
\begin{description}
\item [Ref.~\cite{Bazavov:2014soa}:] The perturbative truncation error is
  estimated by changing the renormalization scale a factor $\sqrt{2}$ above and
  below the physical scale. Their result
  \begin{equation}
    \label{eq:pot_res1}
    \alpha_{\overline{\rm MS} }(M_Z) = 0.1166_{-0.0008}^{+0.0012}\,,
  \end{equation}
  has an uncertainty between $-0.7\%$ and $+1\%$, that is actually dominated by
  the perturbative truncation uncertainty. This uncertainty is smaller than the
  quoted uncertainty by FLAG ($1.4\%$). 
  
\item[Ref.~\cite{Bazavov:2019qoo}:] The estimate comes from a similar analysis,
  but this time the renormalization scale is varied a factor 2 above and below
  the physical scale. Moreover they also include in their estimate the effect of
  the different treatment of the logarithmic corretions in the perturbative
  expansion (that we have ignored in our analysis). This is very similar to our
  approach, and the uncertainty is similar to our quoted $\delta_{(4)}$ in
  table~\ref{tab:scale_truncation}, with the exception that they do not
  symmetrize the error. Their result
  \begin{equation}
    \label{eq:pot_res2}
    \alpha_{\overline{\rm MS} }(M_Z) = 0.1166_{-0.0006}^{+0.0011}\,,
  \end{equation}
  has a very similar uncertainty than the previous work, see
  Eq.~(\ref{eq:pot_res1}), despite the fact that they reach significantly higher
  energies. Obviously this is a result of the more conservative approach to the
  estimate of the truncation uncertainties. 

  Reference~\cite{Bazavov:2019qoo} claims that if they would have
  followed the same recipe as in reference~\cite{Bazavov:2014soa},
  their updated result Eq.~(\ref{eq:pot_res2}) would have the
  uncertainty reduced by a factor two. 
  
\item[Ref.~\cite{Ayala:2020odx}:] This work uses the same data as
  reference~\cite{Bazavov:2019qoo}, {but they use the known terms in
  the perturbative series to fix the normalisation of the renormalon
  ambiguity and subtract some non-perturbative (\ie\, power)
  corrections}. Their final result 
  \begin{equation}
    \label{eq:pot_res3}
    \alpha_{\overline{\rm MS} }(M_Z) = 0.1181(9)\,,
  \end{equation}
  quote a similar uncertainty as~\cite{Bazavov:2019qoo}, but their
  central value is significantly larger.

\end{description}

In summary, the estimate by FLAG of these uncertainties (1.4\%) is reasonable.
It is basically the same as the difference in central
  values between the two most recent works~\cite{Bazavov:2019qoo} and 
  ~\cite{Ayala:2020odx}, that use the same data but different
  strategies to match with perturbation theory. 
  This uncertainty is also larger than the total quoted uncertainty in
both works~\cite{Ayala:2020odx, Bazavov:2019qoo}. 

Determinations of the strong coupling from the static potential have recently
improved significantly. Also the estimate of perturbative
uncertainties is more conservative compared with previous works. 
These determinations are in very good shape, but the following points
should be better understood:
\begin{itemize}
\item The most important point to be understood is {if the
    perturbative region is reached in current extractions. 
    The significant
  difference in central values between extractions using the same
  dataset but different power corrections (\ie\,
  Refs.~\cite{Ayala:2020odx, Bazavov:2019qoo}) needs a better understanding}. 
  
  This difference between central values is about $1.4\%$, similar to
  the uncertainties that we obtained from a simple scale variation approach. 
  We emphasize that our estimate \emph{does not address the
    issue} of the logarithmic corrections to the perturbative series,
  related with the IR divergences of the static potential.
  
\item Another manifestation of the same problem is the strong dependence on the value of
  $\Lambda^{(N_{\rm f} =0)}$ that has been observed for extractions based on the static
  potential for values of the coupling $\alpha^3_s \lesssim
  0.01$~\cite{Husung:2017qjz}. 
  This effect has been studied in~\cite{Bazavov:2019qoo} and they
  observe a much milder effect. 

  Note however that in this last reference $\Lambda$ is extracted as a
  fit over a range of energy scales. This procedure makes it much more
  difficult to see any dependence in the extracted value of $\Lambda$,
  and in fact means that all the different extractions of $\Lambda$
  (with the exception of one point) use data with $\alpha^3> 0.01$.

\item The effect of the bad sampling of the topological charge has still to be
investigated in detail. The ensembles at the finest lattice spacing have the
topology completely frozen (see~\cite{Bazavov:2017dsy}). The impact of frozen
topology can be studied by computing the scale $r_1$ from ensembles that are
stuck in distinct topological sectors. The numerical results so far are all
compatible within errors. Note however that the values of the topological charge
simulated (basically 2 and 0) are small. A dedicated quenched study could shed
some light on these issues.  

\item A related problem is that the physical volumes are relatively
  small (with $m_\pi L \approx 2.2$), and the effect of this small
  volumes on the determination of the scale and the static force is
  unclear. On general grounds, the static force is a short distance
  observable and one expects finite volume effects to be very small.
  Probably the largest source of finite volume effects are in the
  scale determination $r_1$, but one should keep in mind that the
  uncertainty in the scale has a small effect on the uncertainty of
  $\alpha_s(M_Z)$.  It would be interesting to quantify at which level
  of precision finite volume effects start to be a concern for these
  determinations.
\end{itemize}

\subsubsection{Heavy quark correlators}
\label{sec:heavy-quark-corr-1}

\begin{figure}
  \centering
      \begin{subfigure}[t]{0.48\textwidth}
    \centering
    \includegraphics[width=\textwidth]{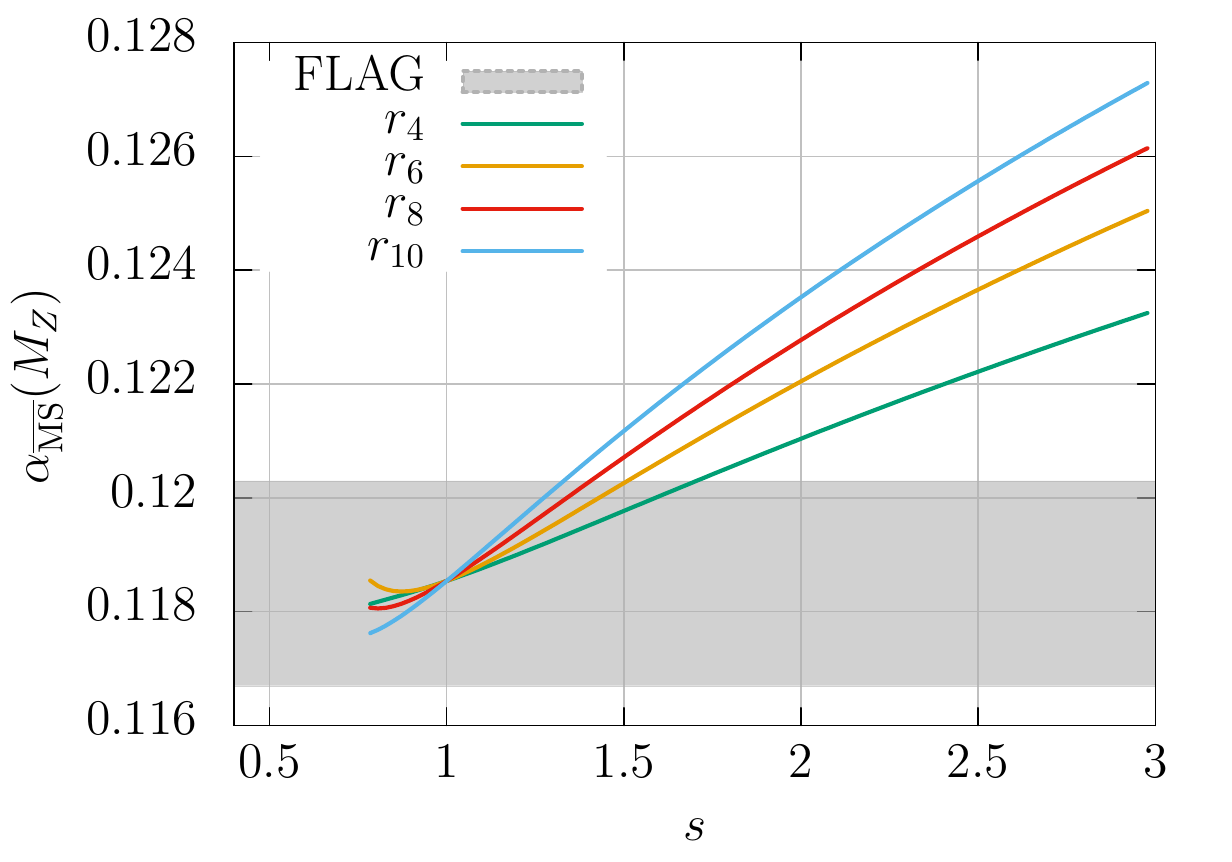}
    \caption{}
  \end{subfigure}
  \begin{subfigure}[t]{0.48\textwidth}
    \centering
    \includegraphics[width=\textwidth]{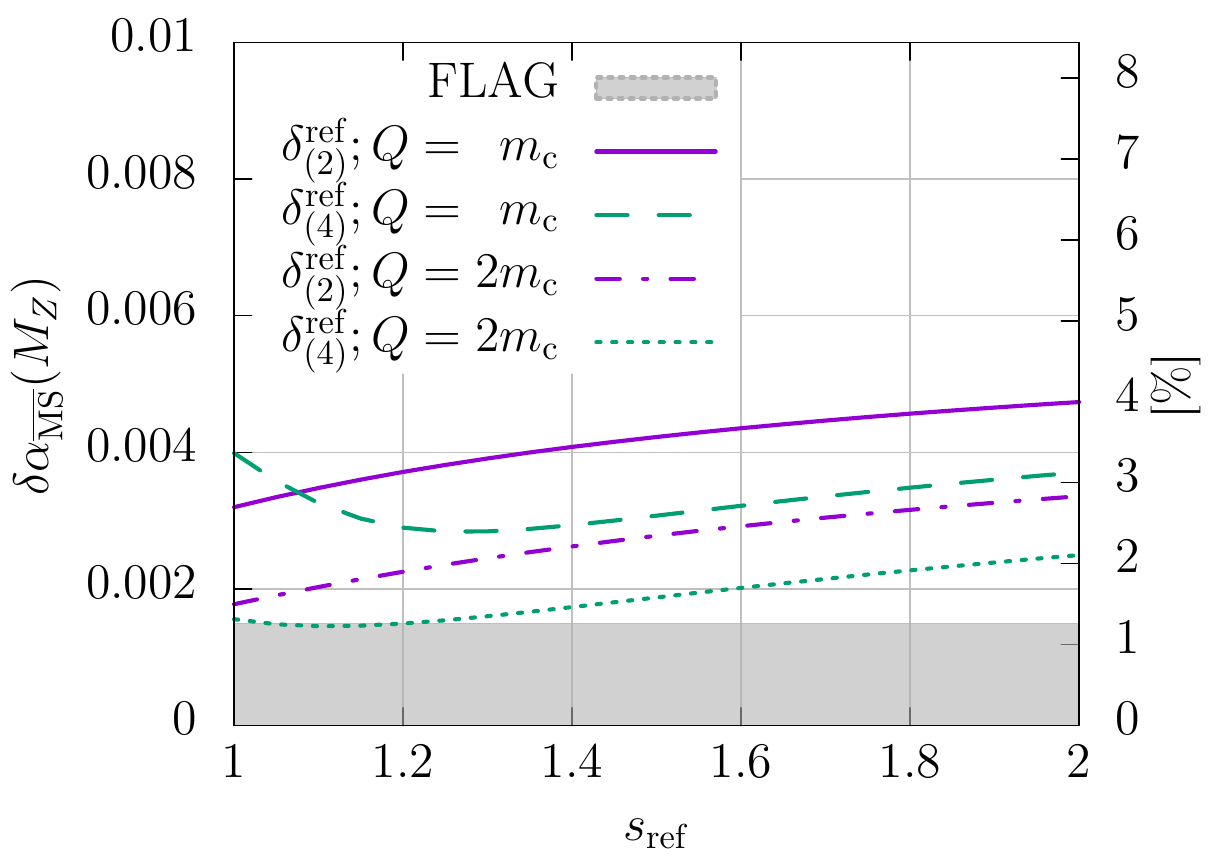}
    \caption{}
  \end{subfigure}
  \caption{Truncation effects in $\alpha_s$ extractions from heavy
    quark correlators.}
  \label{fig:HQtruncation}
\end{figure}

The average in figure~\ref{fig:summary} is the result of the
works~\cite{McNeile:2010ji,Chakraborty:2014aca,Nakayama:2016atf}. Once
again the most delicate sources of uncertainty are the perturbative
error and controlling the continuum extrapolation. It is instructive to
see how different works estimate that 
particular systematic error. 
\begin{description}
\item [JLQCD collaboration~\cite{Nakayama:2016atf}:] The authors use
  a scale variation method, very similar to our procedure (see
  figure~\ref{fig:HQtruncation}). A crucial difference is that they
  vary independently both the renormalization scale in the quark mass
  and in the strong coupling in the range $2-4$ GeV (this typically
  increases the estimates of the uncertainties). Also this work does
  not use the moment $r_4$. This is the quantity with the best
  perturbative behavior (see figure~\ref{fig:HQtruncation}), but they
  find the continuum extrapolation very challenging. Their result
  \begin{equation}
    \alpha_{\overline{\rm MS} }(M_Z) = 0.1177(26)\,,
    \qquad [2.2\%]\,,
  \end{equation}
  claims a $\sim 2\%$ error, mostly dominated by the perturbative
  truncation uncertainties.
  
\item [Ref.~\cite{Maezawa:2016vgv}:] They estimate the truncation uncertainty by
  estimating the $\alpha^4(\mu)$ term in the perturbative expansion of $r_n$.
  They use a range of values  $c_3(1) = \pm 2c_2(1)$ (see Eq.~(\ref{eq:Pinpt})),
  which yields a perturbative truncation uncertainty that is much smaller than
  the one obtained by the scale variation method. The truncation uncertainty
  represents a negligible contribution to the uncertainty of their final result
  \begin{equation}
    \alpha_{\overline{\rm MS} }(M_Z) = 0.11622(84)\,,
    \qquad [0.7\%]\,,
  \end{equation}
  which is 3 times more precise than the JLQCD result.

\item[Ref.~\cite{Petreczky:2019ozv}:] This work can be considered an update
  of~\cite{Maezawa:2016vgv}. Again the perturbative uncertainty is computed by
  estimating the size of the $\mathcal \alpha^4(\mu)$ term in the perturbative
  expansion of $r_n$, but this time they allow a larger range of values for the
  unknown coefficient $c_3(1) = \pm 5c_2(1)$ (see Eq.~(\ref{eq:Pinpt})). Their
  updated result 
  \begin{equation}
    \alpha_{\overline{\rm MS} }(M_Z) = 0.1159(12)\,,
    \qquad [1.0\%]\,,
  \end{equation}
  has in fact a larger uncertainty. 

\item[Ref~\cite{Chakraborty:2014aca}:] The HPQCD collaboration analyze
  data close to the physical charm quark mass ($Q=m_{\rm c}$). Their
  analysis includes higher order terms in the perturbative expansion
  of their data amongst the fitted parameters. These unknown terms (up
  to powers $\alpha^{15}$) are constrained using Bayesian priors. The
  perturbative truncation errors are the main source of uncertainty,
  but their final result
  \begin{equation}
    \alpha_{\overline{\rm MS} }(M_Z) = 0.11822(74)\,,
    \qquad [0.6\%]\,,
  \end{equation}
  claims an uncertainty 4 times smaller than our estimates of the truncation
  uncertainties at the scale $Q=m_{\rm c}$. This uncertainty is estimated by
  varying the number of terms added to the fit. It is not clear to the authors
  why this estimate should be a reliable estimate of the truncation
  uncertainties. In particular these error estimates are substantially
  smaller than the usual estimates coming from scale variation. 
  
\end{description}

FLAG quotes a 1.2\% truncation uncertainty for extractions performed at the
scale $m_{\rm c}$. Our scale variation method tends to point to even larger
values for the truncation uncertainty (see figures~\ref{fig:HQtruncation}):
about 2-3\% for extractions at $m_{\rm c}$ and between 1-2\% for extractions at
$2m_{\rm c}$. 

\begin{figure}
  \centering
  \includegraphics[width=0.8\textwidth]{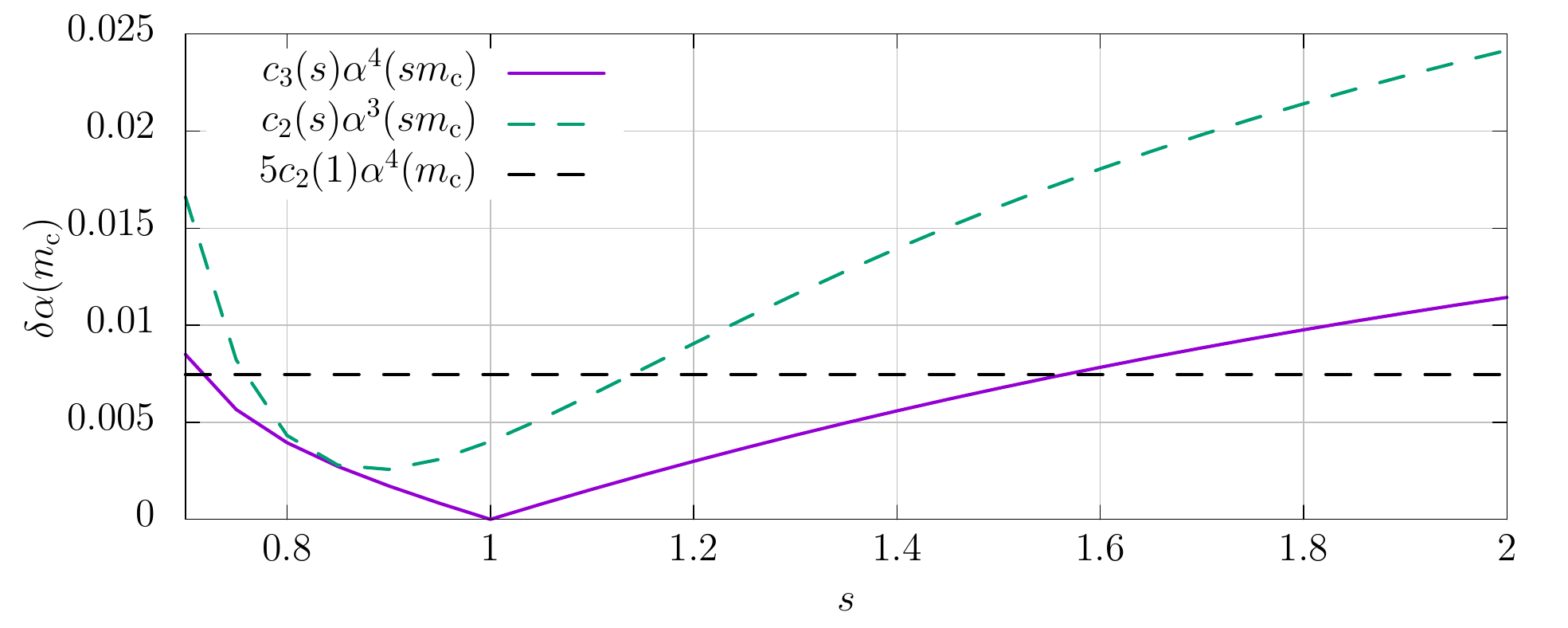} 
  \caption{Truncation effects in $\alpha_s$ extractions using the last
  term in the perturbative series. 
The horizontal dashed line represent the estimate of
Ref.~\cite{Petreczky:2019ozv} $\delta\alpha = 5c_2(1)\alpha^4(m_{\rm c})$. 
This uncertainty is significantly smaller than the last known term in
the series (dashed curve) unless the renormalization scale is chosen
very close to the physical scale (\ie\, $s\approx 1$). 
This estimate is also smaller than the naive estimate coming
\emph{only} from the scale dependence of $c_3$ (solid curve).}
  \label{fig:HQc3}
\end{figure}

It is easy to see why the estimates based on the value of the unknown 
coefficient $c_3(1)$ lead to small uncertainties. 
The last known coefficient in the series is given by
\begin{equation}
  c_2(s) = 0.0796 +  0.588 \log(s)   +  2.052 \log^2(s)\,.
\end{equation}
This coefficient is anomalously small for $s=1$. 
Even when multiplied by a factor 5, it leads to a small estimate for the
coefficient $|c_3|\approx 0.4$. 
On the other hand $c_3(s)$ can be estimated by other means. 
The scale dependence of $c_3(s)$ is fully predicted by the RG
equations
\begin{equation}
  c_3(s) = c_3(1) + 0.865\log(s)   +   2.425\log^2s   +   2.939\log^3 s\,.
\end{equation}
(\ie\, only the non-logarithmic dependence $c_3(1)$ is unknown). This
logarithmic dependence alone predicts a coefficient
$|c_3(s)|\approx 3$ for even modest values of the scale $s=2$. This
value for the $c_3$ term is 7.5 times larger than the estimate used in
the last work Ref.~\cite{Petreczky:2019ozv} and almost 20 times larger
than the estimate of~\cite{Maezawa:2016vgv}. There is an extra
suppression $\alpha_s^4(sQ)$ that makes the uncertainty smaller when
$s>1$, but this effect can only account for a factor between 2.8 (for
$Q=m_{\rm c}$) and 4.5 (for $Q=2m_{\rm c}$): clearly insufficient to
compensate the factor 7.5 or 20 in the estimate of $c_3$. Moreover,
note that this simple estimate of $c_3(s)$ neglects completely
$c_3(1)$. (See Figure~\ref{fig:HQc3}). This explains why uncertainties
based on scale variation are generally larger.
Even assuming that one order more is known, and that
$c_3(1) = 0$, the scale variation approach results in $\approx 1.5\%$
error for $r_4$ at the charm scale. 

Estimates of the
truncation uncertainties based on varying the number of fit parameters
constrained by priors and using Bayesian methods lacks a solid
theoretical basis. We suggest that estimates based on scale variation
should be preferred.

These considerations, together with the complicated scaling violations (see
discussion in section~\ref{sec:heavy-quark-corr}) make it very challenging to
obtain $\alpha_s$ with less than a 1.5\% uncertainty using these methods. Quark
masses significantly larger than the physical charm quark mass would be required
(and one would need to deal with complicated continuum extrapolations). 

There are two interesting directions for future research. First, the next order
in the perturbative relation of the heavy quark moments could be very useful in
future extractions. Second a dedicated pure gauge study, where significantly
larger energy scales could be explored, would shed some light on the
difficulties associated with the truncation uncertainties and the approach to
the continuum in this type of lattice determinations.  

\subsubsection{Observables at the cutoff scale}
\label{sec:observ-at-cutoff}

\begin{figure}
  \centering
      \begin{subfigure}[t]{0.48\textwidth}
    \centering
    \includegraphics[width=\textwidth]{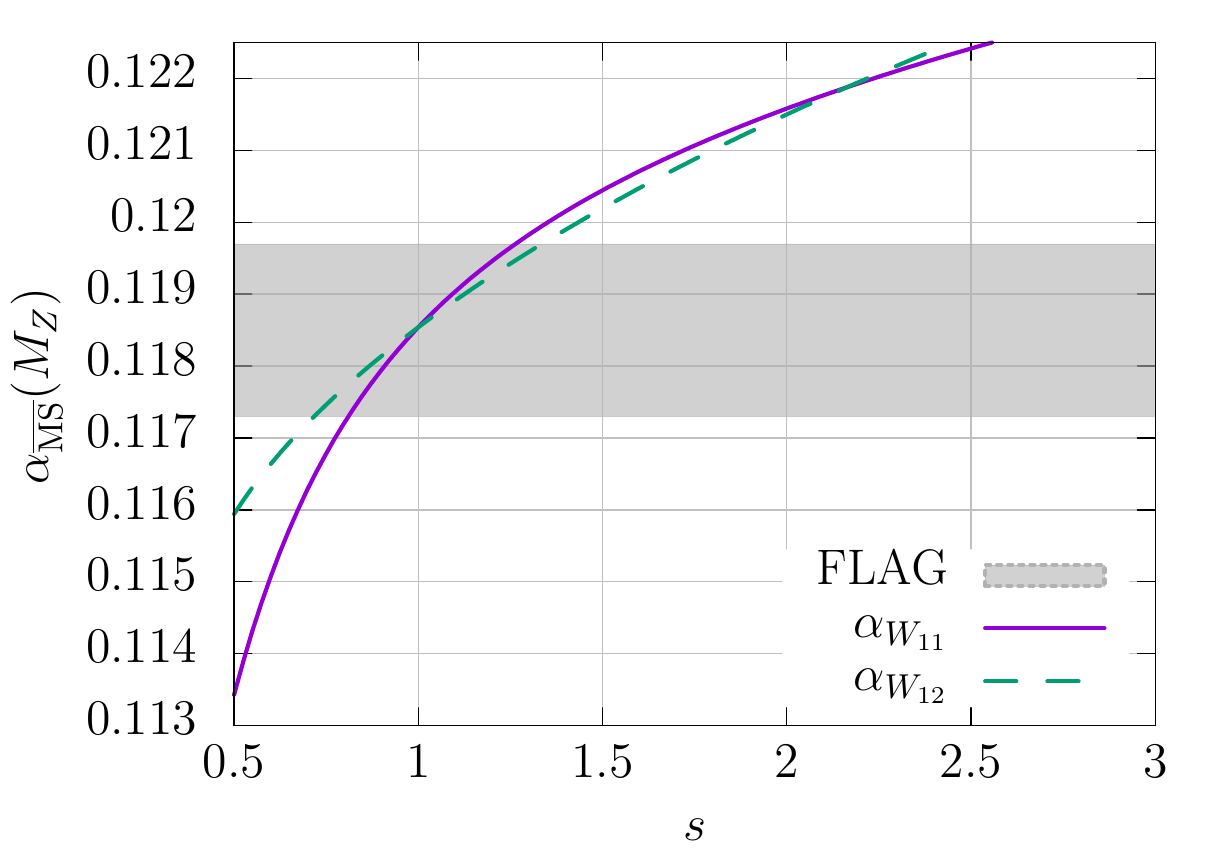}
    \caption{}
  \end{subfigure}
  \begin{subfigure}[t]{0.48\textwidth}
    \centering
    \includegraphics[width=\textwidth]{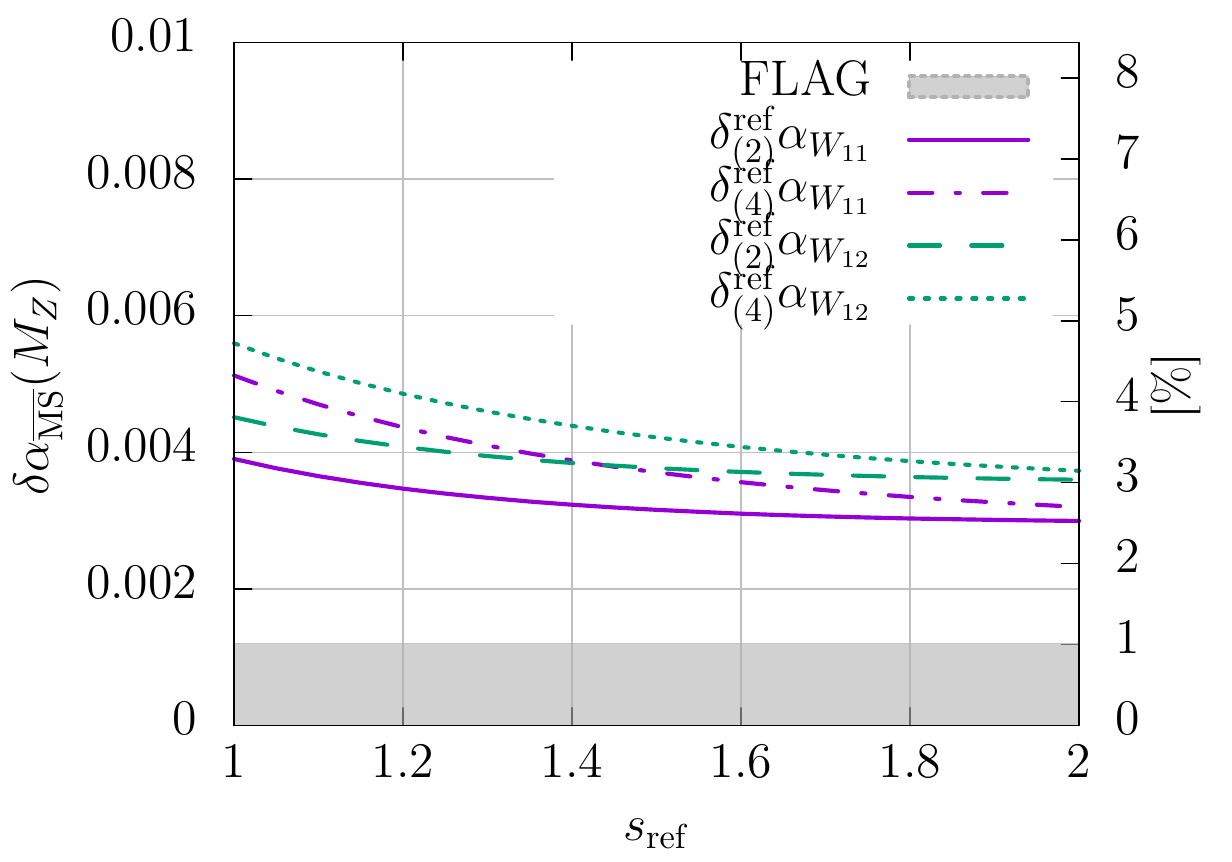}
    \caption{}
  \end{subfigure}
  \caption{Truncation effects in $\alpha_s$ extractions from
    quantities at the scale of the cutoff.}
  \label{fig:Wlooptruncation}
\end{figure}

There are two studies that contribute to this
average~\cite{Maltman:2008bx,McNeile:2010ji}\footnote{Reference~\cite{McNeile:2010ji}
  simply updates the analysis of~\cite{Davies:2008sw} with a more
  precise determination of the scale.}. The main contribution to the
uncertainty in these determinations is purely systematic, with the
perturbative uncertainties playing a leading role. Note that an
advantage for these observables is that there is no need to perform a
continuum extrapolation. 
It is very difficult to obtain an independent estimate of the truncation
uncertainties for these observables. The reason is that the data
\emph{does not follow the known perturbative prediction} in the range
of energy scales reached in the numerical
simulations~\cite{Maltman:2008bx,McNeile:2010ji}. Several terms are
added to the perturbative expansion, up to terms $\alpha^{10}_{\rm V}$ (see
Eq.~(\ref{eq:alphaW})), and the higher-order coefficients are
fixed by fitting the lattice data to the expression
\begin{equation}
  \label{eq:alphaWpt}
  \alpha_{W_{nm}}(1/a) =
  \alpha_{{\rm V} }(\mu) + c_2\alpha_{{\rm V} }^2(\mu)
  + c_3\alpha_{{\rm V} }^3(\mu) +
  \sum_{k=4}^{10} d_i \alpha_{{\rm V} }^i(\mu)\,.
\end{equation}
Here $\alpha_{\rm V}$ is the coupling in the potential scheme (see
Ref.~\cite{Lepage:1992xa}) and  $\mu \approx \pi/a$ with the exact
value depending on the observable). A crucial point in these extractions is the
discretization used. 
We will focus in the same discretization as used in the
works~\cite{Davies:2008sw,McNeile:2010ji}.  

At least one extra term, with an unknown coefficient, is necessary in
order to obtain a satisfactory description of the data. Moreover, the
data is not precise enough to determine all the extra coefficients, so
Bayesian priors are used in order to constrain the size of the
coefficients $d_i \approx 0 \pm 2.5$. The HPQCD
analyses~\cite{McNeile:2010ji} estimate the perturbative truncation error
by varying the number of terms in expression
Eq.~(\ref{eq:alphaWpt}). Firstly let us note that the perturbative
expansion of $\alpha_{W_{nm}}$ is only expected to be asymptotic. In
particular the coefficients in the perturbative expansion should
eventually grow. Constraining the size of all coefficients to the same
size $\mathcal O(1)$ is not justified based on theoretical arguments.

Following the exact same procedure that we used for the other observables, we
can estimate the truncation error by performing a scale variation analysis. For
this purpose, we describe the observable using the \emph{known} perturbative
coefficients in the perturbative expansion in terms of the renormalized $\overline{\rm MS} $ coupling:
\begin{equation}
  \label{eq:alphaWptMSbar}
  \alpha_{W_{nm}}(1/a) =
  \alpha_{{\overline{\rm MS}} }(\mu) + c_2\alpha_{{\overline{\rm MS}} }^2(\mu)
  + c_3\alpha_{{\overline{\rm MS}} }^3(\mu) + \dots\, ,
\end{equation}
with $\mu \approx 2.4/a$ (the exact relation depending on the concrete
quantity). For illustration, let us focus here on the shortest
distance object (the plaquette), and the $1\times 2$ Wilson loop with
tadpole improvement:   
\begin{align}
  \alpha_{W_{11}}(a) &= -\frac{1}{c_1^{(11)}}\log W_{11} = 
  \alpha_V(d/a) + \frac{c_2^{(11)}}{c_1^{(11)}}\alpha_V^2(d/a)
                               + \frac{c_3^{(11)}}{c_1^{(11)}} \alpha_V^3(d/a) + \dots \\
  \alpha_{W_{12}}(a) &= -\frac{1}{c_1^{(12)}}\, \frac{\log W_{12}}{u_0^6} = \alpha_V(d/a) + \frac{c_2^{(12)}}{c_1^{(12)}}\alpha_V^2(d/a)
                                      + \frac{c_3^{(12)}}{c_1^{(12)}} \alpha_V^3(d/a) + \dots \, .
\end{align}
These quantities show a very strong dependence on the scale (see
figures~\ref{fig:Wlooptruncation}), pointing to truncation uncertainties of the
order 3\%.

FLAG uses the HPQCD fit result, $d_4\approx 2$, to estimate the truncation
uncertainty as $2\alpha^4$. This procedure results in a smaller value than the
one obtained form the scale variation approach
\begin{equation}
  \delta\alpha_{\overline{\rm MS} }(M_Z) \approx 0.0012\,,
  \qquad [1\%]\,.
\end{equation}
This last uncertainty is about 2.5 times larger than the estimate of HPQCD. 

A last piece of information comes from comparing the results in
Refs.~\cite{Maltman:2008bx,Davies:2008sw}. They use basically the same dataset
(reference~\cite{Maltman:2008bx} uses a subset of the 22 quantities used
in~\cite{Davies:2008sw}), but analyse it using different perturbative
expressions and deal with the non-perturbative corrections in different ways.
Their results for $\alpha_{\overline{\rm MS} }(M_Z)$ differ by a 0.6\%--1.2\%.

The overall conclusion is that truncation uncertainties based on varying the
number of terms constrained with Bayesian priors underestimate significantly the
true uncertainties. The same phenomena have been observed in the case of
determinations based on currents of heavy quark correlators. Moreover it is
important to recall that the lattice data for these short distance quantities do
not follow the perturbative prediction if only the known coefficients are used,
making it mandatory to fit several additional terms in the perturbative
expansion constrained by Bayesian priors. We think that this method needs
further study. In particular it is mandatory to find a solid estimate of the
truncation uncertainties based on the same techniques that are common in the
estimates of the perturbative QCD uncertainties. A detailed study in pure gauge
could shed some light on the issue. Without these insights the claimed
uncertainties of some of these computations (0.6\%) seem an underestimate of the
true uncertainties. 

\subsection{An opinionated and critical summary}

What can we conclude on the status of the lattice determinations of the strong
coupling? First we should only consider extractions with a dataset that allows a
proper control over the different sources of systematic effects. This is the
approach taken for example by the FLAG review. For the particular case of the
determination of the strong coupling, the most delicate points in the
extractions of the strong coupling is the estimate of the uncertainties related
with the truncation of the perturbative series and the large scaling violations
in short distance quantities. Note that controlling both sources of systematic
uncertainties is challenging: cutoff effects are larger for short distance
quantities, while perturbative truncation errors are larger for low energy
quantities, \emph{the window problem} described in
section~\ref{sec:chall-determ-alph} is again the limiting factor. This is why we
have insisted on these points along the review.  

Many lattice methods do not allow a simultaneous control of these two
sources of systematic effects with current computer resources.
Typically in these cases the extractions of the strong coupling are
performed at a single lattice spacing, and/or the extraction is
performed at energy scales where significant contribution from
non-perturbative effects are present. Our analytic understanding of
these effects is, at best, very limited.  In practical terms we have
to deal with them by fitting these contributions.  This is very
delicate, since distinguishing the perturbative running from the
non-perturbative contributions when the data is available only in a
restricted range of scales is far from easy.  We therefore prefer to
focus on methods where the data follows the perturbative prediction
and non-perturbative corrections are smaller than the uncertainties.

Still, with this ambitious aim, several methods that allow a reliable
extraction the strong coupling have been developed in lattice
QCD. These methods differ in their control over the systematic errors
associated with the continuum extrapolation and the truncation
uncertainties. 

\begin{itemize}
\item Finite size scaling is the only method that offers a \emph{solution} to
  the window problem instead of trying to find a \emph{compromise}. Arbitrarily
  large energy scales can be reached, and the continuum extrapolation can be
  performed by using several lattice spacing at each constant renormalization
  scale. 

  This strategy trades the systematic errors associated with the
  truncation of the perturbative series at relatively low scales with
  the statistical error accumulated when computing the
  non-perturbative running. It remains challenging to obtain precise
  results, but thanks to several recent developments, a sub-percent
  precision has been reached by these kind of determinations. The
  truncation uncertainties are negligible, since perturbation theory
  is typically applied at the electroweak scale. Several observables
  have been studied non perturbatively and in some cases agreement
  with perturbation theory is achieved over a large range of scales
  (from 4 to 140 GeV)~\cite{DallaBrida:2018rfy,Bruno:2017gxd}.
  
  Our reservation with this approach is that until now only two groups have used
  it to determine the strong coupling, and with very similar setups. A new
  independent determination would be welcome.

\item Determinations based on the static potential have several advantages over
  most other determinations. First, the relevant perturbative relations are
  known to N$^3$LO, one order more than what is typically known for other
  observables. Second, the non-perturbative lattice data seems to follow the
  perturbative prediction for energy scales as low as 2 GeV. This energy scale
  can be reached at several values of the lattice spacing and the data can be
  extrapolated to the continuum. 

  At the moment of writing, two new studies~\cite{Bazavov:2019qoo,
    Ayala:2020odx} have been published. 
  One of them  can be 
  considered an update of some of the works evaluated by FLAG. This
  new determination improves significantly in the 
  energy scales that they are able to reach. What is more important, they treat
  the perturbative uncertainties more conservatively than in previous works
  and claim a sub-percent precision in the strong
  coupling. The other recent work (Ref.~\cite{Ayala:2020odx}) uses
  the same dataset but a different treatment of the lattice data and
  matching with perturbation theory.
  They obtain a result for $\alpha_{\overline{\rm MS} }(M_Z)$ about
  1.3\% larger, rising some doubts on the claimed sub-percent accuracy
of these works. 

  According to our analysis using the scale variation approach, the
  uncertainty quoted by FLAG is reasonable, although our method
  neglets the delicate issue of logarithmic corrections to the
  perturbative series.

  As with all large volume determinations, there are reservations with
  this  method beyond the estimates of perturbative uncertainties,
  related with the compromises  
  that are made. Determinations reach high energy scales (up to
  8 GeV), but the energy scales where the continuum limit can be taken
  with several values of the lattice spacing are substantially lower. 
  Some of the volumes simulated are relatively small
  ($m_\pi L \approx 2.2$). Moreover some of these simulations are
  affected by the problem of topology freezing, potentially making
  difficult to asses the statistical errors correctly. The range of
  scales where the methodology can be tested is limited, and we should
  always remember that truncation uncertainties estimated within
  perturbation theory can be misleading (see
  section~\ref{sec:syst-extr-alph}), \emph{even when a conservative
    approach is taken}. These points should (and will) be investigated
  further, but beyond any doubts determinations coming from the static
  potential have reached a considerable level of maturity and
  precision.

\item Some extractions based on currents of heavy quark correlators
  are among the most precise determinations of the strong
  coupling. Scaling violations have a complicated functional form, but
  at least the continuum limit can be explored with several lattice
  spacing. On the other hand, and compared with extractions based on
  the static potential, our current perturbative knowledge in these
  extractions is one order less and cutoff effects seem larger. In
  summary, this method is more challenging 
  than the static potential from both sides of the window
  problem. Most of the determinations are performed at the charm quark
  mass $M_{\rm c}$ GeV. At these low scales, truncation
  uncertainties estimated using the scale variation method are about
  $2-3\%$ in $\alpha_{\overline{\rm MS} }(M_Z)$.  The FLAG 2019 review
  quotes a smaller uncertainty ($1.5\%$). Some
  works claim a sub-percent precision, but we find it difficult to
  consider these estimates reliable. 

  There are several directions in which these extractions can improve. First, it
  should be possible to extend the current perturbative knowledge by one order
  more. Second, this method has never been studied in detail in the pure gauge
  theory, where one could reach energy scales significantly larger than $M_{\rm
  c}$ with full control over the continuum extrapolations. In any case, until
  the perturbative knowledge is known and larger energy scales have been
  studied, it seems difficult to claim a smaller uncertainty than what the FLAG
  review assigns to these extractions.  
    
\item Finally, there is still work to do in order to better understand the
  determinations based on lattice observables defined at the cutoff scale.
  Despite these methods claiming the smallest uncertainties, this claim is not
  backed up by a solid analysis of the truncation uncertainties. A scale
  variation approach points to significantly larger uncertainties. In fact the
  truncation uncertainties in these extractions seem to be significantly larger
  when using several different methods. Our numerical investigation
  suggests that even the FLAG error is still underestimating this source of
  systematic uncertainty. 

  One should also point out that in pure gauge there is a significant
  discrepancy between some recent extractions based on observables defined at
  the cutoff scale and a recent extraction using finite size scaling
  (see~\cite{DallaBrida:2019wur}).
\end{itemize}

Let us end this section with a general comment about figure~\ref{fig:summary}.
Basically every method to extract the strong coupling using lattice QCD seems
to be able to reach a better precision than any phenomenological
extraction. This seems at least in some cases, to be fully justified based on
the general principles on which we have insisted along the review: the most
precise phenomenological determination, the extraction of $\alpha_s$ from
$\tau$ decays, is performed at a \emph{fixed} energy scale
$m_{\tau}\approx 1.7$ GeV. Larger energy scales cannot be probed in these
extractions, while the perturbative expansion of the observable is known at
the same order as in the extractions based on the static potential, where
energy scales $\mu \approx 5$ GeV can be explored at several values of
the lattice spacing. In finite size
scaling the perturbative knowledge is one order less, but perturbation theory
is applied at the electroweak scale and consistency of the results is checked
in the energy range $4-140$ GeV for a one-parameter family of observables.

\subsection{The future of lattice determinations of $\alpha_s$}

Now that we understand the main limitations of the different lattice
methods to extract the strong coupling we are in a good position to
discuss what is needed in order to substantially reduce the current
uncertainty in the strong coupling.

\begin{figure}
  \centering
  \includegraphics[width=\textwidth]{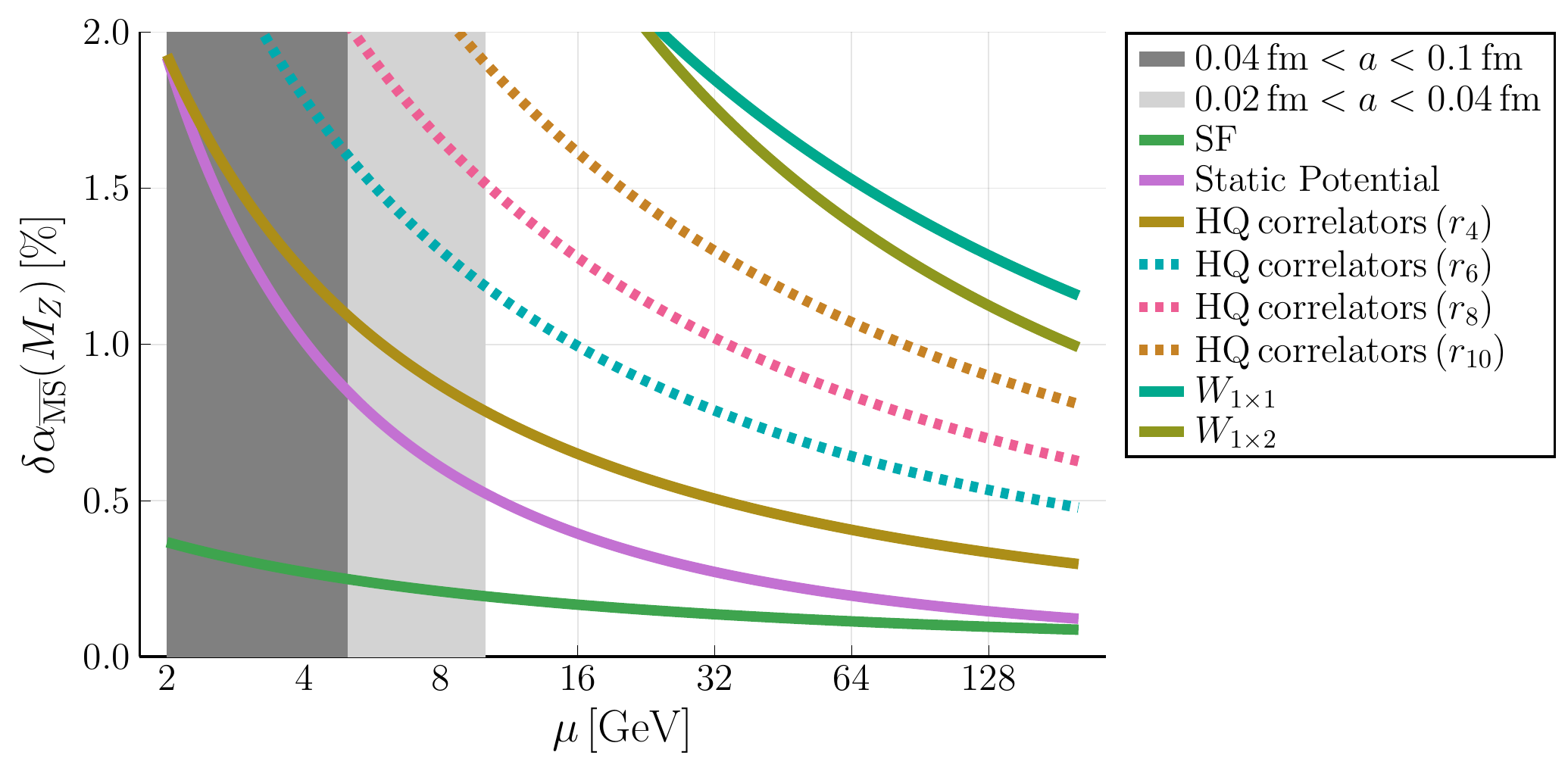}
  \caption{Scale truncation uncertainties for different lattice
    methods (we quote $\delta_{(2)}$, see section~\ref{sec:methods-that-enter}). 
  The dark shaded region is the current accessible range for large
  volume lattice simulations nowadays. 
The light shaded region represents the accessible scales if a
reduction by a factor two in the lattice spacing is possible in the future. 
Note that most methods require a continuum extrapolation with several
lattice spacings: in practice the accessible renormalization scales
for all methods except finite size scaling (SF) is at least factor two/four
smaller than the lattice spacing.}
  \label{fig:scale_errors}
\end{figure}

From a general point of view we must realize that, with the exception
of finite size scaling methods, the limitations of the lattice
determinations of the strong coupling are a direct consequence of the
\emph{window problem}: the fundamental compromise between reaching
large energy scales, where the perturbative behavior is better, and
using low renormalization scales, where the continuum extrapolation is
well under control. 
One should also be aware that truncation uncertainties decrease with
powers of the coupling, and therefore slowly (\ie\, logarithmically)
with the scale (see figure~\ref{fig:scale_errors}). 
This makes the window problem hard to solve by brute force.

What progress can we expect on this front? Since the \emph{window problem} is
basically a computational limitation we expect to see improvements in the
future by just waiting. Computer power has been steadily improving during the
last 50 years, and most probably will continue to do so, with exascale
machines looming on the horizon. Pushing the UV cutoff of a simulation (\ie\
reducing the lattice spacing of the simulation) by a factor two keeping the
physical parameters fixed multiplies the computational cost by a
factor~\footnote{Naively the simulation scales with the lattice volume
  $\propto (L/a)^4$. At the same time simulations at smaller lattice spacing
  show larger autocorrelations, that are expected to scale like $a^2$ in
  absence of topology freezing. On top of these six powers of $a$, the
  integration of the molecular dynamic equations requires to reduce the step
  size at smaller lattice spacing, which brings down another power of $a$ to
  the scaling. } $2^7 = 128$.  We can therefore expect that on exascale
machines the state of the art lattice simulations will be performed on lattice
spacing reaching down to $a\approx 0.02$~fm, assuming that topology freezing
is still under control on such fine lattices. Beyond the trivial observation
that smaller lattice spacing would allow to check the extrapolations
thoroughly, the expected increase in computing power would allow the following
improvements.
\begin{enumerate}
\item Determinations based on QCD vertices and the hadron vacuum
polarization, that are nowadays limited by the size of scaling
violations, will be able to perform a continuum extrapolation at fixed
renormalization scale with several lattice spacing. This will bring
the scaling violations under control.

As of today these methods require to include power corrections as fit
parameters to describe the lattice data. Having access to shorter
distances would clarify whether the perturbative region can be
accessed without having to include these power corrections.

\item Pushing the UV cutoff a factor two higher would allow to match
with the perturbative regime at larger renormalization scales.

For determinations based on the static potential, whose $\beta$-function is
  known up to 4-loops, one expects truncation errors $\mathcal O(\alpha^3)$ to
  be reduced.  Note however that the latest works have reached high energy
  scales $Q\approx 5$ GeV.  At these high energies the running of the coupling
  is really slow, and a factor two in scales only reduces the truncation
  effects in the determination of $\Lambda$ by a factor 2.  But more computer
  power would allow to check the compromises done in the most recent
  computations thoroughly, and reach solid determinations at the
  percent level.

For the determinations based on heavy quark correlators, we also expect a
  reduction of the perturbative truncation errors by a factor of two. Note
  that in this case the $\beta$-function is only known to 3-loops (with
  truncation errors formally $\mathcal O(\alpha^2)$), but the scale typically
  used in the extractions is significantly lower (we use
  $Q \approx \bar m_c$). At these low scales the strong coupling runs
  faster. Using our estimates for the perturbative truncation effects, these
  methods could reach a precision $\sim 1\%$ with a solid and
  conservative estimate of the truncation uncertainties.

  Determinations based on observables defined at the cutoff scale have a
  more uncertain future. Truncation uncertainties estimated using the
  scale variation approach point to a not very well behaved
  perturbative series for these observables. 
  The main benefit would come from the
  possibility of reaching the perturbative domain \emph{without} having to add
  any additional terms to the known perturbative expansion (\ie\,
  reaching comfortably the perturbative region).
  
\item Determinations based on finite size scaling are not limited by the
  systematic error associated with the continuum extrapolation or the
  truncation of the perturbative series, but by statistics. An increase in
  computer power by a factor $2^7$ would help in reducing the statistical
  uncertainty dramatically.  Naively one expects that the non-perturbative
  data would become an order of magnitude more precise~\footnote{As with all
    Monte Carlo methods, the statistical uncertainties of a lattice
    computation decrease $\propto1/\sqrt{N}$ where $N$ is the number of
    configurations. The computer requirements are proportional to $N$.},
  decreasing the uncertainty to the level of $0.07\%$.

  This precision clearly is exagerated. Such level of precision would require
  finer lattices to control the continuum extrapolations. An increase
  in computer power by $2^7$ seems impressive, but this will be eaten by
  just simulating values of the lattice spacing two times smaller. 
  Some compromise between reducing the statistical errors, and
  improving the continuum extrapolations would be needed. 
  Moreover some ingredients of
  the determination (\ie\, the scale setting), that nowadays represent a very
  small contribution to the total uncertainty, would need to be computed on
  large volumes where the uncertainties are not merely statistical.

  Nevertheless, it is clear that such an increase in computer power would
  allow to reduce the uncertainty in $\Lambda$ by a factor two to three. 
  A determination of the strong coupling with an uncertainty $\lesssim 0.3 \%$
  would in principle be possible. 
  Of course at such level of precision one would have to think about
  other problems, like electromagnetic contributions to the scale and
  the running. 
  
\end{enumerate}

Beyond the improvement coming from the increase in computational power, those
determinations that are limited by the truncation of the perturbative series
could benefit from a better perturbative knowledge. This is especially true
for the case of the moments of heavy quark correlators and the observables
defined at the cutoff scale, since the relation to the $\overline{\rm MS} $
scheme is only known to 2-loops, and increasing this to 3-loops seems
feasible\footnote{Note however that for the case of observables defined at the
  cutoff scale, this requires a perturbative computation in \emph{Lattice
    perturbation theory}, that is substantially more complicated.}. This would
naively suppress the perturbative truncation effects by an extra power of
$\alpha$, so potentially a reduction of the error by a factor $3-4$ could be
achieved \emph{if the size of the perturbative coefficients does not
  increase}.  However this assumption is not completely innocent. One should
never forget the asymptotic nature of the perturbative series: \emph{higher
  order is never as good as higher renormalization scales}.

All in all, experience shows that progress in Lattice QCD comes always from
two fronts: computer power, and the new methods and better understanding of
the problems by the community.  We can speculate the improvements that
more computer power would bring, but the most interesting and potentially
better improvements coming from the new methods are much more difficult to
predict. 
{On this front there is a recent proposal~\cite{DallaBrida:2019mqg}
that allows to determine the $N_{\rm f}$-flavor $\Lambda$-parameter
from the pure gauge one. 
The connection between QCD and the pure gauge theory is based on
non-perturbatively decoupling hevy quarks (see
section~\ref{sec:deco-heavy-quark}) using lattice simulations. 
Since the RG equations are solved in the pure gauge theory
(computationally easier), it allows to reach higher energy scales and
improve the statistical precision. 
This idea is discussed in detail in the recent review~\cite{DallaBrida:2020pag}. 
This proposal is promosing, but at the time of writing this review
there are no results for $\alpha_s$ using this method yet.}

\subsubsection{The pure gauge theory as a perfect laboratory}

We have discussed that the limitations usually found in the determination of
the strong coupling are not the same as those of other state of the art
lattice QCD computations. As a consequence of the window problem, the
truncation of the perturbative series and the large scaling violations present
in short distance observables represent the limitation of all lattice methods
to determine the strong coupling, with the exception of finite size
scaling. Statistical accuracy is the main limitation of finite size scaling.

The determination of the $\Lambda$ parameter in the pure gauge theory faces
\emph{exactly} the same limitations. On the other hand, computationally the
pure gauge theory is much more tractable than QCD. Simulations algorithms for
the pure gauge theory are much more efficient, and smaller physical volumes
are acceptable, since finite volume effects are suppressed by the larger mass
of the glueballs.

We do not need to wait 10 years to simulate the pure gauge theory at a lattice
spacing of $a=0.02$ fm. Testing what precision in the strong coupling could
potentially be achieved with 100 times more computer time can be done just by
determining the $\Lambda$ parameter in pure gauge.

In fact the situation in pure gauge is not that clear, with some recent
determinations claiming an uncertainty $\approx 1.5\%$ in the $\Lambda$
parameter, but disagreeing by about a $5\%$~\footnote{The
  determinations~\cite{Gockeler:2005rv,Asakawa:2015vta,Kitazawa:2016dsl},
  based on observables defined at the cutoff scale point to a value
  $r_0\Lambda\approx 0.620$, while the recent
  determination~\cite{DallaBrida:2019wur} based on finite size scaling points
  to a larger value $\approx 0.660$.}. A very detailed and precise study is
been performed using the static potential~\cite{Husung:2017qjz,sommer:19lat},
but at the moment the preliminary results are inconclusive. There is not a
single pure gauge determination using heavy quark correlators.

Given the similarities in the challenges, we hope that the lattice community
takes this discrepancy seriously and does not neglect the issue due to the
lack of phenomenological interest. Pure gauge studies do not enter in the
world average, but in our opinion they will be crucial to help us
understand the problems that we have to face and how to solve them.


\section{Conclusions}

\label{sec:conclusions}

Lattice QCD is, in principle, ideally suited to compute the non-perturbative
quantities that are necessary in order to extract the fundamental parameters of
the standard model. Being a non-perturbative formulation of QCD, it allows a
determination of the value of the strong coupling at energy scales that are
measured in units of some well-known QCD spectral quantity (\ie\ the proton
mass, or the meson decay constants).  

Nevertheless there are considerable challenges in obtaining precise values for
$\alpha_s$. The extraction of the strong coupling requires the application of
perturbation theory, and ideally one would like to use perturbation theory at
very high energy scales, where good convergence is expected. This wish is in
conflict with an intrinsic limitation of any numerical simulation: with finite
computing resources only a limited range of scales can be resolved by a single
lattice QCD simulation. Datasets that have been generated in order to study the
low-energy properties of the strong interactions (\ie\, the spectrum, decay form
factors, the anomalous magnetic moment of the muon, \dots) are limited to reach
a few GeV in energy scales at most. This explains why most lattice QCD
extractions of $\alpha_s$ are limited by the uncertainties associated with the
application of perturbation theory and the truncation of the perturbative series
at relatively low-energy scales.

In writing this review we have focused on explaining our vision of the field. We
have tried to highlight the challenges that lattice determinations of the strong
coupling have to face, and insisted on the specific difference between these and
other lattice computations.

In the last 15 years lattice QCD has made enormous progress. We have witnessed
the surge of dynamical simulations, the values of the lattice spacing has been
pushed down to $a\sim 0.05$ fm, some simulations reach physical volumes of 7 fm
and simulations at the physical point, that once seemed out of reach, are
nowadays common.  

This progress has made it possible to produce solid, first principle predictions
in the low-energy regime of the strong interactions, with a tremendous impact in
flavor physics, searches of beyond the standard model effects and many other
topics in high energy physics. It is this very same progress that has pushed the
lattice determinations of the strong coupling to the top of the podium: as the
overall quality of lattice simulations has improved, the determination of the
strong coupling has become dominated by lattice QCD results. 

But this situation is changing right now. The limitations in current lattice
studies of low-energy hadronic processes are related to the inclusion of charm
effects, electromagnetic corrections, and how to deal with the large
(power-like) finite volume effects that appear in QED. Solving these issues will
have little benefit for the lattice determinations of the strong coupling.

Progress in lattice determinations of $\alpha_s$ will come from \emph{dedicated
approaches}. Mere updates of lattice determinations in new ensembles that are
generated with the aim of studying electromagnetic corrections, or any other
relevant problem of low-energy QCD, will very soon be irrelevant. The lattice
community should embrace this as an opportunity to make real progress. The
current problems of topology freezing, new techniques to simulate very fine
lattices, and more important than anything, the study of alternative techniques
to solve multi-scale problems \emph{should be taken seriously}.  

We hope that this review work will further stimulate the lattice community to
think about these issues. 

The high-luminosity upgrade of the LHC will require high-precision QCD
predictions in order to really benefit from the increase in statistics at the
experiments. With the dominance of lattice determinations of the strong coupling
in the world average, and the relevance of $\alpha_s$ for precision phenomelogy
at the LHC, it is important for the phenomenology community to understand the
problems involved in lattice determinations of the strong coupling, to have some
appreciation of the systematic errors that limit these studies, and to foster
the progress in the lattice community. Providing a self-contained introduction
to the topic for non-lattice specialists has been one of the focal points of our
work. We have given an introduction to lattice QCD with an emphasis in the areas
that are more relevant in the extractions of the strong coupling: the continuum
extrapolation of lattice data, the process of scale setting, and the subtleties
involved in the analysis of lattice QCD data at fine lattice spacings. Moreover
we have dissected all different approaches to extract $\alpha_s$ via lattice
simulations, trying to be as critical as possible. With one important exception,
lattice extractions of $\alpha_s$ based on finite size scaling (see
section~\ref{sec:finite-size-scaling}), all lattice determinations of the strong
coupling represent a compromise between the potentially large cutoff effects
present in every lattice determination of a short distance quantity and the
potentially large systematic effects associated with the use and truncation of
perturbation theory at energy scales of a few GeV. Different lattice methods
take different compromises and suffer from these effects in very different ways.
Our point of view is that only methods that show convincing evidence to have
reached the perturbative running and are able to explore the continuum limit
with several lattice spacing should be used to determine the fundamental
parameters of the standard model. We have studied the perturbative truncation
effects of the most reliable approaches in detail, trying to provide a broad
picture of each of the methods. Our conclusions are that there are at least
three methods that, as of today, allow a reliable extraction of the strong
coupling: finite size scaling (section~\ref{sec:finite-size-scaling}), the
static potential (section~\ref{sec:static-potential}) and moments of heavy quark
correlators (section~\ref{sec:heavy-quark-corr}), although with very different
degrees of reliance. The precision of these extractions, as well as the
potential to improve substantially the current determinations, have been
analyzed. 

On the other hand we have not tried to evaluate each publication. 
The interested reader in this fine detailed perspective should consult
the excellent work of the FLAG review. 

Let us end this review with a comment on the status of the world average. The
most precise lattice determinations show a good agreement. The world average
value of the strong coupling performed by the PDG~\cite{Patrignani:2016xqp}
includes all phenomenological and lattice computations. In the PDG average
individual works are classified according the method of extraction, and
pre-averaged, before determining the final world average. The list of methods
includes several phenomenological procedure (DIS, $\tau$ decays, $e^+-e^-$,
\dots). The rationale behind such a procedure is that \emph{different} works
within the same pre-average have to face the \emph{same} systematic effects. As
we have seen different lattice strategies for the extraction of the strong
coupling are limited by very different reasons. In fact, \emph{once the
continuum extrapolations are under control}, lattice QCD and phenomenological
determinations stand on the same footing, and have to face very similar
challenges (see also~\cite{Salam:2017qdl}). Many lattice QCD extractions using
moments of heavy quark correlators are somehow similar to extractions using
$\tau$ decays: the extraction is performed at a relatively low scale and the
systematics associated with the use of perturbation theory at such low scales
dominate the error budget\footnote{Note however that the perturbative expansion
of the $R_{\tau, V+A}$ is known up to four loops, while the perturbative
expansion of the heavy quark correlators is known only up to 3-loops.}. Lattice
QCD extractions of the strong coupling based on finite size scaling are similar
to phenomenological extractions based on $Z$ boson decays in the sense that they
are performed directly at the electroweak scale and limited by statistics.

We would like to see a world average of $\alpha_s$ that groups different
extractions based on similar systematics. Questions like at What scale is
perturbation theory applied?, What range of scales show agreement with the
perturbative running?, What are the scale variation effects in the extractions?
are the key questions to asses the quality of any determination of $\alpha_s$,
regardless of it being a lattice or a phenomenology extraction. We hope that
this work is also useful to the brave that attempt to produce such a world
average.


\section*{Acknowledgments}
\addcontentsline{toc}{section}{Acknowledgments}

The authors want to thank G. 
Salam for interesting discussions and a critical reading of an earlier
version of this manuscript. 
We also thank him for hospitality in Oxford.

AR has a large debt with the members of the ALPHA collaboration for
a fruitful collaboration in several works, spanning several years,
about many issues covered in this review.
AR also wants to show his gratitude to Rainer Sommer and Stefan Sint
for the many discussions on several topics covered in this review. 

We warmly thank Peter Petreczky for sharing his data
and scripts to produce figures.~\ref{fig:hqcorr}, the authors
of~\cite{Zafeiropoulos:2019flq}, in special Jose
Rodriguez Quintero and Savvas Zafeiropoulos, for sharing and explaining
the data used to produce figure~\ref{fig:alpha_taylor} (b). 
Rainer Sommer for sharing figure~\ref{fig:sign}, and Christopher Kelly
and the RBC/UKQCD collaboration for permission to reproduce
figure~\ref{fig:omega}.

LDD is supported by an STFC Consolidated Grant, ST/P0000630/1, and a
Royal Society Wolfson Research Merit Award, WM140078.  
AR is partially supported by the Generalitat Valenciana
(CIDEGENT/19/040) and was partially supported by the H2020 program in
the Europlex training network, grant agreement No. 813942.

\appendix
\renewcommand*{\thesection}{\Alph{section}}

\section{Challenges in Lattice QCD}
\label{ap:challenges}

\subsection{Topology freezing and large autocorrelation times}

\begin{figure}[h!]
  \centering
  \begin{subfigure}{0.47\textwidth}
    \includegraphics[width=\textwidth]{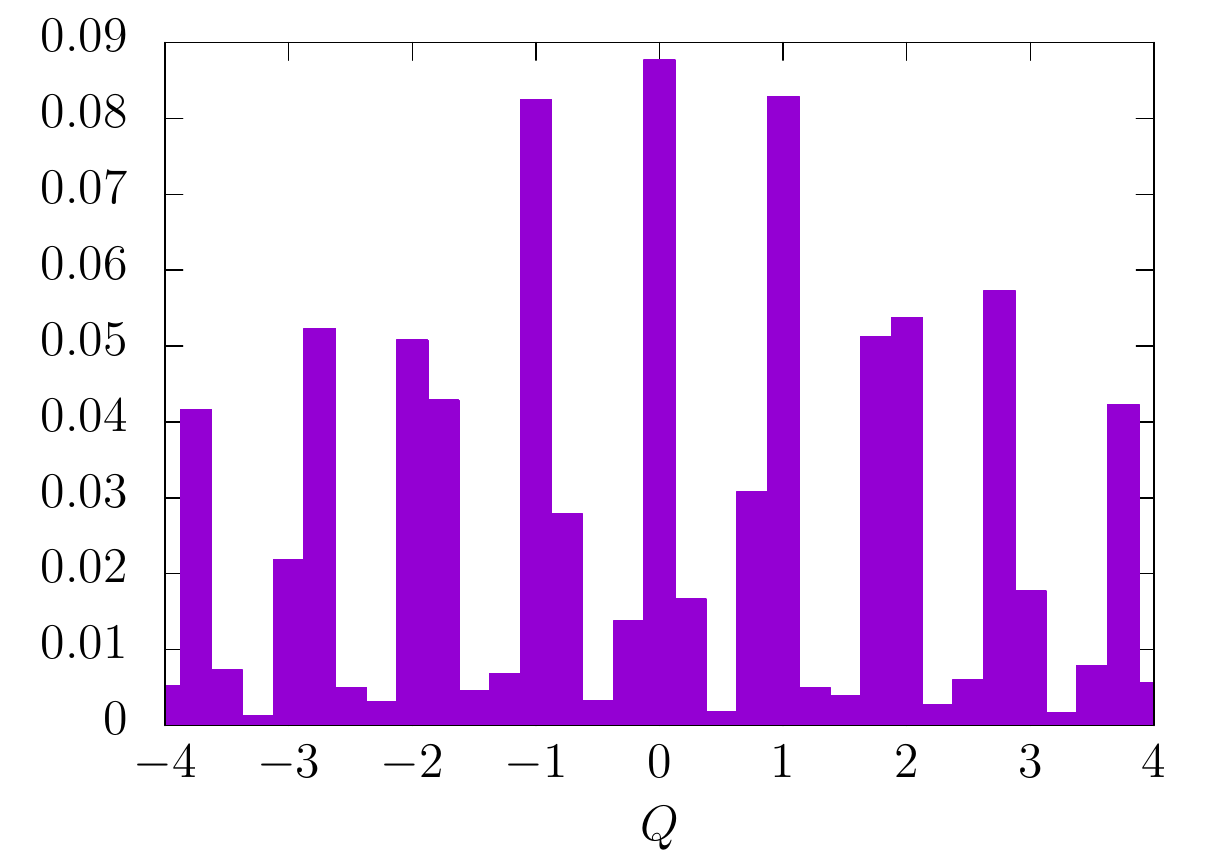}
    \caption{$32^4$ with $a=0.08$ fm.}
  \end{subfigure}
  \begin{subfigure}{0.47\textwidth}
    \includegraphics[width=\textwidth]{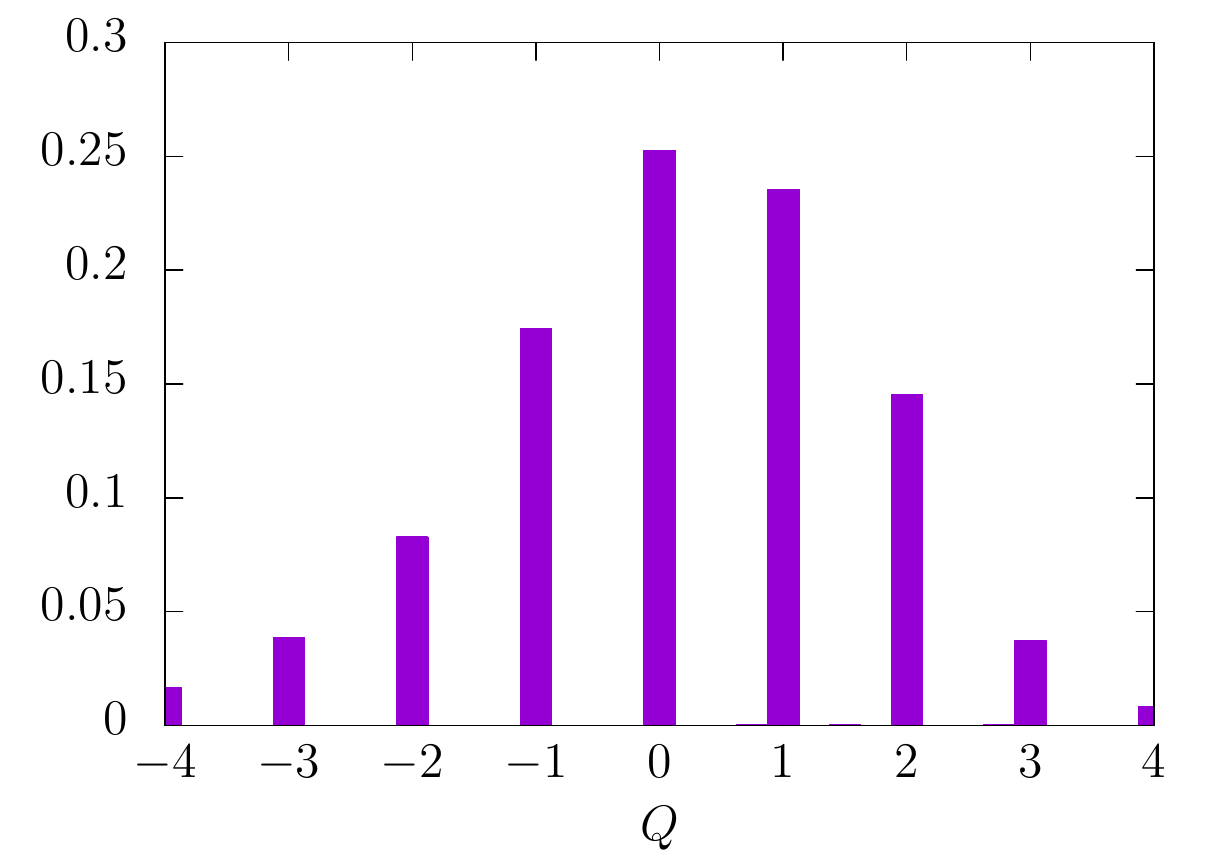}  
    \caption{$48^4$ with $a=0.03$ fm.}
  \end{subfigure}

  \begin{subfigure}{\textwidth}
    \includegraphics[width=\textwidth]{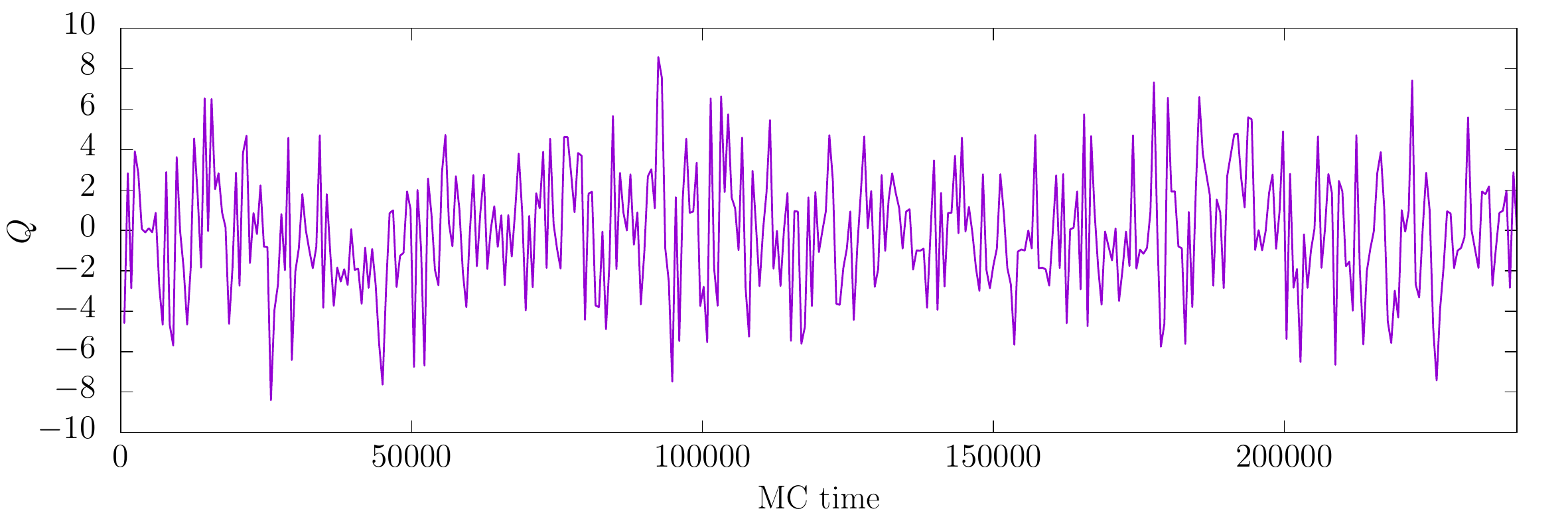}  
    \caption{$32^4$ with $a=0.08$ fm.}
  \end{subfigure}

  \begin{subfigure}{\textwidth}
    \includegraphics[width=\textwidth]{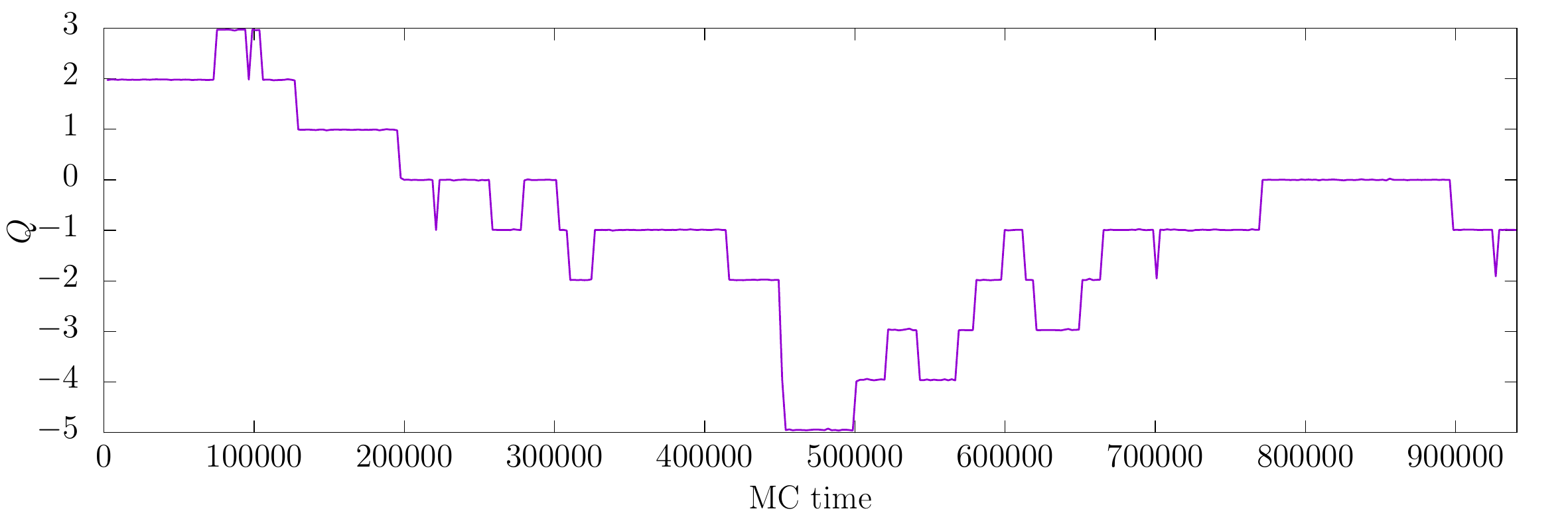}  
    \caption{$48^4$ with $a=0.03$ fm.}
  \end{subfigure}

  \caption{The topological charge is not quantized on the lattice
    (here we use the definition based on the Gradient flow), but
    in practice values for $Q$ in simulations with very fine lattice
    spacing cluster very close to integer values. At the same time the
    transition between different topological sectors becomes less and
    less frequent in simulation time. }
  \label{fig:freeze}
\end{figure}
Most definitions of the topological charge are not quantized on the
lattice. It is only 
when the lattice spacing of a simulation is small enough 
($a< 0.05$ fm), that the values of the topological charge measured on the
lattice cluster close to integer values. The different
topological sectors emerge as the simulation approaches the continuum
limit. As was first realized in~\cite{DelDebbio:2004xh}, at the same
time the \emph{transition} between
different topological sectors also becomes less and less
frequent at small lattice spacings (see
fig.~\ref{fig:freeze})\footnote{This is expected for the kind of 
algorithms used in all lattice QCD simulations, since they are based
on a continous change of the fields (i.e. the HMC). Local update
algorithms (such as those used in the simulations of
fig.~\ref{fig:freeze}) still suffer from topology freezing at small
lattice spacings.}. 

This has implications for the error estimation of observables that
couple strongly with the topological charge. Lattice measurements of
these observables are correlated for very long simulation times, since
they feel the topological sector in which they have been measured. Getting
a solid estimate of the statistical error of such observables is
really difficult since the topological charge is not well sampled.

The most clean solutions to the problem improve the sampling of
different topological sectors by changing the boundary conditions of
the lattice simulation in Euclidean time~\cite{Luscher:2011kk,
  Luscher:2014kea}. In these cases the topological charge is no longer
quantized (even in the continuum), and can fluctuate. In other cases
the problem can be bypassed. One can perform the simulations at fixed
topological sector and deal with power law finite volume effects if
necessary~\cite{Brower:2003yx}. Recently
simulations on very 
large physical volumes so that the finite volume effects of a fixed
topology become irrelevant have been performed
in the pure gauge theory\cite{Giusti:2018cmp}. 

All in all it is pretty uncomfortable that no algorithm is able to
sample correctly the theory at fine lattice spacings. It remains an
algorithmic challenge to find an efficient algorithm for the
simulation of QCD at fine lattice spacings. In the meantime, the
accuracy of the statistical errors quoted for simulations at fine lattice
spacings without the use of some variant of open boundary conditions
has to be taken with a grain of salt. 

\subsection{Signal to noise problem}

\begin{figure}[h!]
  \centering
    \includegraphics[width=\textwidth]{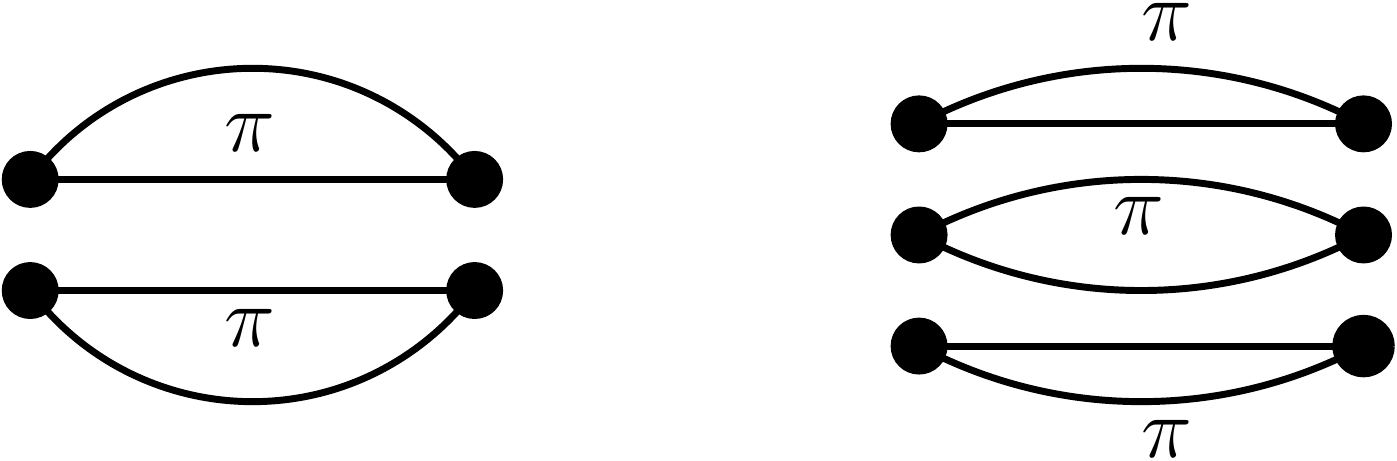}  
    \caption{The variance of a meson at large Euclidean times is
      dominated by two mesons propagating. On the other hand the
      variance of a nucleon
      two-point functions at large Euclidean times is not dominated
      by the propagation of two baryons, but by the propagation of
      three meson states. }
      \label{fig:sign}
  \end{figure}

\begin{figure}[h!]
    \includegraphics[width=\textwidth]{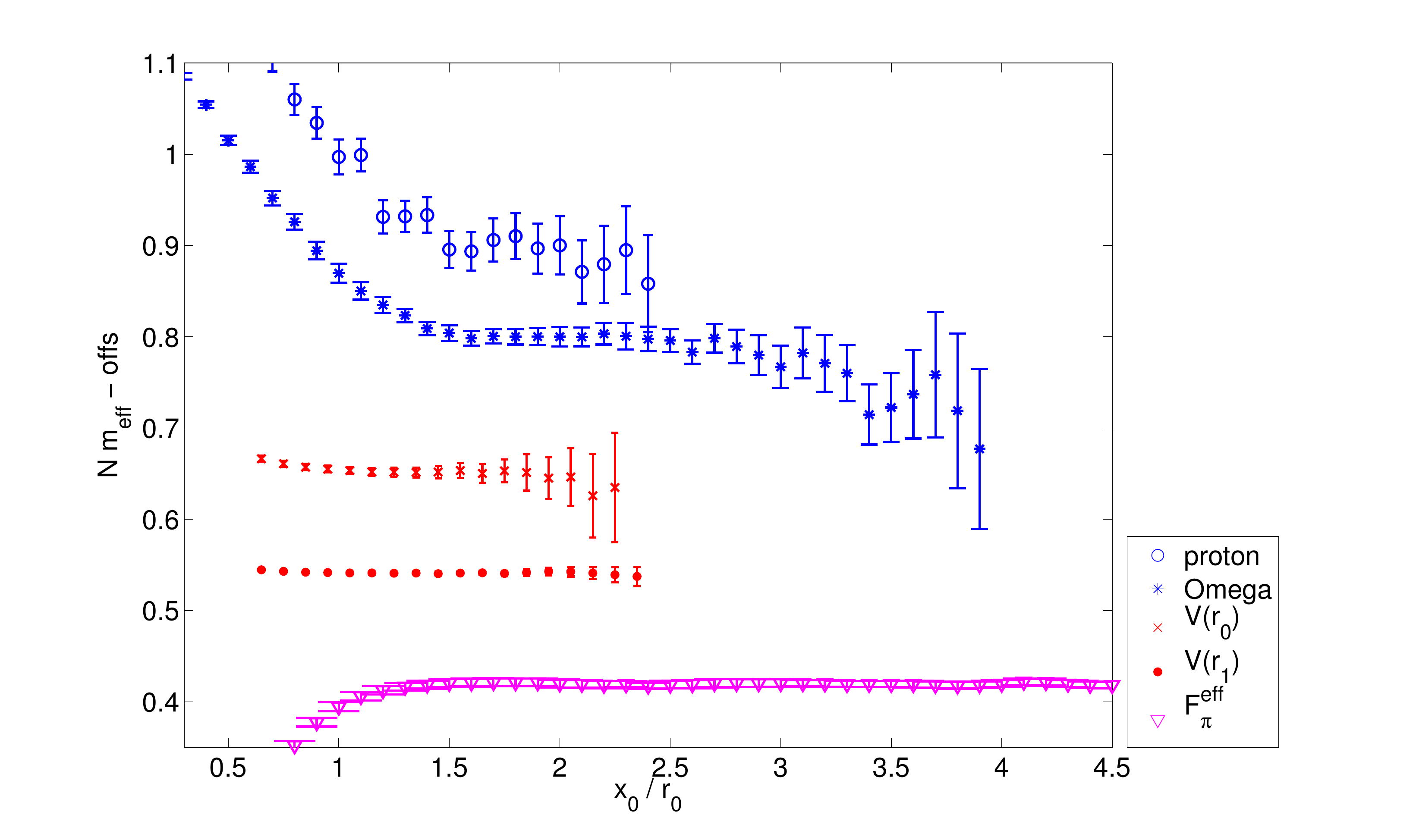}
    \caption{Several meson and baryon correlators, as well as the
      static potential at different distances (shifted vertically). 
      Meson correlators, do
      not have a signal to noise problem, 
   and its value can be determined with high precision at large
   Euclidean times. 
 (Source: $m_p$~\cite{Jager:2013kha},
 $m_{\Omega}$~\cite{Capitani:2011fg},
 $V(\approx r_0), V(\approx r_1)$~\cite{Fritzsch:2012wq},
 $f_\pi$~\cite{Lottini:2013rfa}).}
\label{fig:plateaux}
\end{figure}
  
\begin{figure}[h!]
    \includegraphics[width=\textwidth]{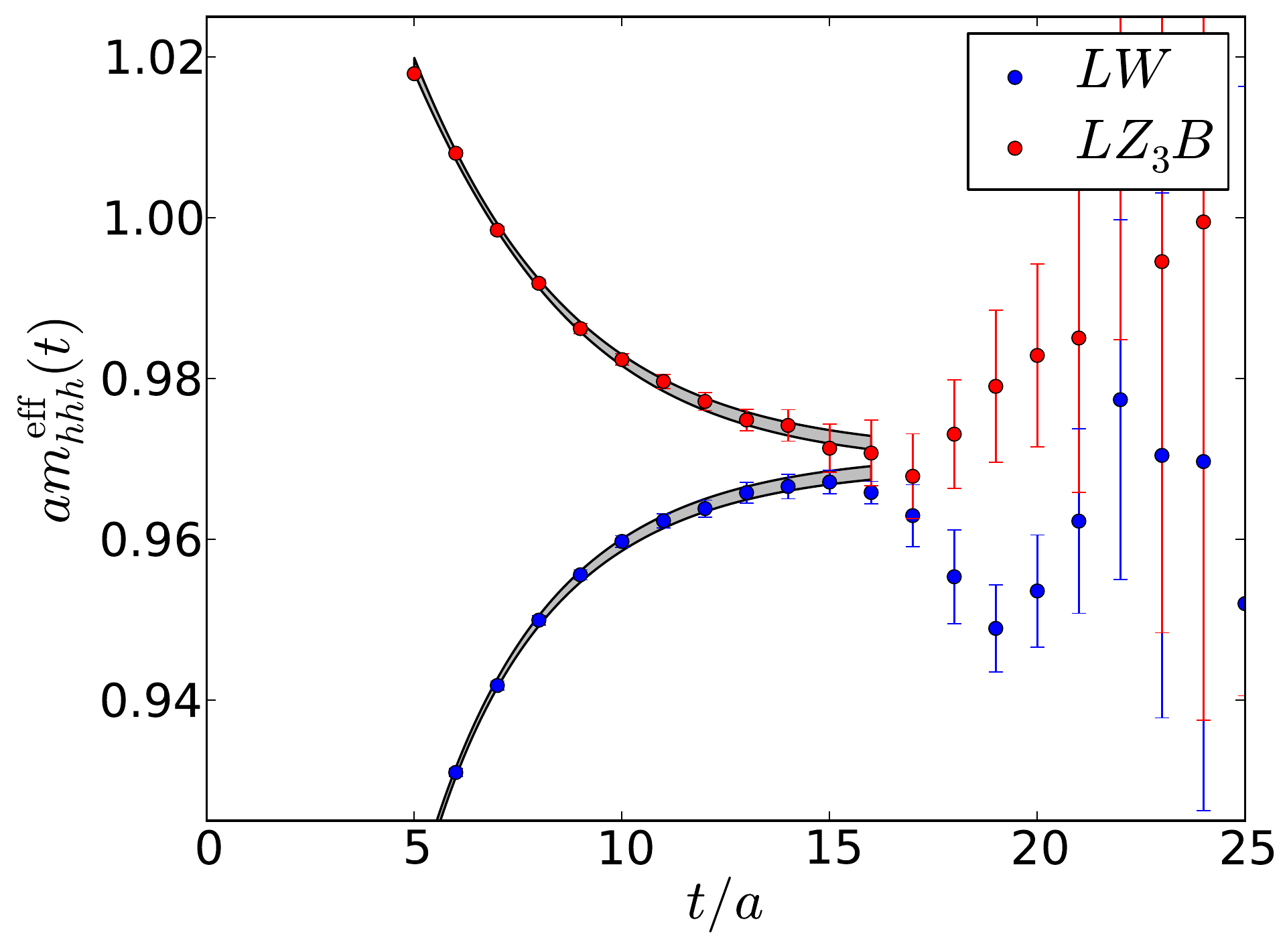}  
    \caption{Precise values of baryon masses require to fit the
      correlators at small Euclidean times, where there are
      significant contributions from excited states.
      (Source~\cite{Blum:2014tka})
    }
    \label{fig:omega}
\end{figure}

Every observable ($O$) in lattice QCD is computed as an avegare over
configurations generated by Monte Carlo sampling
\begin{equation}
  \langle O \rangle = \frac{1}{N}\sum_{t=1}^N O[U_t] + \mathcal
  O({1}/{\sqrt N})\,.
\end{equation}
Here $U_t$ represents the configuration at Monte Carlo time $t$ and
$O[U_t]$ is the value of the observable measured on such a configuration. All
observables carry an statistical uncertainty $\mathcal
O({1}/{\sqrt N})$ due to this stochastic estimation. The error in
the observable is proportional to the variance
\begin{equation}
  (\delta O)^2 \propto \langle O^2 \rangle - \langle O \rangle^2\,.
\end{equation}
The signal to noise problem refers to the particular behavior of the
error in some correlators. If we examine the case of a meson
correlator, like the pion, we have 
\begin{eqnarray}
  \langle [\bar u\gamma_5d(x)\, \bar d\gamma_5u(0)] \rangle^2
  &\simas{x_0\to \infty}& A e^{-2M_\pi x_0} + \dots \\
  \langle [\bar u\gamma_5d(x)\, \bar d\gamma_5u(0)]^2 \rangle
  &\simas{x_0\to \infty}& A' e^{-2M_\pi x_0} + \dots
\end{eqnarray}
The first equation is clear, since this is just the squared $\pi$
propagator. The second equation becomes clear when one realizes that
the Wick contractions in the expectation value
$\langle O^2 \rangle$ can be understood as 2 $\pi$ mesons propagating
between Euclidean times $0$ and $x_0$ (see fig.~\ref{fig:sign}). On the other
hand for the case of a baryon, like the proton $p$, the correlators
have the form ($a,b,c$ are color indices and $C$ the charge
conjugation matrix) 
\begin{eqnarray}
  \langle u^T_a C\gamma_5u_b d_c \epsilon^{abc}(x)\, [u^T_a C\gamma_5u_b d_c \epsilon^{abc}(0)]^\dagger \rangle^2
  &\simas{x_0\to \infty}& A e^{-2M_p x_0} + \dots \\
  \langle \left\{u^T_a C\gamma_5u_b d_c \epsilon^{abc}(x)\, [u^T_a C\gamma_5u_b d_c \epsilon^{abc}(0)]^\dagger\right\}^2 \rangle
  &\simas{x_0\to \infty}& A' e^{-3M_\pi x_0} + \dots
\end{eqnarray}
In this case the dominant contribution at large Euclidean times for
the expectation value $\langle O^2 \rangle$ comes from three $\pi$
meson propagating between Euclidean times 0 and $x_0$ (fig.~\ref{fig:sign}). 
The term with 
two protons propagating also contributes to this expectation value,
but with a term $\propto e^{-2M_p x_0}$ that decays much faster. 
See also~\cite{Luscher:2010ae} for more details.

In summary the ratio signal-to-noise shows a quite different behavior
in both cases (see fig.~\ref{fig:sign})
\begin{eqnarray}
  \frac{C_\pi(x_0)}{\delta C_\pi(x_0)} &\simas{x_0\to\infty}& 1\,,\\
  \frac{C_p(x_0)}{\delta C_p(x_0)} &\simas{x_0\to\infty}& e^{\left(M_p- \frac{3}{2}M_\pi\right)x_0 }\,.\\
\end{eqnarray}
This exponential increase in the signal to noise has in practice
important implications: it is almost impossible to determine the value
of a baryon correlator with high precision at distances of 1 fm and
larger. Precise determinations of baryon masses extract such
quantities from the values of the correlator at relatively small
Euclidean times, where there is significant contamination of excited
estates. The use of different interpolating operators (with different
excited state contamination) and the use of several numerical
techniques are common to deal with this situation (see
figs.~\ref{fig:plateaux} and~\ref{fig:omega}). 

Once more, the situation is far from ideal. Such basic quantities as baryon
masses ideally should be determined without dealing with complicated
systematic effects in the determination of the effective mass
plateau, but currently there no known techniques that allow the
precise extraction of baryon masses at distances $\sim 1$ fm.


\section{Scale variation estimation of truncation errors}
\label{sec:scale-vari-estim}

The definition of a non perturbative coupling in lattice QCD uses an observable
$O(\mu)$ that depends on only one scale ($\mu$) and can be determined non
perturbatively. This observable has a perturbative expansion when $\mu \gg
\Lambda$, that after a proper normalization can be written as
\begin{equation}
  O(\mu) \simas{\mu\to\infty} \alpha_{\overline{\rm MS} }(\mu) + \sum_{k>1}c_k \alpha_{\overline{\rm MS} }^k(\mu)\,.
\end{equation}
The RG evolution of $\alpha_{\overline{\rm MS} }(\mu)$ is given by the $\beta$
function, and can be used to obtain an estimate the truncation effects of the
perturbative series, which is valid under certain assumptions on the size of the
unknown coefficients. The truncated series
\begin{equation}
  \label{eq:Oexp}
  O^{(n)}(\mu) = \alpha_{\overline{\rm MS} }(\mu) + \sum_{k=2}^nc_k \alpha_{\overline{\rm MS} }^k(\mu) + \mathcal O(\alpha_{\overline{\rm MS} }^{n+1})\,,
\end{equation}
can be rewritten, with the same level of precision, as
\begin{equation}
  \label{eq:Otruncated}
  O^{(n)}(\mu,s) = \alpha_{\overline{\rm MS} }(\mu') + \sum_{k=2}^nc_k'(s) \alpha_{\overline{\rm MS} }^k(\mu') + \mathcal O(\alpha_{\overline{\rm MS} }^{n+1})\,,\qquad (s=\mu'/\mu)\,.
\end{equation}
Note that the dependence of $O^{(n)}(\mu, s)$ on $s$ is only due to the
truncation of the perturbative series. 
The coefficients
\begin{equation}
  c_k'(s) = \sum_{l=0}^{k-1} c'_{k,l} \log^l(s)\,,
\end{equation}
can all be determined from $c_k$ that appear in Eq.~(\ref{eq:Oexp}) thanks to
the recursion
\begin{eqnarray}
  c'_{k,0} &=& c_k\,, \\
  c'_{k,l} &=& \frac{2}{l}\sum_{j=0}^{k-1} j(4\pi)^{k-j} b_{k-j-1}c_{j,l-1}'\,,
\end{eqnarray}
where $b_n$ are the coefficients of the $\beta$ function
Eq.~(\ref{eq:BetaFunAsymp}).

Given a value for the observable $O_{\rm ref} = O(\mu_{\rm ref})$
determined on the lattice at some value of the scale $\mu_{\rm ref}$, we can
compare the determination of $\alpha_{\overline{\rm MS} }(M_Z)$ coming
from using Eq.~(\ref{eq:Otruncated}) for different values of $s$, and
use them as an estimate of the truncation uncertainties. 
Note that usually lattice works with $N_{\rm f} = 2+1$ or $N_{\rm f} =2+1+1$
QCD, which means that in order to obtain $\alpha_{\overline{\rm MS}
}(M_Z)$ one needs to match the couplings through the bottom and/or
charm thresholds. 

For numerical results the reader should look at the \texttt{RunDec} package~\cite{Chetyrkin:2000yt, Herren:2017osy}. 
We also provide a freely available
software \texttt{ScaleErrors}\footnote{\label{ftn:scaleerrors}\url{https://igit.ific.uv.es/alramos/scaleerrors.jl}} to
determine the truncation uncertainties in $\alpha_s$ as we detail below.

\subsection{The case of the Schr\"odinger Funcional coupling in full detail}

In this section we describe in full detail our procedure for estimating the
impact of scale variations in the case of works based on finite size
scaling. The
implementation for the other observables is sketched briefly in the rest of the
appendix.

Our starting point is the value of the three flavor theory, that we
assume to be
\begin{equation}
  \label{eq:lam3ref}
\Lambda^{(3)}_{\overline{\rm MS} } = 341\, {\rm MeV}\,,
\end{equation}
and the perturbative expansion of the Schr\"odinger Functional (SF)
coupling, that is measured on a finite volume at the scale $\mu =
1/L$. For $N_{\rm f} =3$ QCD we have
\begin{eqnarray}
  \label{eq:sfptnf3}
  \alpha_{\rm SF}(\mu) &=& \alpha_{\overline{\rm MS} }(s\mu) - [1.37520970   -  1.43239449 \log(s)]\alpha_{\overline{\rm MS} }^2(s\mu) \\
  \nonumber  
                     &+& [0.57120172   - 3.12911612 \log(s)    +  2.05175397 \log(s)^2]\alpha_{\overline{\rm MS} }^3(s\mu)\,.
 \end{eqnarray}
where the $\log$ terms can be computed from
expression~(\ref{eq:Otruncated}) and the perturbative
computation~\cite{Bode:1998hd, Bode:1999sm}. We point that the
routine \texttt{scale\_errors} from the package
\texttt{ScaleErrors.jl} (see footnote~\ref{ftn:scaleerrors}),
takes as input the perturbative coefficients
\begin{equation}
  c_k (1) = \{1.0,-1.3752097,0.571202\} \,,
\end{equation}
and determines the truncation uncertainties exactly as described below.

We proceed as follows:
\begin{enumerate}
\item Chose a reference value for the scale $s_{\rm ref}$ and solve
  the non linear equation
  \begin{equation}
    \frac{\Lambda^{(3)}_{\overline{\rm MS}} }{s_{\rm ref}\mu}= 
  \left[b_0\bar g^2(s_{\rm ref}\mu)\right]^{-\frac{b_1}{2b_0^2}}\,
  e^{-\frac{1}{2b_0\bar g^2(s_{\rm ref}\mu)}}\,
  \exp\left\{-
    \int_{0}^{\bar g(s_{\rm ref}\mu)}{\rm d}x\, \left[\frac{1}{\beta_{\overline{\rm MS} }(x)} +
    \frac{1}{b_0x^3} - \frac{b_1}{b_0^2x}\right]\right\}\,.
  \end{equation}
  in order to obtain $\alpha_{\overline{\rm MS} }(s_{\rm ref}\mu)
  \equiv \bar g^2(s_{\rm ref}\mu)/(4\pi)$. 
  For example, using $\mu = 80$ GeV, $s_{\rm ref} = 2$ and the 5-loop
  $\beta_{\overline{\rm MS} }$ function together
  with our reference value of $\Lambda^{(3)}_{\overline{\rm MS} } =
  341$ MeV we get
  \begin{equation}
    \label{eq:alms1}
    \alpha_{\overline{\rm MS} }(s_{\rm ref}\mu) = 0.09703895\,.
  \end{equation}
  
  \item This value is plugged in Eq.~(\ref{eq:sfptnf3}) in order to
    obtain
    \begin{equation}
      \label{eq:alphasfref}
      \alpha_{\rm SF}(80\, {\rm GeV}) = 0.09287934\,.
    \end{equation}

  \item We can now solve the polynomial equation~(\ref{eq:sfptnf3})
    for different values of $s$ and the l.h.s fixed to the value in
    Eq.~(\ref{eq:alphasfref}). \ie\, Using $s=4$ ($s\mu = 320$ GeV), we solve
\begin{eqnarray}
      0.09287934 &=& \alpha_{\overline{\rm MS} }(s\mu) - [1.37520970   -  1.43239449 \log(s)]\alpha_{\overline{\rm MS} }^2(s\mu) \\
  \nonumber  
                     &+& [0.57120172   - 3.12911612 \log(s)    +  2.05175397 \log(s)^2]\alpha_{\overline{\rm MS} }^3(s\mu)\,,
\end{eqnarray}
To obtain 
    \begin{equation}
    \label{eq:alms2}
      \alpha_{\overline{\rm MS} }(320\, {\rm GeV}) = 0.08802818\,.
    \end{equation}
  \item The two values of the coupling
    Eqs.~(\ref{eq:alms1}),~(\ref{eq:alms2}) should be equivalent,
    except for the truncation uncertainties. 
    In order to compare them, we run the two values to a common scale
    ($M_Z$), crossing the charm and bottom thresholds. 
    We get
    \begin{subequations}
      \label{eq:alvalues}
      \begin{eqnarray}
      \alpha_{\overline{\rm MS} }(80\, {\rm GeV}) = 0.09703895 &\Longrightarrow& \alpha^{(N_{\rm f} = 5)}_{\overline{\rm MS} }(M_Z) = 0.11851821\,,\\
      \alpha_{\overline{\rm MS} }(320\, {\rm GeV}) = 0.08802818 &\Longrightarrow& \alpha^{(N_{\rm f} = 5)}_{\overline{\rm MS} }(M_Z) = 0.11843281\,.
    \end{eqnarray}
    \end{subequations}

    Let us give a few details on how this is done. 
    Using the 5-loop three flavor beta function and the charm quark
    mass $m_{\rm c}^\star = m_{\rm c}(m_{\rm c}) = 1275.0$ MeV, we
    solve the     non-linear equation
    \begin{equation}
      \label{eq:rung}
      \log \left( \frac{80\,{\rm GeV}}{m_{\rm c}^\star} \right) =
      \int_{\bar g(m_{\rm c})}^{\bar g(80\, {\rm GeV})} \frac{{\rm d} x }{\beta_{\overline{\rm MS} }^{(N_{\rm f} = 3)}(x)}\,.
    \end{equation}
    in order to obtain the value of the three flavor coupling at the
    charm scale. 
    We obtain
    \begin{equation}
      \alpha_{\overline{\rm MS} }(80\, {\rm GeV}) = 0.09703895 \Longrightarrow
      \alpha_{\overline{\rm MS} }(m_{\rm c}^\star)  = 0.39566020\,.
    \end{equation}

    We now ``cross the charm threshold'' by using the decoupling
    relations~\cite{Chetyrkin:2005ia,Weinberg:1980wa,Bernreuther:1981sg,Grozin:2011nk,Schroder:2005hy} to determine the four flavor coupling at the scale
    $m_{\rm c}^\star$.
    \begin{equation}
      \alpha_{\overline{\rm MS} }(m_{\rm c}^\star)                = 0.39566020 \Longrightarrow
      \alpha_{\overline{\rm MS} }^{(N_{\rm f} = 4)}(m_{\rm c}^\star) = 0.393947409 \,.
    \end{equation}

    A similar procedure is used to run the coupling to the bottom
    quark threshold $m_{\rm b}^\star = m_{\rm b}(m_{\rm b}) = 4198.0$
    MeV, and convert to the five flavor coupling 
    \begin{displaymath}
       \alpha_{\overline{\rm MS} }^{(N_{\rm f} = 4)}(m_{\rm c}^\star) = 0.393947409 \Longrightarrow
       \alpha_{\overline{\rm MS} }^{(N_{\rm f} = 4)}(m_{\rm b}^\star) = 0.226549731 \Longrightarrow
       \alpha_{\overline{\rm MS} }^{(N_{\rm f} = 5)}(m_{\rm b}^\star) = 0.226311579\,.
    \end{displaymath}

    Finally, one uses Eq.~(\ref{eq:rung}) using the five flavor
    $\beta_{\overline{\rm MS} }$ function to run the coupling to the
    scale $M_Z$.
    
  \item Finally, the difference between values of
    $\alpha_{\overline{\rm MS} }^{(N_{\rm f} = 5)}(M_Z)$ in Eqs.~(\ref{eq:alvalues})
    \begin{equation}
      \delta \alpha_{\overline{\rm MS} }^{(N_{\rm f} = 5)}(M_Z) = 8.5\times 10^{-5}
      \qquad [0.07\%]\,.
    \end{equation}
    can be used as an estimate of the truncation uncertainties. 
    
\end{enumerate}

\subsection{The static potential}
\label{sec:static-potential-app}

We start from the perturbative expression of the potential $V(r)$
\begin{equation}
  V(r) = -\frac{4}{3r} \sum_{n=0} P_n \left( \frac{\alpha_s}{4\pi} \right)^n\,,
\end{equation}
where $\alpha_s = \alpha_{\overline{\rm MS} }(1/r)$ is the strong coupling. 
Note that we ignore the logarithmic corrections $\propto \log\alpha$
due to the IR divergent nature of $V(r)$. 
See section~\ref{sec:static-potential} for more details.
Using the RG equations, we have
\begin{equation}
  \frac{{\rm d} \alpha_s}{{\rm d} r} = \frac{2\alpha_s^2}{r}
  \sum_{i=0}(4\pi)^i b_i\alpha_s^i\, ,
\end{equation}
which we can use to evaluate the derivative of the static potential with respect
to the scale $r$:
\begin{equation}
  F(r) = \frac{{\rm d} V(r)}{{\rm d} r} =
  \frac{4}{3r^2} \sum_{n=0}\frac{P_n}{(4\pi)^n} \alpha_s^{n+1} -
  \frac{4}{3r^2} \sum_{n=0} \frac{2(n+1)P_n}{(4\pi)^n} \alpha_s^{n+2}
  \left[ \sum_{j=0}(4\pi)^{j+1}b_j\alpha_s^j \right]\,.
\end{equation}
Collecting the coefficients in the equation above, we get an expression for the
perturbative expansion of the force, as a function of the number of flavors
$N_f$. We use the force as the observable that determines the strong coupling
constant, with the typical scale associated to the observable being $\mu=1/r$. 
In $N_{\rm f} =3$ QCD the known terms in the perturbative series~\cite{Fischler:1977yf, Peter:1996ig, Smirnov:2009fh, Smirnov:2008pn}
together with expression Eq.~(\ref{eq:Otruncated}) allows to write
\begin{equation}
  \begin{split}
  \alpha_{\rm qq}(\mu) \simas{\mu\to\infty}
  &\alpha_{\overline{\rm MS} }(s\mu) + [-0.0485502   +  1.43239449  \log(s)] \alpha^{2}_{\overline{\rm MS} }(s\mu) \\
  +\, & [  0.687447   +  0.67148339 \log(s)   +  2.05175397 \log(s)^2] \alpha^{3}_{\overline{\rm MS} }(s\mu)\\
  +\, & [  0.818808   +  3.52427341 \log(s)   +  2.6037989 \log(s)^2   +  2.93892108 \log(s)^3 ] \alpha^{4}_{\overline{\rm MS} }(s\mu)\\
  +\, &\dots\,.
  \end{split}
\end{equation}

The scale of fastest apparent convergence is reached at $s^\star =
1.034475$, and the maximum scale reached in current state of the art
determinations is 8 GeV, albeit at a single value of the lattice
spacing.

IR divergences affect the term $\propto \alpha_{\overline{\rm MS} }^4(Q)$. 
If we assume a fixed value for the ultra-soft scale $\nu_{\rm us} = 2$ GeV,
the last term would read instead:
\begin{equation}
  \dots + \alpha^{4}_{\overline{\rm MS} }(sQ) [  0.246933   +  3.52427341 \log(s)   +  2.6037989 \log(s)^2   +  2.93892108 \log(s)^3 ]\,.
\end{equation}
This has an effect in the truncation uncertainties at the $20\%$ level.

\subsection{HQ correlators}

The perturbative expansions for the ratios of moments $\alpha_{\rm HQ,
  n}$ is given by\cite{Chetyrkin:2006xg, Chetyrkin:1997mb, Broadhurst:1991fi}: 

\begin{eqnarray}
  \alpha_{\rm HQ, 4}(\mu) &\simas{\mu \to \infty}& \alpha_{\overline{\rm MS} }(s\mu) - [0.07762325   -  1.43239449 \log(s)]\alpha_{\overline{\rm MS} }^2(s\mu) \\
  \nonumber  
                     &+& [0.07957445   +  .58819524 \log(s)    +  2.05175397 \log(s)^2]\alpha_{\overline{\rm MS} }^3(s\mu) + \dots\,.\\
  \alpha_{\rm HQ, 6}(\mu) &\simas{\mu \to \infty}& \alpha_{\overline{\rm MS} }(s\mu) + [0.77386542   +  1.43239449 \log(s)]\alpha_{\overline{\rm MS} }^2(s\mu) \\
  \nonumber  
                     &-& [0.08560363   -  3.02753059 \log(s)   -  2.05175397 \log(s)^2]\alpha_{\overline{\rm MS} }^3(s\mu) + \dots\,.\\
  \alpha_{\rm HQ, 8}(\mu) &\simas{\mu \to \infty}& \alpha_{\overline{\rm MS} }(s\mu) + [1.08917060   +  1.43239449 \log(s)]\alpha_{\overline{\rm MS} }^2(s\mu) \\
  \nonumber  
                     &+& [0.20034888   +  3.93081340 \log(s)   +  2.05175397\log(s)^2]\alpha_{\overline{\rm MS} }^3(s\mu) + \dots\,.\\
  \alpha_{\rm HQ, 10}(\mu) &\simas{\mu \to \infty}& \alpha_{\overline{\rm MS} }(s\mu) + [1.44848150   +  1.43239449 \log(s)]\alpha_{\overline{\rm MS} }^2(s\mu) \\
  \nonumber  
                     &+& [0.66519861   +  4.96016330 \log(s)   +  2.05175397 \log(s)^2]\alpha_{\overline{\rm MS} }^3(s\mu) + \dots\,.
 \end{eqnarray}
 
\subsection{Wilson loops}

For the couplings defined in Eqs~(\ref{eq:alphaWptMSbar}), defined
from Wilson loops, the perturbative expansion reads~\cite{Davies:2008sw}:  
\begin{eqnarray}
  \alpha_{\rm W_{11}}(\mu) &\simas{\mu \to \infty}& \alpha_{\overline{\rm MS} }(s\mu) - [0.87811924   -  1.43239449 \log(s)]\alpha_{\overline{\rm MS} }^2(s\mu) \\
  \nonumber  
                     &+& [4.20161085   - 1.70505684 \log(s)    +  2.05175397 \log(s)^2]\alpha_{\overline{\rm MS} }^3(s\mu) + \dots\,.\\
  \alpha_{\rm W_{12}}(\mu) &\simas{\mu \to \infty}& \alpha_{\overline{\rm MS} }(s\mu) + [0.79128076   +  1.43239449 \log(s)]\alpha_{\overline{\rm MS} }^2(s\mu) \\
  \nonumber  
                     &+& [3.18658638   + 3.07742188 \log(s)   +  2.05175397 \log(s)^2]\alpha_{\overline{\rm MS} }^3(s\mu) + \dots\,.
 \end{eqnarray} 

The scale of fastest apparent convergence is reached at $s^\star =
1.4252357$, and the maximum scale reached in current state of the art
determinations is $4.4$ GeV.


\addcontentsline{toc}{section}{References}
\bibliography{/home/alberto/docs/bib/math,/home/alberto/docs/bib/campos,/home/alberto/docs/bib/fisica,/home/alberto/docs/bib/computing}

\providecommand{\href}[2]{#2}\begingroup\raggedright\begin{thebibliography}{100}

\bibitem{Aoki:2019cca}
{\bfseries Flavour Lattice Averaging Group} Collaboration, S.~Aoki {\em
  et~al.}, ``{FLAG Review 2019},''
\href{http://arxiv.org/abs/1902.08191}{{\ttfamily arXiv:1902.08191 [hep-lat]}}.

\bibitem{Parker:2018vye}
R.~H. Parker, C.~Yu, W.~Zhong, B.~Estey, and H.~Müller, ``{Measurement of the
  fine-structure constant as a test of the Standard Model},''
  \href{http://dx.doi.org/10.1126/science.aap7706}{{\em Science} {\bfseries
  360} (2018) 191},
\href{http://arxiv.org/abs/1812.04130}{{\ttfamily arXiv:1812.04130
  [physics.atom-ph]}}.

\bibitem{Hanneke:2008tm}
D.~Hanneke, S.~Fogwell, and G.~Gabrielse, ``{New Measurement of the Electron
  Magnetic Moment and the Fine Structure Constant},''
  \href{http://dx.doi.org/10.1103/PhysRevLett.100.120801}{{\em Phys. Rev.
  Lett.} {\bfseries 100} (2008) 120801},
\href{http://arxiv.org/abs/0801.1134}{{\ttfamily arXiv:0801.1134
  [physics.atom-ph]}}.

\bibitem{Politzer:1973fx}
H.~D. Politzer, ``{Reliable Perturbative Results for Strong Interactions?},''
  \href{http://dx.doi.org/10.1103/PhysRevLett.30.1346}{{\em Phys. Rev. Lett.}
  {\bfseries 30} (1973) 1346--1349}.
[,274(1973)].

\bibitem{Gross:1973id}
D.~J. Gross and F.~Wilczek, ``{Ultraviolet Behavior of Nonabelian Gauge
  Theories},'' \href{http://dx.doi.org/10.1103/PhysRevLett.30.1343}{{\em Phys.
  Rev. Lett.} {\bfseries 30} (1973) 1343--1346}.
[,271(1973)].

\bibitem{Callan:1970yg}
J.~Callan, Curtis~G., ``{Broken scale invariance in scalar field theory},''
  \href{http://dx.doi.org/10.1103/PhysRevD.2.1541}{{\em Phys. Rev. D}
  {\bfseries 2} (1970) 1541--1547}.

\bibitem{Symanzik:1970rt}
K.~Symanzik, ``{Small distance behavior in field theory and power counting},''
  \href{http://dx.doi.org/10.1007/BF01649434}{{\em Commun. Math. Phys.}
  {\bfseries 18} (1970) 227--246}.

\bibitem{Bardeen:1978yd}
W.~A. Bardeen, A.~J. Buras, D.~W. Duke, and T.~Muta, ``{Deep Inelastic
  Scattering Beyond the Leading Order in Asymptotically Free Gauge Theories},''
\href{http://dx.doi.org/10.1103/PhysRevD.18.3998}{{\em Phys. Rev.} {\bfseries
  D18} (1978) 3998}.

\bibitem{Boyle:2016wis}
P.~Boyle, L.~Del~Debbio, and A.~Khamseh, ``{Massive momentum-subtraction
  scheme},'' \href{http://dx.doi.org/10.1103/PhysRevD.95.054505}{{\em Phys.
  Rev.} {\bfseries D95} no.~5, (2017) 054505},
\href{http://arxiv.org/abs/1611.06908}{{\ttfamily arXiv:1611.06908 [hep-lat]}}.

\bibitem{Fritzsch:2018kjg}
{\bfseries ALPHA} Collaboration, P.~Fritzsch, R.~Sommer, F.~Stollenwerk, and
  U.~Wolff, ``{Symanzik Improvement with Dynamical Charm: A 3+1 Scheme for
  Wilson Quarks},'' \href{http://dx.doi.org/10.1007/JHEP06(2018)025}{{\em JHEP}
  {\bfseries 06} (2018) 025},
\href{http://arxiv.org/abs/1805.01661}{{\ttfamily arXiv:1805.01661 [hep-lat]}}.

\bibitem{vanRitbergen:1997va}
T.~van Ritbergen, J.~A.~M. Vermaseren, and S.~A. Larin, ``{The Four loop beta
  function in quantum chromodynamics},''
  \href{http://dx.doi.org/10.1016/S0370-2693(97)00370-5}{{\em Phys. Lett.}
  {\bfseries B400} (1997) 379--384},
\href{http://arxiv.org/abs/hep-ph/9701390}{{\ttfamily arXiv:hep-ph/9701390
  [hep-ph]}}.

\bibitem{Czakon:2004bu}
M.~Czakon, ``{The Four-loop QCD beta-function and anomalous dimensions},''
  \href{http://dx.doi.org/10.1016/j.nuclphysb.2005.01.012}{{\em Nucl. Phys.}
  {\bfseries B710} (2005) 485--498},
\href{http://arxiv.org/abs/hep-ph/0411261}{{\ttfamily arXiv:hep-ph/0411261
  [hep-ph]}}.

\bibitem{Baikov:2016tgj}
P.~A. Baikov, K.~G. Chetyrkin, and J.~H. K\"hn, ``{Five-Loop Running of the QCD
  coupling constant},''
  \href{http://dx.doi.org/10.1103/PhysRevLett.118.082002}{{\em Phys. Rev.
  Lett.} {\bfseries 118} no.~8, (2017) 082002},
\href{http://arxiv.org/abs/1606.08659}{{\ttfamily arXiv:1606.08659 [hep-ph]}}.

\bibitem{Luthe:2016ima}
T.~Luthe, A.~Maier, P.~Marquard, and Y.~Schröder, ``{Towards the five-loop
  Beta function for a general gauge group},''
  \href{http://dx.doi.org/10.1007/JHEP07(2016)127}{{\em JHEP} {\bfseries 07}
  (2016) 127},
\href{http://arxiv.org/abs/1606.08662}{{\ttfamily arXiv:1606.08662 [hep-ph]}}.

\bibitem{Herzog:2017ohr}
F.~Herzog, B.~Ruijl, T.~Ueda, J.~A.~M. Vermaseren, and A.~Vogt, ``{The
  five-loop beta function of Yang-Mills theory with fermions},''
  \href{http://dx.doi.org/10.1007/JHEP02(2017)090}{{\em JHEP} {\bfseries 02}
  (2017) 090},
\href{http://arxiv.org/abs/1701.01404}{{\ttfamily arXiv:1701.01404 [hep-ph]}}.

\bibitem{Hasenfratz:1980kn}
A.~Hasenfratz and P.~Hasenfratz, ``{The Connection Between the Lambda
  Parameters of Lattice and Continuum QCD},''
  \href{http://dx.doi.org/10.1016/0370-2693(80)90118-5}{{\em Phys. Lett.}
  {\bfseries 93B} (1980) 165}.
[,241(1980)].

\bibitem{Chetyrkin:2005ia}
K.~G. Chetyrkin, J.~H. Kuhn, and C.~Sturm, ``{QCD decoupling at four loops},''
  \href{http://dx.doi.org/10.1016/j.nuclphysb.2006.03.020}{{\em Nucl. Phys.}
  {\bfseries B744} (2006) 121--135},
\href{http://arxiv.org/abs/hep-ph/0512060}{{\ttfamily arXiv:hep-ph/0512060
  [hep-ph]}}.

\bibitem{Weinberg:1980wa}
S.~Weinberg, ``{Effective Gauge Theories},''
\href{http://dx.doi.org/10.1016/0370-2693(80)90660-7}{{\em Phys. Lett.}
  {\bfseries B91} (1980) 51--55}.

\bibitem{Bernreuther:1981sg}
W.~Bernreuther and W.~Wetzel, ``{Decoupling of Heavy Quarks in the Minimal
  Subtraction Scheme},'' \href{http://dx.doi.org/10.1016/0550-3213(82)90288-7,
  10.1016/S0550-3213(97)00811-0}{{\em Nucl. Phys.} {\bfseries B197} (1982)
  228--236}.
[Erratum: Nucl. Phys.B513,758(1998)].

\bibitem{Grozin:2011nk}
A.~G. Grozin, M.~Hoeschele, J.~Hoff, M.~Steinhauser, M.~Hoschele, J.~Hoff, and
  M.~Steinhauser, ``{Simultaneous decoupling of bottom and charm quarks},''
  \href{http://dx.doi.org/10.1007/JHEP09(2011)066}{{\em JHEP} {\bfseries 09}
  (2011) 066},
\href{http://arxiv.org/abs/1107.5970}{{\ttfamily arXiv:1107.5970 [hep-ph]}}.

\bibitem{Schroder:2005hy}
Y.~Schroder and M.~Steinhauser, ``{Four-loop decoupling relations for the
  strong coupling},''
  \href{http://dx.doi.org/10.1088/1126-6708/2006/01/051}{{\em JHEP} {\bfseries
  01} (2006) 051},
\href{http://arxiv.org/abs/hep-ph/0512058}{{\ttfamily arXiv:hep-ph/0512058
  [hep-ph]}}.

\bibitem{Athenodorou:2018wpk}
A.~Athenodorou, J.~Finkenrath, F.~Knechtli, T.~Korzec, B.~Leder, M.~K.
  Marinkovic, and R.~Sommer, ``{How perturbative are heavy sea quarks?},''
\href{http://arxiv.org/abs/1809.03383}{{\ttfamily arXiv:1809.03383 [hep-lat]}}.

\bibitem{Korzec:2016eko}
T.~Korzec, F.~Knechtli, S.~Cali, B.~Leder, and G.~Moir, ``{Impact of dynamical
  charm quarks},'' {\em PoS} {\bfseries LATTICE2016} (2017) 126,
\href{http://arxiv.org/abs/1612.07634}{{\ttfamily arXiv:1612.07634 [hep-lat]}}.

\bibitem{pdgtbp:2020}
{\bfseries Particle Data Group} Collaboration, P.~Zyla {\em et~al.}, ``{Review
  of Particle Physics},'' {\em to be published in Prog. Theor. Exp. Phys. 2020}
  {\bfseries 083C01} (2020) .

\bibitem{Pich:2013lsa}
A.~Pich, ``{Precision Tau Physics},''
  \href{http://dx.doi.org/10.1016/j.ppnp.2013.11.002}{{\em Prog. Part. Nucl.
  Phys.} {\bfseries 75} (2014) 41--85},
\href{http://arxiv.org/abs/1310.7922}{{\ttfamily arXiv:1310.7922 [hep-ph]}}.

\bibitem{Baikov:2008jh}
P.~A. Baikov, K.~G. Chetyrkin, and J.~H. Kuhn, ``{Order alpha**4(s) QCD
  Corrections to Z and tau Decays},''
  \href{http://dx.doi.org/10.1103/PhysRevLett.101.012002}{{\em Phys. Rev.
  Lett.} {\bfseries 101} (2008) 012002},
\href{http://arxiv.org/abs/0801.1821}{{\ttfamily arXiv:0801.1821 [hep-ph]}}.

\bibitem{Baikov:2012er}
P.~A. Baikov, K.~G. Chetyrkin, J.~H. Kuhn, and J.~Rittinger, ``{Complete ${\cal
  O}(\alpha_s^4)$ QCD Corrections to Hadronic $Z$-Decays},''
  \href{http://dx.doi.org/10.1103/PhysRevLett.108.222003}{{\em Phys. Rev.
  Lett.} {\bfseries 108} (2012) 222003},
\href{http://arxiv.org/abs/1201.5804}{{\ttfamily arXiv:1201.5804 [hep-ph]}}.

\bibitem{Baikov:2012zn}
P.~A. Baikov, K.~G. Chetyrkin, J.~H. Kuhn, and J.~Rittinger, ``{Adler Function,
  Sum Rules and Crewther Relation of Order O($\alpha_s^4$): the Singlet
  Case},'' \href{http://dx.doi.org/10.1016/j.physletb.2012.06.052}{{\em Phys.
  Lett.} {\bfseries B714} (2012) 62--65},
\href{http://arxiv.org/abs/1206.1288}{{\ttfamily arXiv:1206.1288 [hep-ph]}}.

\bibitem{Ball:2011us}
R.~D. Ball, V.~Bertone, L.~Del~Debbio, S.~Forte, A.~Guffanti, J.~I. Latorre,
  S.~Lionetti, J.~Rojo, and M.~Ubiali, ``{Precision NNLO determination of
  $\alpha_s(M_Z)$ using an unbiased global parton set},''
  \href{http://dx.doi.org/10.1016/j.physletb.2011.11.053}{{\em Phys. Lett.}
  {\bfseries B707} (2012) 66--71},
\href{http://arxiv.org/abs/1110.2483}{{\ttfamily arXiv:1110.2483 [hep-ph]}}.

\bibitem{Ball:2018iqk}
{\bfseries NNPDF} Collaboration, R.~D. Ball, S.~Carrazza, L.~Del~Debbio,
  S.~Forte, Z.~Kassabov, J.~Rojo, E.~Slade, and M.~Ubiali, ``{Precision
  determination of the strong coupling constant within a global PDF
  analysis},'' \href{http://dx.doi.org/10.1140/epjc/s10052-018-5897-7}{{\em
  Eur. Phys. J.} {\bfseries C78} no.~5, (2018) 408},
\href{http://arxiv.org/abs/1802.03398}{{\ttfamily arXiv:1802.03398 [hep-ph]}}.

\bibitem{Harland-Lang:2015nxa}
L.~A. Harland-Lang, A.~D. Martin, P.~Motylinski, and R.~S. Thorne,
  ``{Uncertainties on $\alpha _S$ in the MMHT2014 global PDF analysis and
  implications for SM predictions},''
  \href{http://dx.doi.org/10.1140/epjc/s10052-015-3630-3}{{\em Eur. Phys. J.}
  {\bfseries C75} no.~9, (2015) 435},
\href{http://arxiv.org/abs/1506.05682}{{\ttfamily arXiv:1506.05682 [hep-ph]}}.

\bibitem{Salam:2017qdl}
G.~P. Salam, \href{http://dx.doi.org/10.1142/9789813238053_0007}{``{The strong
  coupling: a theoretical perspective},''} in {\em From My Vast Repertoire ...:
  Guido Altarelli's Legacy}, A.~Levy, S.~Forte, and G.~Ridolfi, eds.,
  pp.~101--121.
\newblock 2019.
\newblock
\href{http://arxiv.org/abs/1712.05165}{{\ttfamily arXiv:1712.05165 [hep-ph]}}.
\newblock

\bibitem{Forte:2020pyp}
S.~Forte and Z.~Kassabov, ``{Why $\alpha_s$ Cannot be Determined from Hadronic
  Processes without Simultaneously Determining the Parton Distributions},''
\href{http://arxiv.org/abs/2001.04986}{{\ttfamily arXiv:2001.04986 [hep-ph]}}.

\bibitem{Brida:2016flw}
{\bfseries ALPHA} Collaboration, M.~Dalla~Brida, P.~Fritzsch, T.~Korzec,
  A.~Ramos, S.~Sint, and R.~Sommer, ``{Determination of the QCD
  $\Lambda$-parameter and the accuracy of perturbation theory at high
  energies},'' \href{http://dx.doi.org/10.1103/PhysRevLett.117.182001}{{\em
  Phys. Rev. Lett.} {\bfseries 117} no.~18, (2016) 182001},
\href{http://arxiv.org/abs/1604.06193}{{\ttfamily arXiv:1604.06193 [hep-ph]}}.

\bibitem{DallaBrida:2018rfy}
{\bfseries ALPHA} Collaboration, M.~Dalla~Brida, P.~Fritzsch, T.~Korzec,
  A.~Ramos, S.~Sint, and R.~Sommer, ``{A non-perturbative exploration of the
  high energy regime in $N_{\mathrm{f}}=3$ QCD},''
  \href{http://dx.doi.org/10.1140/epjc/s10052-018-5838-5}{{\em Eur. Phys. J.}
  {\bfseries C78} no.~5, (2018) 372},
\href{http://arxiv.org/abs/1803.10230}{{\ttfamily arXiv:1803.10230 [hep-lat]}}.

\bibitem{Cacciari:2011ze}
M.~Cacciari and N.~Houdeau, ``{Meaningful characterisation of perturbative
  theoretical uncertainties},''
  \href{http://dx.doi.org/10.1007/JHEP09(2011)039}{{\em JHEP} {\bfseries 09}
  (2011) 039},
\href{http://arxiv.org/abs/1105.5152}{{\ttfamily arXiv:1105.5152 [hep-ph]}}.

\bibitem{AbdulKhalek:2019bux}
{\bfseries NNPDF} Collaboration, R.~Abdul~Khalek {\em et~al.}, ``{A First
  Determination of Parton Distributions with Theoretical Uncertainties},''
\href{http://arxiv.org/abs/1905.04311}{{\ttfamily arXiv:1905.04311 [hep-ph]}}.

\bibitem{AbdulKhalek:2019ihb}
{\bfseries NNPDF} Collaboration, R.~Abdul~Khalek {\em et~al.}, ``{Parton
  Distributions with Theory Uncertainties: General Formalism and First
  Phenomenological Studies},''
  \href{http://dx.doi.org/10.1140/epjc/s10052-019-7401-4}{{\em Eur. Phys. J. C}
  {\bfseries 79} no.~11, (2019) 931},
  \href{http://arxiv.org/abs/1906.10698}{{\ttfamily arXiv:1906.10698
  [hep-ph]}}.

\bibitem{Beneke:1998ui}
M.~Beneke, ``{Renormalons},''
  \href{http://dx.doi.org/10.1016/S0370-1573(98)00130-6}{{\em Phys. Rept.}
  {\bfseries 317} (1999) 1--142},
\href{http://arxiv.org/abs/hep-ph/9807443}{{\ttfamily arXiv:hep-ph/9807443
  [hep-ph]}}.

\bibitem{tHooft:1977xjm}
G.~'t~Hooft, ``{Can We Make Sense Out of Quantum Chromodynamics?},''
{\em Subnucl. Ser.} {\bfseries 15} (1979) 943.

\bibitem{DallaBrida:2020pag}
M.~Dalla~Brida, ``{Past, present, and future of precision determinations of the
  QCD parameters from lattice QCD},'' in {\em {38th International Symposium on
  Lattice Field Theory}}.
\newblock 12, 2020.
\newblock \href{http://arxiv.org/abs/2012.01232}{{\ttfamily arXiv:2012.01232
  [hep-lat]}}.

\bibitem{Bietenholz:2010xg}
W.~Bietenholz, U.~Gerber, M.~Pepe, and U.~J. Wiese, ``{Topological Lattice
  Actions},'' \href{http://dx.doi.org/10.1007/JHEP12(2010)020}{{\em JHEP}
  {\bfseries 12} (2010) 020},
\href{http://arxiv.org/abs/1009.2146}{{\ttfamily arXiv:1009.2146 [hep-lat]}}.

\bibitem{PhysRevD.10.2445}
K.~G. Wilson, ``Confinement of quarks,''
  \href{http://dx.doi.org/10.1103/PhysRevD.10.2445}{{\em Phys. Rev. D}
  {\bfseries 10} no.~8, (Oct., 1974) 2445--2459}.

\bibitem{Nielsen:1981hk}
H.~B. Nielsen and M.~Ninomiya, ``{No Go Theorem for Regularizing Chiral
  Fermions},''
\href{http://dx.doi.org/10.1016/0370-2693(81)91026-1}{{\em Phys. Lett.}
  {\bfseries 105B} (1981) 219--223}.

\bibitem{Karsten:1981gd}
L.~H. Karsten, ``{Lattice Fermions in Euclidean Space-time},''
  \href{http://dx.doi.org/10.1016/0370-2693(81)90133-7}{{\em Phys. Lett. B}
  {\bfseries 104} (1981) 315--319}.

\bibitem{Pelissetto:1987ad}
A.~Pelissetto, ``{LATTICE NONLOCAL CHIRAL FERMIONS},''
  \href{http://dx.doi.org/10.1016/0003-4916(88)90299-0}{{\em Annals Phys.}
  {\bfseries 182} (1988) 177}.

\bibitem{Wilson:1975id}
K.~G. Wilson, ``{Quarks and Strings on a Lattice},'' in {\em {New Phenomena in
  Subnuclear Physics: Proceedings, International School of Subnuclear Physics,
  Erice, Sicily, Jul 11-Aug 1 1975. Part A}}, p.~99.
\newblock 1975.
\newblock
[,0069(1975)].

\bibitem{Sheikholeslami:1985ij}
B.~Sheikholeslami and R.~Wohlert, ``{Improved Continuum Limit Lattice Action
  for QCD with Wilson Fermions},''
\href{http://dx.doi.org/10.1016/0550-3213(85)90002-1}{{\em Nucl. Phys.}
  {\bfseries B259} (1985) 572}.

\bibitem{Luscher:1996sc}
M.~L\"uscher, S.~Sint, R.~Sommer, and P.~Weisz, ``{Chiral symmetry and O(a)
  improvement in lattice QCD},''
  \href{http://dx.doi.org/10.1016/0550-3213(96)00378-1}{{\em Nucl. Phys.}
  {\bfseries B478} (1996) 365--400},
\href{http://arxiv.org/abs/hep-lat/9605038}{{\ttfamily arXiv:hep-lat/9605038
  [hep-lat]}}.

\bibitem{Sommer:1997xw}
R.~Sommer, ``{Non-perturbative renormalization of QCD},''
\href{http://arxiv.org/abs/hep-ph/9711243}{{\ttfamily arXiv:hep-ph/9711243
  [hep-ph]}}.

\bibitem{Frezzotti:2003ni}
R.~Frezzotti and G.~Rossi, ``{Chirally improving Wilson fermions. 1. O(a)
  improvement},'' \href{http://dx.doi.org/10.1088/1126-6708/2004/08/007}{{\em
  JHEP} {\bfseries 0408} (2004) 007},
\href{http://arxiv.org/abs/hep-lat/0306014}{{\ttfamily arXiv:hep-lat/0306014
  [hep-lat]}}.

\bibitem{Shindler:2007vp}
A.~Shindler, ``{Twisted mass lattice QCD},''
  \href{http://dx.doi.org/10.1016/j.physrep.2008.03.001}{{\em Phys. Rept.}
  {\bfseries 461} (2008) 37--110},
\href{http://arxiv.org/abs/0707.4093}{{\ttfamily arXiv:0707.4093 [hep-lat]}}.

\bibitem{Kogut:1974ag}
J.~B. Kogut and L.~Susskind, ``{Hamiltonian Formulation of Wilson's Lattice
  Gauge Theories},''
\href{http://dx.doi.org/10.1103/PhysRevD.11.395}{{\em Phys. Rev.} {\bfseries
  D11} (1975) 395--408}.

\bibitem{Susskind:1976jm}
L.~Susskind, ``{Lattice Fermions},''
\href{http://dx.doi.org/10.1103/PhysRevD.16.3031}{{\em Phys. Rev.} {\bfseries
  D16} (1977) 3031--3039}.

\bibitem{Marinari:1981qf}
E.~Marinari, G.~Parisi, and C.~Rebbi, ``{Monte Carlo Simulation of the Massive
  Schwinger Model},''
  \href{http://dx.doi.org/10.1016/0550-3213(81)90048-1}{{\em Nucl. Phys.}
  {\bfseries B190} (1981) 734}.
[,595(1981)].

\bibitem{Sharpe:2006re}
S.~R. Sharpe, ``{Rooted staggered fermions: Good, bad or ugly?},''
  \href{http://dx.doi.org/10.22323/1.032.0022}{{\em PoS} {\bfseries LAT2006}
  (2006) 022},
\href{http://arxiv.org/abs/hep-lat/0610094}{{\ttfamily arXiv:hep-lat/0610094
  [hep-lat]}}.

\bibitem{Bernard:2006vv}
C.~Bernard, M.~Golterman, Y.~Shamir, and S.~R. Sharpe, ``{Comment on `Chiral
  anomalies and rooted staggered fermions'},''
  \href{http://dx.doi.org/10.1016/j.physletb.2007.04.018}{{\em Phys. Lett.}
  {\bfseries B649} (2007) 235--240},
\href{http://arxiv.org/abs/hep-lat/0603027}{{\ttfamily arXiv:hep-lat/0603027
  [hep-lat]}}.

\bibitem{Bernard:2007eh}
C.~Bernard, M.~Golterman, Y.~Shamir, and S.~R. Sharpe, ``{'t Hooft vertices,
  partial quenching, and rooted staggered QCD},''
  \href{http://dx.doi.org/10.1103/PhysRevD.77.114504}{{\em Phys. Rev.}
  {\bfseries D77} (2008) 114504},
\href{http://arxiv.org/abs/0711.0696}{{\ttfamily arXiv:0711.0696 [hep-lat]}}.

\bibitem{Creutz:2007yg}
M.~Creutz, ``{Chiral anomalies and rooted staggered fermions},''
  \href{http://dx.doi.org/10.1016/j.physletb.2007.03.065}{{\em Phys. Lett.}
  {\bfseries B649} (2007) 230--234},
\href{http://arxiv.org/abs/hep-lat/0701018}{{\ttfamily arXiv:hep-lat/0701018
  [hep-lat]}}.

\bibitem{Luscher:1998pqa}
M.~Luscher, ``{Exact chiral symmetry on the lattice and the Ginsparg-Wilson
  relation},'' \href{http://dx.doi.org/10.1016/S0370-2693(98)00423-7}{{\em
  Phys. Lett.} {\bfseries B428} (1998) 342--345},
\href{http://arxiv.org/abs/hep-lat/9802011}{{\ttfamily arXiv:hep-lat/9802011
  [hep-lat]}}.

\bibitem{Ginsparg:1981bj}
P.~H. Ginsparg and K.~G. Wilson, ``{A Remnant of Chiral Symmetry on the
  Lattice},''
\href{http://dx.doi.org/10.1103/PhysRevD.25.2649}{{\em Phys. Rev.} {\bfseries
  D25} (1982) 2649}.

\bibitem{Neuberger:1997fp}
H.~Neuberger, ``{Exactly massless quarks on the lattice},''
  \href{http://dx.doi.org/10.1016/S0370-2693(97)01368-3}{{\em Phys. Lett.}
  {\bfseries B417} (1998) 141--144},
\href{http://arxiv.org/abs/hep-lat/9707022}{{\ttfamily arXiv:hep-lat/9707022
  [hep-lat]}}.

\bibitem{Kaplan:1992bt}
D.~B. Kaplan, ``{A Method for simulating chiral fermions on the lattice},''
  \href{http://dx.doi.org/10.1016/0370-2693(92)91112-M}{{\em Phys. Lett.}
  {\bfseries B288} (1992) 342--347},
\href{http://arxiv.org/abs/hep-lat/9206013}{{\ttfamily arXiv:hep-lat/9206013
  [hep-lat]}}.

\bibitem{Shamir:1993zy}
Y.~Shamir, ``{Chiral fermions from lattice boundaries},''
  \href{http://dx.doi.org/10.1016/0550-3213(93)90162-I}{{\em Nucl. Phys.}
  {\bfseries B406} (1993) 90--106},
\href{http://arxiv.org/abs/hep-lat/9303005}{{\ttfamily arXiv:hep-lat/9303005
  [hep-lat]}}.

\bibitem{Brower:2012vk}
R.~C. Brower, H.~Neff, and K.~Orginos, ``{The Möbius domain wall fermion
  algorithm},'' \href{http://dx.doi.org/10.1016/j.cpc.2017.01.024}{{\em Comput.
  Phys. Commun.} {\bfseries 220} (2017) 1--19},
  \href{http://arxiv.org/abs/1206.5214}{{\ttfamily arXiv:1206.5214 [hep-lat]}}.

\bibitem{Kaplan:2009yg}
D.~B. Kaplan, ``{Chiral Symmetry and Lattice Fermions},'' in {\em {Modern
  perspectives in lattice QCD: Quantum field theory and high performance
  computing. Proceedings, International School, 93rd Session, Les Houches,
  France, August 3-28, 2009}}, pp.~223--272.
\newblock 2009.
\newblock
\href{http://arxiv.org/abs/0912.2560}{{\ttfamily arXiv:0912.2560 [hep-lat]}}.
\newblock

\bibitem{Kennedy:2006ax}
A.~D. Kennedy, ``{Algorithms for dynamical fermions},''
\href{http://arxiv.org/abs/hep-lat/0607038}{{\ttfamily arXiv:hep-lat/0607038
  [hep-lat]}}.

\bibitem{DelDebbio:2007pz}
L.~Del~Debbio, L.~Giusti, M.~Luscher, R.~Petronzio, and N.~Tantalo, ``{QCD with
  light Wilson quarks on fine lattices. II. DD-HMC simulations and data
  analysis},'' \href{http://dx.doi.org/10.1088/1126-6708/2007/02/082}{{\em
  JHEP} {\bfseries 02} (2007) 082},
\href{http://arxiv.org/abs/hep-lat/0701009}{{\ttfamily arXiv:hep-lat/0701009
  [hep-lat]}}.

\bibitem{Vladikas:2011bp}
A.~Vladikas, ``{Three Topics in Renormalization and Improvement},'' in {\em
  {Les Houches Summer School: Session 93: Modern perspectives in lattice QCD:
  Quantum field theory and high performance computing}}, pp.~161--222.
\newblock 3, 2011.
\newblock \href{http://arxiv.org/abs/1103.1323}{{\ttfamily arXiv:1103.1323
  [hep-lat]}}.

\bibitem{Luscher:1985dn}
M.~L{\"u}scher, ``{Volume Dependence of the Energy Spectrum in Massive Quantum
  Field Theories. 1. Stable Particle States},''
  \href{http://dx.doi.org/10.1007/BF01211589}{{\em Commun.Math.Phys.}
  {\bfseries 104} (1986) 177}.

\bibitem{Eidelman:2004wy}
{\bfseries Particle Data Group} Collaboration, S.~Eidelman {\em et~al.},
  ``{Review of particle physics. Particle Data Group},''
  \href{http://dx.doi.org/10.1016/j.physletb.2004.06.001}{{\em Phys.Lett.}
  {\bfseries B592} (2004) 1}.

\bibitem{ramos:19lat}
A.~Ramos, ``{Non-perturbative renormalization by decoupling}.,'' {\em Talk at
  The 37th International Symposium on Lattice Field Theory} (2019) .

\bibitem{Balog:2009np}
J.~Balog, F.~Niedermayer, and P.~Weisz, ``{The Puzzle of apparent linear
  lattice artifacts in the 2d non-linear sigma-model and Symanzik's
  solution},'' \href{http://dx.doi.org/10.1016/j.nuclphysb.2009.09.007}{{\em
  Nucl. Phys.} {\bfseries B824} (2010) 563--615},
\href{http://arxiv.org/abs/0905.1730}{{\ttfamily arXiv:0905.1730 [hep-lat]}}.

\bibitem{Husung:2019ytz}
N.~Husung, P.~Marquard, and R.~Sommer, ``{Asymptotic behavior of cutoff effects
  in Yang-Mills theory and in Wilson's lattice QCD},''
\href{http://arxiv.org/abs/1912.08498}{{\ttfamily arXiv:1912.08498 [hep-lat]}}.

\bibitem{Marciano1993}
W.~J. Marciano and A.~Sirlin, ``{Radiative corrections to pi(lepton 2)
  decays},''
\href{http://dx.doi.org/10.1103/PhysRevLett.71.3629}{{\em Phys. Rev. Lett.}
  {\bfseries 71} (1993) 3629--3632}.

\bibitem{Lubicz:2016xvp}
V.~Lubicz, N.~Carrasco, G.~Martinelli, C.~Sachrajda, N.~Tantalo, C.~Tarantino,
  and M.~Testa, ``{QED corrections to hadronic processes: a strategy for
  lattice QCD},''
\href{http://dx.doi.org/10.22323/1.253.0023}{{\em PoS} {\bfseries CD15} (2016)
  023}.

\bibitem{DiCarlo:2019thl}
M.~Di~Carlo, D.~Giusti, V.~Lubicz, G.~Martinelli, C.~T. Sachrajda,
  F.~Sanfilippo, S.~Simula, and N.~Tantalo, ``{Light-meson leptonic decay rates
  in lattice QCD+QED},''
  \href{http://dx.doi.org/10.1103/PhysRevD.100.034514}{{\em Phys. Rev.}
  {\bfseries D100} no.~3, (2019) 034514},
\href{http://arxiv.org/abs/1904.08731}{{\ttfamily arXiv:1904.08731 [hep-lat]}}.

\bibitem{Sommer:1993ce}
R.~Sommer, ``{A New way to set the energy scale in lattice gauge theories and
  its applications to the static force and alpha-s in SU(2) Yang-Mills
  theory},'' \href{http://dx.doi.org/10.1016/0550-3213(94)90473-1}{{\em Nucl.
  Phys.} {\bfseries B411} (1994) 839--854},
\href{http://arxiv.org/abs/hep-lat/9310022}{{\ttfamily arXiv:hep-lat/9310022
  [hep-lat]}}.

\bibitem{Bernard:2000gd}
C.~W. Bernard, T.~Burch, K.~Orginos, D.~Toussaint, T.~A. DeGrand, C.~E. DeTar,
  S.~A. Gottlieb, U.~M. Heller, J.~E. Hetrick, and B.~Sugar, ``{The Static
  quark potential in three flavor QCD},''
  \href{http://dx.doi.org/10.1103/PhysRevD.62.034503}{{\em Phys. Rev.}
  {\bfseries D62} (2000) 034503},
\href{http://arxiv.org/abs/hep-lat/0002028}{{\ttfamily arXiv:hep-lat/0002028
  [hep-lat]}}.

\bibitem{Sommer:2014mea}
R.~Sommer, ``{Scale setting in lattice QCD},'' {\em PoS} {\bfseries
  LATTICE2013} (2014) 015,
\href{http://arxiv.org/abs/1401.3270}{{\ttfamily arXiv:1401.3270 [hep-lat]}}.

\bibitem{Luscher:2010iy}
M.~L{\"u}scher, ``{Properties and uses of the Wilson flow in lattice QCD},''
  \href{http://dx.doi.org/10.1007/JHEP08(2010)071}{{\em JHEP} {\bfseries 1008}
  (2010) 071},
\href{http://arxiv.org/abs/1006.4518}{{\ttfamily arXiv:1006.4518 [hep-lat]}}.

\bibitem{Narayanan:2006rf}
R.~Narayanan and H.~Neuberger, ``{Infinite N phase transitions in continuum
  Wilson loop operators},''
  \href{http://dx.doi.org/10.1088/1126-6708/2006/03/064}{{\em JHEP} {\bfseries
  0603} (2006) 064},
\href{http://arxiv.org/abs/hep-th/0601210}{{\ttfamily arXiv:hep-th/0601210
  [hep-th]}}.

\bibitem{Luscher:2011bx}
M.~L{\"u}scher and P.~Weisz, ``{Perturbative analysis of the gradient flow in
  non-abelian gauge theories},''
  \href{http://dx.doi.org/10.1007/JHEP02(2011)051}{{\em JHEP} {\bfseries 1102}
  (2011) 051},
\href{http://arxiv.org/abs/1101.0963}{{\ttfamily arXiv:1101.0963 [hep-th]}}.

\bibitem{Borsanyi:2012zs}
S.~Borsanyi, S.~Durr, Z.~Fodor, C.~Hoelbling, S.~D. Katz, {\em et~al.},
  ``{High-precision scale setting in lattice QCD},''
  \href{http://dx.doi.org/10.1007/JHEP09(2012)010}{{\em JHEP} {\bfseries 1209}
  (2012) 010},
\href{http://arxiv.org/abs/1203.4469}{{\ttfamily arXiv:1203.4469 [hep-lat]}}.

\bibitem{Giusti:2018cmp}
{Giusti, Leonardo and L\"uscher, Martin}, ``{Topological susceptibility at
  $T>T_{\rm c}$ from master-field simulations of the SU(3) gauge theory},''
\href{http://arxiv.org/abs/1812.02062}{{\ttfamily arXiv:1812.02062 [hep-lat]}}.

\bibitem{Duane:1987de}
S.~Duane, A.~D. Kennedy, B.~J. Pendleton, and D.~Roweth, ``{Hybrid Monte
  Carlo},''
\href{http://dx.doi.org/10.1016/0370-2693(87)91197-X}{{\em Phys. Lett.}
  {\bfseries B195} (1987) 216--222}.

\bibitem{DelDebbio:2004xh}
L.~Del~Debbio, G.~M. Manca, and E.~Vicari, ``{Critical slowing down of
  topological modes},''
  \href{http://dx.doi.org/10.1016/j.physletb.2004.05.038}{{\em Phys.Lett.}
  {\bfseries B594} (2004) 315--323},
\href{http://arxiv.org/abs/hep-lat/0403001}{{\ttfamily arXiv:hep-lat/0403001
  [hep-lat]}}.

\bibitem{Schaefer:2010hu}
{\bfseries ALPHA} Collaboration, S.~Schaefer, R.~Sommer, and F.~Virotta,
  ``{Critical slowing down and error analysis in lattice QCD simulations},''
  \href{http://dx.doi.org/10.1016/j.nuclphysb.2010.11.020}{{\em Nucl. Phys.}
  {\bfseries B845} (2011) 93--119},
\href{http://arxiv.org/abs/1009.5228}{{\ttfamily arXiv:1009.5228 [hep-lat]}}.

\bibitem{Bruno:2016plf}
M.~Bruno, T.~Korzec, and S.~Schaefer, ``{Setting the scale for the CLS 2+1
  flavor ensembles},'' \href{http://dx.doi.org/10.1103/PhysRevD.95.074504}{{\em
  Phys. Rev.} {\bfseries D95} no.~7, (2017) 074504},
\href{http://arxiv.org/abs/1608.08900}{{\ttfamily arXiv:1608.08900 [hep-lat]}}.

\bibitem{Bruno:2014jqa}
M.~Bruno {\em et~al.}, ``{Simulation of QCD with N$_{f} =$ 2 $+$ 1 flavors of
  non-perturbatively improved Wilson fermions},''
  \href{http://dx.doi.org/10.1007/JHEP02(2015)043}{{\em JHEP} {\bfseries 02}
  (2015) 043},
\href{http://arxiv.org/abs/1411.3982}{{\ttfamily arXiv:1411.3982 [hep-lat]}}.

\bibitem{Wolff:2003sm}
{\bfseries ALPHA} Collaboration, U.~Wolff, ``{Monte {Carlo} errors with less
  errors},'' \href{http://dx.doi.org/10.1016/S0010-4655(03)00467-3,
  10.1016/j.cpc.2006.12.001}{{\em Comput.Phys.Commun.} {\bfseries 156} (2004)
  143--153},
\href{http://arxiv.org/abs/hep-lat/0306017}{{\ttfamily arXiv:hep-lat/0306017
  [hep-lat]}}.

\bibitem{Virotta2012Critical}
F.~Virotta, \href{http://dx.doi.org/http://dx.doi.org/10.18452/16502}{{\em
  Critical slowing down and error analysis of lattice QCD simulations}}.
\newblock PhD thesis, Humboldt-Universität zu Berlin,
  Mathematisch-Naturwissenschaftliche Fakultät I, 2012.

\bibitem{Ramos:2018vgu}
A.~Ramos, ``{Automatic differentiation for error analysis of Monte Carlo
  data},'' \href{http://dx.doi.org/10.1016/j.cpc.2018.12.020}{{\em Comput.
  Phys. Commun.} {\bfseries 238} (2019) 19--35},
\href{http://arxiv.org/abs/1809.01289}{{\ttfamily arXiv:1809.01289 [hep-lat]}}.

\bibitem{Appelquist:1974tg}
T.~Appelquist and J.~Carazzone, ``{Infrared Singularities and Massive
  Fields},''
\href{http://dx.doi.org/10.1103/PhysRevD.11.2856}{{\em Phys. Rev.} {\bfseries
  D11} (1975) 2856}.

\bibitem{Binetruy:1979hc}
P.~Binetruy and T.~Schucker, ``{Gauge and Renormalization Scheme Dependence in
  {GUTs}},''
\href{http://dx.doi.org/10.1016/0550-3213(81)90410-7}{{\em Nucl. Phys.}
  {\bfseries B178} (1981) 293--306}.

\bibitem{Binetruy:1980xn}
P.~Binetruy and T.~Schucker, ``{The Use of Dimensional Renormalization Schemes
  in Unified Theories},''
\href{http://dx.doi.org/10.1016/0550-3213(81)90411-9}{{\em Nucl. Phys.}
  {\bfseries B178} (1981) 307--330}.

\bibitem{Weinberg:1951ss}
S.~Weinberg, ``{New approach to the renormalization group},''
\href{http://dx.doi.org/10.1103/PhysRevD.8.3497}{{\em Phys. Rev.} {\bfseries
  D8} (1973) 3497--3509}.

\bibitem{Wetzel:1981qg}
W.~Wetzel, ``{Minimal Subtraction and the Decoupling of Heavy Quarks for
  Arbitrary Values of the Gauge Parameter},''
\href{http://dx.doi.org/10.1016/0550-3213(82)90038-4}{{\em Nucl. Phys.}
  {\bfseries B196} (1982) 259--272}.

\bibitem{Chetyrkin:1997un}
K.~G. Chetyrkin, B.~A. Kniehl, and M.~Steinhauser, ``{Decoupling relations to O
  (alpha-s**3) and their connection to low-energy theorems},''
  \href{http://dx.doi.org/10.1016/S0550-3213(98)81004-3,
  10.1016/S0550-3213(97)00649-4}{{\em Nucl. Phys.} {\bfseries B510} (1998)
  61--87},
\href{http://arxiv.org/abs/hep-ph/9708255}{{\ttfamily arXiv:hep-ph/9708255
  [hep-ph]}}.

\bibitem{Patrignani:2016xqp}
{\bfseries Particle Data Group} Collaboration, C.~Patrignani {\em et~al.},
  ``{Review of Particle Physics},''
\href{http://dx.doi.org/10.1088/1674-1137/40/10/100001}{{\em Chin. Phys.}
  {\bfseries C40} no.~10, (2016) 100001}.

\bibitem{Herren:2017osy}
F.~Herren and M.~Steinhauser, ``{Version 3 of {\tt RunDec} and {\tt
  CRunDec}},''
\href{http://arxiv.org/abs/1703.03751}{{\ttfamily arXiv:1703.03751 [hep-ph]}}.

\bibitem{Bruno:2014ufa}
{\bfseries ALPHA} Collaboration, M.~Bruno, J.~Finkenrath, F.~Knechtli,
  B.~Leder, and R.~Sommer, ``{Effects of Heavy Sea Quarks at Low Energies},''
  \href{http://dx.doi.org/10.1103/PhysRevLett.114.102001}{{\em Phys. Rev.
  Lett.} {\bfseries 114} no.~10, (2015) 102001},
\href{http://arxiv.org/abs/1410.8374}{{\ttfamily arXiv:1410.8374 [hep-lat]}}.

\bibitem{Sternbeck:2012qs}
A.~Sternbeck, K.~Maltman, M.~Muller-Preussker, and L.~von Smekal,
  ``{Determination of LambdaMS from the gluon and ghost propagators in Landau
  gauge},'' \href{http://dx.doi.org/10.22323/1.164.0243}{{\em PoS} {\bfseries
  LATTICE2012} (2012) 243},
\href{http://arxiv.org/abs/1212.2039}{{\ttfamily arXiv:1212.2039 [hep-lat]}}.

\bibitem{Sternbeck:2010xu}
A.~Sternbeck, E.~M. Ilgenfritz, K.~Maltman, M.~Muller-Preussker, L.~von Smekal,
  and A.~G. Williams, ``{QCD Lambda parameter from Landau-gauge gluon and ghost
  correlations},'' \href{http://dx.doi.org/10.22323/1.091.0210}{{\em PoS}
  {\bfseries LAT2009} (2009) 210},
\href{http://arxiv.org/abs/1003.1585}{{\ttfamily arXiv:1003.1585 [hep-lat]}}.

\bibitem{Takaura:2018lpw}
H.~Takaura, T.~Kaneko, Y.~Kiyo, and Y.~Sumino, ``{Determination of $\alpha_s$
  from static QCD potential with renormalon subtraction},''
  \href{http://dx.doi.org/10.1016/j.physletb.2018.12.060}{{\em Phys. Lett.}
  {\bfseries B789} (2019) 598--602},
\href{http://arxiv.org/abs/1808.01632}{{\ttfamily arXiv:1808.01632 [hep-ph]}}.

\bibitem{Petreczky:2019ozv}
P.~Petreczky and J.~H. Weber, ``{Strong coupling constant and heavy quark
  masses in 2+1 flavor QCD},''
\href{http://arxiv.org/abs/1901.06424}{{\ttfamily arXiv:1901.06424 [hep-lat]}}.

\bibitem{McNeile:2010ji}
C.~McNeile, C.~T.~H. Davies, E.~Follana, K.~Hornbostel, and G.~P. Lepage,
  ``{High-Precision c and b Masses, and QCD Coupling from Current-Current
  Correlators in Lattice and Continuum QCD},''
  \href{http://dx.doi.org/10.1103/PhysRevD.82.034512}{{\em Phys. Rev.}
  {\bfseries D82} (2010) 034512},
\href{http://arxiv.org/abs/1004.4285}{{\ttfamily arXiv:1004.4285 [hep-lat]}}.

\bibitem{Gribov:1977wm}
V.~N. Gribov, ``{Quantization of Nonabelian Gauge Theories},''
  \href{http://dx.doi.org/10.1016/0550-3213(78)90175-X}{{\em Nucl. Phys.}
  {\bfseries B139} (1978) 1}.
[,1(1977)].

\bibitem{Vandersickel:2012tz}
N.~Vandersickel and D.~Zwanziger, ``{The Gribov problem and QCD dynamics},''
  \href{http://dx.doi.org/10.1016/j.physrep.2012.07.003}{{\em Phys. Rept.}
  {\bfseries 520} (2012) 175--251},
\href{http://arxiv.org/abs/1202.1491}{{\ttfamily arXiv:1202.1491 [hep-th]}}.

\bibitem{Martinelli:1994ty}
G.~Martinelli, C.~Pittori, C.~T. Sachrajda, M.~Testa, and A.~Vladikas, ``{A
  General method for nonperturbative renormalization of lattice operators},''
  \href{http://dx.doi.org/10.1016/0550-3213(95)00126-D}{{\em Nucl. Phys.}
  {\bfseries B445} (1995) 81--108},
\href{http://arxiv.org/abs/hep-lat/9411010}{{\ttfamily arXiv:hep-lat/9411010}}.

\bibitem{Chetyrkin:2000dq}
K.~G. Chetyrkin and A.~Retey, ``{Three loop three linear vertices and four loop
  similar to MOM beta functions in massless QCD},''
\href{http://arxiv.org/abs/hep-ph/0007088}{{\ttfamily arXiv:hep-ph/0007088
  [hep-ph]}}.

\bibitem{Boucaud:2005xn}
P.~Boucaud, J.~P. Leroy, A.~Le~Yaouanc, A.~Y. Lokhov, J.~Micheli, O.~Pene,
  J.~Rodriguez-Quintero, and C.~Roiesnel, ``{Non-perturbative power corrections
  to ghost and gluon propagators},''
  \href{http://dx.doi.org/10.1088/1126-6708/2006/01/037}{{\em JHEP} {\bfseries
  01} (2006) 037},
\href{http://arxiv.org/abs/hep-lat/0507005}{{\ttfamily arXiv:hep-lat/0507005
  [hep-lat]}}.

\bibitem{Zafeiropoulos:2019flq}
S.~Zafeiropoulos, P.~Boucaud, F.~De~Soto, J.~Rodríguez-Quintero, and
  J.~Segovia, ``{The strong running coupling from the gauge sector of Domain
  Wall lattice QCD with physical quark masses},''
  \href{http://dx.doi.org/10.1103/PhysRevLett.122.162002}{{\em Phys. Rev.
  Lett.} {\bfseries 122} no.~16, (2019) 162002},
\href{http://arxiv.org/abs/1902.08148}{{\ttfamily arXiv:1902.08148 [hep-ph]}}.

\bibitem{Blossier:2011tf}
B.~Blossier, P.~Boucaud, M.~Brinet, F.~De~Soto, X.~Du, M.~Gravina, V.~Morenas,
  O.~Pene, K.~Petrov, and J.~Rodriguez-Quintero, ``{Ghost-gluon coupling, power
  corrections and $\Lambda_{\bar{\rm MS}}$ from lattice QCD with a dynamical
  charm},'' \href{http://dx.doi.org/10.1103/PhysRevD.85.034503}{{\em Phys.
  Rev.} {\bfseries D85} (2012) 034503},
\href{http://arxiv.org/abs/1110.5829}{{\ttfamily arXiv:1110.5829 [hep-lat]}}.

\bibitem{Blossier:2012ef}
B.~Blossier, P.~Boucaud, M.~Brinet, F.~De~Soto, X.~Du, V.~Morenas, O.~Pene,
  K.~Petrov, and J.~Rodriguez-Quintero, ``{The Strong running coupling at
  $\tau$ and $Z_0$ mass scales from lattice QCD},''
  \href{http://dx.doi.org/10.1103/PhysRevLett.108.262002}{{\em Phys. Rev.
  Lett.} {\bfseries 108} (2012) 262002},
\href{http://arxiv.org/abs/1201.5770}{{\ttfamily arXiv:1201.5770 [hep-ph]}}.

\bibitem{Blossier:2013ioa}
{\bfseries ETM} Collaboration, B.~Blossier, P.~Boucaud, M.~Brinet, F.~De~Soto,
  V.~Morenas, O.~Pene, K.~Petrov, and J.~Rodriguez-Quintero, ``{High statistics
  determination of the strong coupling constant in Taylor scheme and its OPE
  Wilson coefficient from lattice QCD with a dynamical charm},''
  \href{http://dx.doi.org/10.1103/PhysRevD.89.014507}{{\em Phys. Rev.}
  {\bfseries D89} no.~1, (2014) 014507},
\href{http://arxiv.org/abs/1310.3763}{{\ttfamily arXiv:1310.3763 [hep-ph]}}.

\bibitem{Boucaud:2018xup}
P.~Boucaud, F.~De~Soto, K.~Raya, J.~Rodríguez-Quintero, and S.~Zafeiropoulos,
  ``{Discretization effects on renormalized gauge-field Green’s functions,
  scale setting, and the gluon mass},''
  \href{http://dx.doi.org/10.1103/PhysRevD.98.114515}{{\em Phys. Rev.}
  {\bfseries D98} no.~11, (2018) 114515},
\href{http://arxiv.org/abs/1809.05776}{{\ttfamily arXiv:1809.05776 [hep-ph]}}.

\bibitem{Wilson:1974sk}
K.~G. Wilson, ``Confinement of quarks,''
{\em Phys. Rev.} {\bfseries D10} (1974) 2445--2459.

\bibitem{Necco:2001xg}
S.~Necco and R.~Sommer, ``{The N(f) = 0 heavy quark potential from short to
  intermediate distances},''
  \href{http://dx.doi.org/10.1016/S0550-3213(01)00582-X}{{\em Nucl. Phys.}
  {\bfseries B622} (2002) 328--346},
\href{http://arxiv.org/abs/hep-lat/0108008}{{\ttfamily arXiv:hep-lat/0108008
  [hep-lat]}}.

\bibitem{Fischler:1977yf}
W.~Fischler, ``{Quark - anti-Quark Potential in QCD},''
\href{http://dx.doi.org/10.1016/0550-3213(77)90026-8}{{\em Nucl. Phys.}
  {\bfseries B129} (1977) 157--174}.

\bibitem{Billoire:1979ih}
A.~Billoire, ``{How Heavy Must Be Quarks in Order to Build Coulombic q anti-q
  Bound States},''
\href{http://dx.doi.org/10.1016/0370-2693(80)90279-8}{{\em Phys. Lett.}
  {\bfseries 92B} (1980) 343--347}.

\bibitem{Peter:1997me}
M.~Peter, ``{The Static potential in QCD: A Full two loop calculation},''
  \href{http://dx.doi.org/10.1016/S0550-3213(97)00373-8}{{\em Nucl. Phys.}
  {\bfseries B501} (1997) 471--494},
\href{http://arxiv.org/abs/hep-ph/9702245}{{\ttfamily arXiv:hep-ph/9702245
  [hep-ph]}}.

\bibitem{Schroder:1998vy}
Y.~Schroder, ``{The Static potential in QCD to two loops},''
  \href{http://dx.doi.org/10.1016/S0370-2693(99)00010-6}{{\em Phys. Lett.}
  {\bfseries B447} (1999) 321--326},
\href{http://arxiv.org/abs/hep-ph/9812205}{{\ttfamily arXiv:hep-ph/9812205
  [hep-ph]}}.

\bibitem{Brambilla:1999qa}
N.~Brambilla, A.~Pineda, J.~Soto, and A.~Vairo, ``{The Infrared behavior of the
  static potential in perturbative QCD},''
  \href{http://dx.doi.org/10.1103/PhysRevD.60.091502}{{\em Phys. Rev.}
  {\bfseries D60} (1999) 091502},
\href{http://arxiv.org/abs/hep-ph/9903355}{{\ttfamily arXiv:hep-ph/9903355
  [hep-ph]}}.

\bibitem{Smirnov:2009fh}
A.~V. Smirnov, V.~A. Smirnov, and M.~Steinhauser, ``{Three-loop static
  potential},'' \href{http://dx.doi.org/10.1103/PhysRevLett.104.112002}{{\em
  Phys. Rev. Lett.} {\bfseries 104} (2010) 112002},
\href{http://arxiv.org/abs/0911.4742}{{\ttfamily arXiv:0911.4742 [hep-ph]}}.

\bibitem{Anzai:2009tm}
C.~Anzai, Y.~Kiyo, and Y.~Sumino, ``{Static QCD potential at three-loop
  order},'' \href{http://dx.doi.org/10.1103/PhysRevLett.104.112003}{{\em Phys.
  Rev. Lett.} {\bfseries 104} (2010) 112003},
\href{http://arxiv.org/abs/0911.4335}{{\ttfamily arXiv:0911.4335 [hep-ph]}}.

\bibitem{Brambilla:2009bi}
N.~Brambilla, A.~Vairo, X.~Garcia~i Tormo, and J.~Soto, ``{The QCD static
  energy at NNNLL},'' \href{http://dx.doi.org/10.1103/PhysRevD.80.034016}{{\em
  Phys. Rev.} {\bfseries D80} (2009) 034016},
\href{http://arxiv.org/abs/0906.1390}{{\ttfamily arXiv:0906.1390 [hep-ph]}}.

\bibitem{Tormo:2013tha}
X.~Garcia~i Tormo, ``{Review on the determination of $\alpha_s$ from the QCD
  static energy},'' \href{http://dx.doi.org/10.1142/S0217732313300280}{{\em
  Mod. Phys. Lett.} {\bfseries A28} (2013) 1330028},
\href{http://arxiv.org/abs/1307.2238}{{\ttfamily arXiv:1307.2238 [hep-ph]}}.

\bibitem{Bazavov:2014soa}
A.~Bazavov, N.~Brambilla, X.~Garcia~i Tormo, P.~Petreczky, J.~Soto, and
  A.~Vairo, ``{Determination of $\alpha_s$ from the QCD static energy: An
  update},'' \href{http://dx.doi.org/10.1103/PhysRevD.90.074038}{{\em Phys.
  Rev.} {\bfseries D90} no.~7, (2014) 074038},
\href{http://arxiv.org/abs/1407.8437}{{\ttfamily arXiv:1407.8437 [hep-ph]}}.

\bibitem{Ayala:2020odx}
C.~Ayala, X.~Lobregat, and A.~Pineda, ``{Determination of $\alpha(M_z)$ from an
  hyperasymptotic approximation to the energy of a static quark-antiquark
  pair},'' \href{http://dx.doi.org/10.1007/JHEP09(2020)016}{{\em JHEP}
  {\bfseries 09} (2020) 016}, \href{http://arxiv.org/abs/2005.12301}{{\ttfamily
  arXiv:2005.12301 [hep-ph]}}.

\bibitem{sommer:19lat}
R.~Sommer, ``{Yang Mills short distance potential and perturbation theory.}.,''
  {\em Talk at The 37th International Symposium on Lattice Field Theory} (2019)
  .

\bibitem{Allison:2008xk}
{\bfseries HPQCD} Collaboration, I.~Allison {\em et~al.}, ``{High-Precision
  Charm-Quark Mass from Current-Current Correlators in Lattice and Continuum
  QCD},'' \href{http://dx.doi.org/10.1103/PhysRevD.78.054513}{{\em Phys. Rev.}
  {\bfseries D78} (2008) 054513},
\href{http://arxiv.org/abs/0805.2999}{{\ttfamily arXiv:0805.2999 [hep-lat]}}.

\bibitem{Maier:2009fz}
A.~Maier, P.~Maierhofer, P.~Marquard, and A.~V. Smirnov, ``{Low energy moments
  of heavy quark current correlators at four loops},''
  \href{http://dx.doi.org/10.1016/j.nuclphysb.2009.08.011}{{\em Nucl. Phys.}
  {\bfseries B824} (2010) 1--18},
\href{http://arxiv.org/abs/0907.2117}{{\ttfamily arXiv:0907.2117 [hep-ph]}}.

\bibitem{Maier:2007yn}
A.~Maier, P.~Maierhofer, and P.~Marquard, ``{Higher Moments of Heavy Quark
  Correlators in the Low Energy Limit at O($\alpha^2(s)$)},''
  \href{http://dx.doi.org/10.1016/j.nuclphysb.2007.12.035}{{\em Nucl. Phys.}
  {\bfseries B797} (2008) 218--242},
\href{http://arxiv.org/abs/0711.2636}{{\ttfamily arXiv:0711.2636 [hep-ph]}}.

\bibitem{Boughezal:2006px}
R.~Boughezal, M.~Czakon, and T.~Schutzmeier, ``{Charm and bottom quark masses
  from perturbative QCD},''
  \href{http://dx.doi.org/10.1103/PhysRevD.74.074006}{{\em Phys. Rev.}
  {\bfseries D74} (2006) 074006},
\href{http://arxiv.org/abs/hep-ph/0605023}{{\ttfamily arXiv:hep-ph/0605023
  [hep-ph]}}.

\bibitem{Chetyrkin:2006xg}
K.~G. Chetyrkin, J.~H. Kuhn, and C.~Sturm, ``{Four-loop moments of the heavy
  quark vacuum polarization function in perturbative QCD},''
  \href{http://dx.doi.org/10.1140/epjc/s2006-02610-y}{{\em Eur. Phys. J.}
  {\bfseries C48} (2006) 107--110},
\href{http://arxiv.org/abs/hep-ph/0604234}{{\ttfamily arXiv:hep-ph/0604234
  [hep-ph]}}.

\bibitem{Maezawa:2016vgv}
Y.~Maezawa and P.~Petreczky, ``{Quark masses and strong coupling constant in
  2+1 flavor QCD},'' \href{http://dx.doi.org/10.1103/PhysRevD.94.034507}{{\em
  Phys. Rev.} {\bfseries D94} no.~3, (2016) 034507},
\href{http://arxiv.org/abs/1606.08798}{{\ttfamily arXiv:1606.08798 [hep-lat]}}.

\bibitem{Nakayama:2016atf}
K.~Nakayama, B.~Fahy, and S.~Hashimoto, ``{Short-distance charmonium correlator
  on the lattice with Möbius domain-wall fermion and a determination of charm
  quark mass},'' \href{http://dx.doi.org/10.1103/PhysRevD.94.054507}{{\em Phys.
  Rev.} {\bfseries D94} no.~5, (2016) 054507},
\href{http://arxiv.org/abs/1606.01002}{{\ttfamily arXiv:1606.01002 [hep-lat]}}.

\bibitem{Chakraborty:2014aca}
B.~Chakraborty, C.~T.~H. Davies, B.~Galloway, P.~Knecht, J.~Koponen, G.~C.
  Donald, R.~J. Dowdall, G.~P. Lepage, and C.~McNeile, ``{High-precision quark
  masses and QCD coupling from $n_f=4$ lattice QCD},''
  \href{http://dx.doi.org/10.1103/PhysRevD.91.054508}{{\em Phys. Rev.}
  {\bfseries D91} no.~5, (2015) 054508},
\href{http://arxiv.org/abs/1408.4169}{{\ttfamily arXiv:1408.4169 [hep-lat]}}.

\bibitem{Lepage:1992xa}
G.~P. Lepage and P.~B. Mackenzie, ``{On the viability of lattice perturbation
  theory},'' \href{http://dx.doi.org/10.1103/PhysRevD.48.2250}{{\em Phys. Rev.}
  {\bfseries D48} (1993) 2250--2264},
\href{http://arxiv.org/abs/hep-lat/9209022}{{\ttfamily arXiv:hep-lat/9209022
  [hep-lat]}}.

\bibitem{Mason:2005zx}
{\bfseries HPQCD, UKQCD} Collaboration, Q.~Mason, H.~D. Trottier, C.~T.~H.
  Davies, K.~Foley, A.~Gray, G.~P. Lepage, M.~Nobes, and J.~Shigemitsu,
  ``{Accurate determinations of alpha(s) from realistic lattice QCD},''
  \href{http://dx.doi.org/10.1103/PhysRevLett.95.052002}{{\em Phys. Rev. Lett.}
  {\bfseries 95} (2005) 052002},
\href{http://arxiv.org/abs/hep-lat/0503005}{{\ttfamily arXiv:hep-lat/0503005
  [hep-lat]}}.

\bibitem{Hudspith:2018bpz}
R.~J. Hudspith, R.~Lewis, K.~Maltman, and E.~Shintani, ``{$\alpha_s$ from the
  Lattice Hadronic Vacuum Polarisation},''
\href{http://arxiv.org/abs/1804.10286}{{\ttfamily arXiv:1804.10286 [hep-lat]}}.

\bibitem{Shintani:2010ph}
E.~Shintani, S.~Aoki, H.~Fukaya, S.~Hashimoto, T.~Kaneko, T.~Onogi, and
  N.~Yamada, ``{Strong coupling constant from vacuum polarization functions in
  three-flavor lattice QCD with dynamical overlap fermions},''
  \href{http://dx.doi.org/10.1103/PhysRevD.82.074505,
  10.1103/PhysRevD.89.099903}{{\em Phys. Rev.} {\bfseries D82} no.~7, (2010)
  074505}, \href{http://arxiv.org/abs/1002.0371}{{\ttfamily arXiv:1002.0371
  [hep-lat]}}.
[Erratum: Phys. Rev.D89,no.9,099903(2014)].

\bibitem{Nakayama:2018ubk}
K.~Nakayama, H.~Fukaya, and S.~Hashimoto, ``{Lattice computation of the Dirac
  eigenvalue density in the perturbative regime of QCD},''
  \href{http://dx.doi.org/10.1103/PhysRevD.98.014501}{{\em Phys. Rev.}
  {\bfseries D98} no.~1, (2018) 014501},
\href{http://arxiv.org/abs/1804.06695}{{\ttfamily arXiv:1804.06695 [hep-lat]}}.

\bibitem{Sommer:2010ic}
R.~Sommer, ``{Introduction to Non-perturbative Heavy Quark Effective Theory},''
\href{http://arxiv.org/abs/1008.0710}{{\ttfamily arXiv:1008.0710 [hep-lat]}}.

\bibitem{Luscher:1991wu}
M.~L{\"u}scher, P.~Weisz, and U.~Wolff, ``{A Numerical method to compute the
  running coupling in asymptotically free theories},''
\href{http://dx.doi.org/10.1016/0550-3213(91)90298-C}{{\em Nucl.Phys.}
  {\bfseries B359} (1991) 221--243}.

\bibitem{GonzalezArroyo:1981vw}
A.~Gonzalez-Arroyo, J.~Jurkiewicz, and C.~Korthals-Altes, ``{Ground state
  metamorphosis for Yang-Mills fields on a finite periodic lattice},''
{\em Freiburg ASI 1981:0339} (1981) .

\bibitem{vanBaal:1988qm}
P.~van Baal, ``{Gauge theory in a finite volume},''
{\em Acta Phys.Polon.} {\bfseries B20} (1989) 295--312.

\bibitem{tHooft:1981sz}
G.~'t~Hooft, ``Some twisted selfdual solutions for the yang-mills equations on
  a hypertorus,''
{\em Commun. Math. Phys.} {\bfseries 81} (1981) 267--275.

\bibitem{deDivitiis:1994yz}
{\bfseries ALPHA} Collaboration, G.~de~Divitiis {\em et~al.}, ``{Universality
  and the approach to the continuum limit in lattice gauge theory},''
  \href{http://dx.doi.org/10.1016/0550-3213(94)00019-B}{{\em Nucl.Phys.}
  {\bfseries B437} (1995) 447--470},
\href{http://arxiv.org/abs/hep-lat/9411017}{{\ttfamily arXiv:hep-lat/9411017
  [hep-lat]}}.

\bibitem{Ramos:2014kla}
A.~Ramos, ``{The gradient flow running coupling with twisted boundary
  conditions},'' \href{http://dx.doi.org/10.1007/JHEP11(2014)101}{{\em JHEP}
  {\bfseries 1411} (2014) 101},
\href{http://arxiv.org/abs/1409.1445}{{\ttfamily arXiv:1409.1445 [hep-lat]}}.

\bibitem{Luscher:1992an}
M.~L{\"u}scher, R.~Narayanan, P.~Weisz, and U.~Wolff, ``{The Schr{\"o}dinger
  Functional: a renormalizable probe for non-abelian gauge theories},''
  \href{http://dx.doi.org/10.1016/0550-3213(92)90466-O}{{\em Nucl.Phys.}
  {\bfseries B384} (1992) 168--228},
\href{http://arxiv.org/abs/hep-lat/9207009}{{\ttfamily arXiv:hep-lat/9207009
  [hep-lat]}}.

\bibitem{Sint:1993un}
S.~Sint, ``{On the Schr{\"o}dinger functional in QCD},''
  \href{http://dx.doi.org/10.1016/0550-3213(94)90228-3}{{\em Nucl.Phys.}
  {\bfseries B421} (1994) 135--158},
\href{http://arxiv.org/abs/hep-lat/9312079}{{\ttfamily hep-lat/9312079}}.

\bibitem{Capitani:1998mq}
{\bfseries ALPHA} Collaboration, S.~Capitani, M.~L{\"u}scher, R.~Sommer, and
  H.~Wittig, ``{Non-perturbative quark mass renormalization in quenched lattice
  QCD},'' \href{http://dx.doi.org/10.1016/S0550-3213(98)00857-8}{{\em
  Nucl.Phys.} {\bfseries B544} (1999) 669--698},
\href{http://arxiv.org/abs/hep-lat/9810063}{{\ttfamily arXiv:hep-lat/9810063
  [hep-lat]}}.

\bibitem{DellaMorte:2004bc}
{\bfseries ALPHA} Collaboration, M.~Della~Morte {\em et~al.}, ``{Computation of
  the strong coupling in QCD with two dynamical flavors},''
  \href{http://dx.doi.org/10.1016/j.nuclphysb.2005.02.013}{{\em Nucl.Phys.}
  {\bfseries B713} (2005) 378--406},
\href{http://arxiv.org/abs/hep-lat/0411025}{{\ttfamily arXiv:hep-lat/0411025
  [hep-lat]}}.

\bibitem{Aoki:2009tf}
{\bfseries PACS-CS} Collaboration, S.~Aoki {\em et~al.}, ``{Precise
  determination of the strong coupling constant in $N_f = 2+1$ lattice QCD with
  the Schr{\"o}dinger functional scheme},''
  \href{http://dx.doi.org/10.1088/1126-6708/2009/10/053}{{\em JHEP} {\bfseries
  0910} (2009) 053},
\href{http://arxiv.org/abs/0906.3906}{{\ttfamily arXiv:0906.3906 [hep-lat]}}.

\bibitem{Tekin:2010mm}
{\bfseries ALPHA} Collaboration, F.~Tekin, R.~Sommer, and U.~Wolff, ``{The
  running coupling of QCD with four flavors},''
  \href{http://dx.doi.org/10.1016/j.nuclphysb.2010.07.002}{{\em Nucl.Phys.}
  {\bfseries B840} (2010) 114--128},
\href{http://arxiv.org/abs/1006.0672}{{\ttfamily arXiv:1006.0672 [hep-lat]}}.

\bibitem{Fritzsch:2013je}
P.~Fritzsch and A.~Ramos, ``{The gradient flow coupling in the Schr\"odinger
  Functional},'' \href{http://dx.doi.org/10.1007/JHEP10(2013)008}{{\em JHEP}
  {\bfseries 1310} (2013) 008},
\href{http://arxiv.org/abs/1301.4388}{{\ttfamily arXiv:1301.4388 [hep-lat]}}.

\bibitem{Fritzsch:2013yxa}
P.~Fritzsch, A.~Ramos, and F.~Stollenwerk, ``{Critical slowing down and the
  gradient flow coupling in the Schr\"odinger functional},'' {\em PoS}
  {\bfseries Lattice2013} (2013) 461,
\href{http://arxiv.org/abs/1311.7304}{{\ttfamily arXiv:1311.7304 [hep-lat]}}.

\bibitem{Luscher:2014kea}
M.~L\"uscher, ``{Step scaling and the Yang-Mills gradient flow},''
  \href{http://dx.doi.org/10.1007/JHEP06(2014)105}{{\em JHEP} {\bfseries 1406}
  (2014) 105},
\href{http://arxiv.org/abs/1404.5930}{{\ttfamily arXiv:1404.5930 [hep-lat]}}.

\bibitem{DallaBrida:2016kgh}
{\bfseries ALPHA} Collaboration, M.~Dalla~Brida, P.~Fritzsch, T.~Korzec,
  A.~Ramos, S.~Sint, and R.~Sommer, ``{Slow running of the Gradient Flow
  coupling from 200 MeV to 4 GeV in $N_{\rm f}=3$ QCD},''
  \href{http://dx.doi.org/10.1103/PhysRevD.95.014507}{{\em Phys. Rev.}
  {\bfseries D95} no.~1, (2017) 014507},
\href{http://arxiv.org/abs/1607.06423}{{\ttfamily arXiv:1607.06423 [hep-lat]}}.

\bibitem{Blum:2018mom}
{\bfseries RBC, UKQCD} Collaboration, T.~Blum, P.~A. Boyle, V.~Gülpers,
  T.~Izubuchi, L.~Jin, C.~Jung, A.~Jüttner, C.~Lehner, A.~Portelli, and J.~T.
  Tsang, ``{Calculation of the hadronic vacuum polarization contribution to the
  muon anomalous magnetic moment},''
  \href{http://dx.doi.org/10.1103/PhysRevLett.121.022003}{{\em Phys. Rev.
  Lett.} {\bfseries 121} no.~2, (2018) 022003},
\href{http://arxiv.org/abs/1801.07224}{{\ttfamily arXiv:1801.07224 [hep-lat]}}.

\bibitem{Mihaila:2012bt}
L.~Mihaila, J.~Salomon, and M.~Steinhauser, ``{Gauge coupling beta functions in
  the Standard Model},'' \href{http://arxiv.org/abs/1209.5497}{{\ttfamily
  arXiv:1209.5497 [hep-ph]}}.
[PoSLL2012,043(2012)].

\bibitem{Bazavov:2019qoo}
A.~Bazavov, N.~Brambilla, X.~G. Tormo, I, P.~Petreczky, J.~Soto, A.~Vairo, and
  J.~H. Weber, ``{Determination of the QCD coupling from the static energy and
  the free energy},''
\href{http://arxiv.org/abs/1907.11747}{{\ttfamily arXiv:1907.11747 [hep-lat]}}.

\bibitem{Maltman:2008bx}
K.~Maltman, D.~Leinweber, P.~Moran, and A.~Sternbeck, ``{The Realistic Lattice
  Determination of alpha(s)(M(Z)) Revisited},''
  \href{http://dx.doi.org/10.1103/PhysRevD.78.114504}{{\em Phys. Rev.}
  {\bfseries D78} (2008) 114504},
\href{http://arxiv.org/abs/0807.2020}{{\ttfamily arXiv:0807.2020 [hep-lat]}}.

\bibitem{Bruno:2017gxd}
{\bfseries ALPHA} Collaboration, M.~Bruno, M.~Dalla~Brida, P.~Fritzsch,
  T.~Korzec, A.~Ramos, S.~Schaefer, H.~Simma, S.~Sint, and R.~Sommer, ``{QCD
  Coupling from a Nonperturbative Determination of the Three-Flavor $\Lambda$
  Parameter},'' \href{http://dx.doi.org/10.1103/PhysRevLett.119.102001}{{\em
  Phys. Rev. Lett.} {\bfseries 119} no.~10, (2017) 102001},
\href{http://arxiv.org/abs/1706.03821}{{\ttfamily arXiv:1706.03821 [hep-lat]}}.

\bibitem{dEnterria:2015kmd}
D.~d'Enterria and P.~Z. Skands, eds., {\em {Proceedings, High-Precision
  $\alpha_s$ Measurements from LHC to FCC-ee}}, CERN.
\newblock CERN, Geneva, 2015.
\newblock \href{http://arxiv.org/abs/1512.05194}{{\ttfamily arXiv:1512.05194
  [hep-ph]}}.
\newblock
\url{http://lss.fnal.gov/archive/2015/conf/fermilab-conf-15-610-t.pdf}.
\newblock

\bibitem{Husung:2017qjz}
N.~Husung, M.~Koren, P.~Krah, and R.~Sommer, ``{SU(3) Yang Mills theory at
  small distances and fine lattices},''
  \href{http://dx.doi.org/10.1051/epjconf/201817514024}{{\em EPJ Web Conf.}
  {\bfseries 175} (2018) 14024},
\href{http://arxiv.org/abs/1711.01860}{{\ttfamily arXiv:1711.01860 [hep-lat]}}.

\bibitem{Bazavov:2017dsy}
A.~Bazavov, P.~Petreczky, and J.~H. Weber, ``{Equation of State in 2+1 Flavor
  QCD at High Temperatures},''
  \href{http://dx.doi.org/10.1103/PhysRevD.97.014510}{{\em Phys. Rev.}
  {\bfseries D97} no.~1, (2018) 014510},
\href{http://arxiv.org/abs/1710.05024}{{\ttfamily arXiv:1710.05024 [hep-lat]}}.

\bibitem{Davies:2008sw}
{\bfseries HPQCD} Collaboration, C.~T.~H. Davies, K.~Hornbostel, I.~D. Kendall,
  G.~P. Lepage, C.~McNeile, J.~Shigemitsu, and H.~Trottier, ``{Update: Accurate
  Determinations of alpha(s) from Realistic Lattice QCD},''
  \href{http://dx.doi.org/10.1103/PhysRevD.78.114507}{{\em Phys. Rev.}
  {\bfseries D78} (2008) 114507},
\href{http://arxiv.org/abs/0807.1687}{{\ttfamily arXiv:0807.1687 [hep-lat]}}.

\bibitem{DallaBrida:2019wur}
M.~Dalla~Brida and A.~Ramos, ``{The gradient flow coupling at high-energy and
  the scale of SU(3) Yang-Mills theory},''
\href{http://arxiv.org/abs/1905.05147}{{\ttfamily arXiv:1905.05147 [hep-lat]}}.

\bibitem{DallaBrida:2019mqg}
{\bfseries ALPHA} Collaboration, M.~Dalla~Brida, R.~H\"ollwieser, F.~Knechtli,
  T.~Korzec, A.~Ramos, and R.~Sommer, ``{Non-perturbative renormalization by
  decoupling},'' \href{http://dx.doi.org/10.1016/j.physletb.2020.135571}{{\em
  Phys. Lett. B} {\bfseries 807} (2020) 135571},
  \href{http://arxiv.org/abs/1912.06001}{{\ttfamily arXiv:1912.06001
  [hep-lat]}}.

\bibitem{Gockeler:2005rv}
M.~Gockeler, R.~Horsley, A.~C. Irving, D.~Pleiter, P.~E.~L. Rakow,
  G.~Schierholz, and H.~Stuben, ``{A Determination of the Lambda parameter from
  full lattice QCD},'' \href{http://dx.doi.org/10.1103/PhysRevD.73.014513}{{\em
  Phys. Rev.} {\bfseries D73} (2006) 014513},
\href{http://arxiv.org/abs/hep-ph/0502212}{{\ttfamily arXiv:hep-ph/0502212
  [hep-ph]}}.

\bibitem{Asakawa:2015vta}
M.~Asakawa, T.~Hatsuda, T.~Iritani, E.~Itou, M.~Kitazawa, and H.~Suzuki,
  ``{Determination of Reference Scales for Wilson Gauge Action from Yang--Mills
  Gradient Flow},''
\href{http://arxiv.org/abs/1503.06516}{{\ttfamily arXiv:1503.06516 [hep-lat]}}.

\bibitem{Kitazawa:2016dsl}
M.~Kitazawa, T.~Iritani, M.~Asakawa, T.~Hatsuda, and H.~Suzuki, ``{Equation of
  State for SU(3) Gauge Theory via the Energy-Momentum Tensor under Gradient
  Flow},'' \href{http://dx.doi.org/10.1103/PhysRevD.94.114512}{{\em Phys. Rev.}
  {\bfseries D94} no.~11, (2016) 114512},
\href{http://arxiv.org/abs/1610.07810}{{\ttfamily arXiv:1610.07810 [hep-lat]}}.

\bibitem{Luscher:2011kk}
M.~L{\"u}scher and S.~Schaefer, ``{Lattice QCD without topology barriers},''
  \href{http://dx.doi.org/10.1007/JHEP07(2011)036}{{\em JHEP} {\bfseries 1107}
  (2011) 036},
\href{http://arxiv.org/abs/1105.4749}{{\ttfamily arXiv:1105.4749 [hep-lat]}}.

\bibitem{Brower:2003yx}
R.~Brower, S.~Chandrasekharan, J.~W. Negele, and U.~J. Wiese, ``{QCD at fixed
  topology},'' \href{http://dx.doi.org/10.1016/S0370-2693(03)00369-1}{{\em
  Phys. Lett.} {\bfseries B560} (2003) 64--74},
\href{http://arxiv.org/abs/hep-lat/0302005}{{\ttfamily arXiv:hep-lat/0302005
  [hep-lat]}}.

\bibitem{Jager:2013kha}
B.~Jäger, T.~D. Rae, S.~Capitani, M.~Della~Morte, D.~Djukanovic, G.~von
  Hippel, B.~Knippschild, H.~B. Meyer, and H.~Wittig, ``{A high-statistics
  study of the nucleon EM form factors, axial charge and quark momentum
  fraction},'' \href{http://dx.doi.org/10.22323/1.187.0272}{{\em PoS}
  {\bfseries LATTICE2013} (2014) 272},
\href{http://arxiv.org/abs/1311.5804}{{\ttfamily arXiv:1311.5804 [hep-lat]}}.

\bibitem{Capitani:2011fg}
S.~Capitani, M.~Della~Morte, G.~von Hippel, B.~Knippschild, and H.~Wittig,
  ``{Scale setting via the $\Omega$ baryon mass},''
  \href{http://dx.doi.org/10.22323/1.139.0145}{{\em PoS} {\bfseries
  LATTICE2011} (2011) 145},
\href{http://arxiv.org/abs/1110.6365}{{\ttfamily arXiv:1110.6365 [hep-lat]}}.

\bibitem{Fritzsch:2012wq}
{\bfseries ALPHA} Collaboration, P.~Fritzsch, F.~Knechtli, B.~Leder,
  M.~Marinkovic, S.~Schaefer, {\em et~al.}, ``{The strange quark mass and
  Lambda parameter of two flavor QCD},''
  \href{http://dx.doi.org/10.1016/j.nuclphysb.2012.07.026}{{\em Nucl.Phys.}
  {\bfseries B865} (2012) 397--429},
\href{http://arxiv.org/abs/1205.5380}{{\ttfamily arXiv:1205.5380 [hep-lat]}}.

\bibitem{Lottini:2013rfa}
{\bfseries ALPHA} Collaboration, S.~Lottini, ``{Chiral behaviour of the pion
  decay constant in $N_f=2$ QCD},''
  \href{http://dx.doi.org/10.22323/1.187.0315}{{\em PoS} {\bfseries
  LATTICE2013} (2014) 315},
\href{http://arxiv.org/abs/1311.3081}{{\ttfamily arXiv:1311.3081 [hep-lat]}}.

\bibitem{Blum:2014tka}
{\bfseries RBC, UKQCD} Collaboration, T.~Blum {\em et~al.}, ``{Domain wall QCD
  with physical quark masses},''
  \href{http://dx.doi.org/10.1103/PhysRevD.93.074505}{{\em Phys. Rev. D}
  {\bfseries 93} no.~7, (2016) 074505},
  \href{http://arxiv.org/abs/1411.7017}{{\ttfamily arXiv:1411.7017 [hep-lat]}}.

\bibitem{Luscher:2010ae}
M.~Luscher, ``{Computational Strategies in Lattice QCD},'' in {\em {Les Houches
  Summer School: Session 93: Modern perspectives in lattice QCD: Quantum field
  theory and high performance computing}}, pp.~331--399.
\newblock 2, 2010.
\newblock \href{http://arxiv.org/abs/1002.4232}{{\ttfamily arXiv:1002.4232
  [hep-lat]}}.

\bibitem{Chetyrkin:2000yt}
K.~Chetyrkin, J.~H. Kuhn, and M.~Steinhauser, ``{RunDec: A Mathematica package
  for running and decoupling of the strong coupling and quark masses},''
  \href{http://dx.doi.org/10.1016/S0010-4655(00)00155-7}{{\em Comput. Phys.
  Commun.} {\bfseries 133} (2000) 43--65},
  \href{http://arxiv.org/abs/hep-ph/0004189}{{\ttfamily arXiv:hep-ph/0004189}}.

\bibitem{Bode:1998hd}
{\bfseries Alpha} Collaboration, A.~Bode, U.~Wolff, and P.~Weisz, ``{Two loop
  computation of the Schrodinger functional in pure SU(3) lattice gauge
  theory},'' \href{http://dx.doi.org/10.1016/S0550-3213(98)00772-X}{{\em Nucl.
  Phys.} {\bfseries B540} (1999) 491--499},
\href{http://arxiv.org/abs/hep-lat/9809175}{{\ttfamily arXiv:hep-lat/9809175
  [hep-lat]}}.

\bibitem{Bode:1999sm}
{\bfseries ALPHA} Collaboration, A.~Bode, P.~Weisz, and U.~Wolff, ``{Two loop
  computation of the Schrodinger functional in lattice QCD},''
  \href{http://dx.doi.org/10.1016/S0550-3213(00)00187-5,
  10.1016/S0550-3213(01)00045-1, 10.1016/S0550-3213(01)00267-X}{{\em Nucl.
  Phys.} {\bfseries B576} (2000) 517--539},
  \href{http://arxiv.org/abs/hep-lat/9911018}{{\ttfamily arXiv:hep-lat/9911018
  [hep-lat]}}.
[Erratum: Nucl. Phys.B600,453(2001)].

\bibitem{Peter:1996ig}
M.~Peter, ``{The Static quark - anti-quark potential in QCD to three loops},''
  \href{http://dx.doi.org/10.1103/PhysRevLett.78.602}{{\em Phys. Rev. Lett.}
  {\bfseries 78} (1997) 602--605},
  \href{http://arxiv.org/abs/hep-ph/9610209}{{\ttfamily arXiv:hep-ph/9610209}}.

\bibitem{Smirnov:2008pn}
A.~V. Smirnov, V.~A. Smirnov, and M.~Steinhauser, ``{Fermionic contributions to
  the three-loop static potential},''
  \href{http://dx.doi.org/10.1016/j.physletb.2008.08.070}{{\em Phys. Lett. B}
  {\bfseries 668} (2008) 293--298},
  \href{http://arxiv.org/abs/0809.1927}{{\ttfamily arXiv:0809.1927 [hep-ph]}}.

\bibitem{Chetyrkin:1997mb}
K.~Chetyrkin, J.~H. Kuhn, and M.~Steinhauser, ``{Heavy quark current
  correlators to O (alpha-s**2)},''
  \href{http://dx.doi.org/10.1016/S0550-3213(97)00481-1}{{\em Nucl. Phys. B}
  {\bfseries 505} (1997) 40--64},
  \href{http://arxiv.org/abs/hep-ph/9705254}{{\ttfamily arXiv:hep-ph/9705254}}.

\bibitem{Broadhurst:1991fi}
D.~J. Broadhurst, ``{Three loop on-shell charge renormalization without
  integration: Lambda-MS (QED) to four loops},''
  \href{http://dx.doi.org/10.1007/BF01559486}{{\em Z. Phys. C} {\bfseries 54}
  (1992) 599--606}.

\end{thebibliography}\endgroup

\end{document}